\documentclass[reprint,onecolumn,superscriptaddress,nofootinbib,amsmath,amssymb,aps,prd]{revtex4-1}

\usepackage{graphics,graphicx,epstopdf,epsfig,url}
\usepackage{dcolumn}
\usepackage{bm}
\usepackage[normalem]{ulem}
\usepackage{times}
\usepackage{natbib}
\usepackage{color}
\usepackage{multirow}
\usepackage{hyperref}
\usepackage[utf8]{inputenc}
\usepackage{aas_macros}
\usepackage{makecell}

\def\Mpch{~h^{-1} {\rm Mpc}}

\usepackage{hypcap}

\usepackage{comment}

\def\gsim{\;\rlap{\lower 2.5pt
\hbox{$\sim$}}\raise 1.5pt\hbox{$>$}\;}
\def\lsim{\;\rlap{\lower 2.5pt
\hbox{$\sim$}}\raise 1.5pt\hbox{$<$}\;}

\newcommand{\revision}[1]{\textcolor{black}{#1}}
\newcommand{\newrevision}[1]{\textcolor{black}{#1}}

\makeatletter
\def\l@subsubsection#1#2{}
\makeatother

\begin{document}

\preprint{APS/123-QED}
\title{\revision{Towards testing the theory of gravity with DESI: 
summary statistics, model predictions and future simulation requirements}} 
\author{Shadab Alam}
\affiliation{Institute for Astronomy, University of Edinburgh, Edinburgh, UK}

\author{Christian Arnold}
\affiliation{Institute for Computational Cosmology, Department of Physics, Durham University, South Road, Durham DH1 3LE, UK}

\author{Alejandro Aviles}
\affiliation{Departamento de F\'isica, Instituto Nacional de  Investigaciones Nucleares, Apartado Postal 18-1027, Col. Escand\'on, Ciudad de M\'exico,11801, M\'exico}
\affiliation{Consejo Nacional de Ciencia y Tecnologıa, Av. Insurgentes Sur 1582. Colonia Credito Constructor, Del. Benito Ju\'arez C.P. 03940, M\'exico D.F. M\'exico}

\author{Rachel Bean}
\affiliation{Department of Astronomy, Cornell University, Ithaca, NY 14853, USA}
\author{Yan-Chuan Cai}
\affiliation{Institute for Astronomy, University of Edinburgh, Edinburgh, UK}
 %
\author{Marius Cautun}
\affiliation{Institute for Computational Cosmology, Department of Physics, Durham University, South Road, Durham DH1 3LE, UK}
\affiliation{Leiden Observatory, Leiden University, PO Box 9513, NL-2300 RA Leiden, the Netherlands}
 
\author{Jorge L. Cervantes-Cota}
\affiliation{Departamento de F\'isica, Instituto Nacional de  Investigaciones Nucleares, Apartado Postal 18-1027, Col. Escand\'on, Ciudad de M\'exico,11801, M\'exico}

 \author{Carolina Cuesta-Lazaro}
 \affiliation{Institute for Computational Cosmology, Department of Physics, Durham University, South Road, Durham DH1 3LE, UK}

\author{N.~Chandrachani Devi}
 \affiliation{Instituto de Astronomia, Universidad Nacional Aut\'onoma de M\'exico, Apdo. Postal 20-364, M\'exico}
\affiliation{Department of Physics, Manipur University, Canchipur, 795003, Manipur, India}
 %
 
 \author{Alexander Eggemeier}
 \affiliation{Institute for Computational Cosmology, Department of Physics, Durham University, South Road, Durham DH1 3LE, UK}

 \author{Sebastien Fromenteau}
  \affiliation{Instituto de Ciencias F\'isicas, Universidad Nacional Aut\'onoma de M\'exico, Av. Universidad s/n, 62210 Cuernavaca, Mor., M\'exico}
 
 \author{Alma X. Gonzalez-Morales}
  \affiliation{Consejo Nacional de Ciencia y Tecnologıa, Av. Insurgentes Sur 1582. Colonia Credito Constructor, Del. Benito Ju\'arez C.P. 03940, M\'exico D.F. M\'exico}
\affiliation{Departamento de F\'isica, Divisi\'on de Ciencias e Ingenier\'ias, Campus Le\'on, Universidad de Guanajuato, Le\'on 37150, M\'exico}
 
 \author{Vitali Halenka}
 \affiliation{University of Michigan, Department of Physics, 1085 South University, Ann Arbor, 48109, USA}

 \author{Jian-hua He}
 \affiliation{School of Astronomy and Space Science, Nanjing University, Nanjing 210093, China}
 \affiliation{Institute for Computational Cosmology, Department of Physics, Durham University, South Road, Durham DH1 3LE, UK}
  \author{Wojciech A. Hellwing}
 \affiliation{Center for Theoretical Physics, Polish Academy of Sciences, Al. Lotników 32/46, 02-668 Warsaw, Poland}
  \author{C\'esar Hern\'andez-Aguayo}
 \affiliation{Max-Planck-Institut f\"ur Astrophysik, Karl-Schwarzschild-Str. 1, D-85748, Garching, Germany}
 \affiliation{Excellence Cluster ORIGINS, Boltzmannstrasse 2, D-85748 Garching, Germany}
 \affiliation{Institute for Computational Cosmology, Department of Physics, Durham University, South Road, Durham DH1 3LE, UK}
\author{Mustapha Ishak}
\affiliation{Department of Physics, The University of Texas at Dallas, Richardson, Texas 75080, USA}
  \author{Kazuya Koyama}
 \affiliation{Institute of Cosmology and Gravitation, University of Portsmouth, Portsmouth, Hampshire, PO 3FX, UK}
\author{Baojiu Li}
 \affiliation{Institute for Computational Cosmology, Department of Physics, Durham University, South Road, Durham DH1 3LE, UK}
\author{Axel de la Macorra}
\affiliation{Instituto de F\'isica, Universidad Nacional Aut\'onoma de M\'exico, Apdo. Postal 20-364, Ciudad de M\'exico, M\'exico}
%

\author{Jennifer Meneses Rizo}
\affiliation{Instituto de F\'isica, Universidad Nacional Aut\'onoma de M\'exico, Apdo. Postal 20-364, Ciudad de M\'exico, M\'exico}
  
 \author{Christopher Miller}
 \affiliation{University of Michigan, Department of Astronomy, Department of Physics, 1085 South University, Ann Arbor, 48109, USA}
 \affiliation{University of Michigan, Department of Physics, 1085 South University, Ann Arbor, 48109, USA}
 \author{Eva-Maria Mueller}
  \affiliation{Department of Physics, University of Oxford, Denys Wilkinson Building, Keble Road, Oxford OX1 3RH, UK}
  
  \author{Gustavo Niz}
 \affiliation{Departamento de F\'isica, Divisi\'on de Ciencias e Ingenier\'ias, Campus Le\'on, Universidad de Guanajuato, Le\'on 37150, M\'exico}
 
  \author{Pierros Ntelis}
 \affiliation{Aix Marseille Univ, CNRS/IN2P3, CPPM, Marseille, France}

\author{Matias Rodr\'iguez Otero}
\affiliation{Instituto de F\'isica, Universidad Nacional Aut\'onoma de M\'exico, Apdo. Postal 20-364, Ciudad de M\'exico, M\'exico}

 \author{Cristiano G. Sabiu}
\affiliation{Natural Science Research Institute, University of Seoul, 163 Seoulsiripdaero, Dongdaemun-gu, Seoul, 02504, Republic of Korea}

\author{Zachary Slepian}
\affiliation{Einstein Fellow, Lawrence Berkeley National Laboratory, 1 Cyclotron Road, Berkeley, CA 94720, USA}
\affiliation{Department of Astronomy, University of Florida,
211 Bryant Space Science Center, Gainesville, FL 32611, USA}
\author{Alejo Stark}
 \affiliation{University of Michigan, Department of Astronomy, Department of Physics, 1085 South University, Ann Arbor, 48109, USA}
\author{Octavio Valenzuela}
  \affiliation{Instituto de Astronomia, Universidad Nacional Aut\'onoma de M\'exico, Apdo. Postal 20-364, M\'exico}

 %
  \author{Georgios Valogiannis}
 \affiliation{Department of Astronomy, Cornell University, Ithaca, NY 14853, USA}
 \affiliation{Department of Physics, Harvard University, Cambridge, MA 02138, USA}

  \author{Mariana Vargas-Maga\~na}
\affiliation{Instituto de F\'isica, Universidad Nacional Aut\'onoma de M\'exico, Apdo. Postal 20-364, Ciudad de M\'exico, M\'exico}

  \author{Hans A. Winther}
 \affiliation{Institute of Cosmology and Gravitation, University of Portsmouth, Portsmouth, Hampshire, PO 3FX, UK}
 \affiliation{Institute of Theoretical Astrophysics, University of Oslo, 0315 Oslo, Norway}

  \author{Pauline Zarrouk}
 \affiliation{Sorbonne Universit\'e, Universit\'e Paris Diderot, CNRS/IN2P3, Laboratoire de Physique Nucl\'eaire et de Hautes Energies, LPNHE, 4 Place Jussieu, F-75252 Paris, France}
 \affiliation{Institute for Computational Cosmology, Department of Physics, Durham University, South Road, Durham DH1 3LE, UK}
 
  \author{Gong-Bo Zhao}
 \affiliation{National Astronomy Observatories,
Chinese Academy of Science, Beijing, 100101, China}

 \affiliation{School of Astronomy and Space Science, University of Chinese Academy of Sciences, Beijing 100049, China}

 \affiliation{Institute of Cosmology and Gravitation, University of Portsmouth, Portsmouth, Hampshire, PO 3FX, UK}
 
  \author{Yi Zheng}
 \affiliation{School of Physics and Astronomy, Sun Yat-sen University, 2 Daxue Road, Tangjia, Zhuhai, 519082, China}
 \affiliation{School of Physics, Korea Institute for Advanced Study, Hoegiro 85, Seoul 02455, Korea}

  %
  %

\begin{abstract}

Shortly after its discovery, General Relativity (GR) was applied to predict the behavior of our Universe on the largest scales, and later became the foundation of modern cosmology. Its validity has been verified on a range of scales and environments from the Solar system to merging black holes. However, experimental confirmations of GR on cosmological scales have so far lacked the accuracy one would hope for -- its applications on those scales being largely based on extrapolation and its validity there sometimes questioned in the shadow of the discovery of the unexpected cosmic acceleration. Future astronomical instruments surveying the distribution and evolution of galaxies over substantial portions of the observable Universe, such as the Dark Energy Spectroscopic Instrument (DESI), will be able to measure the fingerprints of gravity and their statistical power will allow strong constraints on alternatives to GR. 

In this paper, based on a set of $N$-body simulations and mock galaxy catalogs, we study the predictions of a number of traditional and novel 
\revision{summary statistics} beyond linear redshift distortions in two well-studied modified gravity models -- chameleon $f(R)$ gravity and a braneworld model -- and the potential of testing these deviations from GR using DESI. These 
\revision{summary statistics} employ a wide array of statistical properties of the galaxy and the underlying dark matter field, including two-point and higher-order statistics, environmental dependence, redshift space distortions and weak lensing. We find that they hold promising power for testing GR to unprecedented precision. The major future challenge is to make realistic,  simulation-based mock galaxy catalogs for both GR and alternative models to fully exploit the statistic power of the DESI survey (by matching the volumes and galaxy number densities of the mocks to those in the real survey) and to better understand the impact of key systematic effects. Using these, we identify future simulation and analysis needs for gravity tests using DESI.

\end{abstract}

\pacs{Valid PACS appear here}
\maketitle

\tableofcontents
\newpage
\section{Introduction}
\label{sect:intro}

Since its discovery over a century ago, General Relativity (GR) has been established as our standard theory of gravity, thanks to the many experimental and observational tests of its various predictions. Most of these tests have been laboratory and Solar System tests in the weak-field limit, and its predicted effects from strong gravitational fields have also been detected, through the orbital decay of binary pulsars \cite{Taylor1982} and gravitational waves from merging binary compact objects \cite{Abbott2016, Monitor:2017mdv}. It has passed yet another test recently by the further detection of gravitational waves from a binary neutron star merger (GW170817) associated with a short Gamma-ray burst \cite{Monitor:2017mdv} and various other electromagnetic counterparts \cite{GBM:2017lvd}, which confirmed the equivalence between the speeds of gravity and light with high accuracy. Its application to cosmology involves a huge extrapolation from length scales at which it has been tested accurately to the Universe as a whole, but this was rarely questioned given the success of GR in predicting a diverse set of observations: the Hubble expansion, Big Bang Nucleosynthesis (BBN) and the Cosmic Microwave Background (CMB). The observations of an accelerated rate of the Hubble expansion based on the dimming of distant type Ia supernovae (SNIe) in 1998 \cite{Perlmutter:1998np, Riess:1998cb}, which have since then been supported by various independent probes, however, raised the possibility that GR might have to be replaced by an alternative model on cosmological scales. This is not the only possibility, because the cosmic acceleration may well be driven by a positive cosmological constant $\Lambda$ (as in the standard $\Lambda$CDM paradigm), or some dynamical dark energy component motivated by beyond-standard-model particle physics \cite{Copeland:2006wr}. However, given that there is currently not a widely-accepted theoretical explanation, these different possibilities have all been kept open, and there has been a great body of research on modified gravity (MG) models in recent years (see, e.g., \cite{Clifton:2011jh, joyce, koyama2016,Ishak:2018his} for some recent reviews). From a practical point of view, the studies of modified gravity and dark energy models can be considered as different faces of a common theme: {\it testing GR-based $\Lambda{\rm CDM}$ in cosmology}. This field has now matured enough, thanks to developments of {both} theoretical frameworks, which tells us what happens in cosmology, if there is a deviation from GR, {and} observational techniques, which tells us whether such a deviation is supported by the data.

DESI (Dark Energy Spectroscopic Instrument) will perform one of these advanced wide-area galaxy and quasar redshift surveys. It is a stage-IV ground-based experiment designed to have a sky coverage of 14,000 {\rm deg}$^2$, using four different classes of spectroscopic targets -- luminous red galaxies (LRGs) up to $z\simeq1$, emission line galaxies (ELGs) up to $z\simeq1.7$, quasars and Ly-$\alpha$ features up to $z\simeq3.5$, and bright galaxies at low redshifts ($z_{\rm median}\simeq 0.2$) --
as tracers of the underlying dark matter field. It will measure $\sim30$ million galaxy and quasar redshifts to obtain precise measurements of the Baryon Acoustic Oscillation (BAO) features, Redshift-Space Distortions (RSD), as well as the full galaxy power spectrum, increasing the Dark Energy Task Force (DETF) \cite{Albrecht:2006um} Figure of Merit (FoM) to above $700$ when all these measurements are used \cite{Aghamousa:2016zmz}. Such unprecedented measurement will make DESI an ideal experiment to study fundamental sciences, such as constraints on neutrino mass, parameters of inflation, understanding of the cosmic acceleration, and finally, cosmological tests of gravity theories.

In this paper, we consider two representative examples of modified gravity models: chameleon $f(R)$ gravity \cite{Carroll:2003wy, Carroll:2004de} and the 5D brane-world Dvali-Gabadadze-Porrati (DGP) model \cite{Dvali:2000hr} (we look at the normal branch of the DGP model, nDGP).
Those models can be consider as a subclass of a general class of modified gravity models, named as \textit{Actions of Effective Field Theories} \cite{2020arXiv201006707N}. 
However, the chosen models serve as ideal testbeds of gravity on cosmological scales for a few reasons: (i) they can potentially realise the screening mechanisms \cite{khoury1,khoury2,Vainshtein} to pass local gravity tests and leave observable signatures on much larger, cosmological, scales -- indeed, each of them represents a class of screening mechanisms, the chameleon and Vainshtein mechanism, (ii) the cosmological behavior of these models -- with varying parameters -- qualitatively represents that of various other classes of models, (iii) these models have been studied in great details to date, and simulations for them, which are essential to test them using galaxy surveys, have been developed and matured (see below). As a result, by studying them in detail, we hope to understand and to quantify the constraints that future astronomical observations would place on deviations from GR at astrophysical and cosmological scales, and therefore test the validity of GR in a completely different regime from traditional tests.

In the standard cosmological scenario, the formation and growth of structure is driven by hierarchical gravitational collapse: small structures collapse first, which over time merge and grow larger and more nonlinear. The galaxies that we observe today mostly reside in regions with highly nonlinear density, which cannot be adequately described using linear perturbation theory. Numerical simulations are a more reliable tool for predicting the clustering of matter in such regimes. In the MG models studied here, the enhanced gravity means that structures become more nonlinear. However, as far as simulations are concerned, the main effect these models introduce is screening, which is an inherently nonlinear phenomenon. This is reflected in the highly nonlinear equations governing the behavior of the fifth force, that need to be solved accurately in simulations. For example, linearizing the equations can cause non-negligible errors on length scales as large as $k\sim\mathcal{O}(10^{-2})$ $h{\rm Mpc}^{-1}$ \citep{Li2013} in an otherwise fully consistent simulation. Other approximations that allow to simplify simulations significantly, such as the quasi-static approximation, were shown to be valid in the cosmological regime \cite{Bose_QS2015}. Such considerations have motivated the developments of a wide array of numerical tools to simulate the structure formation in modified gravity models, including complete simulation codes \cite{Oyaizu:2008sr, Li:2009sy, Chan:2009ew, Li:2010re, Li:2010qy, Zhao:2010qy, Davis:2011pj, Li:2011vk, Puchwein:2013lza, Li:2013nua, Li:2013tda, Llinares:2013jza, Barreira:2013eea, MGENZO, Arnold:2019vpg, Hernandez-Aguayo:2020kgq} (see \cite{2015MNRAS.454.4208W} for a detailed comparison of different modified codes) and approximate methods \cite{Khoury:2009tk, Winther:2014cia, Valogiannis:2016ane, Winther:2017jof}. 

Recent progresses in optimizing the simulation codes \cite{Barreira:2015xvp,Bose:2016wms} and in developing faster substitutes \cite{Winther:2017jof} have largely been motivated by the desire to keep up with the rapidly increasing demand to match the size, resolution and variety of observables of ongoing and future galaxy surveys, including DESI. However, to compare theoretical studies with the real observations of galaxy catalogs, there are various intermediate steps, all of which could affect the reliability of the final tests of gravity by introducing sources of uncertainty. 

The main objective of this paper is to identify the key issues in using DESI observations to constrain gravity models, which can help us move to the next stage of preparation for DESI science cases. Redshift-pace distortions provide a promising way to constrain modified gravity models through the measurement of the growth rate. The two MG models considered in this paper have been constrained by RSD measurements from Baryon Oscillation Spectroscopic Survey (BOSS) \cite{Song:2015oza, Barreira:2016ovx, Mueller:2016kpu}. We consider model parameters that are consistent on the 2$\sigma$ level with these measurements. In this paper, we will focus on novel 
\revision{summary statistics} to constrain modified gravity models exploiting information on non-linear scales. A main objective is to quantitatively assess the potential constraining power of these various 
\revision{summary statistics} in testing the two classes of modified gravity models mentioned above. To this end we consider three main categories of 
\revision{summary statistics}: (i) two-point and N-point statistics and marked statistics of galaxy catalogs, (ii) statistics that employ velocity information, including redshift space distortions (RSD) and escape velocity, and (iii) statistics relying on synergy with weak gravitational lensing (WL) data. This enables us to make theoretical predictions of various 
\revision{summary statistics} based on the same set of mock galaxy catalogs, and hence allows a like-for-like comparison. Where possible, we consider different designs in the same 
\revision{class of summary statistics}; an example is the marked galaxy 2-point correlation function (2PCF), which quantifies the correlations of galaxy internal and external properties (marks) \cite{Sheth:2005aj}, where we will test different definitions of marks. Following this philosophy, we have made public the mock galaxy data used in this paper and encourage the community to use these data in their analyses so that their results can be compared with those from this work. The data can be found at this webpage\footnote{\href{https://www.cosma.dur.ac.uk/data/}{\textcolor{black}{https://www.cosma.dur.ac.uk/data/}}}.

The second main objective is to provide a guidance on future simulations and mock requirements for modified gravity tests using galaxy clustering data from the likes of DESI and Euclid. In this paper we will mainly use two sets of simulations: (i) the {\sc elephant} simulations described in \cite{Cautun:2017tkc} with a simulation box size $L_{\rm box}=1024h^{-1}{\rm Mpc}$ and particle number $N_p=1024^3$, and (ii) the {\sc liminality} simulation described in \cite{Shi:2015aya} with $L_{\rm box}=64h^{-1}{\rm Mpc}$ and $N_p=512^3$. The former have been used to construct halo occupation distribution (HOD) galaxy mocks which match the number densities of luminous red galaxies (LRGs) in current galaxy surveys, while the latter is used to build mock galaxies with similar number densities as DESI bright galaxy survey (BGS), using subhalo abundance matching (SHAM).  The simulations considered in this paper do not cover the volume of the full DESI survey, however they do have number densities that are comparable to the full BOSS LRG sample \footnote{\href{https://desi.lbl.gov/trac/attachment/wiki/PublicPages/Instrument/TDR\%20Part\%20I.pdf}{\textcolor{black}{Fig. 3.8 DESI Technical Design Report Part I: Science,Targeting, and Survey Design, July 27, 2015}}}. Recognizing the high cost of large volume, high resolution simulations, this paper is a preliminary study to determine the statistical potential to test gravity for data with the richness of DESI, to help frame simulation specifications and model choices, before embarking on future, heavy computational efforts for survey-tailored full scale simulations.

The final objective is to make an assessment of the statistical and, where possible, systematical uncertainties associated with the individual 
\revision{summary statistics}. For most 
\revision{summary statistics}, we use five independent realizations of simulations and mock galaxy catalogs to estimate the sample variance, while theoretical covariance matrices are used in other cases (e.g., galaxy galaxy lensing). This analysis is intended to set the stage for more comprehensive, quantitative estimate of the impact of systematic errors, beyond the scope of this paper.

The plan of this paper is as follows. In \S~\ref{sect:models} we briefly review the modified gravity models studied in this work, focusing on their key differences from $\Lambda$CDM, the simulations for these models, and the dark matter halo and galaxy catalogs used for the analyses. In \S~\ref{sect:estimators} we have a number of subsections, each of which focuses on a particular estimator and studies the potential of using such 
\revision{summary statistics} from DESI galaxy survey data to constrain the theoretical models.  In \S~\ref{sect:discussion}, the conclusions of the analyses and implications for future work are summarized.

\section{Models, simulations and galaxy catalogs}
\label{sect:models}

In this section we introduce the modified gravity models that are exemplified in this paper -- $f(R)$ gravity and the nDGP model. These are two of the most well-studied modified gravity models, and they represent two main classes of screening mechanisms, thin-shell chameleon \cite{khoury1,khoury2} and Vainshtein \cite{Vainshtein} screening. Cosmological simulations for these models have been carried out by various groups, and we will briefly review the simulations to be used in this paper. Then we will introduce the mock galaxy catalogs constructed using these simulations, which will be used in the calculation of the various 
\revision{summary statistics} in the next section. More details of these models, along with the simulation techniques for them, can be found in \cite{Winther:2015wla}, and more comprehensive reviews of these models and their cosmological tests can be found in the recent review papers, e.g., \citep{koyama2016,Burrage:2017qrf}.

\subsection{Modified gravity models}
\label{subsect:MG}
Ever since the discovery of the accelerated expansion of the Universe, many dark energy and modified gravity models have been proposed, \cite{Clifton:2011jh,joyce,koyama2016,Ishak:2018his}. The recent detection of a binary neutron star merger by gravitational waves (GW170817) associated with various electromagnetic counterparts \cite{Monitor:2017mdv, GBM:2017lvd} has had a substantial impact on this field \cite{Lombriser:2015sxa, Creminelli:2017sry, Sakstein:2017xjx, Ezquiaga:2017ekz}. In the four-dimensional scalar tensor theory, the high-precision measurement of the equivalence between the speeds of photons and gravitational waves leaves very limited scope for a modified gravity model that naturally produces self-acceleration without a cosmological constant \citep{Lombriser:2015sxa} while being compatible with other observations such as the integrated Sachs Wolfe effect (ISW; e.g., \cite{Barreira:2014ija, Barreira:2014jha, Barreira:2015vra, Renk:2017rzu}). The two models that we consider in this paper -- $f(R)$ gravity and nDGP model -- are compatible with this measurement. {The former is a leading example of the class of models in which gravity is modified by a scalar field that has a self-interaction described by a nonlinear potential and interactions with other matter species, while the latter belongs to the category of models where the self-interaction of the scalar field is caused by derivative couplings}. 

Before moving on to the details of these models, it is worthwhile to make a couple of comments:
\begin{enumerate}
    \item in both models, the accelerated expansion of the Universe is not driven by the modification of gravity itself, but has to be explained by an extra cosmological constant or dark energy species. The models may lose some of their original appeals as a consequence. However, they provide useful testbeds for cosmological tests of gravity, which is the prevailing topic of this paper.
    \item while representative, the two models cannot be expected to cover all behaviors of potential MG models. As a few examples, in both models gravity can only be enhanced rather than weakened, and in $f(R)$ gravity the maximum of this enhancement (which is $1/3$) cannot be tuned freely; the expansion history of $f(R)$ models has to be practically indistinguishable from that of $\Lambda$CDM for the model to pass stringent Solar System tests of gravity \cite{Wang:2012kj,Brax:2008hh,Ceron-Hurtado:2016jrp}. Our choices of models are therefore inflexible in some aspects, and such inflexibility is the price to pay for having a manageable number of models to study in greater depth. We believe that an analysis of various 
    \revision{summary statistics} predicted by these two models could offer insight into the cosmological tests of more general models and therefore serve as a starting point to prepare for the tests of gravity using galaxy surveys such as DESI. One alternative to the model-by-model approach here is to use a general parameterization of MG models (see, e.g., \cite{Lombriser:2018guo}, for a review), but this has the disadvantages of having a significantly larger parameter space to explore and -- more importantly -- some of the popular parameterization schemes only mimic the full models in certain (e.g., linear perturbation) regimes and therefore can not be relied on for a fully nonlinear study. Nevertheless, when describing the models below, it is useful to know their links to and position in certain parameterization schemes which are widely used in the literature, for qualitative comparisons.
\end{enumerate}

A popular parameterization scheme for modified gravity is the $\mu$-$\gamma$ parameterization, which we briefly introduce here as we shall try to make connections to the two MG models studied in this paper. In the Newtonian gauge, the line element is given by
\begin{equation}\label{eq:metric}
    {\rm d}s^2 = -\left[1+2\Psi({\bf x},t)\right]{\rm d}t^2 + a^2\left[1-2\Phi({\bf x},t)\right]\delta_{ij}{\rm d}x^i{\rm d}x^j,
\end{equation}
where $a=a(t)$ is the scale factor, $\delta_{ij}$ is the Kronecker $\delta$ and $\Psi, \Phi$, which are functions of space and time, denote the gravitational potentials. We introduce paramtrization functions, $\mu(k,t)$ and $\gamma(k,t)$, in the following linearized Einstein equations, now written in $k$ space,
\begin{eqnarray}
\label{eq:mu-param}k^2\Psi &=& -4\pi\mu(k,t)G\bar{\rho}_ma^2\delta_m,\\
\label{eq:gamma-param}\Psi &=& \gamma(k,t)^{-1}\Phi
\end{eqnarray}
where $\delta_m$ is the matter density contrast and $\bar{\rho}_m$ is the mean matter density ($\delta\rho_m=\bar{\rho}_m\delta_m$); these equations are written in Fourier space so that $\Phi,\Psi, \delta_m$ should be understood as the Fourier transforms of the corresponding fields. Notice that $\mu=\gamma=1$ for GR and deviations of these functions from 1 could be signatures of modified gravity. This model has been constructed under the assumption of statistical cosmic homogeneity and isotropy. This is a well motivated assumption, since even though we cannot prove it, there are good observational evidences \cite{2017JCAP...06..019N,2018arXiv181009362N}.

\subsubsection{f(R) gravity}
\label{subsect:fR}

In this model, the standard Einstein-Hilbert action is extended to include an additional function of the Ricci curvature $f(R)$ (see \cite{DeFelice:2010aj} for a review). This theory is equivalent to the scalar-tensor theory with a potential for the scalar field that is determined by the function $f(R)$, and which realizes the chameleon screening mechanism \citep{khoury1,khoury2} (see \cite{Burrage:2017qrf} for a recent review). Deviations from standard GR can be suppressed in regions with deep gravitational potential by the chameleon mechanism. In regions where the potential is shallow, the theory is in the unscreened regime, in which massive particles experience an additional fifth force mediated by the scalar field $f_R\equiv{\rm d}f(R)/{\rm d}R$, whose strength can be comparable to that of standard gravity. In $f(R)$ gravity, the maximum strength of the fifth force is $1/3$ of Newtonian gravity, while in general chameleon models this can be a free model parameter.

In the quasi-static\footnote{{This means that the time evolution of the gravitational potentials is assumed to
be small compared to the Hubble time so one can assume the derivatives of the potentials to be 
zero for sub-Hubble-horizon scales. For scalar-tensor theories, this approximation also means that one neglects the time derivatives of the fluctuations in the scalar field at scales below the
scalar perturbation sound horizon.}} and weak-field limits, the equations for the Newtonian potential $\Phi$ and the scalar field $f_R$ are given by 
\begin{align}\label{eq:poisson}
\nabla^2 \Phi &= 4 \pi G  a^2 \delta \rho_m - \frac{1}{2} \nabla^2 f_R, \\
\nabla^2 f_R &= -\frac{a^2}{3} \delta R - \frac{8 \pi G a^2}{3} \delta \rho_m,
\end{align}
where $\delta \rho_m$ and $\delta R$ are the matter density perturbations and the perturbation of the Ricci curvature respectively. $\delta R$ can be written as a function of $f_R$, which plays the role of the potential for $f_R$. As a specific example, we consider one of the models proposed by Hu \& Sawicki (HS) \citep{HuSawicky}, in which the functional form of $f(R)$ is given by 
\begin{align}\label{eq:fR}
f(R) = - 6 \Omega_{\Lambda} H_0^2 + |f_{R0}| \frac{\bar{R}^2}{R},
\end{align}
where $H_0$ is the present-day Hubble parameter and $f_{R0} < 0$ is the value of $f_R$ today. Note that the HS $f(R)$ model has another free model parameter, an integer $n$, which has been set to $1$ in this paper; having different values of $n$ could lead to quantitative differences in the model behaviour, but we expect the case of $n=1$ to be representative for the qualitative properties of the fifth force and screening. $f_{R0}$ is conventionally used as a parameter of this model to describe the deviation from $\Lambda$CDM, with smaller $|f_{R0}|$ values denoting weaker deviation from GR, as can be seen from Eqs.~(\ref{eq:poisson},\ref{eq:fR}). For small $|f_{R0}|$ the background expansion history can be approximated by that in $\Lambda$CDM and $\bar{R}$ can be identified with the background Ricci curvature today in $\Lambda$CDM,
\begin{align}
\bar{R}= 3 \Omega_m H_0^2 \left(1+ 4 \frac{\Omega_{\Lambda}}{\Omega_m} \right).
\end{align}
Hereafter we will use this approximation so that there is no difference between $f(R)$ and $\Lambda$CDM in the background. As mentioned above,  $f(R)$ gravity models, and chameleon models in general, have to closely resemble $\Lambda$CDM background expansion history in order to have a sufficiently efficient chameleon screening mechanism to pass the solar system tests of gravity.

The chameleon screening mechanism works because the scalar field has a position-dependent mass $m$, give by
\begin{equation}
    m^2 \simeq \frac{1}{3f_{RR}} \equiv \frac{1}{3}\left|\frac{{\rm d}^2f(R)}{{\rm d}R^2}\right|^{-1},
\end{equation}
so that the fifth force it mediates is of Yukawa type with its potential of the form $\sim\exp(-mr)/r$, where $r$ is the distance between two bodies. In other words, unlike the standard Newtonian force, the fifth force has a finite range characterized by the Compton wavelength $m^{-1}$, beyond which it decays exponentially. In high-density regions, $m$ is large and the fifth force is exponentially suppressed, causing the screening. An important property of $f(R)$ gravity is its prediction of scale-dependent linear growth rate: this is because even in the linear regime, where $m$ can be replaced by its background value $\bar{m}(a)$, the fifth force still has a finite range within which the growth of matter density perturbations is suppressed. 

In the $\mu$-$\gamma$ parameterization framework described in Eqs.~(\ref{eq:mu-param},\ref{eq:gamma-param}), $f(R)$ gravity can be described as
\begin{eqnarray}
\label{eq:mu-exp}\mu &=& \frac{4+2\omega}{3+2\omega},\\
\label{eq:gamma-exp}\gamma &=& \frac{1+\omega}{2+\omega}.
\end{eqnarray}
with
\begin{equation}\label{eq:omega-exp}
    \omega = \omega(k,a) = \frac{3a^2\bar{m}^2}{2k^2},
\end{equation}
where $\bar{m}=\bar{m}(a)$ is the background scalar field mass mentioned above. These equations indicate that, in the large-scale limit where $a\bar{m}\gg k$ or $\omega\gg1$, $\mu,\gamma\rightarrow1$ and GR is recovered, while $\mu\rightarrow4/3$ in the opposite limit (therefore the scale dependence). As mentioned earlier, this paramterization only works for the linear perturbations regime. For models with larger $|f_{R0}|$, such as $|f_{R0}|=10^{-4}$ used in this paper, chameleon screening is inefficient and this parameterization can be a good description, but for smaller values of $|f_{R0}|$ we expect it to miss some important effects of screening \cite{Zhao:2010qy}.

\subsubsection{DGP model}
\label{subsect:DGP}

The DGP model \citep{Dvali:2000hr} is a braneworld model where standard model particles are confined to a four-dimensional brane in a five-dimensional spacetime. This model has one parameter $r_c$ of length dimension, below which gravity becomes four dimensional. On scales smaller than $r_c$ and in the quasi-static and weak-field limits, the equations for the Newtonian potential and the scalar field $\varphi$ that represents the brane-bending mode is given by
\begin{eqnarray}\label{eq:poisson_dgp}
\nabla^2 \Phi &=& 4 \pi G  a^2 \delta \rho_m + \frac{1}{2} \nabla^2 \varphi, \\
\label{DGPeq} \nabla^2 \varphi + \frac{r_c^2}{3 \beta a^2}\Big[(\nabla^2 \varphi)^2- (\nabla_i \nabla_j \varphi)^2 \Big] &=& \frac{8 \pi G a^2}{3 \beta} \delta \rho_m,
\end{eqnarray}
where 
\begin{align}\label{eq:beta_dgp}
\beta = \beta(a) = 1 + 2 H r_c \left(1 + \frac{\dot{H}}{3 H^2} \right).
\end{align}
Note that we assumed the normal branch of the DGP model. This branch requires an additional dark energy to explain the cosmic acceleration, {but does not suffer from the instabilities of the self-accelerating branch, see, e.g., \cite{Charmousis:2006pn,Gregory:2007xy,Gorbunov:2005zk}}. In order to make the comparison between $\Lambda$CDM and $f(R)$ models easier, we tune the dark energy equation of state so that the background expansion history is identical with that of $\Lambda$CDM \cite{Schmidt:2009sv}. 

In the DGP model, massive particles also feel a fifth force -- as can be seen from Eq.~(\ref{eq:poisson_dgp}) -- whose potential is governed by the scalar field $\varphi$. The model realizes the so-called Vainshtein screening mechanism \citep{Vainshtein}, by which the fifth force can be suppressed in regions where the second derivatives of the scalar field $\varphi$ ($\nabla^2\varphi$) is large. This can be seen from Eq.~\eqref{DGPeq}: in regions where $\nabla^2\varphi$ is small, nonlinear terms such as $(\nabla^2\varphi)^2$ and $(\nabla_i\nabla_j\varphi)^2$ are subdominant so that $\nabla^2\varphi\sim\nabla^2\Phi$ for $\beta\sim\mathcal{O}(1)$; while in regions where $\nabla^2\varphi$ is large, the nonlinear terms are dominant and so $|\nabla^2\varphi|\ll|\nabla^2\Phi|$.

Unlike in $f(R)$ gravity, in the DGP model the linear growth rate is scale-independent as the scalar field is massless. Detailed comparisons between nDGP and $f(R)$ gravity, in particular Vainshtein and chameleon screening mechanisms can be found in \cite{Schmidt:2010jr,Falck:2015rsa}. {This feature can again be seen if we map this model to the $\mu$-$\gamma$ parameterization above, which corresponds to Eqs.~(\ref{eq:mu-param},\ref{eq:gamma-param}, \ref{eq:mu-exp},\ref{eq:gamma-exp}) with
\begin{equation}\label{eq:omega-exp-ndgp}
    \omega = \omega(a) = \frac{3}{2}\left[\beta(a)-1\right].
\end{equation}
}

\subsection{N-body simulations and halo/galaxy catalogs}
\label{sec:simulations}

In this subsection we outline the simulations used in the analysis and present some of the statistics derived from the simulated dark matter halo and mock galaxy catalogs.

\subsubsection{N-body simulations}
\label{sec:Nbody}

The $f(R)$ simulations were performed using an optimized version of the {\sc ecosmog} code \citep{Li:2011vk,Bose:2016wms}, and the nDGP simulations were done using an optimized version of the {\sc ecosmog-v} code \cite{Li:2013nua,Barreira:2015xvp}. Both codes are extensions to the publicly-available $N$-body and hydrodynamical simulation code {\sc ramses} \citep{ramses}, with new routines added to solve the scalar field and modified Einstein equations in the MG models. These codes are parallelized using {\sc mpi} and use the adaptive-mesh-refinement (AMR) technical to achieve high resolution in overdense regions where the requirement for the force resolution is high and the screening effect is strong. The simulations start with a uniform (domain) grid with $N^{1/3}_{\rm dc}$ cells a side which covers a cubic box of size $L_{\rm box}$. The cells are refined (split into eight daughter cells) if the number of particles contained in them grows over some pre-set threshold ($N_{\rm ref}$), in such a way as to hierarchically refine the domain grid by adding higher-resolution meshes.

\begin{table}
\begin{tabular}{@{}|l|l|l|}
\hline\hline
parameter & physical meaning & value \\
\hline
$\Omega_m$  & present fractional matter density & $0.281$ \\
$\Omega_{\Lambda}$ & $1-\Omega_m$ & $0.719$ \\
$h$ & $H_0/(100$~km~s$^{-1}$Mpc$^{-1})$ & $0.697$ \\
$n_s$ & primordial power spectral index & $0.971$ \\
$\sigma_{8}$ & r.m.s. linear density fluctuation & $0.820$ \\
\hline
$|f_{R0}|$ & Hu \& Sawicki $f(R)$ parameter & $0$ (GR) $10^{-6}$ (F6), $10^{-5}$ (F5), $10^{-4}$ (F4) \\
$H_0r_c$ & nDGP parameter & $5.0$ (N5), $1.0$ (N1) \\
\hline
$L_{\rm box}$ & simulation box size & 1024~$h^{-1}$Mpc\\
$N_{\rm p}$ & simulation particle number & $1024^3$\\
$m_{\rm p}$ & simulation particle mass & $7.78\times 10^{10}h^{-1}M_{\odot}$\\
$N_{\rm dc}$ & domain grid cell number & $1024^3$\\
$N_{\rm ref}$ & refinement criterion & 8 \\
\hline\hline
\end{tabular}
\caption{The parameters and technical specifications of the N-body simulations of this work. Note that the refinement criterion $N_{\rm ref}$ is the same for refinement levels, and that $\sigma_8$ is for the $\Lambda$CDM model and only used to generate the initial conditions -- its value for $f(R)$ gravity is different but is irrelevant here.}
\label{table:simulations}
\end{table}

The cosmological and technical parameters of the simulations are given in Table~\ref{table:simulations}. The former are chosen as the best-fit $\Lambda$CDM parameters of the WMAP9 cosmology \citep{WMAP9}. The simulations were started at $z_{\rm ini}=49$, from initial conditions generated using Zel'dovich approximation\footnote{\textcolor{black}{We note that at $z_{\rm ini}=49$ the Zel'dovich approximation can lead to an error of a few percent in the generated initial condition, e.g., for the power spectrum of the density field we find the measured $P(k)$ from our initial conditions agrees better with the linear theory prediction with $\sigma_8=0.842$. This suggests that future simulations should use initial conditions generated either at higher $z_{\rm ini}$ or using the second-order Lagrangian perturbation theory \cite{Crocce:2006ve}. For this work, we simply note that the small inaccuracy should not affect our main conclusions since we are mainly looking at relative differences from $\Lambda$CDM.}} with the publicly available {\sc Mpgrafic} code \citep{Prunet:2008fv}. Because the $f(R)$ and nDGP model parameters are chosen such that they only deviate from $\Lambda$CDM non-negligibly at late times, at $z_{\rm ini}$ the modified gravity effect can be neglected, and so all our simulations started from exactly the same initial condition. In order to estimate the effect of sample variance, we have used five independent realizations of boxes, whose initial conditions differ only in their random phases of the density field. We shall refer to these different realizations as `Box 1' to `Box 5'. {For $f(R)$ gravity we ran three variants of the HS model, with $\log\left(|f_{R0}|\right)=-6, -5, -4$ (with increasing deviation from GR), which we shall refer to as F6, F5 and F4 respectively in what follows; note that GR, or $\Lambda$CDM, is a special subcase of $f(R)$ gravity with $f_{R0}=0$. For nDGP we consider two variants with $H_0r_c=5.0, 1.0$ (again with increasing deviation from GR) and refer to them as N5 and N1 respectively; $\Lambda$CDM is a special case of nDGP with $H_0r_c=\infty$.}

To gain some quick insight into the qualitative behavior of the different models, we show the predictions of some cosmological quantities here. 

\begin{figure*}[!tb]
     \centering
     \begin{tabular}{cc}
        \includegraphics[width=0.5\textwidth,angle=0]{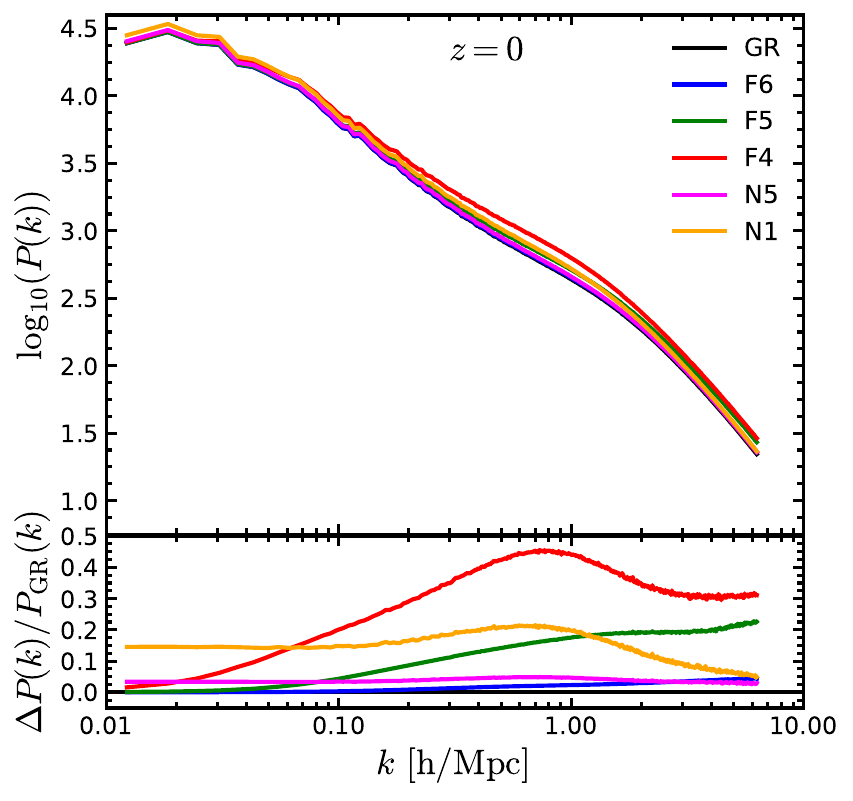}  & 
        \includegraphics[width=0.5\textwidth,angle=0]{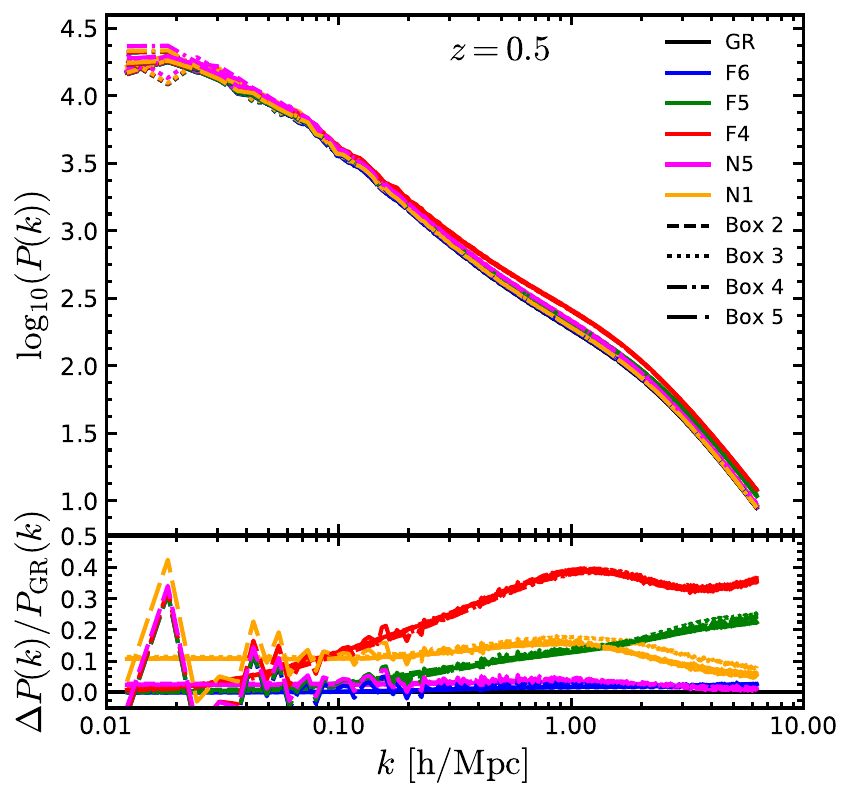} 
     \end{tabular}
\caption{(Color Online) The matter power spectra of the different models compared in this work (upper panels) and their relative differences of the modified gravity models with respect to $\Lambda$CDM (lower panels). {\it Left panels}: results at $z=0$. {\it Right panels}: results at $z=0.5$, for all five boxes. This plot sets the convention of line styles that will be used in the rest of this paper: black, blue, green, red, magenta and orange respectively represent GR, F6, F5, F4, N5 and N1, while solid, dashed, dotted, dash-dotted and long-dash-dotted lines represent results from Boxes 1, 2, 3, 4 and 5. This convention will be used in other plots unless otherwise stated. Note that for $z=0$ we only show the results for Box 1, while for $z=0.5$ we show all five boxes -- the upper panel shows the average $P(k)$ from these boxes, while in the lower panel we plot individual curves of $\Delta P/P_{\rm GR}$ for the five boxes, in order to show the agreement between the different realizations. All power spectra have been measured using the publicly-available code {\sc powmes} \citep{powmes}}
\label{fig:pofk}
\end{figure*}

Figure~\ref{fig:pofk} shows the matter power spectra of the MG models at two redshifts, $z=0$ (left) and $z=0.5$ (right), as well as their relative difference with respect to $\Lambda$CDM (bottom subpanels). For $z=0$, we show the results for Box 1 only, while for $z=0.5$ we show the results for all boxes {(the different line styles are for individual realizations, to highlight the good agreement between them). The line styles and color scheme in Fig.~\ref{fig:pofk} (see the legend and caption for more details) will be used in other plots across this paper.}

Figure \ref{fig:pofk} confirms that in $f(R)$ gravity the linear growth rate is scale dependent while in nDGP it is scale independent, as can be seen from the bottom subpanels at $k\lesssim0.1h{\rm Mpc}^{-1}$, {where linear theory works relatively well for both models}. The amount of deviation from $\Lambda$CDM in the MG models follows the expected order, and in all models it increases with time as the effect of enhanced gravity accumulates. In both F4 and N1, the enhancement of $P(k)$ starts to decrease at $k\sim0.8h{\rm Mpc}^{-1}$. For F4 this is not a signature of chameleon screening --- but is related to the internal structures of halos \citep{Li2013} --- as can be realized from the facts that F5 and F6, which both have stronger screening effect, actually do not show a similar decrease of $\Delta P/P_{\rm GR}$ at that scale. For N1, in contrast, the decrease of $\Delta P/P_{\rm GR}$ is a real signal of Vainshtein screening, which very efficiently suppresses the fifth force near and inside halos.

\subsubsection{Halo catalogs and halo mass functions}
\label{sec:halocat}

Dark matter halos and the self-bound substructures associated with them are identified using the publicly-available {\sc rockstar} halo finder\footnote{\href{https://bitbucket.org/gfcstanford/rockstar}{https://bitbucket.org/gfcstanford/rockstar}.}\citep{Behroozi2013}.  {\sc rockstar} uses the six-dimensional phase-space information from the dark matter particles to identify halos. {Note that, in principle, the presence of the fifth force in $f(R)$ gravity\footnote{In nDGP, the fifth force is strongly suppressed in halos and we can safely neglect its effect.} would require a modification to the unbinding procedure in {\sc rockstar}, but the effect is expected to be small \cite{Li2010} and so we use identical versions of {\sc rockstar} for GR and MG simulations, and we also use the same halo mass definition, $M_{200c}$, which is the mass enclosed in $R_{200c}$, the radius from halo center within which the mean mass density is 200 times the critical density $\rho_{\rm crit}(z)$}. In this paper, we make use of only independent (`main') halos, and not their substructures, partly because of the relatively low resolution of our simulations.

\begin{figure*}
     \centering
     \begin{tabular}{cc}
        \includegraphics[width=0.5\textwidth,angle=0]{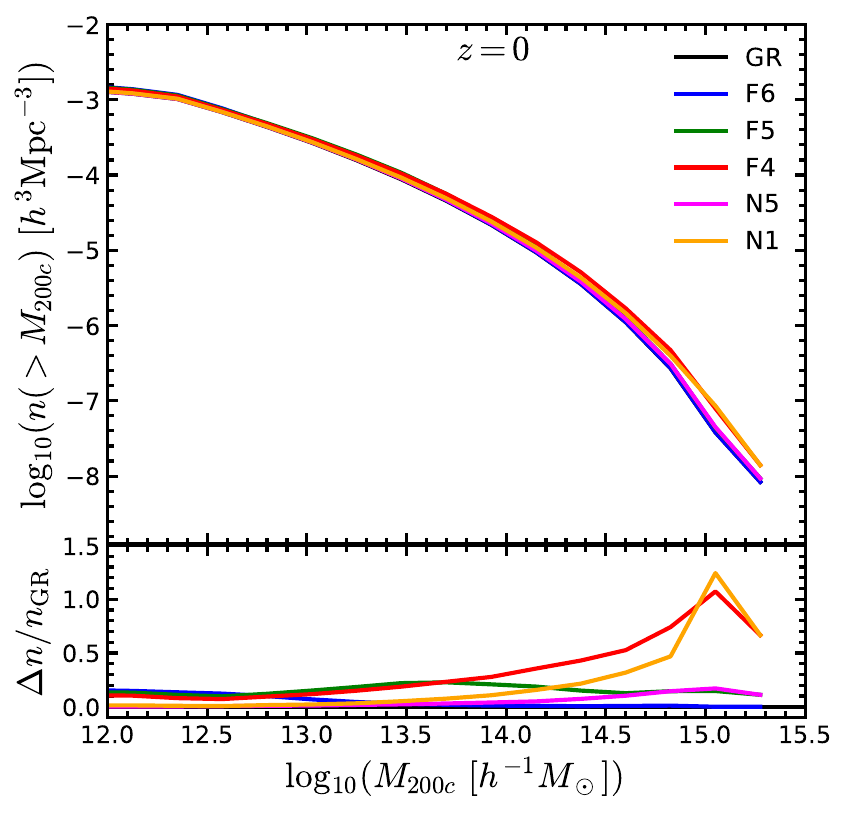}  & 
        \includegraphics[width=0.5\textwidth,angle=0]{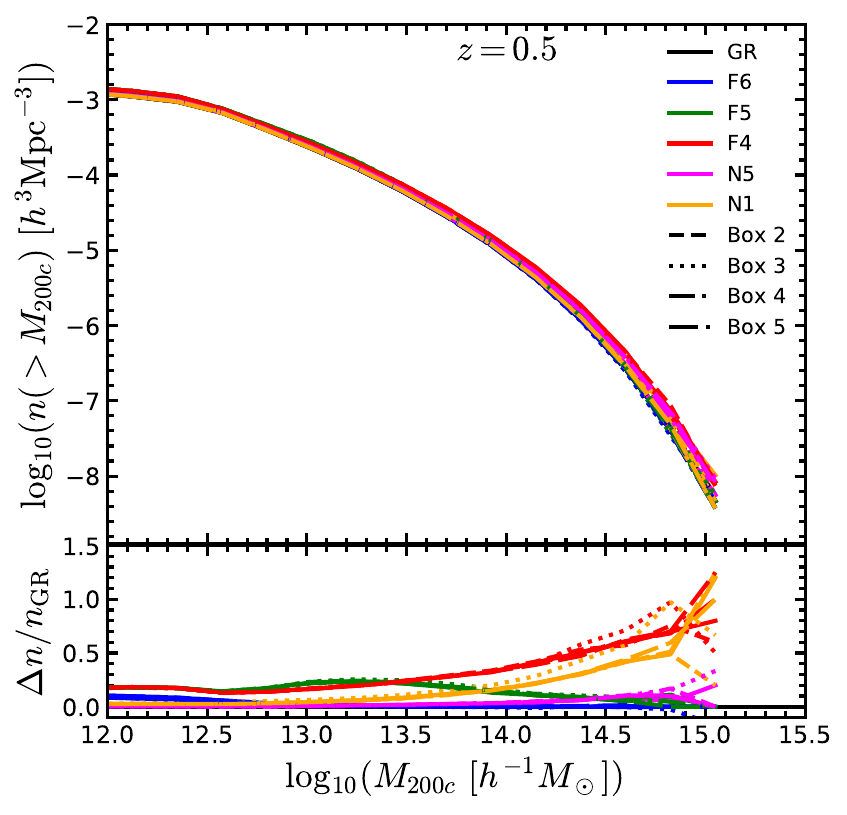} 
     \end{tabular}
     \caption{(Color Online) The cumulative halo mass function for the various models simulated here, at $z=0.0$ (left panel) and $z=0.5$ (right panel). The bottom panels show the relative difference between the various modified gravity models and $\Lambda$CDM. For the $z=0.5$ plot we have used all 5 realizations, and it shows that the model differences are not sensitive to the specific realization. For the $z=0$ plot we only used Box 1. The line colors and styles are the same as in Fig.~\ref{fig:pofk}.}
     \label{fig:hmf}
\end{figure*}

Figure \ref{fig:hmf} shows the cumulative halo mass function (HMF) of all models at $z=0$ (left panel) and $z=0.5$ (right panel). The bottom subpanels show the relative differences between the MG models and GR. In the case of $z=0.5$, there is good agreement between the five realizations (the different line styles) again. \textcolor{black}{For both redshifts we have compared the simulation HMF with the analytical fitting formula of Ref.~\cite{Tinker:2008ff} and found very good agreements above $10^{13}h^{-1}M_\odot$; this comparison is not shown here to avoid the plot becoming two crowded.}

As the halo catalogs are the starting point of the mock galaxy catalogs to be described below, it is useful to note some main features and their physical origins. Although $\Delta n/n_{\rm GR}$ is smaller at higher redshift, the qualitative features are the same in both redshifts. Of the $f(R)$ variants, in F6 the difference is strongly suppressed by the chameleon mechanism except for the smallest halos for which the screening is weak; this feature remains in F5, though the deviation from GR now starts at higher halo mass; for F4, the screening is essentially non-existent, leading to a significant increase in the number density of the most massive halos resolved in the simulations ($M_{200c}>10^{14.5}h^{-1}M_\odot$). Due to the faster mergers of small halos to form larger ones, F4 actually produces fewer halos in the mass range $10^{13}\sim10^{13.5}h^{-1}M_\odot$ than F5. The nDGP models are qualitatively similar to F4, but with smaller differences from GR. {The Vainshtein mechanism does not prevent more massive halos from forming in N1 and N5 as compared with GR, because the growth of halos is largely determined by how much matter the halos can accrete from their surroundings: while the Vainshtein mechanism is efficient in suppressing the fifth force close to and inside the halos, gravity can still be stronger than in GR within regions of size $\mathcal{O}(10)h^{-1}$Mpc from halos, which means that the largest structures end up growing more by accreting more matter from further away.}

As will be discussed next, the differences in the HMFs of the different models means that we have to slightly tune the galaxy populating scheme to obtain galaxy catalogs with the same desired clustering properties.

\subsubsection{Mock \newrevision{HOD} galaxy catalogs }
\label{sec:HOD}

To map the halo catalogs to a corresponding galaxy distribution, we populate halos with galaxies using the Halo Occupation Distribution (HOD) method \citep{Berlind:2002rn,Zheng:2004id}, in which it is assumed that the probability for a halo to host a certain number of galaxies can be computed through a simple functional dependence on the mass of the host halo. We use the form of the HOD suggested by \cite{Zheng2007}, in which the mean number of central galaxies, $\left< N_{{\rm cen}} (M) \right>$, and the mean number of satellite galaxies, $\left< N_{{\rm sat}} (M) \right>$, in a halo of mass $M$, are given respectively by:
\begin{eqnarray} \label{eq:HOD_eqns}
\left< N_{{\rm cen}} (M) \right> &=& \frac{1}{2} \left[ 1 + {\rm erf} \left( \frac{\log M - \log M_{{\rm min}}}{\sigma_{\log M}} \right) \right], \nonumber \\ 
\left< N_{{\rm sat}} (M) \right> &=& \left<N_{{\rm cen}}\right> \left( \frac{M-M_0}{M_1} \right)^{\alpha},
\end{eqnarray}
where $M_{{\rm min}}$, $M_0$, $M_1$, $\sigma_{\log M}$ and $\alpha$ are free parameters of the HOD model. Once their values have been specified, the mean number of galaxies in a halo of mass $M$ is then given by $\left<N(M)\right> = \left<N_{{\rm cen}} (M) \right> + \left< N_{{\rm sat}} (M) \right>$. From Eq.~\eqref{eq:HOD_eqns}, it can be seen that $M_{{\rm min}}$ and $M_0$, respectively, denote the threshold halo mass required to host at least one central or one satellite galaxy. When placing HOD galaxies in halos, central galaxies are assumed to reside at the center of potential of their host halo. Satellites, on the other hand, are distributed between $\left[0,R_{200c}\right]$ 
of the host halo center, according to a Nararro-Frenk-White (NFW, \cite{Navarro:1995iw,Navarro:1996gj}) profile with the concentration of the host halo computed by {\sc rockstar}. \textcolor{black}{This naturally takes into account the effect of the fifth force on halo density profiles, which can be substantial for $f(R)$ gravity \cite{Mitchell:2019qke}; for the DGP model, though, the effect on halo concentration is quite small \citep[e.g.,][]{Mitchell:2021b} but still taken into account}. Furthermore, central galaxies are assigned the center-of-mass velocity of the host halo, $V_{{\rm CM}}$; the velocity of a satellite galaxy is $V_{{\rm CM}}$ plus a perturbation along the $x, y$ and $z$ axes sampled from a 3D Gaussian distribution with a dispersion equal to the root-mean-squared (RMS) velocity dispersion of the host halo \textcolor{black}{as calculated by \textsc{rockstar}, which again takes into account of the modified gravity effects in the $f(R)$ and DGP models. We note that this way of modelling satellite galaxies necessarily incurs approximations, not least because in reality the satellite velocity distribution is more complicated \citep[e.g.,][]{Hikage:2015wfa}; this could also suppress the correlation between central and satellite galaxies which encodes the memory of the infall history of the latter. Other methods to set up satellite velocities are possible, e.g., by assigning the velocity of a dark matter particle randomly selected near the satellite position to the said satellite, but this is beyond the scope of this work and deserves a dedicated study using future high-resolution simulations}.

If the HOD catalogs in the MG models had been constructed using the same HOD parameters as in the $\Lambda$CDM model, there would generally be a difference of order $10$-$20$\% in the resulting number density and two-point correlation function (2PCF) of HOD galaxies, reflecting the MG effects on the halo abundance and clustering. Since there is only one observed Universe, if we do not know which cosmological model is the correct one, a more conservative way is to demand that all models make predictions that are consistent with observations. For this consideration, we have tuned the HOD parameters for the MG models to ensure that their resulting galaxy catalogs have roughly the same number densities and clustering properties as the corresponding $\Lambda$CDM catalogs. The assumption that MG models have different HOD parameters from GR is reasonable, because the evolution of the matter field and the assembly histories of galaxies are generally different in these models. This tuning of HOD parameters to fix galaxy clustering actually can help to remove one source of contamination when it comes to the model differences predicted by the various 
\revision{summary statistics} to be studied below. As a result of this tuning, some 
\revision{summary statistics}, such as the projected two point correlation functions, by construction cannot be used to discriminate the MG models from GR, and we need to find other ways to use the galaxy catalogs.

In practice, our tuning of MG HOD parameters was carried out using a Nelder-Mead simplex search through the 5-dimensional HOD parameter space. {From a MG and a $\Lambda$CDM HOD catalogs, the projected galaxy 2PCFs, $w_p(r_p)$, were measured\footnote{{This was obtained by projecting the 3D redshift-space 2PCF $\xi(r_p,r_\pi)$, with $r_p, r_\pi$ being respectively the galaxy pair separations transverse and parallel to the line of sight, using a projection depth of $90h^{-1}$Mpc. $\xi(r_p,r_\pi)$ was measured using the Correlation Utilities and Two-point Estimation ({\sc cute}) code \cite{Alonso2012}, with the distant-observer approximation. For the main results in this paper we have used the $z$ axis of the simulation box as the line-of-sight direction, but we have checked the HOD tuning when using the $x$ and $y$ axes as the lines of sight, and found very similar values of the best-fit HOD parameters.}} with the comoving projected separation range of $0.1<r_p<50h^{-1}{\rm Mpc}$.} The RMS difference between two models was calculated with an identical weight {of 1.0 for all $\log\left(r_p\right)$ bins}. To ensure that the two models have similar galaxy number density $n(z)$, the fractional difference in the respective $n(z)$ values was also used in the calculation of the RMS difference, with a (somewhat arbitrary) weight of 8.0. The code then walked through the 5D HOD parameter space to look for the smallest RMS difference, {and the search stopped if the value dropped below $1.5\%$ (with the exception of N1 for which the minimum value the code found was $2.2\%$). As the 
\revision{summary statistics} to be studied later in this paper are all evaluated at $z=0.5$, we only did this tuning to produce $z=0.5$ HOD galaxy catalogs.}

{This is a simplified and less rigorous approach in several ways. First, unlike a Markov chain Monte Carlo approach, the search only led to the `best-fit' HOD parameters rather than their posterior distributions. Second, the fitting of HOD parameters did not involve real data; instead, we adopted the best-fit HOD parameter values taken from \cite{Manera2013} for the CMASS data in the GR halo catalogs (all five realizations) to produce the $\Lambda$CDM HOD catalogs, and then tuned the HOD parameters for the MG models to match the $\Lambda$CDM results. Finally, often the HOD parameters are constrained simultaneously with cosmological parameters using a combination of different probes, while here we used a single probe (the projected galaxy 2PCF) to fix the HOD parameters and study other probes afterwards, as we wanted to use the same HOD catalogs to study a variety of probes. Note that for each MG model a single set of `best-fit' HOD parameters were tuned and used in producing the HOD catalogs for all five realizations. To check the sensitivity of the physical results presented below to the way in which the best-fit HOD parameters were obtained, we made two additional checks by: (1) tuning the parameters so that the MG models match the $\Lambda$CDM prediction of the real-space 3D galaxy 2PCF $\xi_{gg}(r)$ instead of the projected 2PCF, and (2) tuning the parameters individually for each simulation realization of each MG model. In both cases, we found little difference from the default case in terms of the halo occupancy properties and the physical results of the various cosmological probes.}

\begin{figure*}[!tb]
     \centering
     \begin{tabular}{cc}
        \includegraphics[width=0.495\textwidth,angle=0]{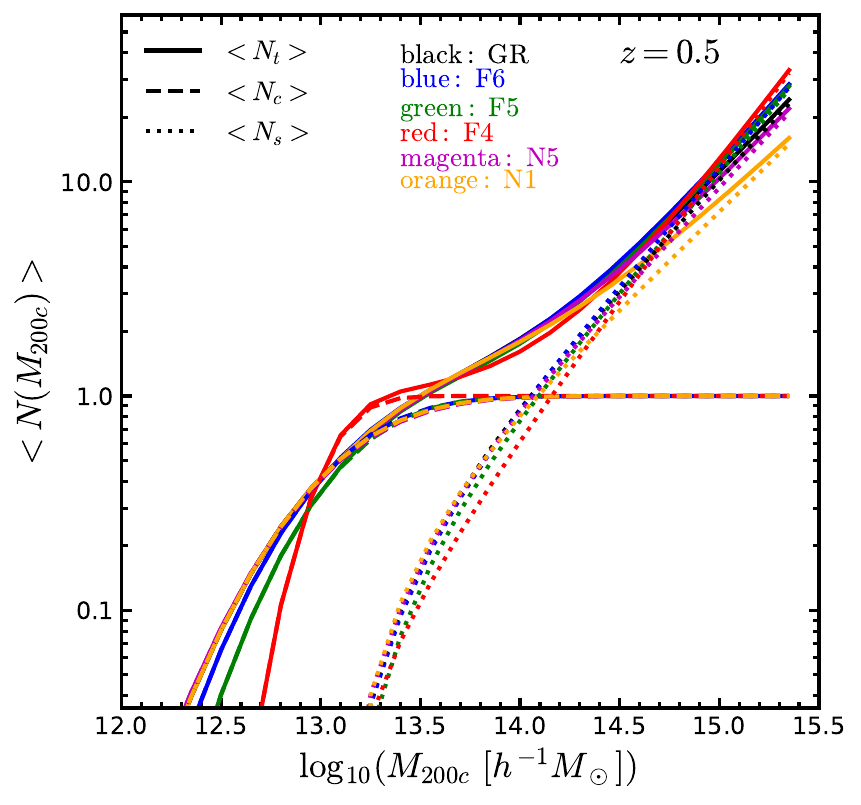}  & 
        \includegraphics[width=0.515\textwidth,angle=0]{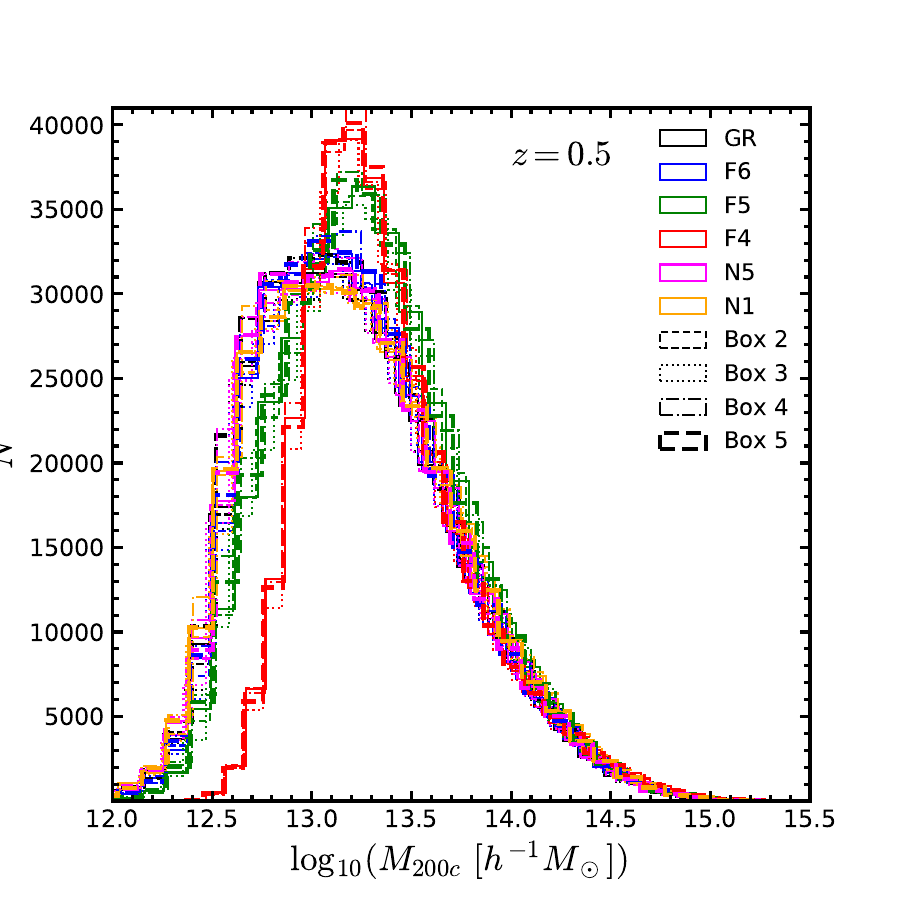} 
     \end{tabular}
\caption{(Color Online) {\it Left panel}: the mean galaxy number as a function of host halo mass for central (dashed), satellite (dotted) and all (solid) galaxies for $\Lambda$CDM and the five MG models. The figure is made using Eq.~\eqref{eq:HOD_eqns} with the tuned HOD parameters (cf.~Table \ref{tab:hod}) for the individual models. {\it Right panel}: histograms of galaxy numbers as a function of the host halo mass for all models and all realizations; this is essentially the product of the mean occupancy number (as shown in the left panel) and the halo mass function. The line colors and styles in the right panel are the same as in Fig.~\ref{fig:pofk}. All results are at $z=0.5$.}
\label{fig:HOD}
\end{figure*}

\begin{figure*}[!tb]
     \centering
     \begin{tabular}{cc}
        \includegraphics[width=0.494\textwidth,angle=0]{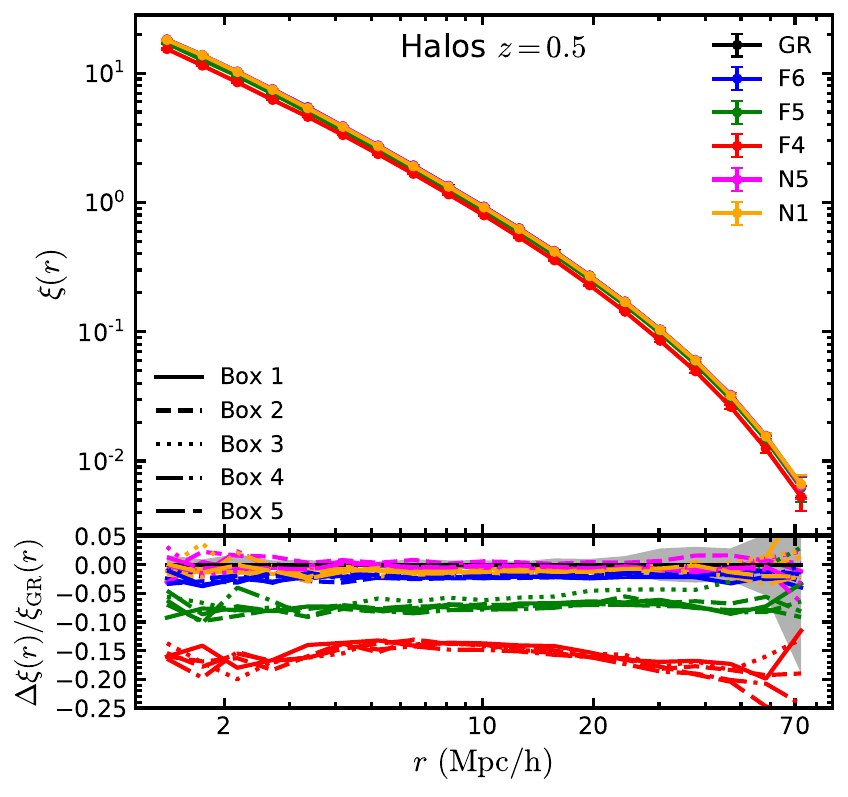}  & 
        \includegraphics[width=0.5\textwidth,angle=0]{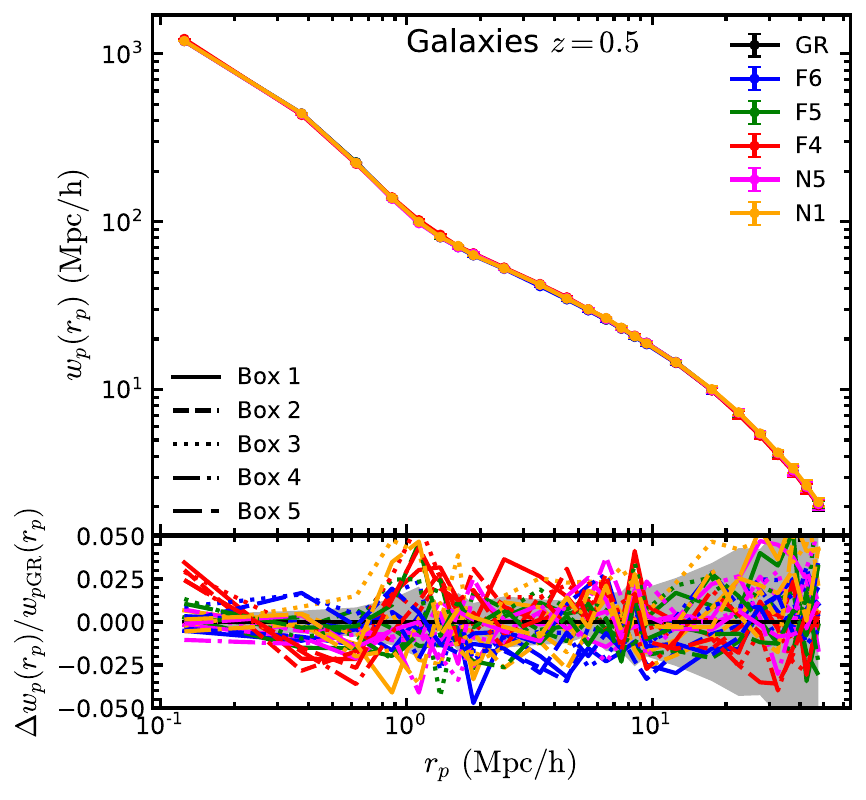} 
     \end{tabular}
\caption{(Color Online) {\it Left panels}: the 3D halo two-point correlation functions, $\xi_{hh}(r)$, for the six models, averaged over all five realizations, as a function of halo separation $r$ (upper panel) and the relative differences with respect to $\Lambda$CDM (lower panel; the five simulation realizations are shown individually using different line styles as in Fig.~\ref{fig:pofk}). Note that only halos more massive than $10^{13}h^{-1}M_\odot$ are used, so that the halo number densities are different in the different models. {\it Right panels}: the same as the left panels, but for the projected correlation functions of HOD galaxies, $w_p(r_p)$. All results are at $z=0.5$. The error bars in the top panels and the shaded regions in the bottom panels are the standard deviation of 5 $\Lambda$CDM realizations.}
\label{fig:h2pcf}
\end{figure*}

The left panel of Fig.~\ref{fig:HOD} shows the mean halo occupancy numbers for the different models in Box 1 (with other boxes being in good agreement), where a complicated pattern can be observed. For example, in F4 $N_{\rm cen}$ is substantially higher than in other all models at $10^{13}\lesssim{M}/(h^{-1}M_\odot)\lesssim10^{13.5}$ but decays much faster at $M\lesssim10^{13}h^{-1}M_\odot$. $N_{\rm cen}$ in the two nDGP variants both agree very well with that in $\Lambda$CDM, because for massive halos $N_{\rm cen}$ is equal to 1 anyway, while for smaller halos these models have very similar HMFs, cf.~Fig.~\ref{fig:hmf}. The right panel of Fig.~\ref{fig:HOD} shows the histograms of the numbers of HOD galaxies in host halos of different masses, which is essentially the product of the mean occupancy number multiplied by the host halo mass function. We have plotted the results from all five boxes using different line styles, and a good agreement among them (for a given model) is visible. As expected, more galaxies reside in more massive halos in F4 and F5 than in the other models.

\begin{table*}
\centering
\begin{tabular}{cccccccc} \hline \hline
           \textcolor{black}{Gravity Model} & $\log(M_{\rm min}/[h^{-1} {\rm M_\odot}])$ & $\log(M_1/[h^{-1} {\rm M_\odot}])$ & $\log(M_0/[h^{-1} {\rm M_\odot]})$ & ~~~~~~~$\sigma_{\log M}$~~~~~~~ & ~~~~~~$\alpha$~~~~~~ & \revision{$n_g/\left[h^{-1}{\rm Mpc}\right]^{-3}$} & 
           ~~~\revision{$f_{\rm sat}$}~~~ \\ \hline
GR & 13.09 & 14.00 & 13.077 & 0.596 & 1.0127 & \revision{$3.196\times10^{-4}$} & \revision{$10.88\%$} \\
F6 & 13.090 & 14.011 & 13.035 & 0.552 & 1.0766 & \revision{$3.192\times10^{-4}$} & \revision{$10.88\%$} \\
F5 & 13.132 & 14.049 & 13.061 & 0.512 & 1.1008 & \revision{$3.215\times10^{-4}$} & \revision{$11.18\%$} \\
F4 & 13.030 & 14.132 & 12.939 & 0.260 & 1.2393 & \revision{$3.177\times10^{-4}$} & \revision{$11.24\%$} \\
N5 & 13.098 & 14.017 & 13.054 & 0.610 & 0.9930 & \revision{$3.204\times10^{-4}$} & \revision{$10.90\%$} \\
N1 & 13.126 & 14.036 & 13.111 & 0.633 & 0.8967 & \revision{$3.204\times10^{-4}$} & \revision{$11.20\%$} \\
\hline\hline
\end{tabular}
\caption{The tuned HOD parameter values (columns 2-6) for the GR, $f(R)$ (F6, F5, F4) and nDGP (N5, N1) models at $z=0.5$ and  different realizations (Boxes 1-5). Note that all GR simulations use the same set of HOD parameters, which are taken from the best-fit parameters using CMASS data \cite{Manera2013}. The galaxy number density for all HOD catalogs is around $n_g=3.2\times10^{-4}[h^{-1}{\rm Mpc}]^{-3}$\revision{, and the number densities for each model, averaged over all 5 realizations, are shown in the second last column. The last column shows the satellite fractions for all models, $f_{\rm sat}$, again averaged over all 5 realizations. }}
\label{tab:hod}
\end{table*} 

The right panel of Fig.~\ref{fig:h2pcf} shows the projected galaxy 2PCFs $w_p(r_p)$ of the HOD galaxies for all models (top subpanel; average over five boxes), and the relative differences of the MG models from GR (bottom subpanel; individual boxes). {This verifies that the HOD parameter tuning for the different MG models has served its purpose of making $w_p(r_p)$ in the different models agree within $\sim1.5$-$2.5\%$, and that the same HOD parameters applied to different simulation realizations do give convergent results of $w_p(r_p)$.}

As a comparison, we also show, in the left panel of Fig.~\ref{fig:h2pcf}, the same results but the 2PCFs of dark matter halos with $M_{200c}\geq10^{13}h^{-1}M_\odot$. Here we see that $f(R)$ models generally have weaker clustering than GR, since for the same $M_{200c,{\rm min}}$ the model with stronger gravity would have a higher halo number density. This means that some of its halos correspond to initial density peaks too small to form halos with $M\geq M_{200c,{\rm min}}$ in a weaker gravity model. Note again the good agreement of halo 2PCFs in the different realizations.

It is interesting to see that very different gravity models can give nearly identical HOD galaxy number densities and clustering, showing the flexibility of the 5-parameter HOD model used here. This highlights the fact that galaxy-halo connection can bring a main theoretical uncertainty in using galaxy clustering to test models. 

\textcolor{black}{Note also that in screened MG models one may expect an additional assembly bias as the strength of gravity can vary from region to region, and that might lead to corresponding variations of $N_{\rm cen}$ and $N_{\rm sat}$. For example, Ref.~\cite{2015MNRAS.451L..45H} found a comparable redshift-space distortion signal due to assembly bias, modelled by galaxy color reshuffling, of the order of $10$-$20\%$ in the galaxy cluster environments, $\simeq1$-$10h^{-1}\mathrm{Mpc}$, which is comparable to the signal predicted by $f(R)$ models as shown below. However, Ref.~\cite{2016arXiv160102693M} find that the matter 2-point correlation function in the presence of assembly bias can be recovered at $2\%$ using matter-galaxy cross-correlation modelling in galaxy-galaxy lensing at scales larger than $r \geq 1\ h^{-1}\mathrm{Mpc}$. Therefore, we shall not pursue this extra complication here, but will leave a dedicated study of the assembly bias effect in modified gravity models to future work.}

\subsubsection{\newrevision{Additional mock galaxy catalogs used in this work}}
\label{sect:SHAM_SHYBONE}

\newrevision{Due to the relatively low resolution of the \textsc{elephant} simulations used in the previous subsection (\S~\ref{sec:HOD}), the halo catalogs are incomplete for halos of small mass (e.g., below $10^{12.5}$--$10^{13}h^{-1}M_\odot$), which makes them not ideal for building mock HOD galaxies 
with number densities above a few times $10^{-4}[\Mpch]^{-3}$ (such as the ones described in \S~\ref{sec:HOD} and used for most of the analyses in this paper). In addition, the low force resolution of the simulations also means that the spatial clustering of halos and HOD galaxies could be inaccurate below a few Mpc. As a complementation, therefore, we have employed some additional mock galaxy catalogs in this work. These additional mocks are from simulations with much higher resolutions, which means that they offer the possibility to study the effect of modified gravity in high-density galaxy catalogs (up to $10^{-2}[\Mpch]^{-3}$); 
they also employ more realistic methods of assigning satellite galaxies that can be important for small scales ($\sim1\Mpch$ and smaller).}

\newrevision{The first set of additional mock galaxy catalogs are constructed by using the subhalo abundance matching (SHAM) technique \cite{Conroy:2005aq,Rachel,moster}, and these are described in Ref.~\cite{He:2018oai}. The SHAM technique assumes that galaxies reside in subhalos, and that through a monotonic relation between a property of a subhalo and an observed property of a galaxy, subhalos selected in a simulation correspond to galaxies observed in a galaxy survey. The clustering of subhalos from the simulation therefore can be compared to the observed clustering of galaxies directly. An advantage of SHAM is that there is no ambiguity of galaxy bias. Moreover, the method can fully explore nonlinear effects, such as the FoG, on small scales as it utilizes N-body simulations directly. However, straightforward as the SHAM method may seem to be, it is indeed non-trivial to practically implement it and effectively control systematics.}

\newrevision{On the observational side, instead of photometrically selected samples in optical bands, stellar-mass-selected samples should be used. State-of-the-art hydrodynamic simulations (e.g., \cite{Schaye:2014tpa,Chaves-Montero:2015iga}) suggest that it is the stellar mass of a galaxy, rather than its $r$-band luminosity, that has a {tighter} correlation with $v_{\rm peak}$ (the peak value of the maximum circular velocity over a subhalo's merger history) of a subhalo. However, unlike its luminosity, a galaxy's stellar mass cannot be directly measured but has to be derived from a stellar population synthesis model. A challenge we face here is the uncertainty in the estimation of stellar mass. There are two main sources of the systematics: one is in theory, especially the stellar initial mass function (IMF); the other is the uncertainty in determining the total flux of a galaxy. In order to mitigate these systematics, we have constructed volume-limited samples that are complete in stellar mass, with galaxies being selected in terms of number densities rather than stellar mass cut. The idea here is to keep the ranking orders of galaxies, {and as demonstrated in \cite{He:2018oai} this method can effectively minimize the impact of systematics in the estimation of stellar mass on the measured RSD multipoles, particularly for samples with higher number densities, $n_g\gtrsim1\times10^{-2}[\Mpch]^{-3}$.}}

\newrevision{On the theoretical side, a high mass resolution of the simulation is crucial for SHAM. A satellite subhalo close to the host halo center may substantially lose mass due to tidal stripping. Sometimes even a massive halo at an earlier time in a simulation can be completely disrupted by tidal stripping and can not be resolved at a later time, leading to a phenomenon called ``orphan galaxies" \cite{moster}, which can result in an under-estimation of the galaxy clustering on small scales \cite{moster}. 
}

\newrevision{We only have SHAM mock galaxy catalogs for the GR and F6 models \cite{He:2018oai}. These are constructed on halo/subhalo catalogs and merger history obtained by applying the \textsc{rockstar} halo finder on GR and F6 simulations run with the \textsc{ecosmog} code. The simulations follow $512^3$ dark matter particles in a cubic box of size $64h^{-1}\mathrm{Mpc}$, with a mass resolution of $\simeq1.52\times10^8h^{-1}M_\odot$. The domain grid size used here is $512^3$ cells, but the code adaptively refines this grid so that in dense regions the mesh cell size can be as small as $\simeq1h^{-1}\mathrm{kpc}$. These lead to a high force resolution that allows us to look at the clustering of SHAM galaxies down to $\simeq1h^{-1}\mathrm{Mpc}$, and the force resolution enables the SHAM mocks to reach a number density of $10^{-2}[\Mpch]^{-3}$, which is impossible for the HOD catalogs described above. Further details of these mock catalogs can be found in Ref.~\cite{He:2018oai}.}

\newrevision{Our second set of additional mock galaxy catalogs are obtained from full hydrodynamical simulations of galaxy formation in modified gravity models \cite[the \textsc{shybone} simulations;][]{Arnold:2019vpg,Hernandez-Aguayo:2020kgq}. These simulations employ the IllustrisTNG galaxy formation model \citep{2017MNRAS.465.3291W,Pillepich:2017jle} and include runs for both GR and modified models (F6, F5, N5, N1). The simulations have been run using the highly parallel and optimised hydrodynamical cosmological simulation code \textsc{arepo} \citep{2010MNRAS.401..791S}, which has been suitably adapted to include a modified gravity solver, for the HS $f(R)$ gravity and the DGP model respectively, that accurately calculates the fifth force in high-density regions using adaptive mesh refinement. The simulations took place in a box of side length 62$h^{-1}{\rm Mpc}$,  following $512^3$ dark matter particles and the same number of initial gas resolution elements (which are Voronoi cells in \textsc{arepo}); all runs begin at redshift $z=127$, from the same initial condition for all gravity models. The cosmological parameters are ($h$, $\Omega_{\rm M}$, $\Omega_{\rm B}$, $\Omega_{\Lambda}$, $n_{\rm s}$, $\sigma_8$) $=$ ($0.6774$, $0.3089$, $0.0486$, $0.6911$, $0.9667$, $0.8159$), where note that the $\sigma_8$ value is for $\Lambda$CDM at $z=0$. The mass resolution for DM particles is $m_{\rm DM}=1.28\times10^8h^{-1}M_{\odot}$ and the average gas cell mass is $m_{\rm gas}\approx2.5\times10^7h^{-1}M_{\odot}$. The group catalogs were constructed using the \textsc{subfind} code \cite{Springel:2000qu} inbuilt in \textsc{arepo}, which uses the friends-of-friends (FOF) algorithm combined with an unbinding method to identify bound structures within a FOF group.}

\newrevision{The IllustrisTNG model used in the \textsc{shybone} simulations is a realistic and highly sophisticated description of the simplified galaxy formation physics, including star formation, cooling, and stellar and black hole (BH) feedback. It adopts the Eddington ratio as the criterion for deciding the accretion state of BHs, and employs a kinetic AGN feedback model that produces a BH-driven wind, which is responsible for the quenching of star formation in galaxies residing in high- and intermediate-mass halos, and for the production of red and passive galaxies at late times. The standard IllustrisTNG subgrid model has been used in the $f(R)$ and nDGP simulations without any further tuning --- although in principle a retuning is needed for any new cosmological model, it has been checked explicitly that the default IllustrisTNG model still predicts baryonic observables in good agreement with observations even for stronger MG models such as F5 and N1.}

\begin{figure*}[!tb]
     \centering
     \begin{tabular}{cc}
        \includegraphics[width=1.0\textwidth,angle=0]{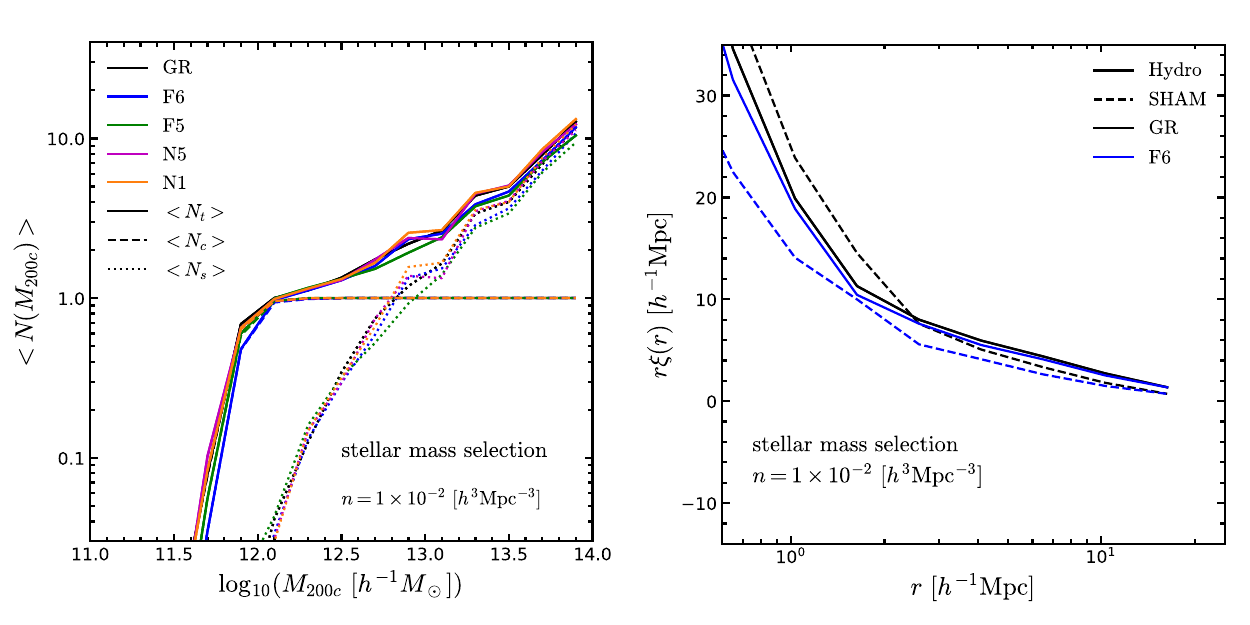} 
     \end{tabular}
\caption{\newrevision{(Color Online) {\it Left panels}: the HOD of the stellar-mass-selected galaxies from the \textsc{shybone} hydrodynamical simulations for GR (blue), F6 (blue), F5 (green), N5 (magenta) and N1 (orange). The dashed lines are for central galaxies ($N_c$), dotted lines for satellite galaxies ($N_s$) and solid lines the sum of the two ($N_t$). {\it Right panels}: the real-space galaxy correlation functions ($\xi(r)$) of the GR (black) and F6 (blue) models, as predicted from SHAM (dashed lines) and the \textsc{shybone} simulations (solid lines) for stellar-mass-selected galaxies. In all cases the mock galaxy number density is $10^{-2}[\Mpch]^{-3}$.}}
\label{fig:sham_shybone_plot}
\end{figure*}

\newrevision{The left panel of Fig.~\ref{fig:sham_shybone_plot} shows the halo occupancy numbers for different gravity models as predicted by the \textsc{shybone} simulations, which indicate that galaxies in these simulations occupy halos of mass down to $10^{11.7}$--$10^{12}h^{-1}M_\odot$. The right panel of the figure compares the real-space galaxy correlation function $\xi(r)$ (which is multiplied by $r$ to increase readability) for GR and F6, as predicted by SHAM and the \textsc{shybone} simulations. We can see that the two approaches give very different predictions of $\xi(r)$, as well as very different predictions in the relative difference between F6 and GR, though both of them predict that F6 has a smaller $\xi(r)$ than GR. We will comment on this again later in \S\ref{subsect:subhalo}.}

\newrevision{These additional mock galaxy catalogs, based on SHAM and the \textsc{shybone} simulations respectively, will only be used to study the small-scale clustering of galaxies in \S~\ref{subsect:subhalo} below, but not for other summary statistics analysed in the next section. This is because their box sizes ($62$--$64h^{-1}\mathrm{Mpc}$) are very small and so they offer very little statistical power. In contrast, the other summary statistics to be studied below mostly involve: large length scales, which are beyond the probe of these small boxes; or large cosmological objects such as galaxy clusters and cosmic voids, for which there are few in the box; or higher-order statistics, for which a larger sample size is even more important; or smoothed (galaxy) density fields with large smoothing lengths. These summary statistics will be more reliably studied using larger (yet still high-resolution) simulations than used in this paper, and we hope to revisit them in future works.}

\subsubsection{Fast simulations and mocks}
\label{sect:FastSim}

A rigorous analysis of the 
\revision{summary statistics} and their capability to constrain models requires to generate a substantial number of realizations of simulations to explore the statistical properties of the 
\revision{summary statistics} and characterize the significance of the differences observed. Furthermore, for comparing models one also needs the covariance matrices for the different 
\revision{summary statistics}. Since full MG simulations introduced above are expensive, it is not practical to use them for that purpose. Thus, it is helpful to use approximate methods to generate a large number of simulations. 

One of the existing fast simulation codes is {\sc mg}-{\sc cola} {(COmoving Lagrangian Acceleration)} \cite{Winther:2017jof}, which is an approximation method based on second order Lagrangian Perturbation Theory (2LPT), to generate quick mock catalogs for the different MG models considered. {\sc mg}-{\sc cola} has flexibility to allow a simulation to be run with varying numbers of time steps: the use of very few time steps gives the predictions of 2LPT, while the use of more time steps can improve the accuracy towards that of a full simulation. This flexibility, however, means that {\it a priori} we do not know the smallest number of time steps to be used in order to meet our target accuracy. Therefore, in order to use {\sc mg}-{\sc cola} simulations, we need to calibrate them using full simulations to make sure that the setup can reproduce the results of the latter on the scales of interest, and this step should be performed separately for each 
\revision{summary statistic} of interest to us. Apparently, this will be a substantial effort that goes beyond the scope of this paper. 

As a quick check, we have run 20 simulations for each of the six models considered in this paper, using the same specifications as given in Table \ref{table:simulations} except that the particle number is 8 times higher ($2048^3$). With 50 code time steps (30 steps for $30\leq z \leq 0.5$ and 20 up to $z=0$), each {\sc mg}-{\sc cola} simulation takes about 300 core hours, which is a factor of $\mathcal{O}(100)$ faster than full simulations which take about 350 coarse and $\sim10,000$ fine time steps. The huge saving of computing time partly comes from the fact that full simulations spend a substantial fraction of their running time solving the nonlinear field equations for the scalar fields on refinements. The HMFs of these fast simulations agree with those from the full simulations within $\sim20\%$ between $10^{12.5}\sim10^{14.5}h^{-1}M_\odot$, but the agreement worsens rapidly at $>10^{14.5}h^{-1}M_\odot$. The two-point correlation functions of halos from the fast simulations agree with those of the full simulations within $15\%$ between $3\sim30h^{-1}$Mpc, and the discrepancy increases quickly for $r>30h^{-1}$Mpc. The disagreements between $3\sim30\Mpch$ are at a similar level to the model differences that we are interested in tests, and so are not suitable for estimating the observables or their covariances. This suggests that the fast simulations may need to be run at higher time, space or mass resolutions in order to agree with full simulations, especially for probes that employ small-scale information such as within dark matter halos. Future developments in this area, e.g., in optimising existing codes, developing possible new fast codes, or proposing (semi)analytical methods for covariance estimation, will therefore be of great importance and much welcomed effort.

\section{{Summary statistics}}
\label{sect:estimators}

Having introduced the theoretical models, simulations, and halo and mock galaxy catalogs in the previous section, we can now move on to showing how the various 
\revision{summary statistics} studied in this paper predict different signals for the different {gravity} models.  Because of the restrictions in the number densities and volumes of our mock galaxies, and due to the various systematic effects which are not accounted for, we shall not conduct a rigorous statistical analysis of the future constraints from DESI, but instead focus on understanding the model behaviors and signatures, quantifying roughly the significance at which different models can be statistically distinguished, and identifying future simulation and analysis needs.

In this section we consider a diverse range of 
\revision{summary statistics}. In \ref{subsect:RSD}, we consider the use of redshift-space clustering on large and small scales. In \ref{subsect:marked}, we investigate how clustering in under-dense and over-dense regions can be contrasted through both the explicit consideration of statistics in dense and void environments and the application of environment-dependent, or ``marked" statistics. In \ref{subsect:beyond2pt}, the applications of a range of real- and Fourier-space statistics beyond the two-point correlation function are studied, including the three-point correlation function, the bispectrum, hierarchical clustering, Minkowski functionals and phase space statistics. In \ref{subsect:lensing}, the use of lensing information in addition to spectroscopic clustering information is considered both in clustered and void environments.

\subsection{{Redshift-space galaxy clustering}}
\label{subsect:RSD}

Peculiar velocities of galaxies induce anisotropies in redshift space and leave distinctive imprints on the clustering pattern at different regimes. On large and linear scales, galaxies infall into high-density regions such as clusters, producing a squashing effect of these regions along the line-of-sight: this is the Kaiser effect \cite{Kaiser:1987}. On smaller (nonlinear) scales, the random motions of galaxies in virialized objects produce the Fingers-of-God (FoG) effect where the density field becomes stretched and structures appear elongated along the line of sight \cite{FoG:Jackson}.

{In this section we consider the potential constraints from redshift space distortions (RSD) on large scales, section \ref{subsect:LRSD}, and small scales, in section \ref{subsect:subhalo}.}

\subsubsection{Large-scale RSD}
\label{subsect:LRSD}

In linear theory, the amplitude of the RSD is related to the distortion parameter $\beta$, defined as
 \begin{equation}\label{eq:beta}
 \beta(z) \equiv \frac{f(z)}{b(z)}\,,
 \end{equation}
where $f$ is the linear growth rate and $b$ is the linear galaxy bias, as functions of redshift.

The linear growth for the matter fluctuations in different gravity models can be obtained by solving the equation of the linear growth factor, $D_+$,
\begin{equation}\label{eq:Dp}
D''_+ + \left[2 - \frac{3}{2}\Omega_{\rm m}(a) \right]D'_+ - \frac{3}{2}\mu(k,a)\Omega_{\rm m}(a) D_+ = 0\,,
\end{equation}
where $^\prime$ denotes a derivative with respect to $\ln a$ and {$\mu$, introduced in Eq.~(\ref{eq:mu-exp}), is the ratio between the effective Newton constant $G_{\rm eff}$ and the true one $G$:}
\begin{equation}\label{eq:Geff}
\mu(k,a)=\frac{G_{\rm eff}}{G} = 
      \begin{cases}
          1 & \text{for GR},\\
          1 + \frac{k^2}{3(k^2 + a^2m^2_{f_R})} & \text{for $f(R)$ gravity},\\ 
          1 + \frac{1}{3\beta_{\rm DGP}(a)} & \text{for nDGP},
      \end{cases}
\end{equation}
where $k$ is the wavenumber of a perturbation mode, $m_{f_R}$ is the mass of the scalaron field defined by $m^2_{f_R} \simeq \left[3|f_{RR}|\right]^{-1}$ and $\beta_{\rm DGP}(a)$ is given by Eq.~\eqref{eq:beta_dgp}. 
Note that $G^{f(R)}_{\rm eff}$ is a function of time and scale, which means that the linear growth of structure for $f(R)$ gravity is scale dependent, while for GR and nDGP is scale independent. The linear growth rate, $f$, is defined as 
\begin{equation}\label{eq:f_lin}
f \equiv \frac{{\rm d}\ln D_+}{{\rm d}\ln a}\,.
\end{equation}

Large-scale redshift distortions have been studied with a large variety of tracers, including luminous red galaxies, e.g., \citep{Cabre:2008sz,Sanchez:2009}, cosmic voids \citep{Hamaus:2015yza,Hamaus:2017dwj,Cai:2016jek} and quasi-stellar-objects (QSOs) \citep{Hou:2018dr14, Gil-marin:2018dr14, Zarrouk:2018dr14}, and been successfully used to extract cosmological information by assuming a standard cosmological model, $\Lambda$CDM, based on GR. Current studies of modified gravity have used redshift-space distortions to put constraints on the $\beta$ parameter, Eq.~\eqref{eq:beta}, see, e.g., Ref.~\citep[][]{Hernandez-Aguayo:2018oxg}. In Ref.~\cite{Wright:2019qhf}, the authors studied a coupled model of $f(R)$ gravity and massive neutrinos to break the degeneracy between the enhancement of the growth of large-scale structure produced by MG models and the suppression due to the free-streaming of massive neutrinos at late times. 

\begin{figure*}[!tb]
\begin{center}
{\includegraphics[width=0.49\textwidth]{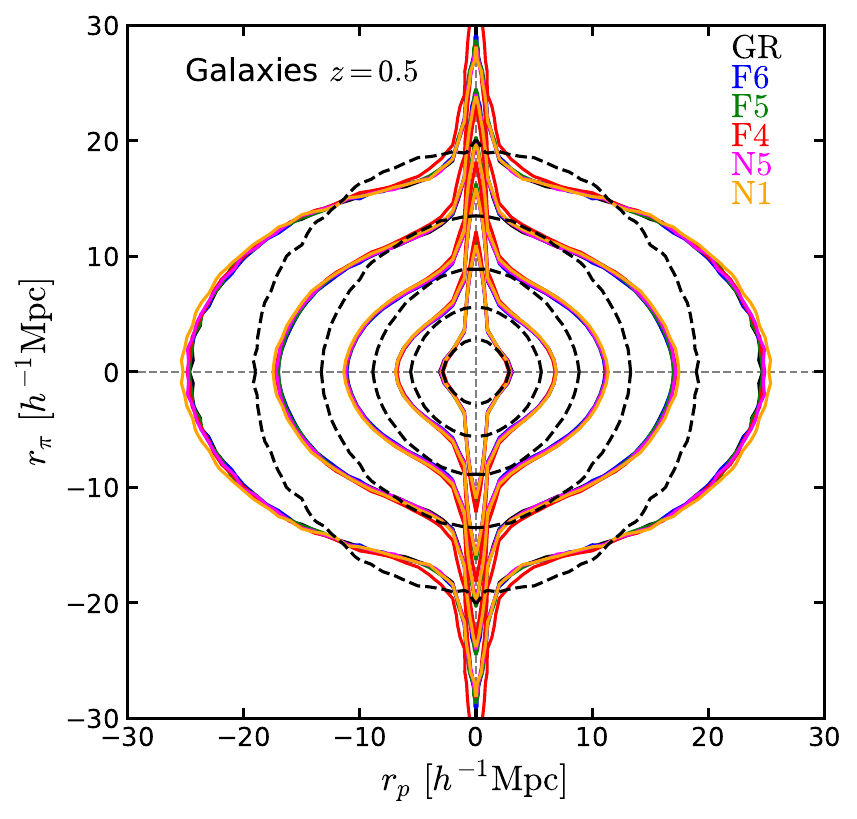}
\includegraphics[width=0.49\textwidth]{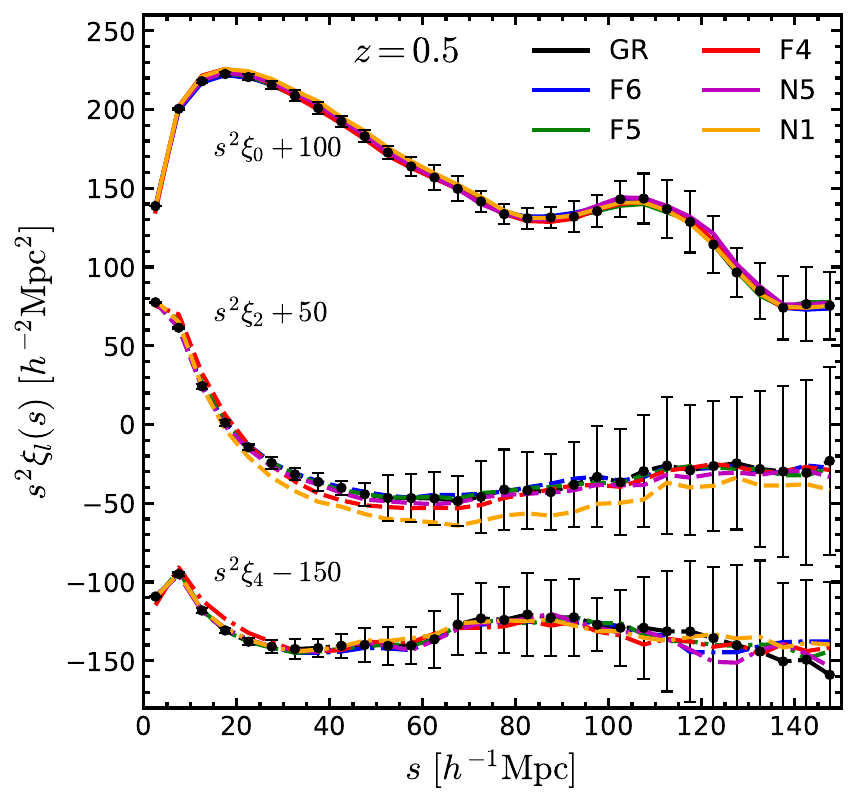}}
\caption{{\it Left panel:} The two dimensional galaxy correlation function $\xi(r_p,r_\pi)$ measured from our mock catalogues at $z=0.5$ as a function of separation across $(r_p)$ and along $(r_\pi)$ the line-of-sight. Contours show lines of constant $\xi(r_p,r_\pi)$ at $\xi(r_p,r_\pi)$ = 5, 2, 1, 0.5, 0.25. The correlation functions correspond to the average of fifteen measurements obtained by projecting five realizations over the three LOS directions. The black dashed contours correspond to the real-space measurements of the correlation function at the same values of its counterpart in redshift-space, for simplicity we show the GR case only {for these contours}.
{\it Right panel}: Monopole, quadrupole and hexadecapole moments of the correlation function, Eq.~\eqref{eq:xi_l}, for our six gravity models at $z=0.5$. The moments have been shifted by a factor of $100$, $50$ and $-150$ for better visualization. The error bars correspond to the standard deviation over fifteen GR measurements.
Different colour line correspond to different gravity model as specified in the legend.}
\label{fig:2pcf_zspace}
\end{center}
\end{figure*}

The effects of redshift-space distortions can be measured using the redshift-space correlation function of galaxies, $\xi(r_p,r_\pi)$, which is the excess probability of finding a pair of galaxies at separations transverse ($r_p$) and parallel ($r_\pi$) to the line of sight (LOS). The left panel of Fig.~\ref{fig:2pcf_zspace} shows $\xi(r_p,r_\pi)$ as a function of separation $(r_p,r_\pi)$ at $z=0.5$, using the HOD catalogs described in \ref{sec:HOD} for the different gravity models. For comparison, the black dashed curve corresponds to the spherical two-dimensional correlation function in real-space of the $\Lambda$CDM (GR) model. We clearly see that along the LOS at $r_p \lesssim 2\Mpch$ the clustering is enhanced by the FoG effect, while at $r_p > 2\Mpch$ the clustering pattern is squashed by the Kaiser effect. 

Given the symmetry along the line of sight, the transverse and parallel separations, $(r_p,r_\pi)$, can be expressed as a distance in redshift space, $s$, and the cosine of the angle between $s$ and the LOS direction, 
\begin{equation}
s = \sqrt{r^2_\pi + r^2_p}\,, \qquad \mu = \frac{r_\pi}{s}\,.
\end{equation}
The resulting anisotropic correlation function, $\xi(s,\mu)$, can then be decomposed into multipole moments,
\begin{equation}\label{eq:xi_l}
\xi_l(s) = (2l + 1) \int^1_{0}{\xi(s,\mu)} P_l(\mu)~{\rm d}\mu\,,
\end{equation}
where $P_l(\mu)$ are the Legendre polynomials. In linear theory, the $l=0$, $2$ and $4$ moments are non-zero with $P_0(\mu) = 1,\, P_2(\mu) = (3\mu^2 + 1)/2,\, P_4(\mu) = (35\mu^4 - 30\mu^2 +3)/8$, corresponding to the monopole, quadrupole and hexadecapole moments. 

The right panel of Fig.~\ref{fig:2pcf_zspace} shows the multipole moments, $\xi_l(s)$, of the correlation functions measured from the same galaxy catalogues. From the monopole, $\xi_0(s)$, we observe that the position of the baryon acoustic oscillations (BAO) peak, at $s_{\rm BAO} \simeq 105\Mpch$ or $150~{\rm Mpc}$, is not affected by modified gravity.
Higher-order multipole moments such as the quadrupole $\xi_2(s)$ and the hexadecapole $\xi_4(s)$ encode the anisotropies induced by redshift distortions. In the case of the quadrupole, $\xi_2(s)$, N1 shows the strongest deviation with respect to GR, especially on scales $s>20\Mpch$, followed by F4 and N5. Our measurements of the hexadecapole are almost indistinguishable among all models studied here. Comparing the plot with the left panel of Fig.~4 of \cite{Hernandez-Aguayo:2018oxg}, in which the HOD parameters were tuned to match the real-space galaxy two point correlation functions in MG and GR, we can see that the results are almost identical. 

\begin{figure*}[!tb]
\begin{center}
{\includegraphics[width=0.49\textwidth]{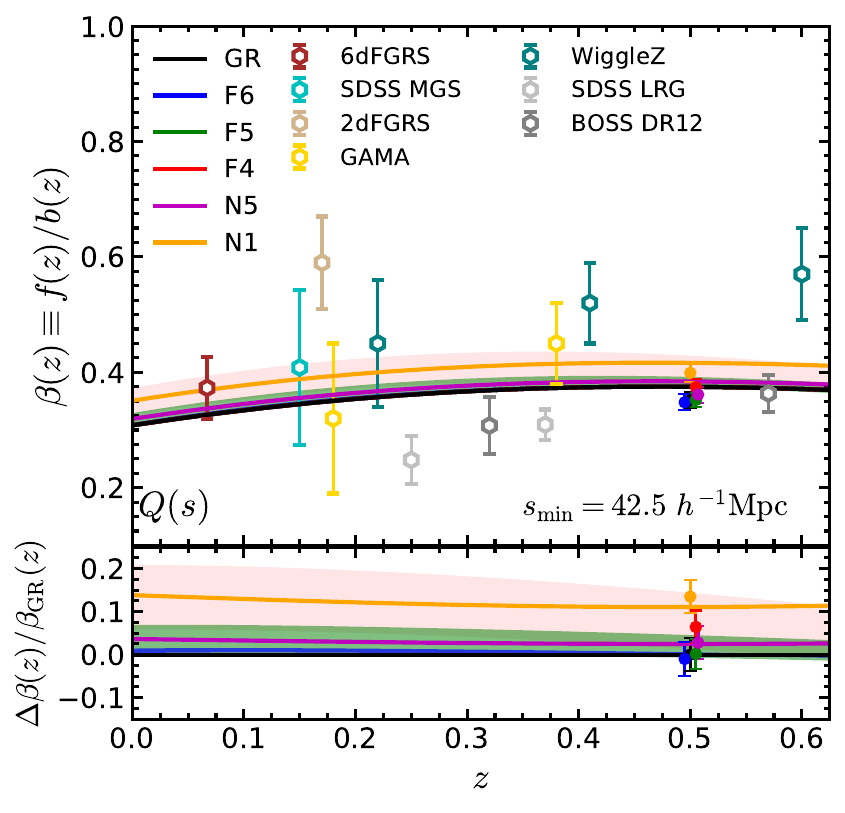}
\includegraphics[width=0.49\textwidth]{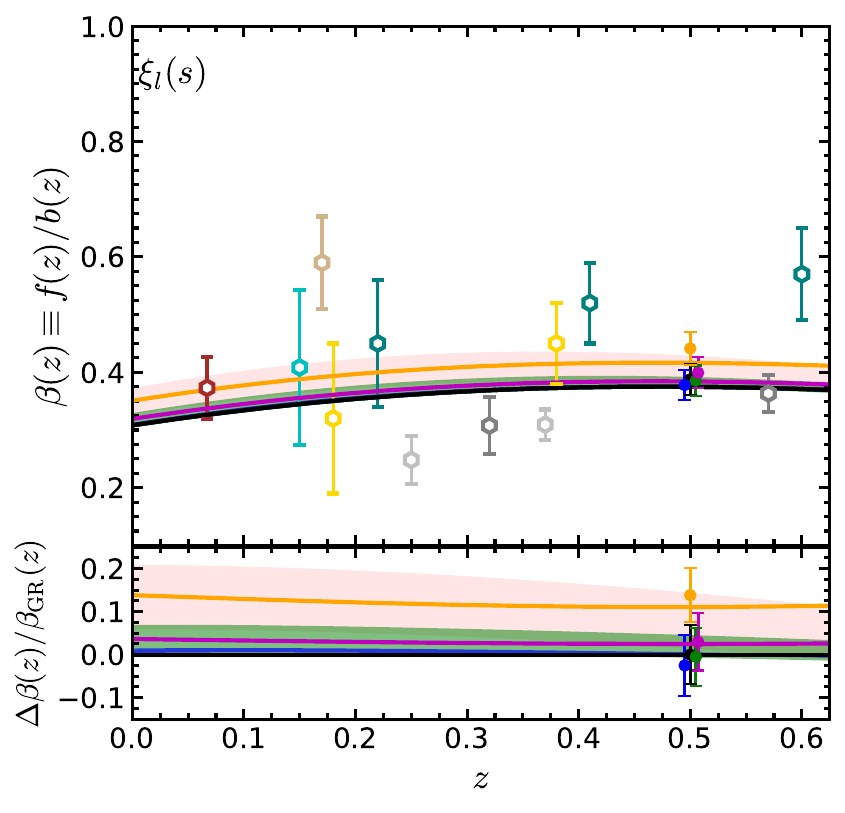}}
\caption{Evolution of $\beta$ as a function of redshift. The curves show the theoretical predictions for the gravity models as shown in the legend, for $f(R)$ gravity models the theoretical predictions are shown as a shaded region for wavenumbers $0.01 \leq k/[h\,{\rm Mpc}^{-1}] \leq 0.1$. Open coloured symbols correspond to current constraints from observational galaxy survey data: 6dFGRS \cite{6dFGRS:2012px}, SDSS MGS \cite{SDSS-MGS:2014opa}, 2dFGRS \cite{2dFGRS:2004fs}, GAMA \cite{GAMA:2013nif}, Wiggle Z \cite{WiggleZ:2011rj}, SDSS LRG \cite{SDSS-LRG:2011cs} and BOSS DR12 \cite{BOSS-DR12:2015sqa}.  Filled symbols, at $z=0.5$, denote the $\beta$ estimates obtained from the simulations using the 
\revision{summary statistics}---\textit{Left panel}: $Q(s)$ Eq.~\eqref{eq:Qs} using linear theory; \textit{Right panel}: $\xi_l(s)$ Eq.~\eqref{eq:xi_l}  using non-linear theory with $s_{\rm min} = 42.5\,h^{-1}\mathrm{Mpc}$ at $z=0.5$. The lower subpanels show the relative difference between the modified gravity models and GR. Error bars correspond to the $1\sigma$ confidence level.}
\label{fig:fits_beta}
\end{center}
\end{figure*}

{To investigate the impact of MG on RSD more quantitatively, we discuss two methods to constrain the distortion parameter, $\beta$. The first is based on the Kaiser linear model \cite{Kaiser:1987}, considering 
\revision{summary statistic}, $Q(s)$, to obtain the distortion parameter $\beta$ \revision{as usually done in the literature}, \citep[see e.g.,][]{Hawkins:2002sg}:
\begin{equation}\label{eq:Qs}
Q(s) \equiv \frac{\xi_2(s)}{\xi_0(s) - \bar{\xi}_0(s)} = \frac{(4/3)\beta + (4/7)\beta^2}{1 + (2/3)\beta + (1/5)\beta^2}\,
\end{equation} 
The second method is based on an extension of the Galilean-invariant renormalized perturbation theory \citep[gRPT;][]{Crocce:2005xy}, where the anisotropic correlation function is obtained as the inverse Fourier transform of the power spectrum.} 

To estimate $\beta(z)$ from $Q(s)$ using the linear theory model, we use a $\chi^2$-test by minimizing the $\chi^2$ defined as
\begin{equation}\label{eq:chi2}
\chi^2(\beta) = \sum_i \left[\frac{Q(s_i) - Q^{\rm{th}}(s_i;\beta)}{\sigma_{Q_i}}\right]^2\,,
\end{equation}
where $Q(s)$ is the average of the linear 
\revision{summary statistic} given by Eq.~\eqref{eq:Qs} from the 15 redshift-space HOD catalogues constructed from the 5 real-space HOD catalogs, $\sigma_{Q}$ is the standard deviation in the same catalogs, and $Q^{\rm{th}}(s;\beta)$ is the theoretical prediction of $Q(s)$ given by the third expression of Eq.~\eqref{eq:Qs}. We searched in a grid of values in $\beta \in [0,1]$, with a step size of $\Delta\beta = 10^{-4}$, for the theoretical 
\revision{summary statistic} and identified the value of $\beta$ that minimizes $\chi^2$ as $\chi^2_{\rm min} = \chi^2(\beta_{\rm fit})$. As we vary only one parameter, the $1\sigma$ error on $\beta$ corresponds to $\Delta\chi^2 \equiv \chi^2 - \chi^2_{\rm min} = 1$. The fit was performed in the range of scales $s = 42.5 - 147.5 \Mpch$, \revision{to avoid contamination due to non-linear effects and potential assembly bias that would complicate the measurements}.

To obtain the constraints of $\beta$ using the gRPT model, we used Bayesian statistics and maximize the likelihood,
\begin{equation}
\mathcal{L}({\bm \xi}|{\bm \lambda})\propto \exp\left[-\frac{1}{2}\left({\bm \xi}-{\bm \xi}_{\rm model}({\bm \lambda})\right)^{\rm T}{\Psi}\left({\bm \xi}-{\bm \xi}_{\rm model}({\bm \lambda})\right)\right],
\label{eqn: gaussian-likelihood}
\end{equation}
where the $\Psi = \rm C^{-1}$ is the inverse of the covariance matrix. We applied the Gaussian recipe \citep{Grieb:2016cov} to estimate the covariance matrix, which is then rescaled by the number of simulations. The parameters that enter the default fitting are $\{f, b_1, b_2, \gamma^-_3, a_{\rm vir}\}$, {where $b_1$ and $b_2$ are local galaxy biases to linear and second order; $\gamma^-_3$ is a non-local bias coefficient, see, e.g. \citep{Chan:2012jj}; and 
$a_{\rm vir}$ is a free parameter that describes the kurtosis of the velocity distribution on small scales. Two additional parameters, $\{q_\parallel, q_\perp\}$, relating fiducial and real distances,  are needed when applying the Alcock–Paczynski \revision{(AP)} test \cite{Alcock:1979mp}}. \revision{However, we fixed the AP parameters to make a fair comparison with the linear Kaiser model.} We marginalize over the nuisance parameters to find the probability distribution of the distortion parameter $\beta = f/b_1$. For more details of the method, we refer the readers to  \cite{Sanchez:2016sas,Hernandez-Aguayo:2018oxg}.

In the upper panels of Fig.~\ref{fig:fits_beta}, we compare the theoretical predictions for $\beta(z)$ to current observational measurements of the distortion parameter from galaxy surveys, including the 6dFGRS at $z\simeq 0.067$ \cite{6dFGRS:2012px}, the SDSS MGS at $z\simeq0.15$ \cite{SDSS-MGS:2014opa}, the 2dFGRS at $z\simeq0.17$ \cite{2dFGRS:2004fs}, GAMA at $z\simeq0.18$ and $0.38$ \cite{GAMA:2013nif}, Wiggle Z at $z\simeq0.22$, $0.41$ and $0.6$ \cite{WiggleZ:2011rj}, the SDSS LRG at $z\simeq0.25$ and $0.37$ \cite{SDSS-LRG:2011cs} and BOSS DR12 at $z\simeq0.32$ and $0.57$ \cite{BOSS-DR12:2015sqa}. 

In the left-hand panel, we also include the best-fit $\beta$ values for all gravity models extracted from the simulations at $z=0.5$ using the linear Kaiser method, with $1\sigma$ error bars. The extracted estimates consistently underestimate the $\beta$ value for all gravity models compared with the theoretical prediction. This is due non-linearities that produce smaller values of $Q(s)$. In general, the linear Kaiser model fails to model RSD in configuration space even on large scales $(s_{\rm min} = 42.5\Mpch)$. In the right-hand panel of Fig.~\ref{fig:fits_beta}, the constraints on $\beta$ using the gRPT model are shown, where we observe a good agreement between the best-fit values and the theoretical predictions for all models, and the additional details of the RSD model corrects the inaccuracy of the Kaiser model on linear scales.

In the lower subpanels of Fig.~\ref{fig:fits_beta}, we plot the relative differences of the MG models with respect to GR. Despite the different best-fit $\beta$ values in the linear and non-linear methods, the relative differences between models predicted by them are almost the same. However, the difference of the F6 and F5 models with respect to GR is $\lesssim 1\%$, making these three models statistically indistinguishable from each other, while F4 keeps a difference of $\sim 5\%$ with respect to GR; this reflects the fact that the growth of structure in $f(R)$ gravity is not enhanced on large scales, beyond the Compton wavelength of the scalaron field. On the other hand, N5 and N1 models show a difference of respectively $\sim 2.5\%$ and $\sim 12\%$ with respect to GR, due to the long-range nature of the fifth force. 

We conclude that with current data it is difficult to distinguish between the various gravity models simply by using constraints on $\beta$. Future data from surveys like DESI will likely improve on this situation, though tests of models like F5, N5 and F6 may still remain a challenge \citep[see e.g.][]{Hernandez-Aguayo:2020oiw}. \revision{As an example, we can rescale our error size  by the square root of the inverse volume ratio, $\sigma^\prime=\sigma\sqrt{V_{\rm mock}/V_{\rm DESI}}$, where $V_{\rm DESI} = 20~({\rm Gpc}/h)^3$ and $V_{\rm mock} = (1.024~{\rm Gpc}/h)^3$. This would lead to a new error size of $\sigma^\prime=0.23\sigma$ which would help to distinguish between gravity models on large scales. As we discuss next, the power of RSD to distinguish between gravity models can also be improved by the inclusion of smaller scale information.}

\subsubsection{Small-scale redshift-space galaxy clustering}
\label{subsect:subhalo}

We have seen in the previous subsection that large-scale RSD can be a useful tool to test gravity theories which strongly affect structure formation on large scales, such as nDGP, while for models such as $f(R)$ gravity, where gravity is modified on small nonlinear scales, the constraints are generally weaker. However, this conclusion should be read with the particular context of the analysis in mind. Neither the perturbative theoretical templates for RSD nor the numerical results from our HOD catalogs are accurate enough for the highly nonlinear regime, where the FoG effect due to the virial motions of small galaxies dominates the anisotropies in galaxy clustering and can potentially be affected by an enhanced gravity force. For this reason, it is 
\revision{important} to explore this regime in greater details using different techniques. 
\revision{In this subsection, we visit this topic by using two alternative approaches, subhalo abundance matching (SHAM) technique \cite{Conroy:2005aq,Rachel,moster} and hydrodynamical simulations of galaxy formation in modified gravity models \cite{Arnold:2019vpg,Hernandez-Aguayo:2020kgq} \newrevision{(cf.~\S~\ref{sect:SHAM_SHYBONE})}, to predict galaxy clustering, in particular small-scale RSD, in the $f(R)$ and nDGP models studied above. Part of this subsection consists of a review of the results presented in Ref.~\cite{He:2018oai}. We note that, in considering halo catalogs generated by HOD, SHAM and hydrodynamical simulations, our intention in this paper is not to explicitly undertake apple-to-apple comparisons of the various simulations, i.e., it is rather to enumerate and present the possible summary statistics that can be derived from upcoming galaxy clustering survey data to test gravity, than to conduct a rigorous quantitative comparison of different summary statistics under exactly the same controlled conditions. The latter will be left for future works.}

\begin{figure*}
\includegraphics[width=4.0in,height=4.0in]{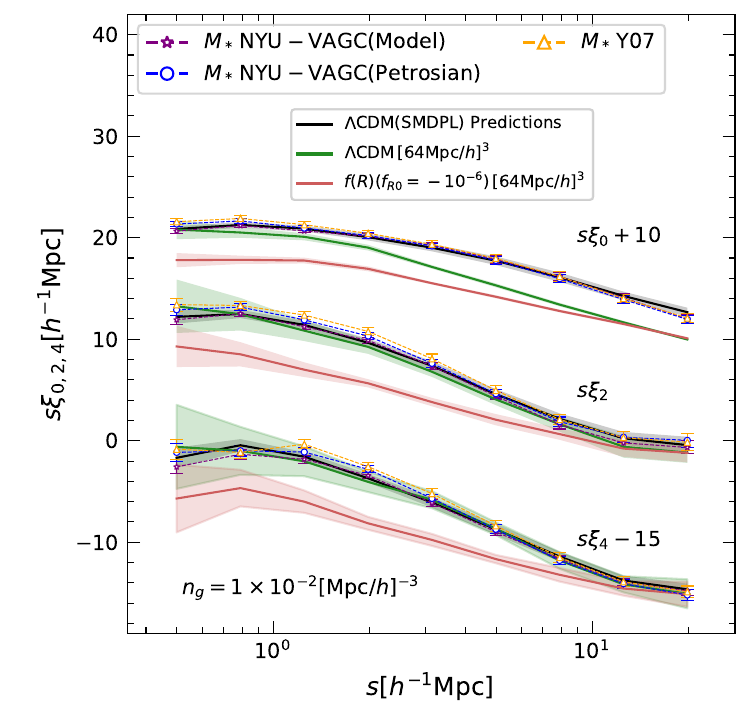}\caption{A comparison of RSD measurements using three different stellar mass models: a template-fit method adopted in the NYU catalog with the SDSS model magnitude (stars with dashed line), the same template-fit method but with the SDSS Petrosian magnitude (circles with dashed lines), and a single-color method (triangles with dashed lines). Different stellar mass models give rise to convergent results. The error-bars in the above plot are estimated using a jackknife re-sampling technique with $133$ realizations. Here the error-bars represent $3\sigma$ statistical error. The predicted RSD multipoles (monopole $\xi_0$, quadrupole $\xi_2$ and hexadecapole $\xi_4$) in $\Lambda$CDM from SHAM mock are in good agreement with observations. Note that the SHAM mock is based on the SMDPL simulation~\cite{Klypin:2014kpa} and has the same geometry as the real data. The plot is reproduced from~\cite{He:2018oai}. The green and red lines show the predicted RSD multipoles for $\Lambda$CDM and $f(R)$ gravity using a suite of simulations with $64[\Mpch]^3$ box size. The shaded regions give the $1\sigma$ uncertainty around the expectation value, derived from $500$ realizations with line-of-sight along different directions of the simulation box. Despite the deficit of power in the monopole due to the small box size of the simulations, the higher order multipoles, such as quadrupole $\xi_2$ and hexadecapole $\xi_4$, show significant differences, indicating that even a model like F6 can be testable given the accuracy of the clustering measurement of DESI.}\label{SDSSdata}
\end{figure*}

Figure~\ref{SDSSdata} shows a comparison between the observed redshift space galaxy clustering and the predictions in $\Lambda$CDM for galaxy samples with 
$n_g=1\times 10^{-2}[h^{-1}{\rm Mpc}]^{-3}$. The SHAM predictions in $\Lambda$CDM agree very well with the observation especially for the Fingers-of-God on small scales. To demonstrate the robustness of SHAM in constraining MG models, we also present its predictions for the F6 model in Fig.~\ref{SDSSdata}. The predictions are based on set of high-resolution $f(R)$ and GR simulations presented in Ref.~\cite{Shi:2015aya}, using the ``effective halo'' technique developed in Refs.~\cite{He:2015oua,He:2015bua,He:2016uxb}. As we can observe from this figure, the $f(R)$ result deviates substantially from both the $\Lambda$CDM predictions and observational data for all three RSD multipoles. 

\begin{figure*}
\includegraphics[width=5.0in,height=3.5in]{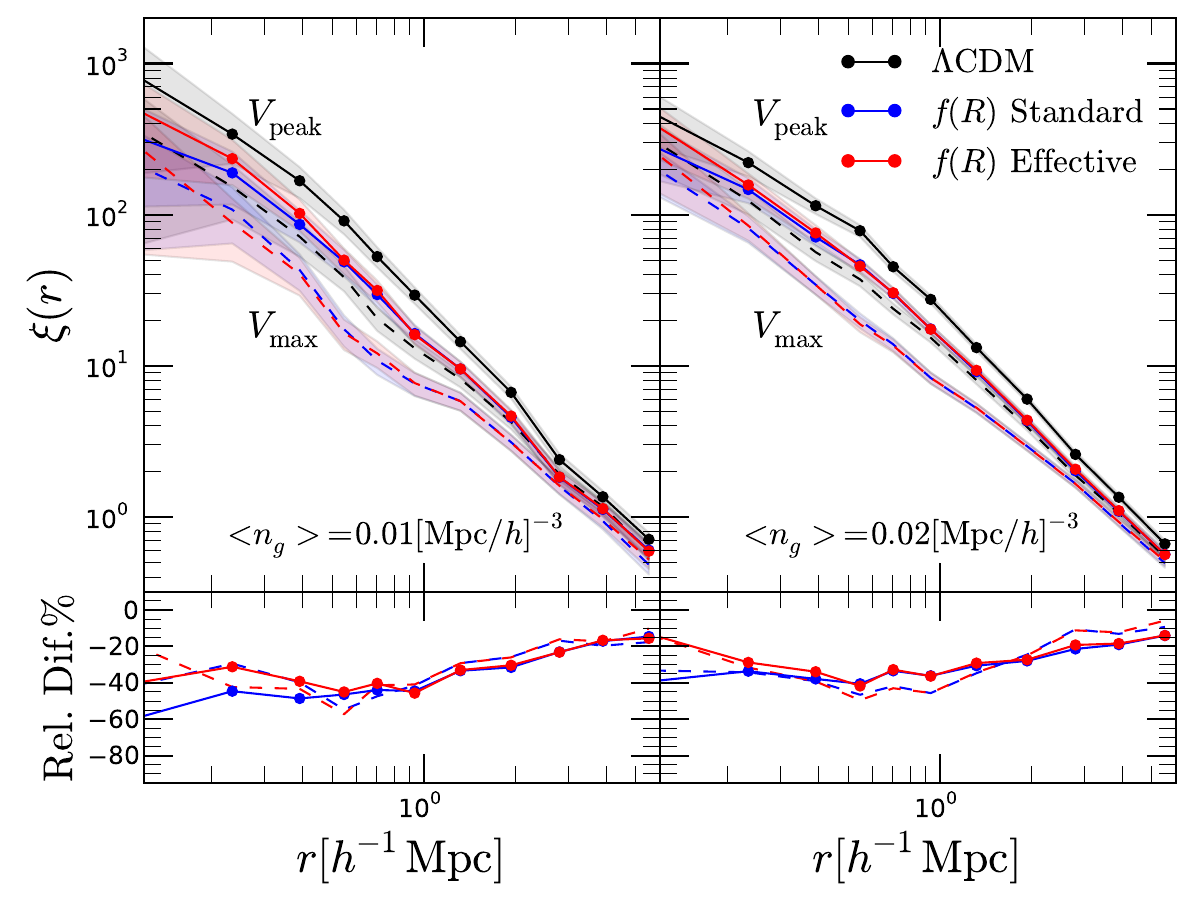}\caption{The predicted three-dimensional two-point galaxy correlation functions from the SHAM model (upper panels). The shaded regions represent the $1\sigma$ Poisson errors. The lower panels show the fractional differences between the $f(R)$ model and $\Lambda$CDM. The left panels show the results for a galaxy density $\langle n_g\rangle=0.01[\Mpch]^{-3}$ and the right panels are for $\langle n_g\rangle=0.02[\Mpch]^{-3}$ . For comparison, the dashed lines show the results obtained using the current maximum circular velocity $v_{\rm max}$. The plot is reproduced from Ref.~\cite{He:2016uxb}} \label{frsham_real}
\end{figure*}

To further understand the substantial differences between F6 and GR in Fig.~\ref{SDSSdata}, we have plotted in Fig.~\ref{frsham_real} the real space 2PCFs of the SHAM mock galaxies, for two sample number densities: $n_g=1\times10^{-2}[\Mpch]^{-3}$ (left) and $2\times10^{-2}[\Mpch]^{-3}$ (right). In both cases there is a $20\sim40\%$ difference at $r\lesssim6\Mpch$, which partially explains the behavior of Fig.~\ref{SDSSdata} (note the difference in the galaxy velocities between the two models also contributes to the model difference in redshift space in Fig.~\ref{SDSSdata}). The clustering is actually weaker in F6, which is because on small scales the enhanced gravitational force makes structures grow faster, which means that lower initial density peaks, which would not have become massive enough to host halos for the chosen $n_g$ cutoff in GR, have indeed become galaxy bearing. These halos are intrinsically less clustered than the ones which form from higher initial density peaks. This is clearly a feature that can only be probed in the nonlinear regime due to the short-range nature of the fifth force in this model. 

\revision{While the above result seems to suggest that small-scale RSD can be a promising tool to constrain gravity models, the SHAM prediction for F6 is based on an assumption on the relationship between the peak circular velocity and the effective mass of a halo. In obtaining Figs.~\ref{SDSSdata} and \ref{frsham_real}, it is assumed that the circular velocity profile $v_{\rm circ}(r)$ is related to the effective mass $M_{\rm eff}$ by the usual $v_{\rm circ}^2=GM_{\rm eff}(<r)/r$. This relation is applicable for relaxed halos with constant (effective) masses and density profiles. However, in $f(R)$ gravity, the chameleon screening efficiency becomes weaker with time\footnote{\revision{This is because as time progresses, the background value of the scalar field increases, so that screening a halo of the same mass becomes harder.}}, and so the effective mass of a halo, $M_{\rm eff}$, will change from the true halo mass $M$ in the fully-screened regime to $4M/3$ when the halo becomes unscreened. Such a change could happen rapidly, and so the above relationship does not always hold: when $M_{\rm eff}$ has changed, it will take time for $v_{\rm circ}$ to adapt, which means that using $M_{\rm eff}$ to estimate $v_{\rm circ}$ can lead to overestimate of the modified gravity effect. Note that in this way the SHAM approach will have a different halo population in F6 from in GR, because even for halos of the same mass, their effective masses $M_{\rm eff}$ can be different depending on the environments (a consequence of the chameleon screening). This can naturally lead to different clustering predictions between F6 and F5. Furthermore, in low-density regions where chameleon screening is inefficient, small halos are likely to be unscreened and have higher effective masses than their GR counterparts, but this does not necessarily translate into higher stellar masses, since there is less baryonic matter in these regions to start with. All these complicated effects cannot be expected to be fully accounted for in the simple SHAM approach.}

\begin{figure*}[!tb]
\begin{center}
{
\includegraphics[width=0.49\textwidth]{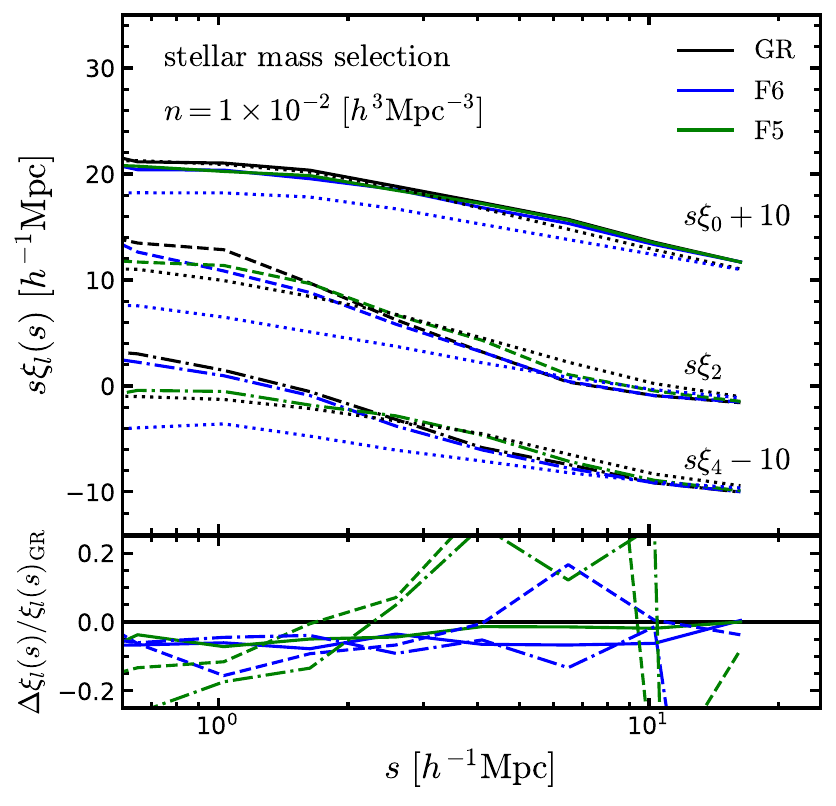}
\includegraphics[width=0.49\textwidth]{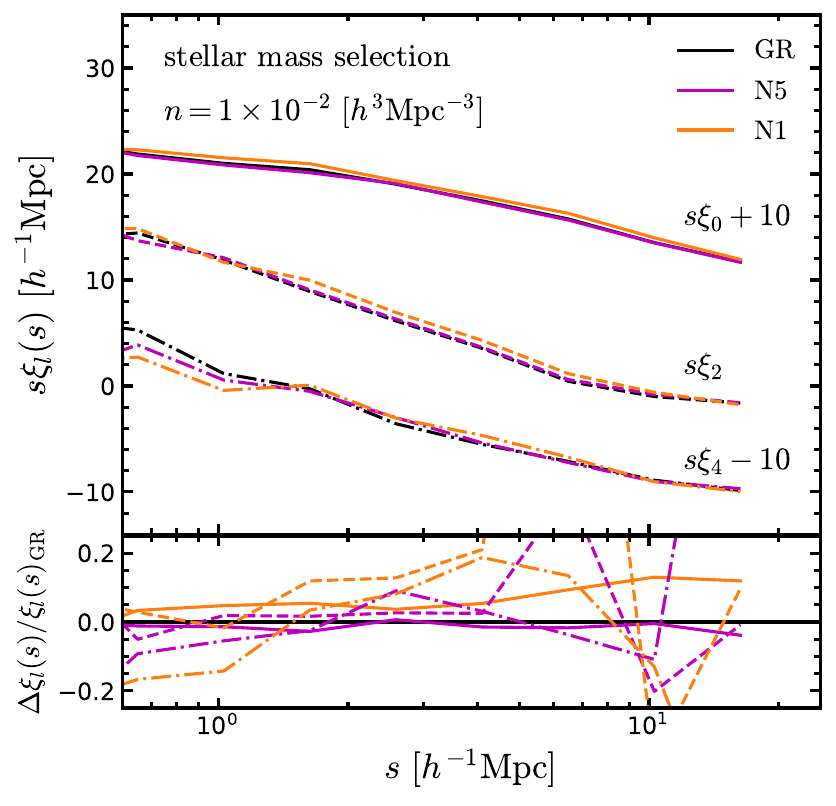}
}
\caption{\textcolor{black}{Small-scale redshift space galaxy clustering measured from the full-hydrodynamical {\sc shybone} simulations \citep{Arnold:2019vpg,Hernandez-Aguayo:2020kgq} for the F6 and F5 models ({\it left panel}), and the N5 and N1 models ({\it right panel}). In the {\it left panel} we additionally show the measurements from the SHAM catalog (dotted lines) for comparison. The lower subpanels show the relative difference of the measurements from the MG full-physics simulations with respect to their GR counterpart.}}
\label{fig:hydro_RSD}
\end{center}
\end{figure*}

\revision{Therefore, as a cross comparison of the different approaches to model small-scale RSD, we have measured the small-scale redshift space clustering of the {\sc shybone} simulations \cite{Arnold:2019vpg,Hernandez-Aguayo:2020kgq}, which are the first realistic full-physics hydrodynamical galaxy formation simulations of MG ($f(R)$ gravity and nDGP) models, employing the IllustrisTNG subgrid physics model \cite{Pillepich:2017jle} \newrevision{(see \S~\ref{sect:SHAM_SHYBONE} and \cite{Arnold:2019vpg,Hernandez-Aguayo:2020kgq} for more details)}. We select galaxies according to their stellar masses, to match the number density of the SHAM catalogs ($n_g=1\times10^{-2}[\Mpch]^{-3}$). Figure \ref{fig:hydro_RSD} shows the measurements of the redshift space clustering multipoles of the $f(R)$ gravity (left panel) and nDGP (right panel) models, and we have also included the SHAM RSD measurements in the left panel of Fig.~\ref{fig:hydro_RSD} for comparison.} 

\revision{Of the three multipoles, we can see that the monopole (solid lines) is least affected by MG effects, we find a difference of around $5\%$ on most scales for the F6, F5 and N1 models; the N5 model is almost indistinguishable from GR. On the other hand, MG effects on the quadrupole (dashed lines) and the hexadecapole (dash-dotted lines) can produce difference of {$\simeq$}$20\%$--$30\%$ on scales $s=2$--$8\Mpch$ for the F5 and N1 models. This trend is not surprising, given that the monopole is largely determined by the real-space galaxy correlation function, while the quadrupole and hexadecapole depend more sensitively on the pairwise velocity fields, and a modified gravity force is expected to affect the velocity field first and more, since it is the first integral of the acceleration field while the position is the second integral.}

\revision{Interestingly, the predictions by the galaxy formation simulations differ significantly from those by SHAM (for F6), suggesting that the complications mentioned above can bear a non-negligible systematic impact on the SHAM predictions. \newrevision{This observation is consistent with the right panel of Fig.~\ref{fig:sham_shybone_plot}, which indicates the differences between these two approaches already appear in the real-space clustering, is not completely down to different mappings from real to redshift space.} Of course, even hydrodynamical simulations are not immune of systematic effects, e.g., different subgrid physics models can give quantitatively different results. Nevertheless, the results of Fig.~\ref{fig:hydro_RSD} confirm that RSD on small scales ({$\lesssim${$10h^{-1}{\rm Mpc}$}}) can be strongly modified in models such as $f(R)$ gravity, for which the effect on larger scales is generally small, (Fig.~\ref{fig:2pcf_zspace}; right panel). In particular, because the quadrupole and hexadecapole are mostly sensitive to the velocity field (e.g., \cite{Cuesta-Lazaro:2020ihk}), we expect them to be less affected by the complications caused by baryonic effects (e.g., \cite{Hellwing:2016ucy}).}

Small-scale RSD can thus be a promising tool when using future galaxy surveys such as DESI, in particular its Bright Galaxy Survey (BGS), to constrain gravity models. The BGS can obtain redshifts of ``bright'' galaxies up to two magnitudes fainter than the limit of the Sloan Digital Sky Survey (SDSS) main galaxy redshift survey \cite{Strauss:2002dj}. In its current design, the BGS will observe
approximately 17 million galaxies over 14,000 deg$^2$, in two completeness
tiers: $r< 19.5$ galaxies with completeness of $\approx 90\%$ and
$19.5 < r < 20.0$ galaxies with completeness of $\approx 75\%$. The exceptionally high sampling density allows the best achievable measurements of RSD with unparalleled accuracy. 

\subsubsection{Redshift-space distortions around voids}
\label{subsect:rsdvoid}

Cosmic voids are regions in our Universe that are underdense in terms of tracer numbers and matter. 
Redshift space distortions around voids can be used to probe the growth rate of structures around these regions \citep{Cai2016,Hamaus2016}. The void-galaxy correlation function is distorted in redshift space because of the peculiar motions of galaxies. While such peculiar motions respond only to the Newtonian potential in GR, in MG models they can also be affected by the fifth force, causing the distortion patterns to change -- a diagnostic that can then be used to
distinguish the model from GR. Since the fifth force is expected to be unscreened in voids \citep{Hui2009,Clampitt2013,Lam2015,Cai2015}, the effect should be larger around voids. 

\begin{figure*}[!tb]
\scalebox{0.85}{
\hspace{-1.0cm}
\includegraphics{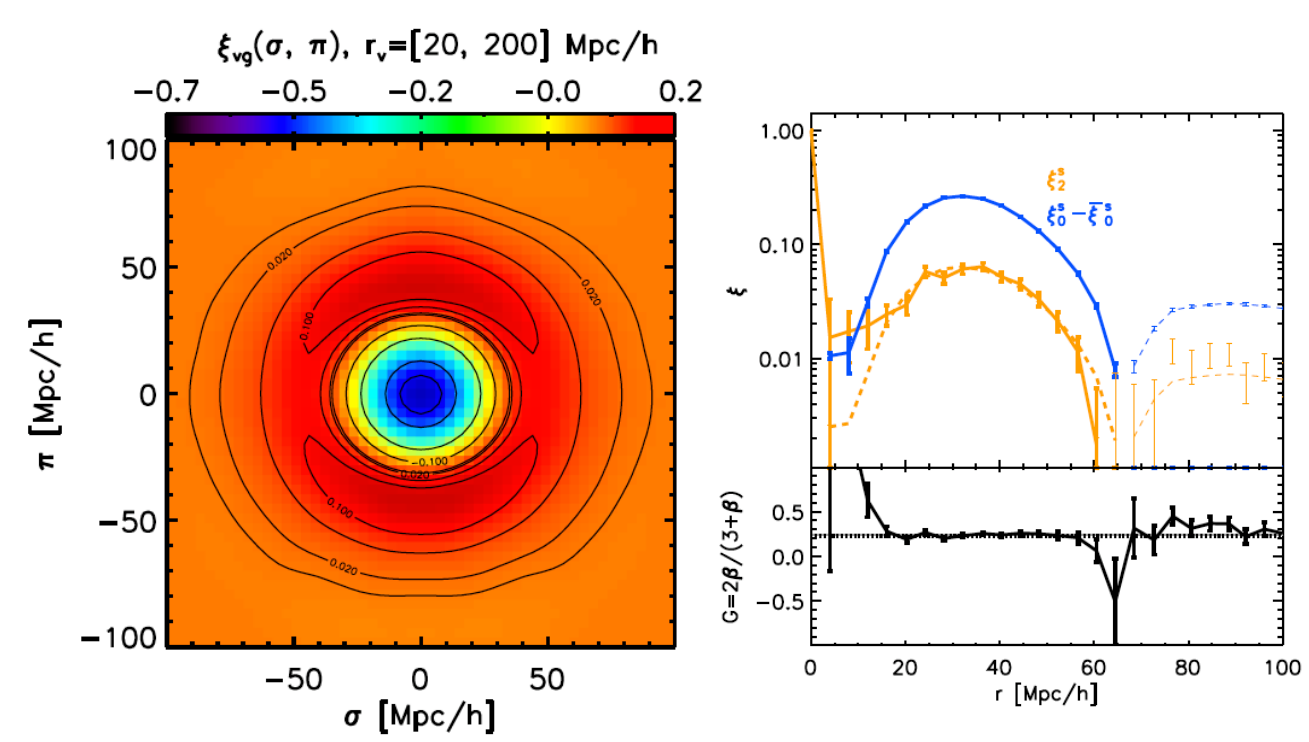}}
\vspace{-0.3cm}
\caption{Left: the void-galaxy correlation function in redshift space for the GR model for voids identified by {\sc zobov} with the HOD galaxies in 5 boxes of the $1024h^{-1}{\rm Mpc}$-aside simulation. Voids with effective radius $r_{\rm v}>20h^{-1}{\rm Mpc}$ are used.Right: the upper panel shows the monopole (blue) and quadrupole (orange) moments of the redshift-space correlation function. Thin lines indicate the absolute values of negative parts. The lower panel shows the ratio of the quadrupole versus monopole. The black curve is from taking all measurements of voids along three different major axes of the simulation box, with the error bars showing the error on the mean. The dotted line is the true answer for the growth rate parameter $G$ given by the cosmology and galaxy bias. The black shaded region shows the 1$\sigma$ range around the best-fit value.}
\label{VoidRSD1}
\end{figure*}

We conduct an analysis for the redshift-space distortions around voids using the redshift-space void-galaxy correlation function $\xi^s_{\rm vg}$, using the mock galaxy catalogs from the GR and modified gravity simulations described in previous section. Voids are identified in the redshift-space galaxy fields using the {\sc zobov} void finder, which makes use of Voronoi tessellation of the galaxy field \citep{Neyrinck2008} (details for the definitions of {\sc zobov} voids including void center and radius are described in Section~3.3 of \citep{Cautun:2017tkc}). We then measure the void-galaxy correlation function $\xi^s_{\rm vg}$ in redshift space. An example for the GR model is shown in the left panel of Fig.~\ref{VoidRSD1}. The extracted monopole $\xi^s_{0}$ and quadrupole $\xi^s_{2}$ moments are shown in the top of the right panel of Fig.~\ref{VoidRSD1}. All voids with radius $r_{\rm v}$ greater than $20h^{-1}{\rm Mpc}$ are used for the analysis.

In GR, Following the linear model of \citep{Cai2016}, the ratio between the quadrupole and monopole is a constant $G=2\beta/(3+\beta)$, where $\beta=f/b$ is the distortion parameter introduced above. We can estimate $\beta$ using the 
\revision{summary statistic}
\begin{equation}
\label{Eq:Estimator}
\tilde G(\beta) = \frac{\xi_{2}^s(r)}{\xi_{0}^s(r)-\frac{3}{r^3}\int_0^r \xi^s_0(r')r'^2dr'} = \frac{2\beta}{3+\beta}.
\end{equation}
The multipoles of the redshift-space correlation function can be obtained by
\begin{equation}
\xi^s_{\ell}(r)=\int_0^1\xi^s_{\rm vg}(r, \mu)(1+2\ell)P_{\ell}(\mu)d\mu,
\end{equation}
where $P_{0}(\mu)=1$ and $P_{2}(\mu)=\frac{1}{2}(3\mu^2-1)$, and $\mu=\cos\theta$ where $\theta$ is the {the angle between the line connecting a galaxy-void pair and} the LOS. In linear theory, the model has only one free parameter, $\beta$.

We follow the same procedure as in \citep{Cai2016} for constructing the covariance matrix and for the parameter fitting. The correlation functions from all the 5 boxes of simulation at $z=0.5$ are treated as independent and used for the fitting. We have also viewed the simulation box along three different major axes of the box to further increase the data size, although we do not expect the volume to increase by three times since they are not independent.

We appreciate that the growth rate in non-GR models may be scale dependent even in the linear regime. In this study, we do not explore this subtlety and simply recover the effective growth rate parameters $\beta_{\rm effective}$ for all models. We have also treated the linear bias as scale independent and taken the measurement from the galaxy 2PCF versus the dark matter correlation function between 20 to 50 $h^{-1}{\rm Mpc}$. The linear galaxy biases for all different models are very close to each other, $b\sim1.9$ at $z=0.5$. 

We find that the best-fit growth rate parameter for the GR model agrees with the true answer within the 1$\sigma$ range. However, with the same fitting procedure, we recover similar growth rates for F6 and N5, not distinguishable from GR. Both the F5 and N1 models may be distinguished from GR at the 2-3$\sigma$ level, with the F5 model having a slightly lower best-fit value of $\beta$ and the N1 model having a higher value of $\beta$ than GR, which is somewhat unexpected as the linear growth rate in both these two models should be higher than in the GR model with the same expansion history. This may have been complicated by non-linearity and scale-dependent galaxy bias being different among different models. To fully exploit the information from these measurements, more accurate models of RSD which work in the non-linear regime are needed, e.g., \cite{Nadathur2019, Paillas2021}.

We have also found that error budget for $\beta$ does not goes down with the square root of the volume, which suggests that viewing the simulation along three different axes does not give us more independent data. Having a higher number density of galaxies will help to resolve smaller voids and increase their number for a fixed volume. This will be guaranteed for the DESI BGS survey, where the number density of galaxies will be at least an order of magnitude higher than the current CMASS mocks. This may help to further reduce the errors and increase the distinguishing power between the MG models and GR. Void RSD is a relatively unexplored probe, and its potential in testing gravity will require further investigations in the future.

\revision{Another interesting avenue to explore with voids is to compare the distribution functions for the sizes of voids (VSD) among different models. In theory, this has been shown to be powerful for distinguishing gravity models with chameleon screening mechanism (see, e.g., \cite{Clampitt2013, Lam2015, Cai2015}). However, we do not find the VSDs found with our galaxy samples distinguishable among different models (see, e.g., \cite{Cautun:2017tkc} for a study of the VSD in different models that used HOD catalogues similar to the ones used in this paper). The main reason for this is that we have made sure that the clustering of our HOD galaxy samples are similar to each other, which has forced the VSDs to be similar, as VSDs are correlated with galaxy clustering. Also, we note that a fraction of the CMASS-like HOD galaxies are hosted by massive halos, which are likely to live in high-density regions where the fifth force in MG is screened: this may again indicate the benefit of using denser/fainter galaxy samples to find voids, which will be left for future studies.}

\subsubsection{Discussion}

Redshift-space distortions, through combining information from both large and small scales, can be a potentially powerful tool to discriminate between different gravity scenarios. Numerical, semi-analytic and analytic techniques may each be valuable to accurately predict RSD in gravity models that deviate from GR. In the $\Lambda$CDM model, various RSD models have been proposed. Substantial progress has been made based on perturbation theory such as the one given by \cite{Scoccimarro:1999Bispectrum, Scoccimarro:2004tg}, and the Taruya-Nishimichi-Saito model \cite{Taruya:2010mx} and the effective theory of large-scale structure \cite{Lewandowski:2015ziq} (see \cite{delaBella:2018fdb, Bose:2019psj, Bose:2019ywu} for comparisons of these models). These need to be combined with perturbative galaxy bias models such as the one studied in \citep{Chan:2012jj} (see \cite{Desjacques:2016bnm} for a review). 

\textcolor{black}{In configuration space, the redshift-space correlation function is often modelled by the Gaussian streaming model (GSM) \cite{Reid:2011ar} combined with Lagrangian perturbation theory \cite{Carlson:2012bu}. GSM models assume the pairwise velocity distribution of galaxies can be modelled as a Gaussian. However, N-body simulations have demonstrated that gravitational clustering introduces non-negligible skewness and kurtosis \cite{Scoccimarro:2004tg}, that play an important role in shaping the redshift-space galaxy clustering multipoles on small scales. Various models aimed at incorporating these properties have been proposed in recent years, e.g.,  \cite{Bianchi:2014kba,Bianchi:2016qen,Uhlemann:2015hqa,Kuruvilla:2017kev,Cuesta-Lazaro:2020ihk}, which can improve the accuracy of redshift-space clustering predictions on smaller scales (e.g., $\lesssim20h^{-1}$Mpc) than the Gaussian case, but their accuracy is often limited by the ability of perturbation theory to reproduce the pairwise velocity moments accurately. In the near future, this limitation might be overcome by calibrating non-Gaussian models to N-body simulations.}

\textcolor{black}{More generally, N-body simulations are being used extensively not only to test the accuracy of perturbation theory approaches, but also to directly constrain gravity by means of either hybrid techniques, such as the one presented in \cite{Song:2018afp} where the redshift-space power spectrum is computed on large scales using perturbation theory and a template approach is taken for the small scales, or by taking simulations a step further to create emulators for clustering 
\revision{summary statistics} \cite{Zhai:2018plk,Kobayashi:2020zsw}. The main challenges for emulator approaches are their expensive computational requirements, and their reliability on accurate models of the galaxy-halo connection to produce unbiased constrains of gravity from data. }

The advantage of the perturbation-theory-based RSD models is that they can be extended to different gravity models. This has been done for several MG models including $f(R)$ gravity and nDGP models \cite{Taruya:2013quf, Taruya:2014faa, Bose:2016qun, Bose:2017dtl, Bose:2018orj, Valogiannis:2019nfz}. This enables one to directly constrain modified gravity parameters such as $|f_{R0}|$ by taking into account consistently the scale dependence of the growth rate and non-linear interactions due to screening \cite{Song:2015oza}. The disadvantage is that perturbative approaches are not able to model the RSD effect accurately on small scales and a careful analysis is required to determine the scales that can be used using galaxy mocks (as we shall discuss in the following subsection). On the other hand, the emulator approaches can exploit the information on much smaller scales but they require a large number of high-resolution N-body simulations in a given gravity model, {although {\sc cola} approaches have brought some alleviation to such a requirement}  \cite{Winther:2019mus,Ramachandra:2020lue}. It is important to develop more efficient and optimal RSD models to distinguish different gravity models for surveys like DESI.

\subsection{Density-dependent environmental effects on galaxy clustering}
\label{subsect:marked}

We have seen above that many MG models have environmentally-dependent behaviors, such as stronger deviation from GR in under dense regions. The mock galaxy catalogs used for this paper have been produced using HOD parameters tuned for galaxies from all environments and so, while the different gravity models match each other in their overall projected 2PCFs, it is natural to ask whether stronger differences can be found in subsets of galaxies from specific environments. In this section, we investigate this possibility through a variety of statistics in contrasting environments. We consider statistics that utilize density transformations, or ``marked tracers'', to test modified gravity with simulated data.

A general marked correlation function can be defined as \citep{Sheth:2005aj}:
\begin{equation}\label{eq:Mr}
{\mathcal{M}(r)} \equiv \frac{1}{n(r)\bar{m}^2} \sum_{ij} m_i m_j = \frac{1 + W(r)}{1 + \xi(r)}\,,
\end{equation}
where the sum is over all tracer pairs with a given separation, $r$, $n(r)$ is the number of such pairs, $m_i$ the mark for the $i$th tracer and $\bar{m}$ the mean mark for the entire sample. In the second equality, $\xi(r)$ denotes the standard two-point correlation function and $W(r)$ is the `weighted' correlation function at a separation $r$. 

In the following discussion we consider 
\revision{summary statistics} based on number counts of halos and galaxies in \ref{sub:density} and the gravitational potential that accentuate under-dense regions in \ref{sub:potential}. In addition to the 
\revision{summary statistics} derived from the simulations,  we also discuss how marked correlation functions can be predicted from analytical estimates in section \ref{sub:analyticalmCF}. Finally we consider correlation functions that highlight over-dense regions, in \ref{sub:rho-wp}.

\begin{figure*}[!tb]
\begin{center}
{\includegraphics[width=0.49\textwidth]{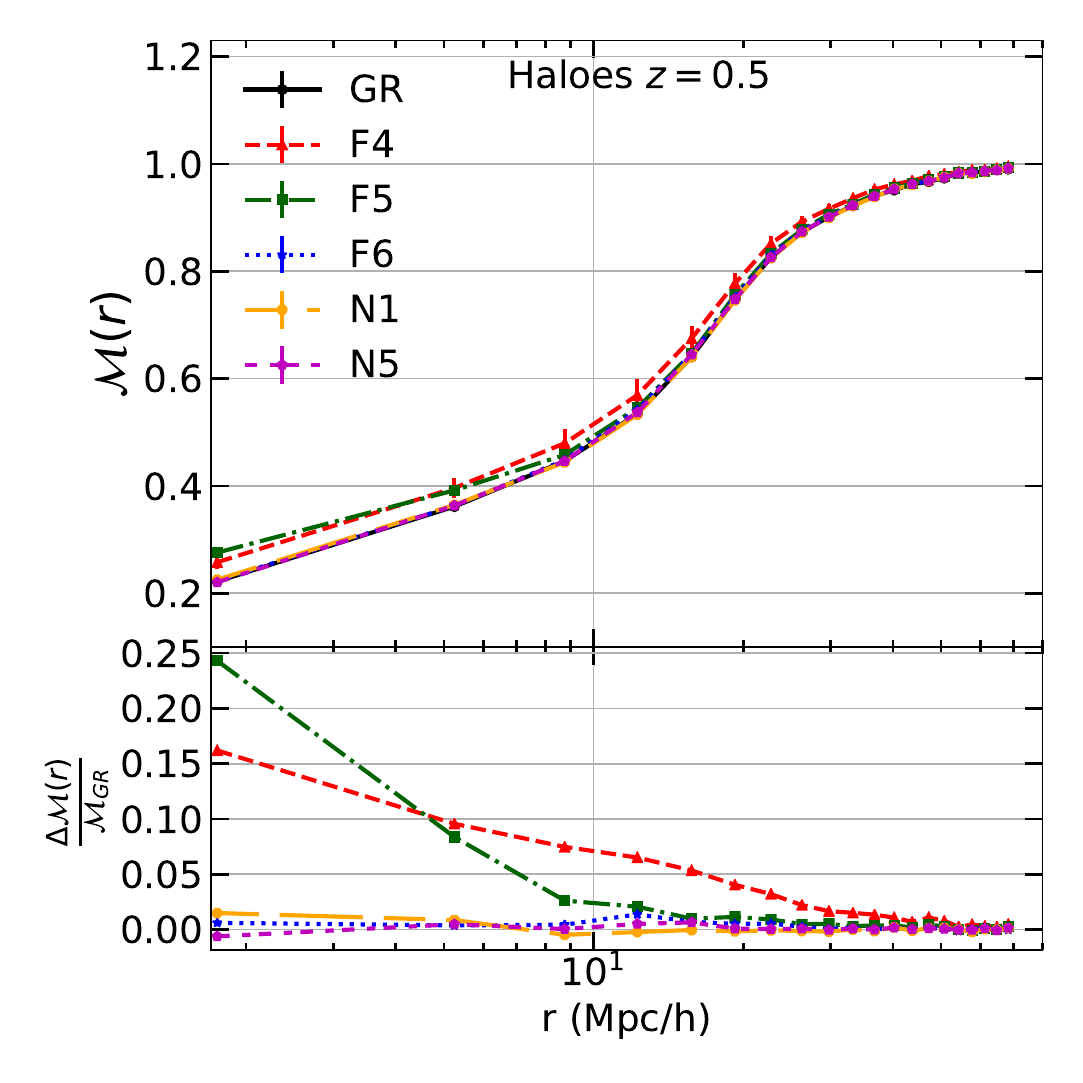}
\includegraphics[width=0.49\textwidth]{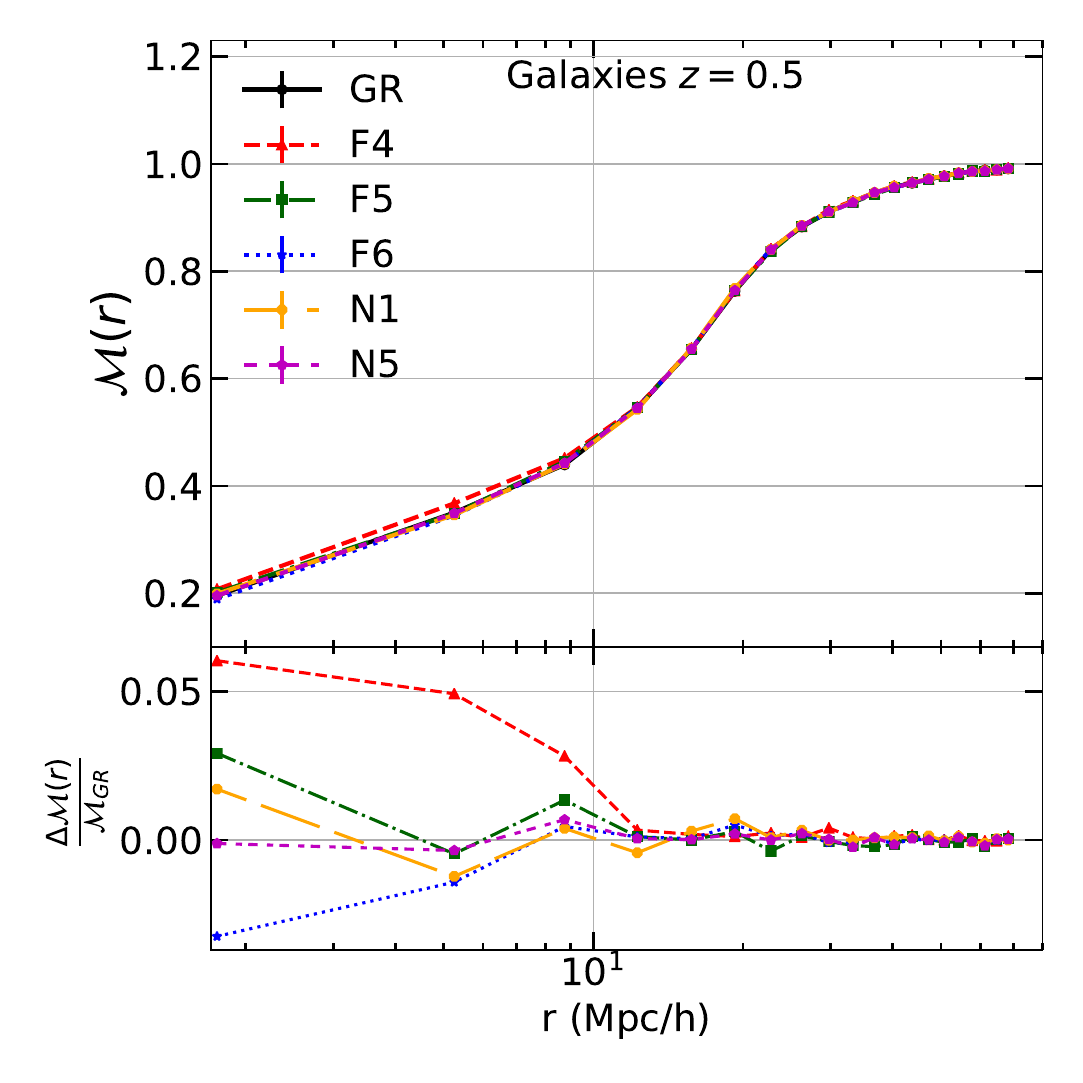}
}
\caption{(Color Online) The {tracer}-number-density marked correlation statistic for {\it halos} (left) and {\it galaxies} (right) with $\rho_{*}=4,p=10$, for GR (black), $f(R)$ with $|f_{R0}|=10^{-4}$ (red), $|f_{R0}|=10^{-5}$ (green), $|f_{R0}|=10^{-6}$ (blue) and nDGP with $r_c H_0=1$ (orange), $r_c H_0=5$ (magenta). The curves correspond to the averages over the 5 realizations of HOD catalogs, while the bottom panels are the fractional deviation ${\Delta \mathcal{M}}/{\mathcal{M}}={\mathcal{M}_{\rm MG}}/{\mathcal{M}_{\rm GR}}-1$ for each MG model. The error bars correspond to the standard deviations over the 5 boxes. \revision{The results shown are all in real space.}}
\label{fig:marked1}
\end{center}
\end{figure*}

\begin{figure*}[!tb]
\begin{center}
{
\includegraphics[width=0.49\textwidth]{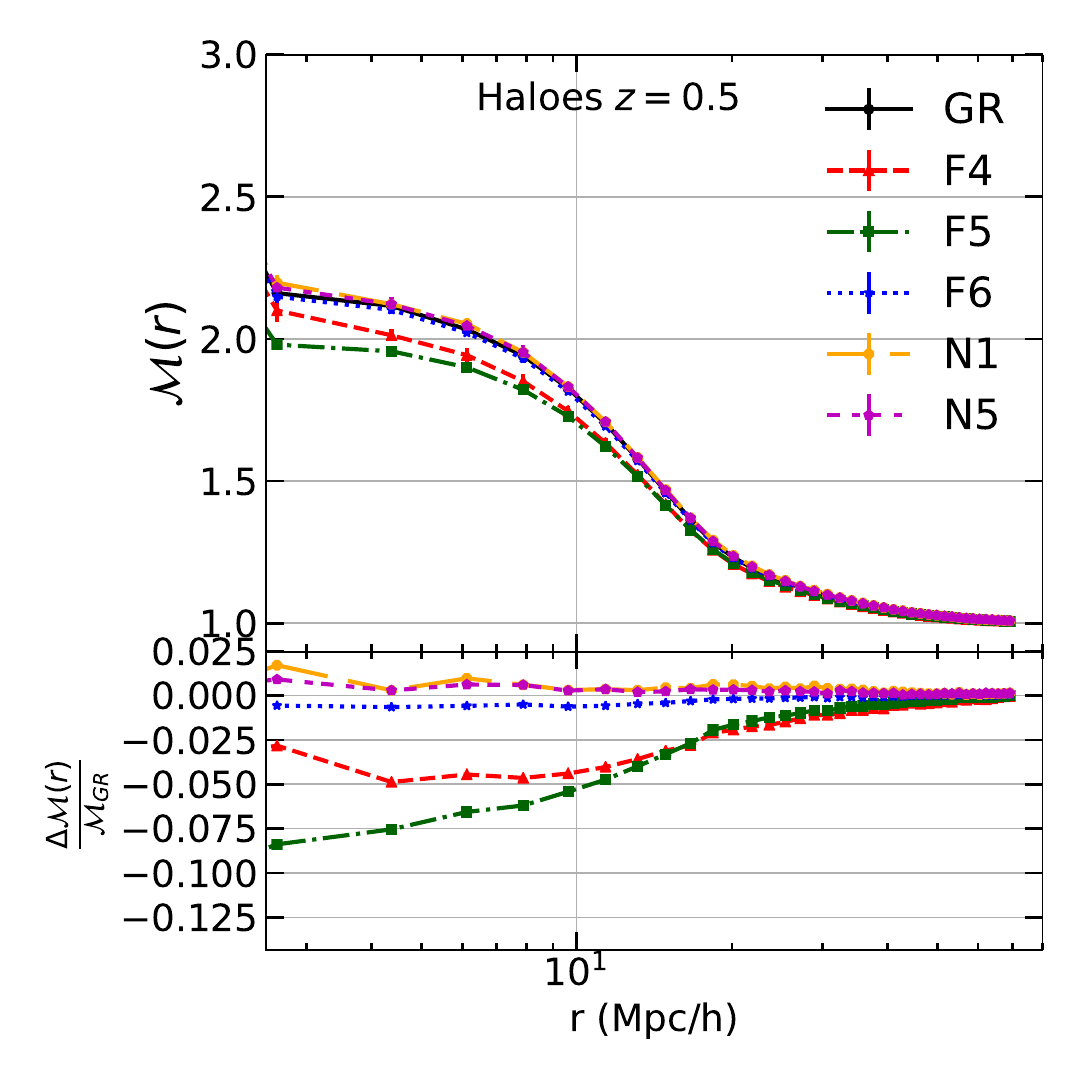}
\includegraphics[width=0.49\textwidth]{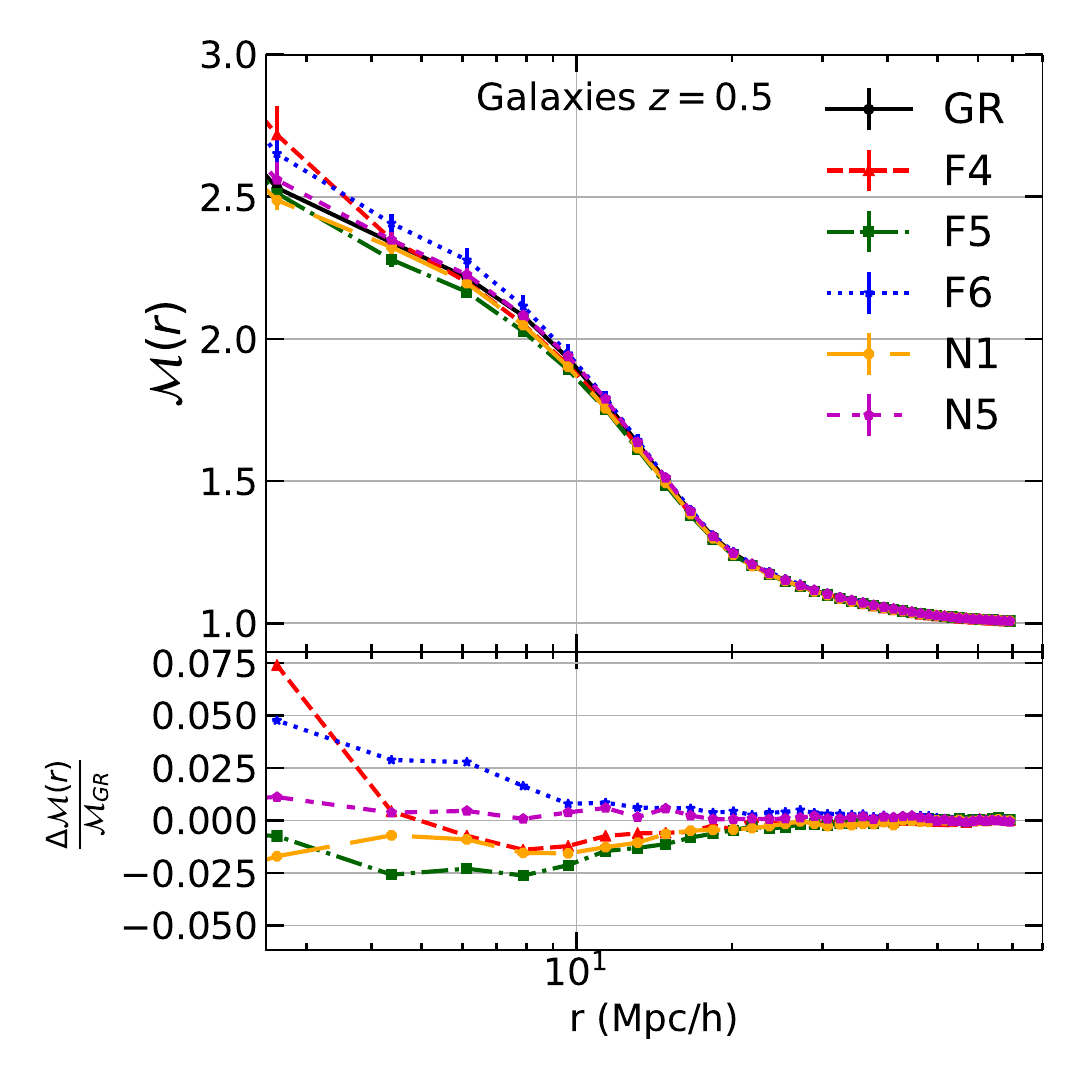}
}
\caption{(Color Online) Comparison of the \textcolor{black}{tracer}-number-density marked correlation statistic for {\it dark matter halos} (left) and {\it galaxies} (right) for the case of up-weighting high density regions ($\rho_{*}=1,p=-1$), for GR (black), $f(R)$ with $|f_{R0}|=10^{-4}$ (red), $|f_{R0}|=10^{-5}$ (green), $|f_{R0}|=10^{-6}$ (blue) and nDGP with $r_c H_0=1$ (orange), $r_c H_0=5$ (magenta). The curves correspond to the averages over the 5 realizations. The bottom panels show the fractional deviation ${\Delta \mathcal{M}}/{\mathcal{M}}={\mathcal{M}_{\rm MG}}/{\mathcal{M}_{\rm GR}}-1$ for each MG model. The error bars correspond to the standard deviations over the 5 boxes.}
\label{fig:mCF_negative_p}
\end{center}
\end{figure*}

\begin{figure*}[!tb]
\begin{center}
{
\includegraphics[width=0.49\textwidth]{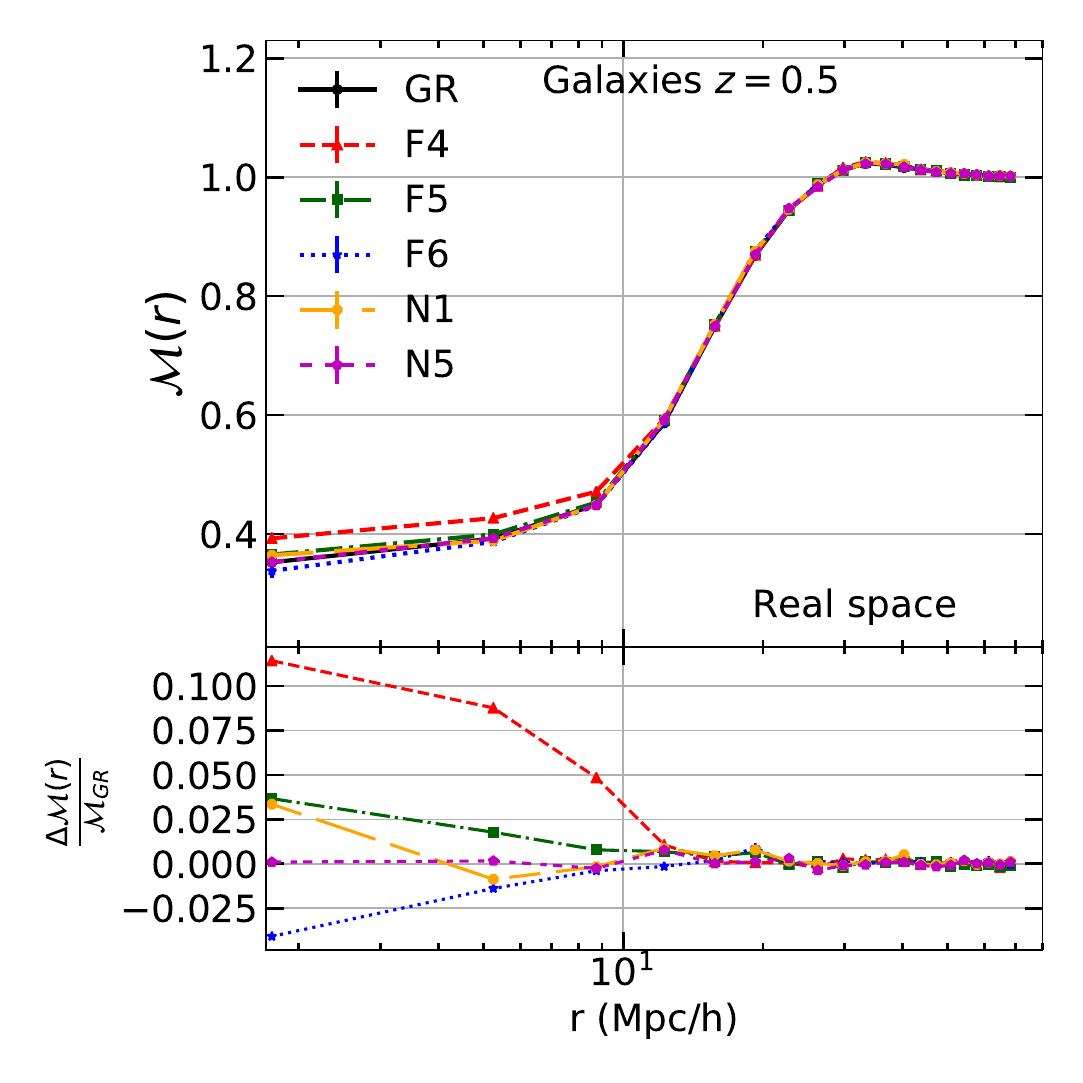}
\includegraphics[width=0.49\textwidth]{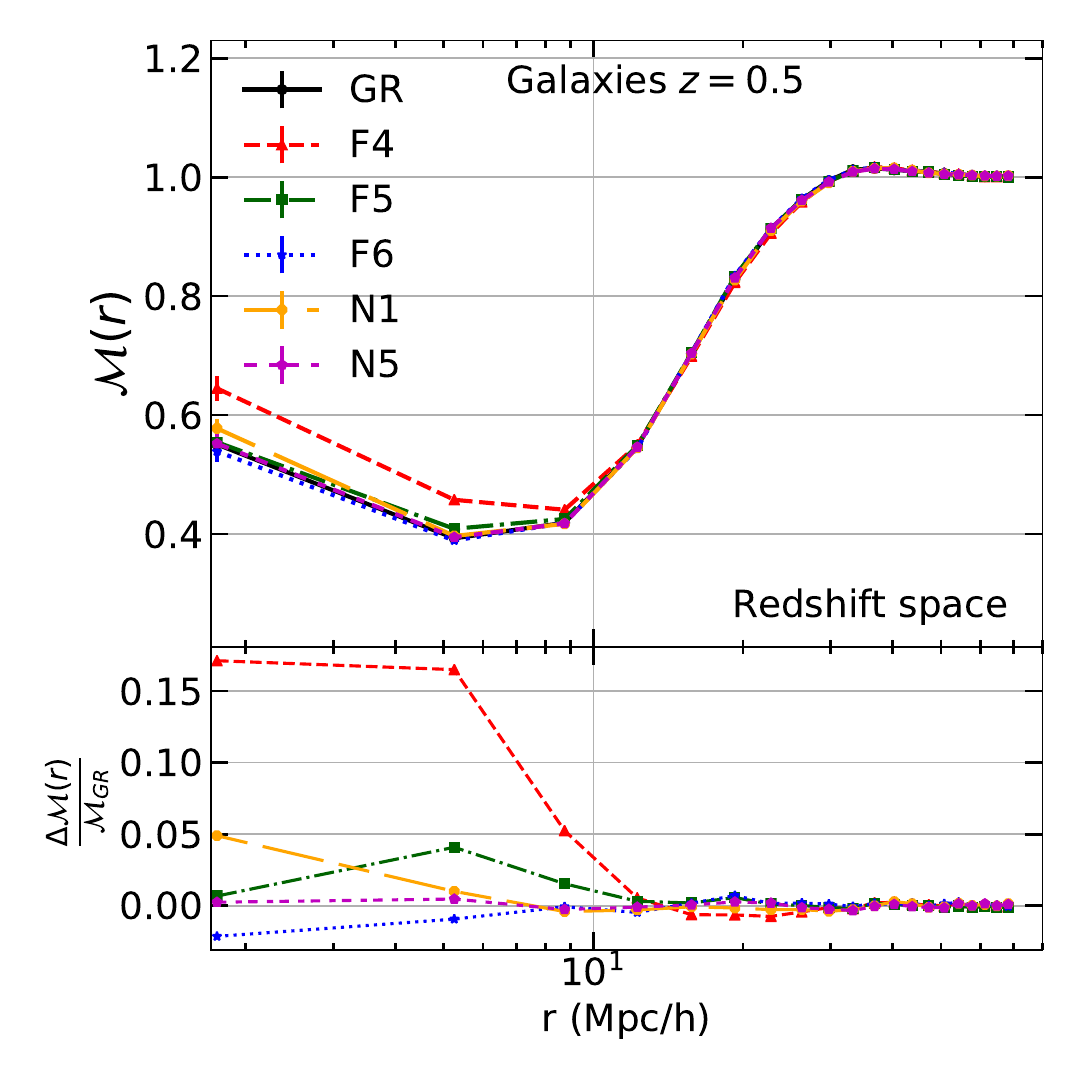}
 }
\caption{(Color Online) Comparison of the galaxy-number-density marked correlation statistic for {\it galaxies} with $\rho_{*}=4,p=10$, for GR (black), $f(R)$ with $|f_{R0}|=10^{-4}$ (red), $|f_{R0}|=10^{-5}$ (green), $|f_{R0}|=10^{-6}$ (blue) and nDGP with $r_c H_0=1$ (orange), $r_c H_0=5$ (magenta) when using SPH spline kernel interpolation for density estimate in real space (left) and redshift space (right). The curves correspond to the averages over the 5 realizations (in redshift space we consider 3 directions of each realization as independent samples), while the bottom panels show the fractional deviation ${\Delta \mathcal{M}}/{\mathcal{M}}={\mathcal{M}_{\rm MG}}/{\mathcal{M}_{\rm GR}}-1$ for each MG model. The error bars correspond to the standard deviations over the 5 boxes.}
\label{fig:mCFg_SPH}
\end{center}
\end{figure*}

\subsubsection{Marks defined using tracer number densities}
\label{sub:density}

Screening in modified gravity models, which occurs in regions of high density, has led to the consideration of density transformations that up-weight lower density, unscreened, regimes as a mechanism to enhance the modified gravity signatures. Such transformations include: the logarithmic transform of the density field \citep{2009ApJ...698L..90N,Wang:2011fj,2012PhRvL.108g1301C}, for which the transformed field becomes more Gaussian, facilitating easier information extraction from the 2-point function; a clipped density field \citep{2011PhRvL.107A1301S,Simpson:2013nja,Lombriser:2015axa} in which density peaks are all allocated a common value $\delta_0$,
\begin{equation}\label{clipstr}
\delta'=\delta_c=  \left\{\def\arraystretch{1.2} 
  \begin{array}{@{}c@{\quad}l@{}}
   \delta  & \text{if $ \delta < \delta_0$ }\\
    \delta_0 & \text{if $\delta>\delta_0.$}\\
  \end{array}\right.
\end{equation}
and more general ``marked" density transformations that up-weight low densities. An example of such a mark is the one proposed in \cite{White:2016yhs}, 
\begin{equation}\label{markdel}
\delta'=m(\delta)=\left(\frac{\rho_*+1}{\rho_*+\rho}\right)^p=\left(\frac{\rho_*+1}{\rho_*+(\delta+1)}\right)^{p}, 
\end{equation}
where $\rho_*$ and $p$ are free parameters and $\rho$ the dark matter mass or tracer number density field in a grid cell, in units of the mean density $\bar{\rho}$. Marked statistics have been previously explored in the context of MG in \citep{Valogiannis:2017yxm}, using CDM simulations produced in \cite{Valogiannis:2016ane}, and in \cite{Armijo:2018urs,Hernandez-Aguayo:2018yrp}. Tested on the $f(R)$ and symmetron \cite{Olive:2007aj,PhysRevLett.104.231301} MG models, the density-marked statistic was found to boost the signal-to-noise ratio encoded in the two-point statistics, providing thus additional discriminatory power with respect to $\Lambda$CDM.  

To calculate the marked correlation function for our models, we measure the galaxy number density using the nearest-grid-point scheme, by dividing the simulation box into cells of the same size, and then counting the number of galaxies inside each cell and using this to assign a density to the cell. Hence, we can compute the overdensity, $\delta$, as
\begin{equation}\label{eq:drho}
1 + \delta \equiv \frac{n_g}{\bar{n}_g}\,, 
\end{equation} 
where $n_g$ is the number of galaxies in each cell and $\bar{n}_g$ is the mean number of galaxies in cells of a given size over the simulation volume. To compute the galaxy density, we have used $60^3$ cells of size $\sim 17~h^{-1}$ Mpc. \newrevision{We have checked that the relative model differences are stable against the number of cells, provided that we use at least $30^3$--$40^3$ cells (see Appendix~\ref{App:A} for details).}

In the left-hand panel of Fig.~\ref{fig:marked1}, we show the marked correlation statistic Eq.~(\ref{eq:Mr}) for halos of mass $M_{200c}>10^{13}h^{-1} M_\odot$, evaluated using galaxy-number-weighted density estimates in Eq.~(\ref{markdel}), with $\rho_{*}=4,p=10$, for GR and all the modified gravity models considered. In addition, the right-hand panel of Fig.~\ref{fig:marked1} shows the same marked statistic, for the same models using the same parameters for the mark $(\rho_*,p)$, but calculated using the simulated galaxy catalogs as described in \S~\ref{sec:simulations}. The correlation function calculations were performed using the publicly available code Super W Of Theta ({\sc swot}) \citep{2012A&A...542A...5C}.

For the halo marked correlation functions, the fractional deviations from GR are more pronounced for the F4 and F5 models, as expected, reaching $\simeq20\%$ in the lowest $r$ bin. {The deviations in the F6, N5 and N1 cases are at the (sub)percent level at all scales.} The predicted differences are overall less pronounced in the marked correlations of galaxies, with the various models predicting deviations not larger than $\simeq5\%$ {even} on the smallest scales. This is {possibly} a consequence of the HOD parameters used to populate galaxies in each halo being tuned to match the unmarked correlation functions. An explanation of this behavior is presented in \S~\ref{sub:m_discussion}.

For completeness, we also studied the case with negative $p$. Figure \ref{fig:mCF_negative_p} shows the marked correlation using the galaxy-number-weighted density estimate for halos and galaxies with $\rho_*=1, p=-1$. For dark matter halos, the deviations from GR are much smaller compared with the case with $\rho_*=4, p=10$, confirming our expectation that up-weighting low-density regions enhances MG signals. On the other hand, the galaxy marked correlation function shows similar small deviations to the case with $\rho_*=4, p=10$, with no clear trend among the MG models in their relative difference to GR. {This, together with Fig.~\ref{fig:marked1}, implies that if we 
use the galaxy number density with the mark of Eq.~\eqref{markdel}, 
it is difficult to 
substantially enhance the model differences, at least for low-density galaxy samples as used here.} This difficulty may be alleviated if one uses external information such as the gravitational potential to define marks as discussed in \S~\ref{sub:potential}, {or if the tracer number density is higher (which will be left for future investigations when higher-resolution simulations become available), or if other definitions of marks are used (which is beyond the scope of this work). However, even if some other mark definitions can lead to stronger model difference, we still need to ensure that their predictions are stable, and not too sensitive to modelling systematics such as the galaxy-halo connection.}

In the discussion so far, we have used the NGP assignment scheme to calculate the galaxy density field. We have also checked that the conclusions for Figs.~\ref{fig:marked1} and \ref{fig:mCF_negative_p} also apply if we use the matter density to define the marks in Eq.~\eqref{markdel}. For completeness, we further checked if the results are sensitive to the details of the density assignment method used. To this end we have tested two other ways to estimate the density, smoothed particle hydrodynamics (SPH) and space tessellation, respectively. Since their results are again consistent with each other, here we only discuss the results from the former case.

In Fig.~\ref{fig:mCFg_SPH}, we estimate the local galaxy number density $n_g$ at each galaxy's position by the spline kernel interpolation \cite{Springel:2005mi},
\begin{equation*}
\label{eqn:rho_R_spline}
\rho_R=\sum_1^{20}W(r_i,h),
\end{equation*}
where $W(r_i,h)$ is the SPH spline kernel, $r_i$ is the distance to $i$th galaxy and the smoothing length $h$ being the distance to the 20th nearest galaxy \cite{Park2007} ,
\begin{equation*}
\label{eqn:kernel}
W(r,h)=\frac{8}{\pi h^3} \left\{
\begin{array}{ll}
1-6\left(\frac{r}{h}\right)^2 + 6\left(\frac{r}{h}\right)^3, &
0\le\frac{r}{h}\le\frac{1}{2} ,\\
2\left(1-\frac{r}{h}\right)^3, & \frac{1}{2}<\frac{r}{h}\le 1 ,\\
0 , & \frac{r}{h}>1 .
\end{array}
\right. 
\end{equation*}
The smoothing scale $h$ for density estimation in this method is varying across the simulation box.

The left panel of Fig.~\ref{fig:mCFg_SPH} shows the galaxy-number-density-weighted marked correlation functions for HOD galaxies in real space at $z=0.5$, using the same parameters $(\rho_*,p)$ as in Fig.~\ref{fig:marked1}. The deviations from GR show the same trend as in Fig.~\ref{fig:marked1}, but the amplitudes are slightly larger, now reaching $\sim10\%$ for F4 on small scales; for the two DGP variants the deviation from GR is negligible. The right panel of Fig.~\ref{fig:mCFg_SPH} represents the monopole of galaxy marked correlation functions in redshift space. Rather than averaging the nominator and denominator of Eq.~\eqref{eq:Mr} over five boxes and taking the ratio, we directly measure the marked correlation function by Eq.~\eqref{eq:Mr} and then take the average. There are some quantitative differences from the real-space case, with the deviations from GR generally slightly larger, but the overall behaviors are similar to the left panel. The only MG model that is significantly different from GR is F4. This shows that the result of marked correlation functions could be sensitive to the scheme of the mark evaluation, adding to the complexity in making theoretical predictions.

\subsubsection{\textcolor{black}{Marked correlation function with gravitational potential 
\revision{summary statistics}}}
\label{sub:potential}

\begin{figure*}[!tb]
\begin{center}
\includegraphics[width=0.485\textwidth]{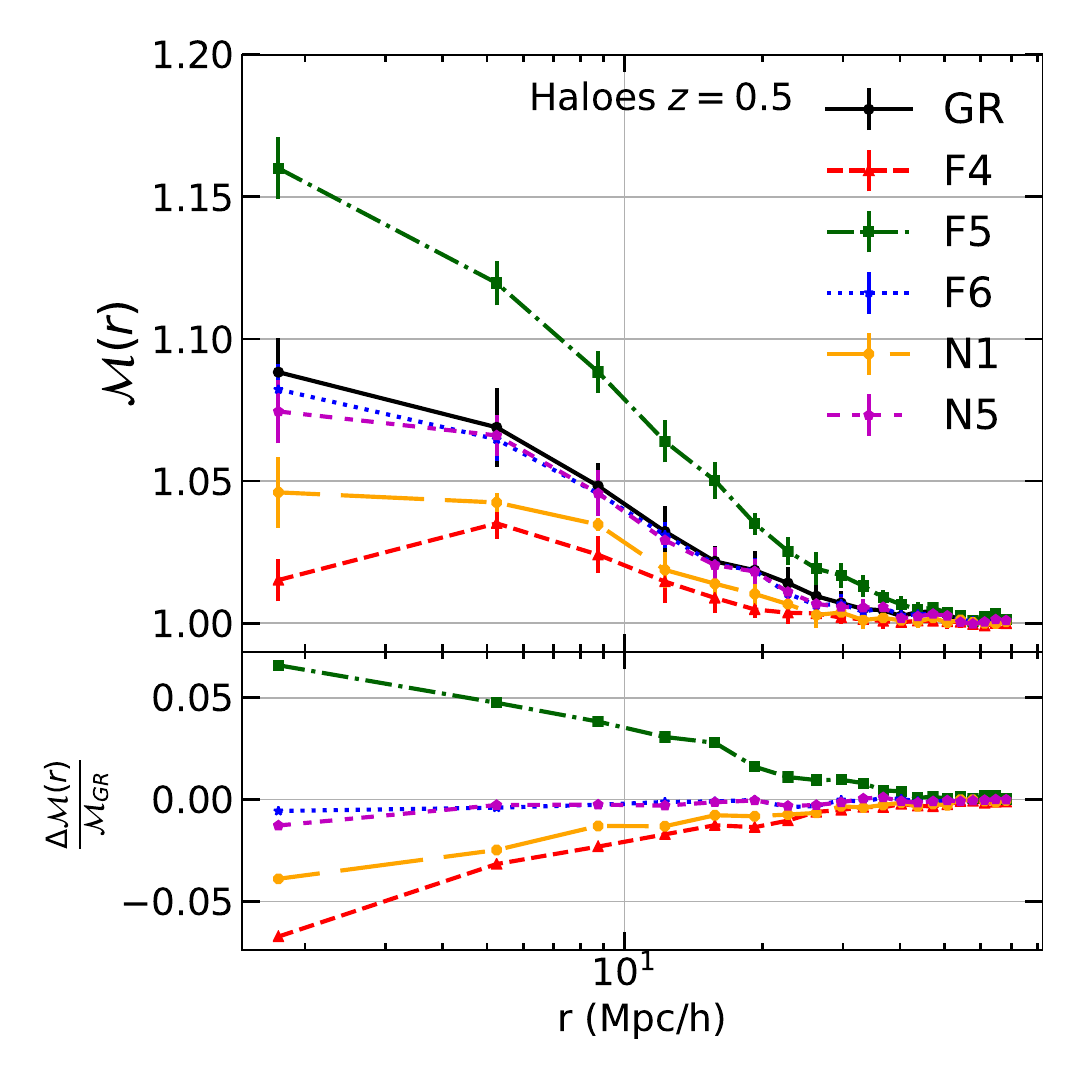}
\includegraphics[width=0.475\textwidth]{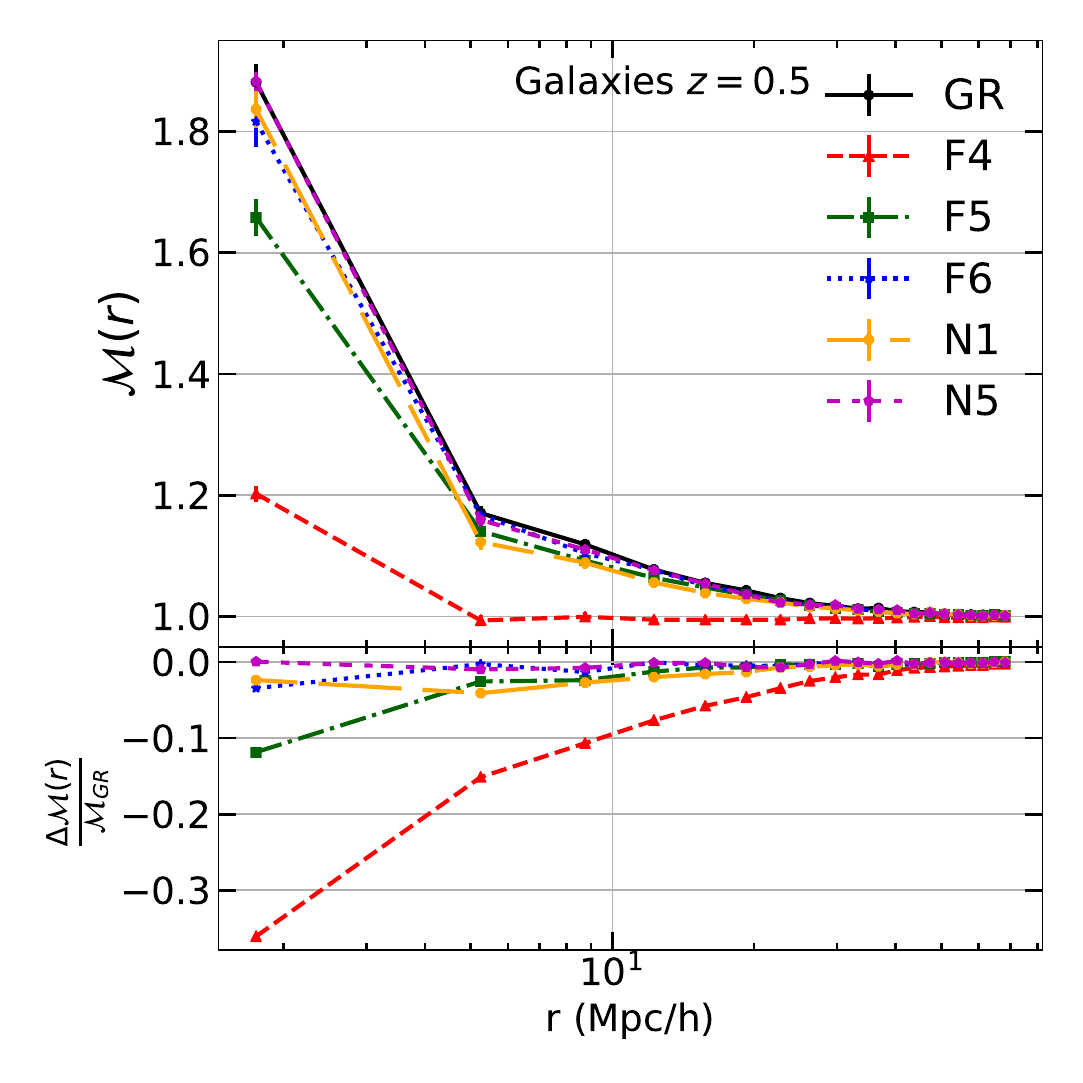}
\caption{Gravitational potential marked correlation function with $\Phi_*=-5.295$ and $\sigma_\Phi=0.1$ at $z=0.5$ for halos (left panel) and galaxies (right panel). The curves correspond to the averages over the 5 realizations. Different colors and line styles correspond to different models as labeled. The lower panels show the relative difference between the MG models and GR, and the error bars correspond to the standard deviation over the 5 realizations.}
\label{fig:mCFg_phi}
\end{center}
\end{figure*}

{In the above we have used the galaxy density field, calculated in various ways, to define the mark. Due to the projected galaxy 2PCFs in the different models all matching each other, we have seen that the resulting marked 2PCFs show very mild difference which is generally of the same order as the residual mismatch in the projected galaxy 2PCFs, making it hard to be tied to any MG effect. This implies that a better definition of the mark may be obtained using quantities other \revision{than} the density field. In the mean time, we know that in the MG models studied here, whether strong deviations from GR happen for a galaxy depends on certain properties of its environment. Here we try a mark defined using the Newtonian gravitational potential, $\Phi_N$, of the galaxy's host dark matter halo, which is one measure of this environment (e.g., \cite{Cabre:2012tq,StarkMG,Shi:2017pyd}). As Eq.~(\ref{eq:PhiN}) below shows, $\Phi_N$ is closely related to the halo mass $M_{200c}$; the abundance and clustering of the latter are strongly affected by MG (cf.~Figs.~\ref{fig:hmf}, \ref{fig:h2pcf}), and the way galaxies populate in halos of different masses is also sensitive to gravity (cf.~Fig.~\ref{fig:HOD}).} 

In many cosmological models, dark matter halos follow a NFW density profile \cite{Navarro:1995iw,Navarro:1996gj}:
\begin{equation}\label{eq:rho_nfw}
\rho_\textrm{NFW} = \frac{\rho_{\rm s}}{(r/r_{\rm s})(1+r/r_{\rm s})^2},
\end{equation}
where $r_{\rm s}$ is the characteristic radius where the profile has a slope of $-2$ and $\rho_{\rm s}$ is the density at $r_s$. {The NFW profile also works well for many MG models, including those studied here (e.g., \cite{Lombriser:2012nn,Barreira:2014zza,Barreira:2014kra,Shi:2015aya,Mitchell:2019qke})}. The gravitational potential \revision{at $r=r_{200c}$} is accordingly given by \cite{Navarro:1996gj,Cole:1996nfw,Lokas:2000mu}:
\begin{equation}\label{eq:PhiN}
\Phi_N = - \frac{G M_{200c}}{r_{200c}} \frac{\ln(1 + c)}{\ln(1 + c) - c/(1 +c)}\,,
\end{equation}
where $G$ is Newton's gravitational constant and $c$ is the concentration parameter defined as $c\equiv r_{200c}/r_{\rm s}$.

The new mark we define here is a Gaussian function of $\Phi_N$,
\begin{equation}\label{eq:mGphi}
m = \frac{1}{\sqrt{2\pi}\sigma_{\Phi}} \exp\left[- \frac{(\log_{10}(|\Phi_N|) - \Phi_*)^2}{2\sigma_{\Phi}^2} \right]\,,
\end{equation}
where $\Phi_*$ and $\sigma_\Phi$ are free parameters which control the amplitude and width of the Gaussian. {This mark allows one to up-weight galaxies hosted by halos of certain mass, and as an illustration here we choose host halos with $M_{200c}/[h^{-1} M_\odot] \in[10^{13}, 10^{14}]$, which is where most of the galaxies in our HOD catalogues reside and where the different gravity models differ most substantially (cf.~the right panel of Fig.~\ref{fig:HOD}). As above, we consider the marked 2PCFs for halos and galaxies separately, using parameter values $\Phi_\ast = -5.295$ and $\sigma_\Phi = 0.1$ for both tracers: halos and galaxies.}

The results for halos are shown in the left-hand panel of Fig.~\ref{fig:mCFg_phi} where we find that F5 differs most strongly from GR, with a maximum positive difference of $\sim 7\%$, whereas the other MG models all produce a negative difference with respect to GR. The distinct behavior of F5 is likely a consequence of the MG effect on the halo mass function (cf.~right panel of Fig.~\ref{fig:hmf}) --- the halo catalogs for all models shown in this plot are cut at $10^{13}h^{-1}M_\odot$, so that the numbers of halos in the different models are quite different. The result  \newrevision{(shown in Fig.~\ref{McutvsNh} of Appendix \ref{App:B})} is qualitatively different if one cuts the halo catalogs to have the same halo number in all models, with the differences from the GR prediction becoming less pronounced in the case of the two largest deviation f(R) models, F4 and F5. Nevertheless, for the rest of the models we find that the fractional deviation w.r.t. to the GR mCF does not change by more than $1\%$ compared to when using a fixed halo mass cut.

The right panel of Fig.~\ref{fig:mCFg_phi} shows the gravitational potential marked correlation function at $z=0.5$ for galaxies. We note that F4 differs most from GR, giving a maximum difference of $\sim 36\%$, while F6 and F5 give a monotonically increasing difference of $\sim 3\%$ and $\sim 12\%$, respectively. The nDGP models (N1 and N5) are both very close to GR with a relative difference of $<2\%$. We have measured the marked 2PCFs in redshift space, and found very similar result. We have also checked the HOD catalogs for which the HOD parameters were tuned so that the 3D galaxy 2PCFs $\xi_{gg}(r)$ match between the MG and GR models, and they again showed very similar features, with slightly smaller (larger) differences of $f(R)$ gravity (nDGP) from GR. While accurately measure the Newtonian potentials of galaxy host halos is nontrivial, methods to estimate halo masses from observations do exist and have been constantly improved (e.g,. \cite{Henden:2019muv,Bradshaw:2019nkx,Yang:2007yr,Lim:2017yr,Wang:2020awd}). The result here suggests that using information other than the density field itself to define the mark can be a potential way to increase the constraining power of the marked CF, and it will be worthwhile to pursue this direction further. In particular, it will be interesting to investigate how the results presented in Fig.~\ref{fig:mCFg_phi} are affected by uncertainties in the halo mass or potential estimation.

\subsubsection{Analytical predictions for the marked correlation function}
\label{sub:analyticalmCF}

In the discussion of marked correlation functions so far, the results have been obtained numerically from mock galaxy catalogs. It would be useful to have a theoretical template to make analytical predictions, for example based on perturbation theory, which can be used to check consistency at large scales and to more efficiently explore the model parameter space.

In this subsection we compute the White marked correlation function [eq.~(\ref{markdel})] using the MG Lagrangian Perturbation Theory (LPT) of \cite{Aviles:2017aor,Aviles:2018saf,Valogiannis:2019xed}. Since we are up-weighting low-density regions it is expected that higher than linear order corrections will be highly suppressed, and therefore we focus on the Zel'dovich approximation. The LPT considers the mapping $\mathbf{x}(t) = \mathbf{q} + \mathbf{\Psi}(\mathbf{q},t)$ between initial (Lagrangian) and final (Eulerian) coordinates, and performs a formal expansion on the Lagrangian displacement $\mathbf{\Psi}$. To linear order
\begin{equation}
\mathbf{\Psi}^{(1)}(\mathbf{k},t)= i \frac{\mathbf{k}}{k^2} D_+(k,t) \delta_L(\mathbf{k},t_0), 
\end{equation}
with $\delta_L(\mathbf{k},t_0)$ the linear overdensity field, evaluated at $t_0$, and $D_+$ the linear growth factor introduced in Eq.~(\ref{eq:Dp}). Remember that while $D_+$ is scale-independent in $\Lambda$CDM, it can have a scale dependence in general MG models.

In the majority of MG models which are viable to explain the accelerated expansion of the Universe, including those considered in this work, the evolution of perturbations at a sufficiently early time is indistinguishable from GR, hence the linear power spectrum in MG can be obtained by the relation $P_L^\text{MG}(k,t) = [D_+(k,t)/D_+(k,t_0)]^2 P_L^\text{GR}(k,t_0)$. Otherwise, it can be obtained from an Einstein-Boltzmann code.
 
To model halos, we use the biasing prescription of Ref.~\cite{Matsubara:2008wx,Carlson:2012bu} that evolves initially Lagrangian biased tracers, and consider linear $b_1$ and second order $b_2$ local biases. However, the $k$ dependence of the $\mu$ function
in Eqs.~(\ref{eq:mu-exp},\ref{eq:Dp}) implies that even linear local bias becomes scale dependent. In order to model this, one can perform an expansion of the function $\mu$ in powers of $k^2$ and substitute the linear local bias by $b_1 \longrightarrow b_1 + b_{\nabla^2\delta} k^2 + \cdots$ with $\nabla^2\delta$ the curvature bias operator \cite{Desjacques:2016bnm}. In addition to the biasing expansion, we expand the mark as $m(\delta_R) \simeq 1 + B_1 \delta_R + \frac{1}{2}B_2 \delta_R^2$, with $\delta_R(\mathbf{q})$ obtained by convolving the matter density field with a Gaussian kernel $W_R \propto e^{-|\mathbf{q}|^2/2R^2}$, and we choose the smoothing scale as $R=6  \,h^{-1}\text{Mpc}$. 
 
The analytical prediction is obtained following the standard methods of Convolution Lagrangian Perturbation Theory  \cite{Carlson:2012bu,White:2016yhs,Aviles:2018thp,Aviles:2019fli} 
\begin{align} \label{eq:CLPmCFW}
&1+ W(r) = \int \frac{d^3 q}{(2\pi)^{3/2}(\text{det}A)^{1/2}} e^{-\frac{1}{2}(r_i-q_i)(r_j-q_j)A_{ij}^{-1}} 
         \Bigg\{  1 + b_1^2 \xi_L   - 2 b_1  U_i g_i - (b_2+b_1^2) U_iU_j G_{ij} \nonumber\\
 &\quad  - 2 b_1b_2\xi_L U_i g_i + B_1^2  \xi_{RR}  - 2B_1 U_i^R g_i- (B_2+B_1^2)U_i^{R} U_j^{R} G_{ij} - 2 B_1B_2 \xi_{RR} U^{R}_i g_i + 2b_1B_1 \xi_{R} \nonumber\\
 &\quad   - 4b_1B_1 U_i U_j^{R} G_{ij}   - 2b_1^2B_1\xi_L  U_i^{R} g_1 - 2B_1^2b_1\xi_{RR} U_i g_i  - 2(b_2 + b_1^2)B_1 \xi_{R} U_i g_i -  2(B_2 + B_1^2)b_1  \xi_{R} U_i^{R} g_i  \nonumber\\
 &\quad   + 2(1+b_1+B_1)b_{\nabla^2\delta}\nabla^2\xi_L(q) + b_{\nabla^2\delta}^2 \nabla^4\xi_L(q)\Bigg\}.
\end{align}
Here, $\Delta_i = \Psi_i(\mathbf{q})-\Psi_i(0)$, and 
\begin{align}
  &A_{ij}(\mathbf{q}) = \langle \Delta_i \Delta_j\rangle,\quad U_i(\mathbf{q}) = \langle \delta(\mathbf{q}) \Delta_i \rangle, \quad U_i^R(\mathbf{q}) = \langle \delta_R(\mathbf{q}) \Delta_i \rangle \nonumber\\
  &\xi_L(q) = \langle \delta(\mathbf{q})\delta(0)\rangle, \quad \xi_R(q) = \langle \delta_R(\mathbf{q})\delta(0)\rangle, \quad \xi_{RR}(q) = \langle \delta_R(\mathbf{q})\delta_R(0)\rangle,
\end{align}
and the tensors are defined as $g_i = A^{-1}_{ij}(q_j-r_j) $, $G_{ij} = A^{-1}_{ij} - g_ig_j$. The (unmarked) correlation function for tracer type $X$ is given by $\xi_X(r) = W(r;B_i=0)$. 
Eqs.~(\ref{eq:Mr},\ref{eq:CLPmCFW}) provide a good approximation for large scales as long as the RMS of matter Lagrangian displacements is smaller than the smoothing scale $R$. In other words, as larger the smoothing scale is, or as higher the evaluation redshift is, more accurate predictions are expected \cite{Aviles:2019fli}.

\begin{figure*}[!tb]
	\begin{center}
	\includegraphics[width=0.49\textwidth]{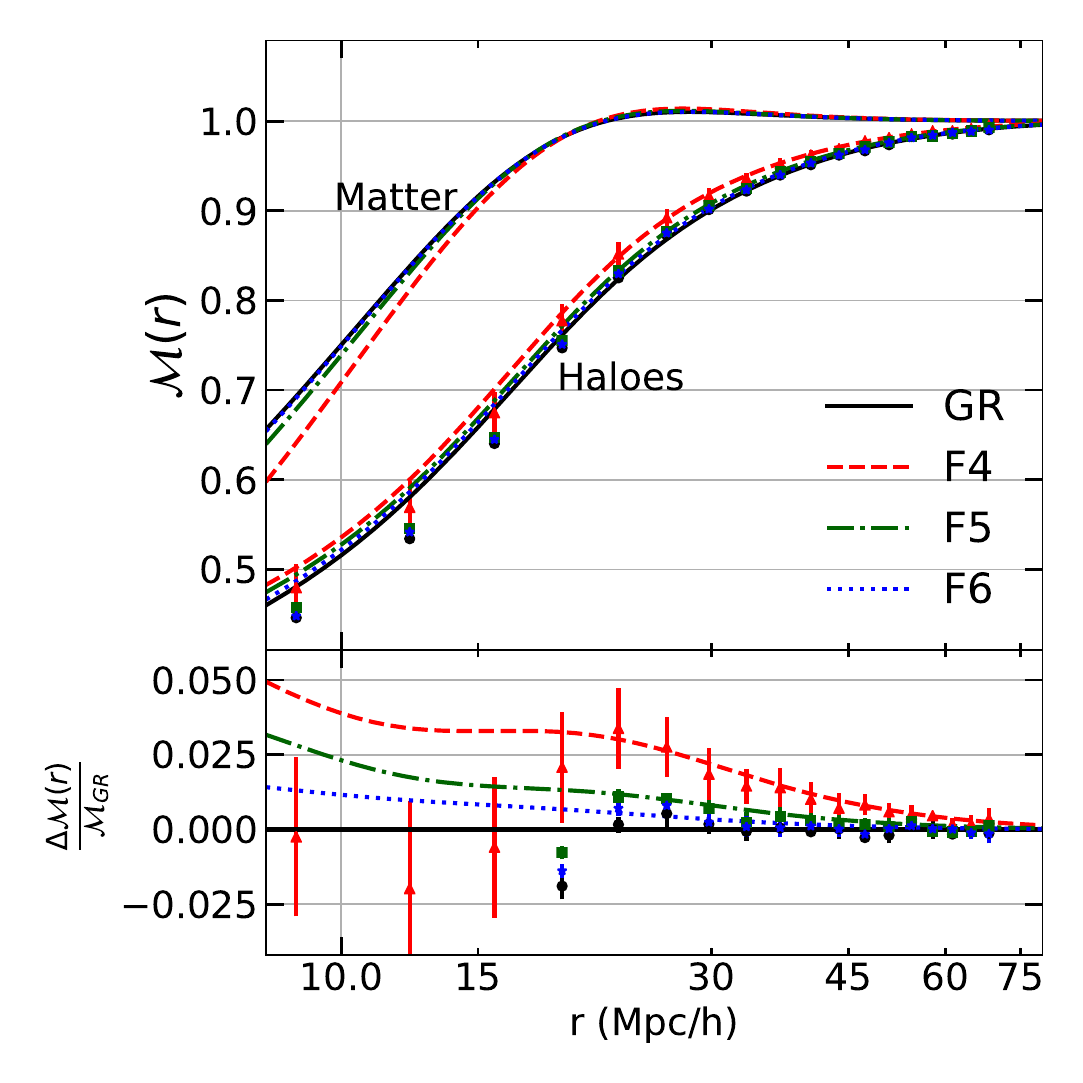}
	\caption{The Zel'dovich-approximation prediction of the marked correlation function Eq.~\eqref{markdel} with $\rho_*=4$ and $p=10$, for models F4 (shown in dashed red curves), F5 (dot-dashed green), F6 (dotted blue), and GR (solid black) at redshift $z=0.5$. 
	The data points are the simulation results taken from 
	Fig~\ref{fig:marked1}: black data points with error bars are for GR, blue for F6, green for F5, and red for F4. The lower panel shows the relative differences in the marked correlation function for \revision{tracers} (and the data) with respect to the Zel'dovich approximation in GR.
	\label{fig:LPTplot}}
	\end{center}
\end{figure*} 

By applying the above formalism to $\Lambda$CDM and the three variants of $f(R)$ gravity considered here, we find that the trends of the marked correlation function for matter and biased tracers are qualitatively different, as shown in  Fig.~\ref{fig:LPTplot}. In the matter case, the F4, F5, and F6 marked correlation functions fall below that of GR, a behavior that has been observed recently  in simulations 
\citep{Valogiannis:2017yxm}. This can be interpreted by considering the mean mark $\bar{m} = (1+b_1)B_1 \sigma_R^2$, which shows
that for the unbiased case $\bar{m}^\text{MG} < \bar{m}^\text{GR}$, simply because $\sigma_R^\text{MG} > \sigma_R^\text{GR}$ and $B_1<0$. For the biased tracer case, we have used halos instead of HOD galaxies. A key factor here is to identify the halo bias, because its effect is to reduce the marked correlations \cite{White:2016yhs} -- the larger the bias the larger the reduction. A local bias is typically controlled by two parameters, the density threshold for collapse, $\delta_c(M)$, and the variance $\sigma(M)$ (roughly through their ratio). The former (latter) is smaller (bigger) in $f(R)$ gravity, resulting in smaller bias values for our MG models than for GR \cite{Aviles:2018saf}. As a result, the halo bias may bring the marked correlation above that of GR, as is the case for the halo masses interval presented here. These results can be understood on physical grounds as even though low density regions in GR correspond to even more underdense regions in $f(R)$, the halos are more efficiently formed due to a larger gravitational strength.
 
To compare with simulations, we let free the bias parameters and fit them on scales $r>20 \,h^{-1} \text{Mpc}$, finding a good agreement within the error bars of the data. Our best fit is given by linear local Lagrangian biases $b_1^\text{GR}=1.15 $, $b_1^\text{F6}=1.12 $, $b_1^\text{F5}= 1.07$, and $b_1^\text{F4}=0.90$, and second order biases $b_2^\text{GR}=0.8 $, $b_2^\text{F6}=0.7 $, $b_2^\text{F5}= 0.5$, and $b_2^\text{F4}=0.3$, while the curvature bias does not contribute significantly over this region. In  Fig.~\ref{fig:LPTplot}, we also show the number weighted density estimates, as presented in the left panel of Fig.~\ref{fig:marked1}. 
\revision{This analysis suggests the Lagrangian perturbation approach may serve to construct theoretical templates for future gravity tests against survey data, as it has been observed also in \cite{Aviles:2019fli}. However, even at large scales our modeling accuracy is very sensitive to the expansion of the mark, and in some applications one needs to go to very high perturbative order \cite{Philcox:2020fqx,Philcox:2020srd}. This becomes prohibitive for an analytical treatment, at least one is able to construct semi-analytical templates valid at any order in PT, as it is done in \cite{Philcox:2020srd} for the marked power spectrum.}

\subsubsection{Clustering in over-dense environments}
\label{sub:rho-wp}

\begin{figure}
\begin{center}
    \includegraphics[width=0.9\textwidth]{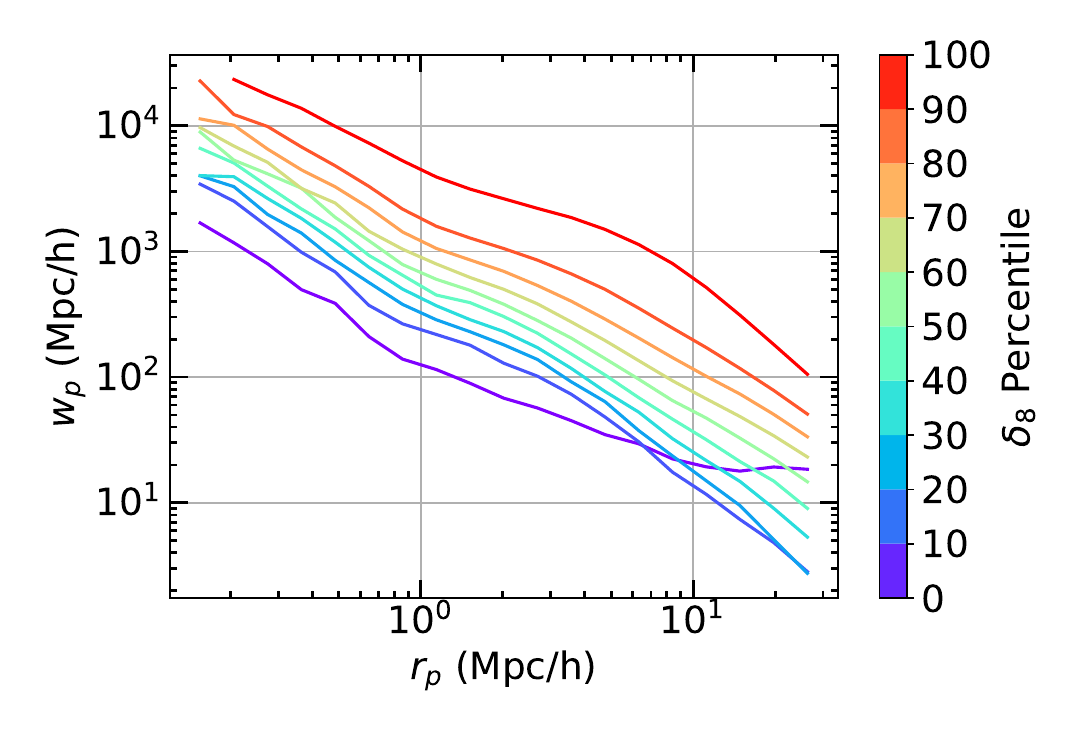}
	\caption{The projected correlation function, $w_p(r_p)$, estimated for GR HOD catalogues at redshift $z=0.5$. The different colored line show the $w_p$ for over-density ($\delta_8$) percentiles indicated by the colorbar. At small scale ($r_p<5\Mpch$) the clustering amplitude increases with $\delta_8$ monotonically. At larger scale the clustering amplitude shows non-monotonic behavior with $\delta_8$ as expected \cite{Alam2017}.}
\label{fig:rho-wp}
\end{center}
\end{figure}

\begin{figure}
\begin{center}
    \includegraphics[width=0.9\textwidth]{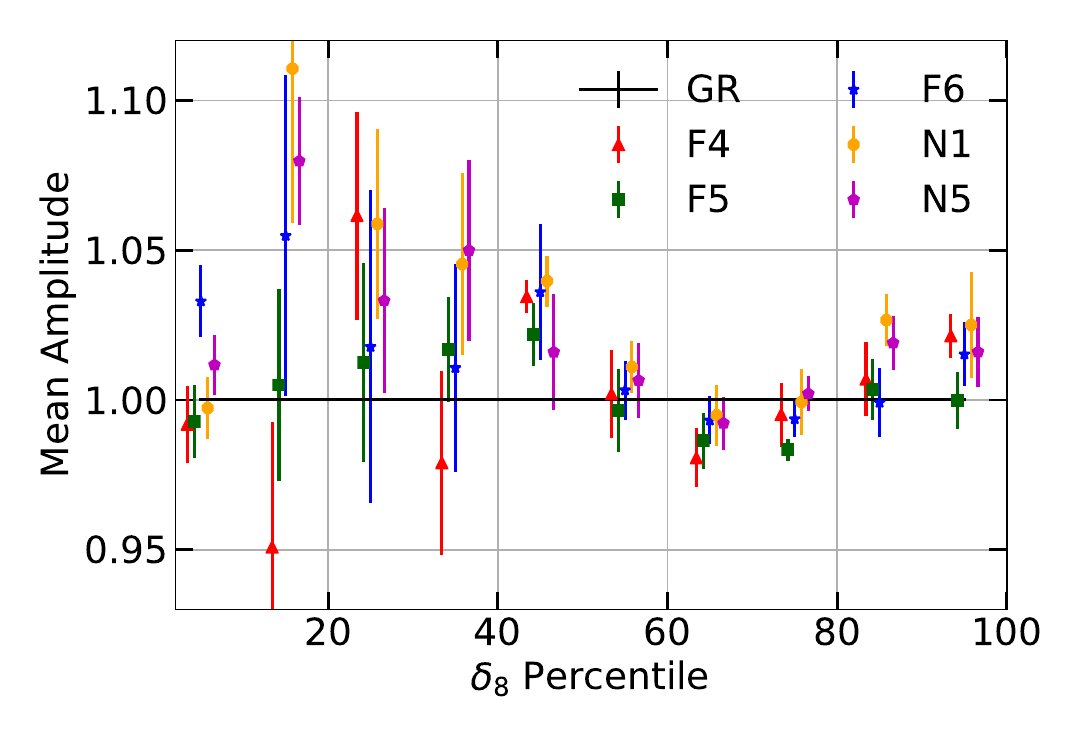}
	\caption{The mean amplitude as a function of over-density ($\delta_8$). The different MG models are shown with different colors, and the {1-$\sigma$} errors are computed as the error on the mean of the 5 HOD catalogs. The F4, F5, and N1 models show large $\chi^2$ values and should be detectable at better than $3\sigma$ significance but F6 and N5 are very close to GR and do not show any statistically significant difference. Note the points representing different models at same over-density are slightly shifted horizontally to avoid clutter and increase visibility.}
\label{fig:rho-amp}
\end{center}
\end{figure}

\begin{table*}
\centering
\begin{tabular}{ccccc} \hline \hline
        model & ~~~~ $r_p>0.5 h^{-1}{\rm Mpc}$ ~~~~& ~~~~ $r_p>1 h^{-1}{\rm Mpc}$  ~~~~ & ~~~~ $r_p>4 h^{-1}{\rm Mpc}$ ~~~~ & $r_p>8 h^{-1}{\rm Mpc}$  ~~~~ \\ \hline
F4 & 355 & 106 & 191 & 81  \\
F5 & 854 & 52 & 25 & 30   \\
F6 & 16 & 28 & 13 & 18   \\
N1 & 28 & 26 & 67 & 42 \\
N5 & 12 & 15 & 18 & 23 \\
\hline\hline
\end{tabular}
\caption{The $\chi^2$ with 10 degrees of freedom for the mean amplitude as a function of the minimum $r_p$ used in the calculation. The different rows corresponds to different MG models and the different columns for different minimum scale. See the text for more details and discussion.
}
\label{tab:rho-wp}
\end{table*}

The idea behind using marked correlation function to test models is that different gravity models can show clustering patterns that are density dependent in different ways. Actually, instead of estimating the marked correlation functions as done above, one can more directly estimate the clustering in different density environments and compare them to distinguish between different gravity models. Although our simulation have matched the overall projected 2PCFs, we may still find the signature of different gravity models and screening mechanisms by looking at $w_p(r_p)$ for galaxies in various environments. 

Fig.~\ref{fig:rho-wp} shows the projected correlation functions computed for the GR galaxy catalogs in different $\delta_8$ percentile bins, where $\delta_8$ is the environmental density contrast averaged over a $8\Mpch$ scale using a modified tessellation scheme. Although the results here are for periodic boxes, the method can account for survey mask and have been applied for real surveys \cite{Alam2017}. Briefly, we count the number of randoms nearest to each galaxy and turn that into an estimate of volume occupied by the galaxy. This automatically accounts for the survey mask and incompleteness (see \cite{Alam2017} for more details). This only requires a random catalog sampling the volume of the survey which will be anyway produced as part of the standard survey pipeline and does not require any extra information. One should think of $\delta_8$ as over-density of galaxy field here. Once the over-density is assigned for each galaxy, we rank the galaxies based on their $\delta_8$ values and split the catalog into 10 sub-samples each containing 10\% of the whole sample. We then measure the two-point correlation functions using the Landy-Szalay estimator and project out the line-of-sight information to obtain projected correlation functions. The measurement of projected correlation function ($w_p$) is obtained with 20 logarithmic bins in $r_p$  covering $0.1-30\Mpch$. We then define the `amplitude' as the ratio of projected correlation function between a MG model and GR, averaged over a range of scale above a chosen $r_p$. 

Fig.~\ref{fig:rho-amp} shows the mean amplitude as the function of $\delta_8$ percentile for all the MG models at $z=0.5$ and with minimum $r_p$ of $4\Mpch$. The curves is the mean of the amplitude values from the five simulation boxes, and the errorbar denotes the error on the mean of five boxes. One can see that in high-density environments (i.e., the five largest values of $\delta_8$) the different models all agree with each other very well, confirming the effect of screening. For smaller values of $\delta_8$, the amplitude generally show more deviations from 1: there are large scatters for the three $f(R)$ variants, making it hard to see any clear trend, while nDGP shows a clearer pattern. This is a consequence of the range of scale ($[4,30]\Mpch$) used to calculate the amplitude, because the effect of $f(R)$ gravity -- especially for F6 and F5 -- is restricted to smaller scales not used in the calculation, while the nDGP model deviates from $\Lambda$CDM on larger scales which are covered by the calculation. We show the results for \revision{$r_p>4\Mpch$} in Fig.~\ref{fig:rho-amp} here because, as discussed below, this is the optimal choice for N1, but we have checked other minimum values of $r_p$ and found that decreasing it generally leads to larger amplitude values for the $f(R)$ variants, as we show more quantitatively next.

In order to combine the information from all the $\delta_8$ subsamples, we also need to estimate the covariance between the amplitude. We generate 100 jackknife realizations for each model and each $\delta_8$ subsample, and use these jackknife realizations to estimate the correlation matrix of the mean amplitude with $\delta_8$. The correlation matrix is then scaled by the error estimated from the mean of 5 boxes to get the covariance matrix $\mathbb{C}(i,j)$, where $i,j$ represent the different $\delta_8$ subsamples. This covariance matrix is then employed to estimate the $\chi^2$ for each MG model with respect to GR, according to
\begin{equation}
    \chi^2 = \sum_{i,j=1}^{10}\left(A_i-A^{\rm GR}_i\right)\mathbb{C}(i,j)\left(A_j-A^{\rm GR}_j\right),
\end{equation}
where $A_i$ denotes the amplitude value of the $i$th $\delta_8$ subsample, calculated using the projected 2PCFs measured in the range of $r_p$ between a varying minimum value and $30\Mpch$. 
Our estimated $\chi^2$ for different models and different choices of the minimum $r_p$ are given on Table~\ref{tab:rho-wp}. As mentioned above, the $f(R)$ models show increases in $\chi^2$ as we include smaller scales. F4, which is the model with very little screening and the largest Compton wavelength of the background scalar field, can be detected with high confidence independent of choice of scale; F5 can be detected with high confidence if at least scales up to $1\Mpch$ is used and F6 can not be detected independent of scale used. We note that for nDGP models, including smaller scales first increases the $\chi^2$ up to an optimal scale and then decreases it: N1 can be detected with an optimal choice of minimum scale as $4\Mpch$ while N5 can not be detected independent of scale used. 

Given that the number density of galaxies in future surveys such as DESI will be higher and the volume will be much larger than the simulations used, the statistical error from DESI will be much smaller. However, at the same time the real data also have systematic errors not included in the analysis here. For this particular test, the most important systematics will come from fibre collision and completeness uncertainty, both of which are major concerns for DESI main science cases and huge amounts of work have been put in to keep them under control. These conclusions about detecting the difference should hold for DESI in presence of systematics as long as they do not dominate the total error budget. 

We have checked this 
\revision{summary statistic} using the mock galaxy catalogs where the HOD parameters were tuned to match the real space 3D galaxy 2PCFs in different models, and found the results change slightly for the lowest five bins of $\delta_8$, which suggests that the uncertainty associated with HOD modelling can be a theoretical systematic which needs better control. On the other hand, we also note the MG models can be distinguished from GR using the scales above $\sim1\Mpch$, and hence these detection should be largely free from the baryonic effects.

\subsubsection{Discussion}\label{sub:m_discussion}

While the halo and HOD galaxy catalogs might be several steps away from the dark matter distribution, the behavior of the marked correlation functions above can be qualitatively understood. A marked correlation function quantifies the correlation between the marks. In the case of the mark in Eq.~\eqref{markdel}, for a tracer $i$ -- which can be a galaxy or a halo -- with $m_i<\bar{m}$, it is likely to find another tracer $j$ nearby with $m_j<\bar{m}$ because these tracers are in dense regions; for a tracer $i$ with $m_i>\bar{m}$, which means that it is in a low-density region, it is less likely to find a neighbor $j$. This leads to $\mathcal{M}<1$. A similar reasoning can explain why $\mathcal{M}>1$ in the case of gravitational potential mark.

The model differences between $f(R)$ gravity and $\Lambda$CDM can be explained as follows. An enhanced gravity means that more halos can form in regions where these tracers have low densities in the $\Lambda$CDM counterpart, while in high-density regions the increase in halo number density is less significant due to either more efficient screening (e.g., in F5 and F6) or more frequent mergers (e.g., in F4). For the mark in Eq.~\eqref{markdel}, this means that for a halo $i$ with $m_i>\bar{m}$ (i.e., in a low-density region), it is more likely to find neighbors with $m_j>\bar{m}$ either from the same cell or from a neighboring cell, leading to a larger $\mathcal{M}(r)$ in MG than in GR. In the cases of F5 and F4, the difference is significant and can be observed to larger halo separations (cf.~Fig.  \ref{fig:marked1}). For the gravitational potential mark, again, the enhanced gravity produces more halos with $M_{200c}>10^{13}h^{-1}M_{\odot}$ which correspond to smaller halos in GR; these halos are less biased and more uniformly distributed; they also have relatively low marks, so that they increase the likelihood of finding a halo $i$ with $m_i<\bar{m}$ near a halo $j$ with $m_j>\bar{m}$, thereby reducing $\mathcal{M}(r)$. The strange behavior of F5 (left panel of Fig.  \ref{fig:mCFg_phi}) is probably a consequence of this model having many more halos of $M\sim10^{13}h^{-1}M_\odot$, which significantly decreases $\bar{m}$, leading to an overall increase in $\mathcal{M}(r)$. This suggests that it may possible or even preferrable to define marks in ways so to pick out the region of the HMF with the most significant model differences.

We also noticed above that for the mark in Eq.~\eqref{markdel} the marked correlation functions for galaxies show much smaller model differences, and a less clear trend, than for the gravitational potential mark. This is probably because in the former case the marks themselves are obtained from the galaxy field, which has been tuned to match in all the different models. In other words, with a similar galaxy number density and spatial distribution, the marks for the HOD galaxies in the various models are similar, and so are their correlations. In contrast, for the gravitational potential mark, the mark itself encodes external information beyond what is contained in the galaxy distribution, and what we see in the right panel of Fig.~ \ref{fig:mCFg_phi} is the correlation of this information. This suggests that it can prove useful to study the many possible ways to define the marks for marked correlation functions, by using complementary observational information and by varying the parameter values of the marks.

The nDGP models are very difficult to be distinguished from $\Lambda$CDM either using halos or galaxies by the marked correlation function. In these models, deviations from $\Lambda$CDM appear only at the high-mass end of the mass function (Fig.~\ref{fig:hmf}). In addition, the halo correlation function is very close to $\Lambda$CDM (Fig.~\ref{fig:h2pcf}), despite the fact that the underlying dark matter clustering is enhanced (Fig.~\ref{fig:pofk}). The difference of the HOD parameters is less prominent and the galaxy numbers as a function of the host halo mass is also very similar to $\Lambda$CDM (Fig.~\ref{fig:HOD}). As a consequence, irrespective of tracers and marks, the difference between nDGP and $\Lambda$CDM is always small. The environmentally-dependent clustering of galaxies shown in Figs.~\ref{fig:rho-wp} and \ref{fig:rho-amp}, and Table \ref{tab:rho-wp}, on the other hand, seems to be more promising than the marked statistics for both the $f(R)$ and nDGP models: a possible explanation to this is that here we have singled out the subsamples of tracers from low-density regions---where the effect of modified gravity is strongest---for clustering measurement, while in the marked statistics analysis the tracers from high-density environments still contribute, though with a smaller weight, which weakens the model difference.

\subsection{{Beyond two-point statistics}}
\label{subsect:beyond2pt}

In cosmological perturbation theory, {see, e.g.,}   \cite{Bernardeau:2001qr}, the velocity-velocity and density-velocity couplings render the evolution equations nonlinear, generating non-Gaussian late-time density fields from a Gaussian random field as the initial condition, and the non-Gaussianities may be enhanced in models with a stronger gravity. Furthermore, primordial non-Gaussianities due to the detailed properties of inflation are generically expected at some level. Both of these features indicate that the galaxy distribution cannot be fully characterized by the two-point correlation function. Higher-order correlation functions contain additional information regarding the nature of gravitational interactions, making them useful to test deviations from GR {(see, e.g., \cite{Borisov:2008xn,Bose:2019wuz})}, in particular for theories where nonlinear interactions play a significant role in screening modifications at short scales. Although there are several theoretical and observational studies showing the potential of such 
\revision{summary statistics} for LSS analysis (see, for example, the review \cite{Bernardeau:2001qr} or the studies summarized in \cite{Guo:2014nka, 2016A&A...592A..38S, Sugiyama:2018yzo,Sugiyama:2020uil}), it is still relatively poorly explored compared to its lower order cousin, the 2PCF.

In \ref{subsect:threept} and \ref{subsec:bispectrum} we respectively discuss the information to investigate modifications to gravity in the three-point correlation function (3PCF) and its Fourier space counterpart, the bispectrum. {For the former we focus on real space while for the latter on the monopole in redshift space (see Refs.~\cite{Sugiyama:2018yzo,Sugiyama:2020uil} for complementary studies of the bispectrum)}. In \ref{subsec:hierarchical_clustering} we consider constraints using hierarchical clustering, while in \ref{subsect:MF} we discussed the application of Minkowski functionals, and in \ref{subsect:stackedcl} we close with a consideration of the use of phase space information from stacked clusters.

\subsubsection{Three-point correlation functions}
\label{subsect:threept}

Although three-point statistics in configuration and Fourier spaces have theoretically the same information, in practice their implementations are different. Measurement in configuration space is conceptually more straightforward, and does not require special effort to deal with gridding, numerical Fourier transforms or shot noise corrections \cite{Hoffman2018}. Moreover, large surveys such as BOSS and DESI bias the clustering due to the targeting algorithm. There are methods to mitigate this effect, which may be more easily implemented in configuration space than Fourier space (e.g., \cite{Burden:2016cba, Guo:2012, Bianchi:2018rhn, Smith:2018tyi}, though this is still a subject under study. Therefore in this subsection we focus on the three-point correlation function (3PCF) first.

Large surveys such as DESI will render a brute-force 3PCF calculation computationally challenging. However, a few efficient algorithms have lately been developed that should help surmount this obstacle (for example \cite{2001misk.conf...71M,2005ApJ...635..743C,2005ApJ...635..743C,Scoccimarro:2015bla,2015MNRAS.454.4142S,2016MNRAS.455L..31S, 2019ApJS..242...29S, 2019ascl.soft08004S, Nunez:2020vdm}); of particular interest is the one recently proposed by Slepian and Eisenstein \cite{2015MNRAS.454.4142S}, which scales as $\mathcal{O}(N^2)$ for the isotropic case, with $N$ the number of objects. The isotropic 3PCF, $\xi^{(3)}$, which does not track the line of sight, accounts for triangular configurations parametrized by 3 variables, which can be chosen to be the two sides of the triangle, $|\vec{r}_1|=r_1$ and $|\vec{r}_2|=r_2$, and the opening angle between those sides, $\cos\ \theta_{12} = \hat{r_1}\cdot\hat{r_2}$. The key idea to reduce the computational load is to use a Legendre polynomial basis \cite{Szapudi:2004gg}, namely
\begin{equation}\label{multi-dec}
\xi^{(3)}(r_1,r_2,\hat{r}_1\cdot\hat{r}_2)=\sum_\ell \xi^{(3)}_\ell(r_1,r_2)P_\ell(\hat{r}_1\cdot\hat{r}_2),
\end{equation}
with $P_\ell$ the Legendre polynomials. While this basis does not have a straightforward geometrical interpretation, one should bear in mind that each multipole contains information about all possible shapes because the coefficients are averages over all triangles weighted by the Legendre polynomials. For example, the monopole contribution is averaged over all triangles of sides $r_1$ and $r_2$ with an equal (constant) weight on $\hat{r}_1\cdot\hat{r}_2$. The multipole basis reduces the computational cost because the Legendre polynomial can be factored into spherical harmonics of one unit vector each, avoiding the need to explicitly construct the opening angle between the two galaxies at $\vec{r}_1$ and $\vec{r}_2$ from the vertex. 

The multipole representation has additional benefits beyond the computational acceleration it offers. For instance, it displays more information on the isotropic 3PCF than using particular shapes (such as the equilateral or specific triangles with two sides fixed). In the case of a converging series, only a few multipole moments $\xi^{(3)}_\ell$ are needed to capture most of the 3PCF (studied in \cite{2015MNRAS.454.4142S}, seen in Fig.  8 for a triangle with $r_1, r_2 = 70, 40h^{-1}$Mpc, and further discussed in \cite{SlepianEisenstein3PCFmodel}). Numerical estimates suggest that the series easily converges away from the diagonal ($r_1\neq r_2$), while for $r_1=r_2$ it converges on scales larger than a given value, which for our catalogs is around 10 $h^{-1}{\rm Mpc}$ by using the first 10 multipoles. Therefore, it is expected that the first ten multipoles capture most of the information up to some small regions in the $r_1$-$r_2$ plane. Another benefit of this Legendre basis is the simplicity of calculating the edge corrections due to the survey boundary (for details see \cite{2015MNRAS.454.4142S}).

For these reasons, in this subsection we shall mostly focus on the algorithm for the isotropic 3PCF proposed in Ref.~\cite{2015MNRAS.454.4142S}\footnote{The code is publicly available in the n-body kit \cite{Hand:2017pqn}.}. However, for completeness we will also show some results from considering a few triangle configurations using a full 3PCF algorithm near the end. One can indeed calculate the full 3PCF for any triangular shape using the multipole expansion Eq.~(\ref{multi-dec}). For fixed values in $r_1$ and $r_2$, all the remaining 3PCF data can be compressed into a single-valued function of the opening angle between $\vec{r}_1$ and $\vec{r}_2$ (or equivalently the side $\vec{r}_3$). This approach of arbitrarily choosing two sides of the triangle has often been followed in the literature to compare three point statistics. However, each coefficient $\xi_\ell^{(3)}(r_1,r_2)$ is a function of two  independent scales, and thus contains all the relevant information about interactions in a particular gravitational theory. In practice, the largest deviations of alternative models to GR may not be found where the strongest 3PCF signal is. Therefore, it is wise to study both the full 3PCF and particular triangular shapes when testing gravity. Moreover, we find that the previous Legendre expansion contains interesting information about gravitational clustering at each multipole, suggesting that a comparison between GR and modified gravity models for each multipole may be more useful than using the full correlation function, where often the lower multipoles comprise the dominant portion of the signal.

We run the 3PCF algorithm, up to a maximum multipole of $\ell=9$, over the five realizations of each of the different models described in \S~\ref{sect:models}. For all models we use the same random catalogs based on an equal box size as the data, and 600,000 random points. To avoid Poisson noise due to the randoms yet also elude the need to compute a large number of random pairs, we use 50 realizations of the random catalogs and average them over the data using the Szapudi-Szalay estimator for the 3PCF \cite{SzapudiSzalay}
\begin{equation}
\xi^{(3)}=\frac{\overline{NNN}}{RRR},
\end{equation}
where $N=D-R$ and once the product is expanded, the $XXX$ (with $X$ either $D$ or $R$) refer to the histogram of distances of triplets where each vertex belongs to the corresponding $X$-space. The overbar denotes an average over the different random catalog realizations. 

\begin{figure*}[!tb]
\begin{center}
{\includegraphics[width=0.99\textwidth]{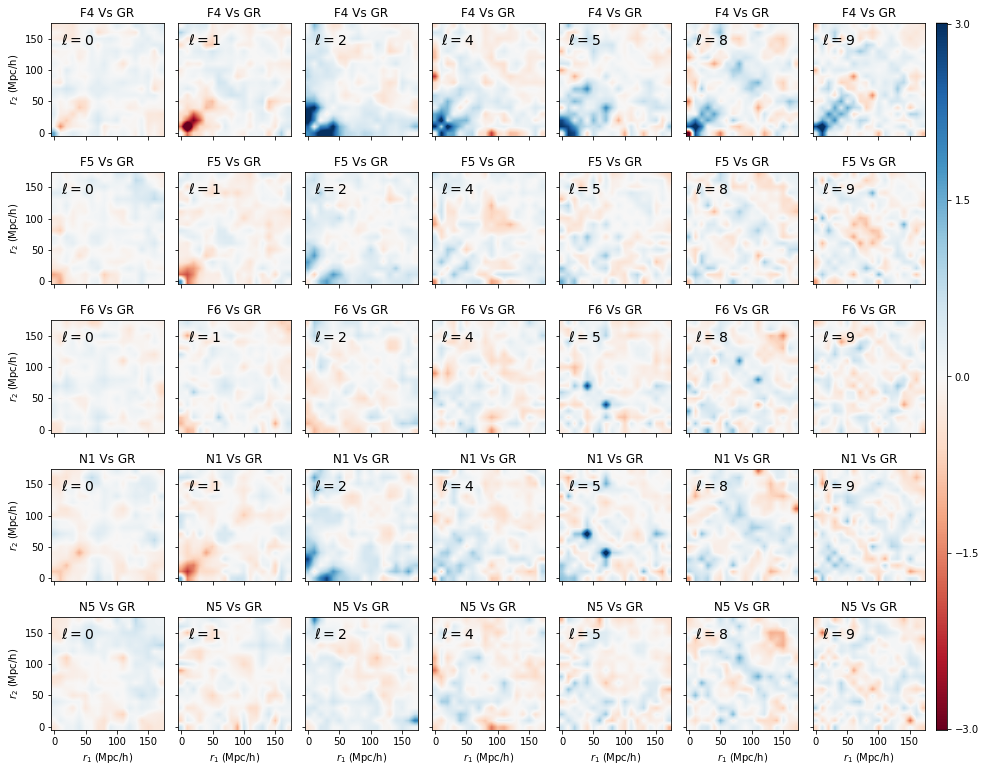}
}
\caption{3PCF comparison of the models using the variable $\Delta_\sigma^{(3)}\xi_\ell$ defined in Eq.~(\ref{def:delta_xi}) and the multipole decomposition. Saturated regions may reach values beyond the colorbar shown. {We use 18 bins of $10h^{-1}{\rm Mpc}$ each for the triangle sides.}}
\label{fig:3pcfmodels}
\end{center}
\end{figure*}


To compare the correlation function coefficients $\xi^{(3)}_\ell$ between models, we use the variable 
\begin{equation}\label{def:delta_xi}
\Delta_\sigma\xi_\ell^{(3)}=\frac{\xi^{(3)}_\ell{(X)}-\xi^{(3)}_\ell{(\textrm{GR})}}{\sqrt{\sigma^2_\ell{(X)}+\sigma^2_\ell{(\textrm{GR})}}},
\end{equation}
where we have taken the mean over the 5 realizations of the modified gravity models, $X$, and General Relativity, $\textrm{GR}$, to obtain $\xi_\ell^{(3)}{(X)}$ and $\xi_\ell^{(3)}{(\textrm{GR})}$ respectively, together with their standard deviations $\sigma^2_\ell{(X)}$ and $\sigma^2_\ell{(\textrm{GR})}$, which we have assumed to be uncorrelated. This expression is the square root contribution of each bin to the overall $\chi^2$, using GR as our fiducial model. The results for the 5 models are shown in Fig.~\ref{fig:3pcfmodels}. To compare the complete 3PCF, $\xi^{(3)}$, we use a similar expression to Eq.~(\ref{def:delta_xi}), but with the complete 3PCFs and their propagated errors in place of the expressions with $\ell$'s.

There is a complex structure in the resulting values for the variable $\Delta_\sigma^{(3)}\xi_\ell$, showing departures of each modified gravity model from GR on all scales and multipoles. Although each model shows a different pattern, there are some generic features shared by all. In most of the parameter space explored in Fig.~\ref{fig:3pcfmodels}, the differences are roughly of the same order as the scatter between model realizations, which translates into $\Delta \xi^{(3)}_\ell\sim 1$. However, there are departures with $\Delta \xi^{(3)}_\ell> 1$ in different regions of the $r_1$-$r_2$ plane (see the high intensity or strongly saturated regions of Fig.~\ref{fig:3pcfmodels}). These regions come in two classes: isolated (almost point-like) areas where at least one of the scales ($r_1$ or $r_2$) is large, and the small-scale sector where both sides of the triangle are smaller than roughly $40h^{-1}$Mpc. In the later case, we expect these strong departures from GR due to the nonlinear evolution of modes, hence all models (and multipoles) have some degree of departure from GR on these scales, which are better appreciated for F4 and N1. In contrast, for the case of the isolated points, these regions only show up in particular multipoles and models, and one would need more realizations of each model to assess whether they are physical. Actually, if one constructs the HOD catalogs differently (for examples, by choosing the HOD parameters independently for each model realization, or by matching the full 2PCF between $\Lambda$CDM and the MG models) these highly-saturated points of Fig.~\ref{fig:3pcfmodels} on large scales may change {their} saturation, suggesting that some of them are not truly physical. However, under these different HOD prescriptions the overall patterns remain invariant, supporting evidence for using the 3PCF to test gravity. It is important to stress that using an equivalent expression to Eq.~(\ref{def:delta_xi}) for the 2PCF, we find that all models depart from GR with significance less than $\Delta\xi^{(2)}<0.4$ on scales between $10$ and $150h^{-1}$Mpc\footnote{Note that this range is different from the one within which the 2PCFs are used for the tuning of the HOD parameters.}. This suggests strong departures of the 3PCF compared with those found in the 2PCF; however, further tests are needed to check if the 3PCF offers additional or complementary information for testing the gravity models.

The 3PCF monopole is slightly suppressed with respect to the higher multipoles, although one should have in mind that each multipole contribution to the full 3PCF is weighted by the Legendre polynomials, hence the full 3PCF is usually dominated by the first few multipoles. Moreover, largest departures from GR are found for F4, F5, and N1, as expected (see Fig.   \ref{fig:3pcfmodels}), and are particularly noticeable in the quadrupole. For these three models, in the small-scale region of the plots (where both sides of the triangle are smaller than roughly $40h^{-1}$Mpc), the monopole and dipole show less clustering with respect to GR, manifested by a redder color, but an enhancement of clustering (bluer color) in the same regions for the higher multipoles. As expected, models F6 and N5 are generically closer to GR in the whole parameter space. However, for particular shape configurations, specially in the small-scale region, their departures from GR may be larger than for other models, as we will discuss later. Furthermore, for certain multipoles there is no monotonic trend in the different MG models in certain regions of $r_1$-$r_2$, e.g., at $r_1,r_2<50h^{-1}$Mpc the quadrupole is stronger in F4, F5,  but weaker in F6, than in GR, which is possibly a residual of the HOD tuning; this implies that understanding the impact of the uncertainties in the galaxy-halo connection will be important for using the 3PCF multipoles to test models.

We calculate particular cases of the full 3PCF using the Legendre basis to exemplify in a different way its complex dependence on the triangle sides $r_1$, $r_2$ and the opening angle $\theta_{12}$. When computing the full 3PCF for a specific bin pair, the signal shows convergence in the Legendre Series in most of the parameter space, except for the first couple of bins along the diagonal. The 3PCF reaches the highest values often when $\theta=0$, and in a few other cases when $\theta=\pi$, points where the Legendre polynomials also peak (taking on values $\pm 1$ for all $\ell$). The propagated error, coming from the standard deviation among realizations, grows towards $\theta=0$, implying that for small angles our measurement of the deviation with respect to GR (the $\Delta\xi^{(3)}$ variable) would shrink to zero while for larger angles it would be enhanced. Moreover, one should bare in mind that the strength of the signal in $\Delta\xi^{(3)}$ depends on a large difference in the 3PCFs and a small combined variance, hence tiny fluctuations in the variance using the multipole basis may lead to larger changes in $\Delta\xi^{(3)}$. As a result, this would either enlarge or reduce the significance of the departure from GR, or introduce a small shift on the angle $\theta$ where the departure occurs. Therefore, further realizations (and possibly higher multipoles) may be needed to increase the stability of our 
\revision{summary statistic} $\Delta\xi^{(3)}$ for large scales. While bearing this in mind, we explore some triangle configurations.

\begin{figure*}[!tb]
\begin{center}
{\includegraphics[width=0.445\textwidth]{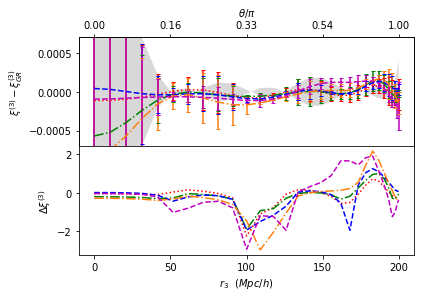}}
{\includegraphics[width=0.54\textwidth]{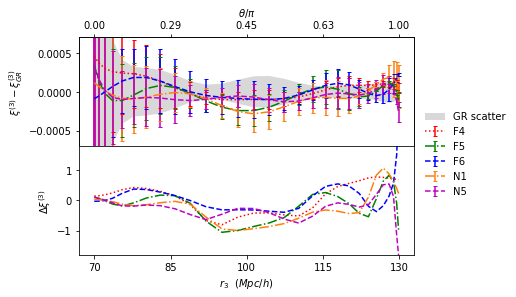}}
\caption{Full 3PCFs for $r_1=100h^{-1}$Mpc and $r_2=r_1$ (\emph{Left}) or $r_2=30h^{-1}$Mpc (\emph{Right}). The \emph{top panels} show differences of the MG 3PCF with respect to GR as a function of the third triangle side $r_3$, or equivalently, the opening angle $\theta_{12}$ between $r_1$ and $r_2$. The \emph{bottom panels} show the model differences using the analog of the variable $\Delta_\sigma\xi^{(3)}_{\ell}$ but for the full 3PCF. We use the first $\ell<10$ multipole moments in Fig.~\ref{fig:3pcfmodels} to construct the complete 3PCF.
}
\label{fig:3pcfBAO}
\end{center}
\end{figure*}

Figure \ref{fig:3pcfBAO} shows the dependence on the opening angle (or equivalently the distance $r_3$ to close the triangle) by the full 3PCF, for cases where at least one of the sides is around the BAO scale  ($r_1=100h^{-1}$Mpc). For the other side, $r_2$, we consider two cases: one along the diagonal ($r_2=r_1$), and another far from the diagonal ($r_2=30h^{-1}$Mpc). In both cases, the MG departures from GR show an oscillatory behavior, which is remnant of the multipole expansion, affecting both the signal and the error bars. In the diagonal case, the largest differences amongst models are found at certain values of the opening angle, with a particularly strong enhancement for the equilateral shape ($\theta=\pi/3$). Actually, for other bin pairs along the diagonal ($r_2=r_1$) away from the BAO scale, we often find stronger departures from GR around the equilateral shape. {In agreement with these findings, it is worth mentioning that for the diagonal and in the non-BAO region the authors of \cite{Sugiyama:2020uil} show that the 3PCF exhibits a trough in the gravitational shift and tidal force terms. In contrast, the off-diagonal cases show a non-trivial small-to-large scale mixing with no particular features over the different MG models.} Further realizations would be desirable to assess strong departures from GR. A careful exploration of the non-diagonal regions may shed more light on the correlations between overdensities and underdensities, which should strongly differ from GR when screening mechanisms are at work.

For all MG models, the small-scale region (where all triangle sides are $\lesssim 40h^{-1}$Mpc) contain important deviations from GR, as one would expect due to the nonlinear evolution of matter perturbations. This is clearly shown in Fig.~\ref{fig:3pcfmodels}, especially for the F4 model. Consequently, in the remaining part of this subsection we focus on the small-scale limit for both $r_1$ and $r_2$, where the overall variation with respect to GR is larger and the structure is richer. However, here we are reaching scales where the density of objects, together with the simulation resolution and the multipole truncation, disfavors the convergence of the Legendre series Eq.~(\ref{multi-dec}), especially in the diagonal case $(r_1 = r_2)$. Regardless of the convergence behavior, the multipole moments of the 3PCF nonetheless remain well-defined in and of themselves, hence one may still employ this decomposition truncated at a given $\ell$ as a discriminator to compare models. The behaviors are now rather particular to each pair of bins, with drastic changes from adjacent cells. This rapid variation implies the need for a careful choice of bin size, simulation resolution, and HOD properties. Again, further studies with higher simulation resolution, larger volume or more realizations of each model, and different HOD constructions may be required to fully assess the physical significance of the 3PCF structures due to MG on the shortest scales using the multipole expansion.

One may worry that the MG signal is degenerated with certain systematics, such as the  allocation algorithm used by the DESI instrument which artificially modifies the clustering \cite{Aghamousa:2016zmz}. However, given the complexity in the structures of the 3PCF, it is hard for other effects to mimic a true MG signal such as the ones reviewed here. For example, in the case of previously mentioned DESI  assignment effect, by using a subsample of the mocks described in Refs~\cite{Burden:2016cba,White:2013psd}, the 3PCF was computed and compared with the full sample without allocation with the yearly passes, and a very strong difference was found along the diagonal in the multipole basis for modes higher than the monopole (details will be found in a future publication). The signal is clearly stronger than the MG one, but with a very different structure from the one appreciated in Fig.~\ref{fig:3pcfmodels}. Therefore, it is advisable to consider the full 3PCF information, and not particular shapes, where there could be degeneracies.

\begin{figure}
\includegraphics[width=0.32\columnwidth,trim=0 0 0 0, clip]{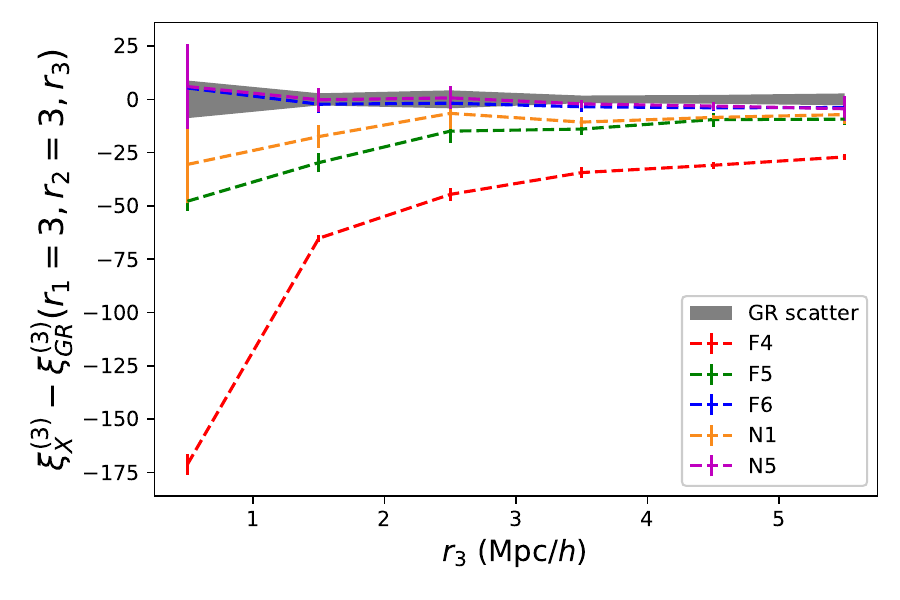}
\includegraphics[width=0.32\columnwidth,trim=0 0 0 0, clip]{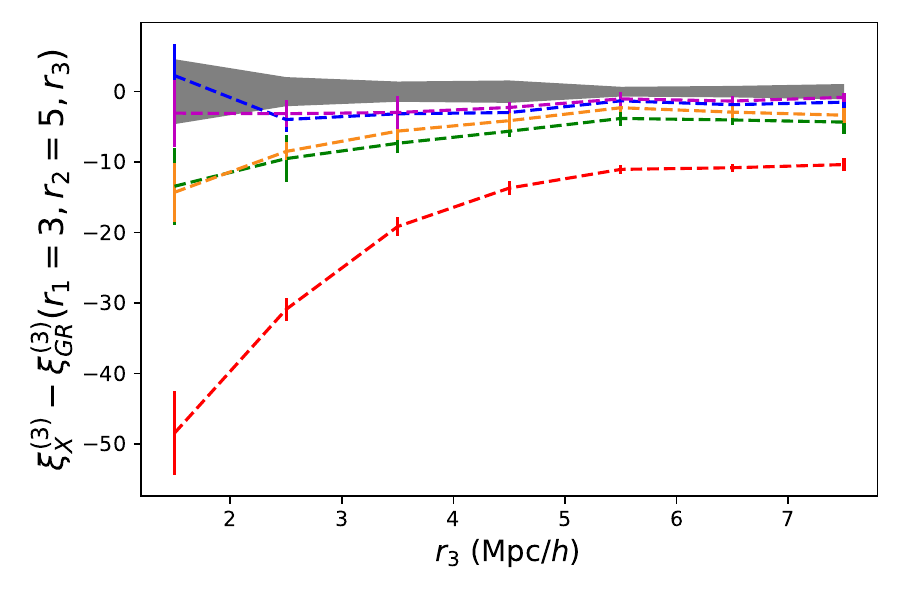}
\includegraphics[width=0.32\columnwidth,trim=0 0 0 0, clip]{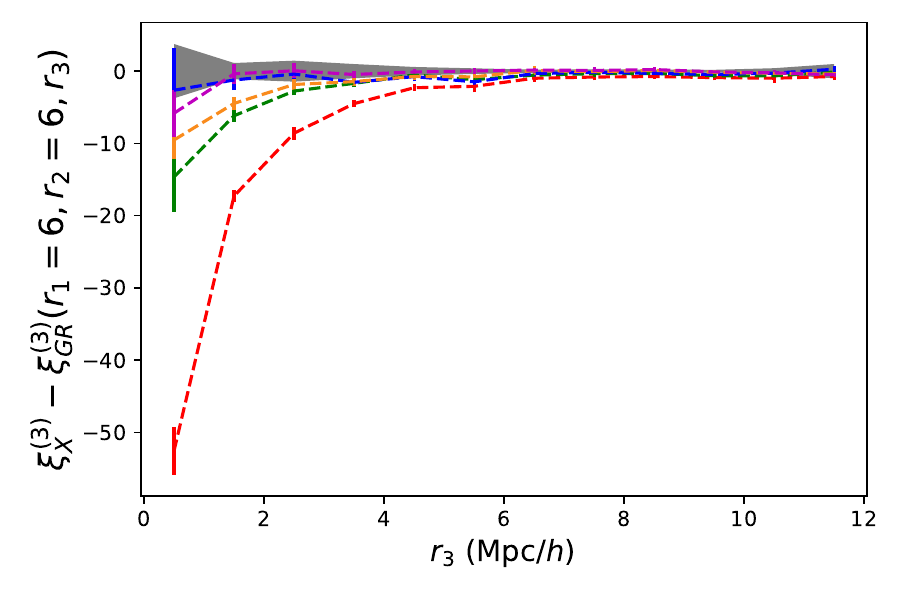}
\caption{The 3-point correlation function difference between GR and several modified gravity scenarios. {\em Left:} A triangular configuration with $r_1=r_2=3\Mpch$. {\em Center:} with $r_1=3\Mpch$ and $r_2=5\Mpch$. {\em Right:} $r_1=r_2=6\Mpch$. The shaded regions are the standard deviations of the five realizations of the HOD catalogs for GR. All results are for $z=0.5$.
}
\label{fig:3pcfz} 
\end{figure}

Given the poor convergence of the series on small scales, one may also try to run algorithms calculating the full 3PCF. Even though they are computationally more costly with respect to the code presented previously, for triangular configurations below $10\Mpch$, they can indeed be realistically run even for large surveys such as DESI. Therefore, following Ref.~\cite{2016A&A...592A..38S}, we focus on the small-scale 3PCF in redshift space, using a parallel {\sc kd}-{\sc tree} algorithm\footnote{We use the public code, KSTAT, https://bitbucket.org/csabiu/kstat2.} for efficient neighbor matching. We computed all triplets in various triangular configurations without using any approximation. Using a random catalog 20 times larger than the galaxy data, we measured all possible triangular configurations with scales $0<r_1<r_2<r_3<20\Mpch$. In Fig.~\ref{fig:3pcfz}, we show the differences between the 3PCFs of GR and the various MG models, focusing in three particular triangular configurations. One may notice that apart from the closest models to GR (F6 and N5), all others deviate significantly. The difference with respect to GR tends to decrease for larger values of $r_1,r_2$ and $r_3$, consistent with the discussion above on the 3PCF multipoles. We also checked these results for mock galaxy catalogs where the HOD parameters were tuned by matching the 3D galaxy 2PCFs of the different models, and found them to be very insensitive to the HOD details, which confirms that the 3PCFs encode important information useful for testing gravity. The big challenges in this regime are those shared by the 2PCF on the same small scales, namely to model the signal and to clearly understand the involved systematics.

The analysis in this subsection has demonstrated that the 3PCF of galaxies contains independent information to the traditional 2-point statistics, and has the capacity to distinguish MG from GR and break degeneracies on a wide range of scales. The use of the fast algorithm in \cite{2015MNRAS.454.4142S} allows for an efficient analysis of DESI-like surveys and simulations, with an easy-to-picture comparison of models based on the Legendre series coefficients. Moreover, based on this algorithm, some authors have constructed and optimized the anisotropic 3PCF in redshift space \cite{Friesen:2017acf,Slepian:2017lpm}, which could be used in large galaxy surveys such as DESI, and capture the RSD physics to further test gravity. Other ideas, such as the recent squeezed 3PCF construction \cite{Yuan:2018qek}, may be worth exploring to test gravity in galaxy surveys as well. Finally, the potential of the 3PCF goes beyond the models and conditions explored here, and may be used to characterize other possible deviations from $\Lambda$CDM and, given the richness in the pattern of differences, it may also provide a platform to understand other systematics such as fibre collision.

\subsubsection{Galaxy bispectrum}
\label{subsec:bispectrum}

The galaxy bispectrum is the counteraprt of the 3PCF of the galaxy field in Fourier space, and forms a Fourier transform pair with the configuration-space 3PCF that was discussed in Sec.~\ref{subsect:threept}. In principle, these two measures would carry the same information, but in practice this is not guaranteed as our analyses are restricted to a finite range of scales, and configuration- and Fourier-space statistics are impacted differently by systematic effects. In addition, modelling approaches of configuration- and Fourier-space quantities tend to differ and come with their own unique challenges, but models of the bispectrum have received more attention in the recent literature, putting them generally into a more mature state (see, e.g., \cite{HasRasTar0817,DesJeoSch1218,BosTar1018,EggScoSmi0619,Bose:2019wuz, Sugiyama:2018yzo, Sugiyama:2020uil}). For these reasons, the bispectrum can provide us with a valuable complementary point of view when studying the impacts of the MG dynamics in the $f(R)$ and DGP models, and our aim of this subsection is therefore to present measurements of the bispectrum from the various galaxy mock catalogs introduced in Sec.~\ref{sec:simulations}, both in real and in redshift space.

The bispectrum $B({\bm k}_1,{\bm k}_2,{\bm k}_3)$ is defined as the correlation of three density modes,
\begin{equation}
  \left<\delta({\bm k}_1)\delta({\bm k}_2)\delta({\bm k}_3)\right> \equiv (2\pi)^3\,B({\bm k}_1,{\bm k}_2,{\bm
    k}_3)\,\delta_D({\bm k}_1+{\bm k}_2+{\bm k}_3)\,,
\end{equation}
where the three wave vectors ${\bm k}_1$, ${\bm k}_2$ and ${\bm k}_3)$ form a closed triangle. In the absence of redshift-space distortions (or statistical anisotropies in general), the bispectrum is fully defined in terms of the length of the three triangle sides, otherwise two additional variables are required to capture the orientation of the triangle with respect to the LOS (or the direction along which isotropy is broken). We still lack a thorough understanding of whether particular triangle configurations are prominently affected in theories of modified gravity, and so in the following we are going to consider all possible triangles between the two extreme scales $k_{\mathrm{min}}$ and $k_{\mathrm{max}}$, given a specified bin width $\Delta k$ for each side. A detection of a strong configuration dependence could be regarded as compelling motivation to further investigate higher-order statistics, as this would allow us to disentangle the MG signal from other potential cosmological effects, which might be degenerate in two-point statistics and other alternative measures.

\begin{figure*}
  \centering
  \includegraphics[width=\textwidth]{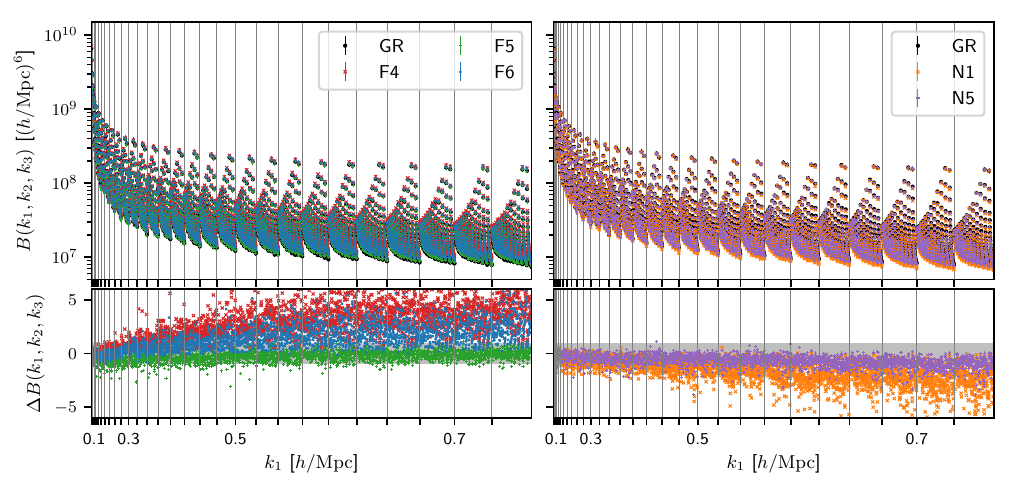}
  \caption{Upper panels: real-space bispectrum measurements for $f(R)$ models (left) and nDGP models (right) compared to GR. Each data point corresponds to one of the 2825 triangle configurations, ordered according to when they appear in the measurement loop. Vertical lines are spaced by the bin width $\Delta k \approx 0.025\,h/\mathrm{Mpc}$ and indicate the value of $k_1$, i.e., the largest triangle side. Lower panels: difference between modified gravity models and GR quantified in terms of the variable $\Delta B$ given in Eq.~(\ref{eq:bispectrum.DeltaB}). The grey band marks the $1$-$\sigma$ area where deviations are compatible with the combined standard deviation of GR and a given modified gravity model.}
  \label{fig:bisp_rs}
\end{figure*}

For our measurements we use an implementation of the bispectrum estimator presented in Ref.~\cite{Scoccimarro:2015bla} with fourth-order density interpolation on two interlaced cubic grids \cite{Sefusatti16} of $N = 380$ cells per side. Starting from $k_{\mathrm{min}} = 0.025\,h/\mathrm{Mpc}$, we loop through all configurations satisfying $k_1 \geq k_2 \geq k_3$ and $k_1 \leq k_2+k_3$ (the triangle closure condition) with bin width $\Delta k = 4 k_f \approx 0.025\,h/\mathrm{Mpc}$, where $k_f$ denotes the fundamental mode. We correct each measurement for Poissonian shot noise, finding that for the galaxy catalogs studied here the redshift-space bispectrum becomes shot noise dominated for scales $k > 0.75\,h/\mathrm{Mpc}$, which is why we choose that value for $k_{\mathrm{max}}$. This procedure yields a total of $2825$ distinct triangle configurations.

In the upper-left and upper-right panels of Fig.~\ref{fig:bisp_rs}, we show the raw real-space measurements for the $f(R)$ and nDGP models, compared to GR. The $x$-axis of the plots reflects the ordering of the triangle configurations corresponding to when they appear in the loop, while vertical lines and axis labels indicate the increasing values of $k_1$ from the left- to the right-hand side. The lower panels show the difference between the modified gravity models and GR in terms of the variable defined in Eq.~(\ref{def:delta_xi}), i.e.,
\begin{equation}
  \label{eq:bispectrum.DeltaB}
  \Delta B(k_1,k_2,k_3) \equiv \frac{B_X(k_1,k_2,k_3) -
    B_{\mathrm{GR}}(k_1,k_2,k_3)}{\sqrt{\sigma^2_X(k_1,k_2,k_3) + \sigma^2_{\mathrm{GR}}(k_1,k_2,k_3)}}\,,
\end{equation}
where $\sigma_X$ and $\sigma_{\mathrm{GR}}$ are the standard deviations obtained from the five HOD catalogs for either a MG model or GR. A general trend revealed by the plots is an enhancement of the bispectrum signal for the $f(R)$ models relative to GR, whereas the modified dynamics in nDGP lead to a suppression. These effects are growing towards smaller, more non-linear scales, and the deviations from GR are strongest for the F4 and N1 models, for which $\Delta B$ takes values of $\sim 3$ and $-1.8$ when averaged over all configurations in the interval $k_1 \in [0.3, 0.75]\,h/\mathrm{Mpc}$ where the effect is most significant. However, even for the F6 model we get $\Delta B \sim 1.6$ in the same interval, which is a factor $\sim 25$ larger than the analogue quantity we would instead obtain for the power spectrum. The behaviour of the F5 model, on the other hand, is qualitatively different from F4 and F6, and the deviations from GR are noticeably smaller. We do not find the same trend when measuring the bispectrum from the \revision{underlying halo catalogs, where we have used either all identified halos in the simulation boxes, or the 340,000 most massive ones, so that their number density is kept fixed across the various gravity models \newrevision{(see Fig.~\ref{fig:halo_bispectra})}. This implies that the difference between F5 and F4/F6 is likely driven by the HOD modelling.} It is interesting to note that the 3PCF analysis in Sec.~\ref{subsect:threept} does not \revision{give an indication of suppressed deviations between GR and F5 compared to F6 (see, e.g., Figs.~\ref{fig:3pcfmodels} and \ref{fig:3pcfz}), while our findings here (sign and relative strength of deviations) are qualitatively consistent with the measurements of the reduced cumulants in Sec.~\ref{subsec:hierarchical_clustering} below (in particular, cf. Fig.~\ref{fig:s345_compare_gal}).} A more detailed future investigation is needed to check if this non-monotonic behavior is sensitive to the HOD model employed. Since the galaxy 2PCFs in the different models have been matched to each other by the HOD tuning, the strong difference in their bispectrum hints that the latter can offer independent information of and constraints on the MG dynamics, consistent with the findings of the 3PCF subsection above.

\begin{figure*}
  \centering
  \includegraphics[width=\textwidth]{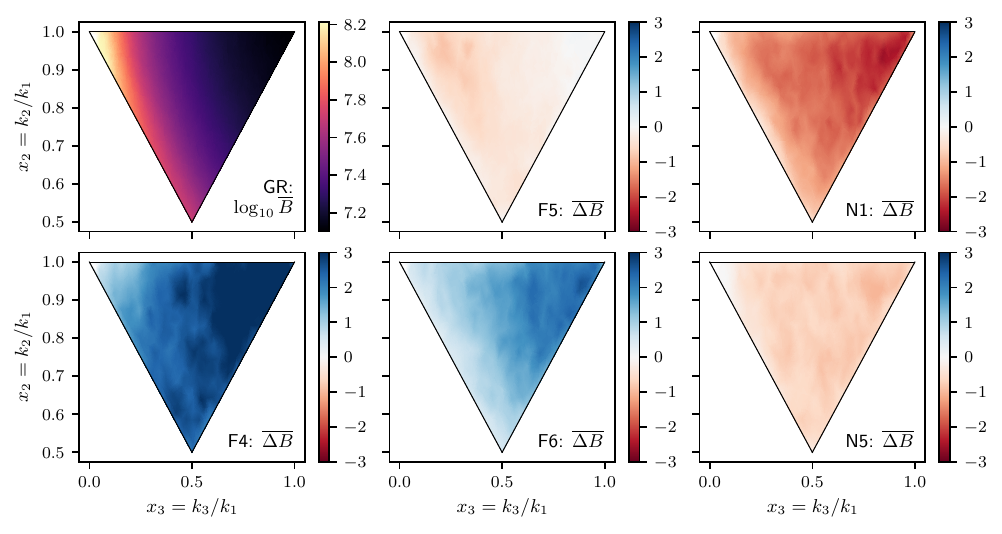}
  \caption{Configuration dependence of the real-space bispectrum, obtained by integrating the raw measurements
    over $k_1$ (see Eq.~\ref{eq:bispectrum.Bx2x3}) between $k_1 = 0.3\,h/\mathrm{Mpc}$ and
    $0.75\,h/\mathrm{Mpc}$. Configurations outside of the triangular area are forbidden according to the closure
    condition and the top left, top right and bottom corners correspond to squeezed, equilateral and folded
    configurations, respectively.}
  \label{fig:bisp_x2x3_rs}
\end{figure*}

In order to highlight the configuration dependence of the difference between modified gravity models and GR, we average the bispectrum (or $\Delta B$) over the largest triangle side $k_1$, while keeping the ratios $x_2 = k_2/k_1$ and $x_3=k_3/k_1$ fixed:
\begin{equation}
  \label{eq:bispectrum.Bx2x3}
  \overline{B}(x_2,x_3) = \frac{1}{k_u-k_l} \int_{k_l}^{k_u} B(k, k x_2, k x_3)\,\mathrm{d}k\,,
\end{equation}
where we again use $k_u = 0.75\,h/\mathrm{Mpc}$ and $k_l = 0.3\,h/\mathrm{Mpc}$ for the upper and lower limits, respectively, which yields the results shown in Fig.~\ref{fig:bisp_x2x3_rs} --- note that in case of GR we plot $\overline{B}(x_2,x_3)$, while all other panels display the difference $\overline{\Delta B}(x_2,x_3)$. The overall amplitude and sign of $\overline{\Delta B}$ can readily identified from Fig.~\ref{fig:bisp_rs}, but now we see more clearly that for the three models with the strongest deviations from GR, i.e., F4, N1 and F6, differences are maximized for configurations that are mostly equilateral (towards $x_2 = 1 = x_3$), whereas collinear configurations ($k_1 = k_2 + k_3$ or $x_2+x_3 = 1$), which include squeezed and folded triangles, tend to be less affected. Again, F5 is qualitatively different from the other two $f(R)$ models with a preference for nearly squeezed configurations, and for N5 we do not identify any clear configuration dependence.

As discussed in Sec.~\ref{subsect:RSD}, modified gravity not only impacts the clustering of galaxies, but also their infall and virial velocities, and as such alters the redshift-space distortions of clustering statistics. These distortions are present in any real measurement, so it is interesting to explore the difference between GR and MG models for the bispectrum in redshift space. For that we use the distant observer approximation, adopting the same LOS for all galaxies in the simulation volume, and focus on the bispectrum monopole, which averages over all orientations of the triangle with respect to the LOS, i.e.,
\begin{equation}
  B^{(s)}_0(k_1,k_2,k_3) = \int_0^{2\pi} \frac{\mathrm{d}\phi}{2\pi} \int_0^1 \mathrm{d}\mu\,B^{(s)}({\bm
    k}_1,{\bm k}_2,{\bm k}_3)\,,
\end{equation}
where $\mu = \cos{\theta}$, and $\phi$ and $\theta$ are the angles describing the orientation. \newrevision{Fig.~\ref{fig:triangle_plots_bisp_zs} in Appendix \ref{App:C} offers a quick visualization of the configuration dependence of the redshift-space bispectrum monopole, i.e., the $z$-space counterpart of Fig.~\ref{fig:bisp_x2x3_rs}.}

\begin{figure*}
  \centering
  \includegraphics[width=\textwidth]{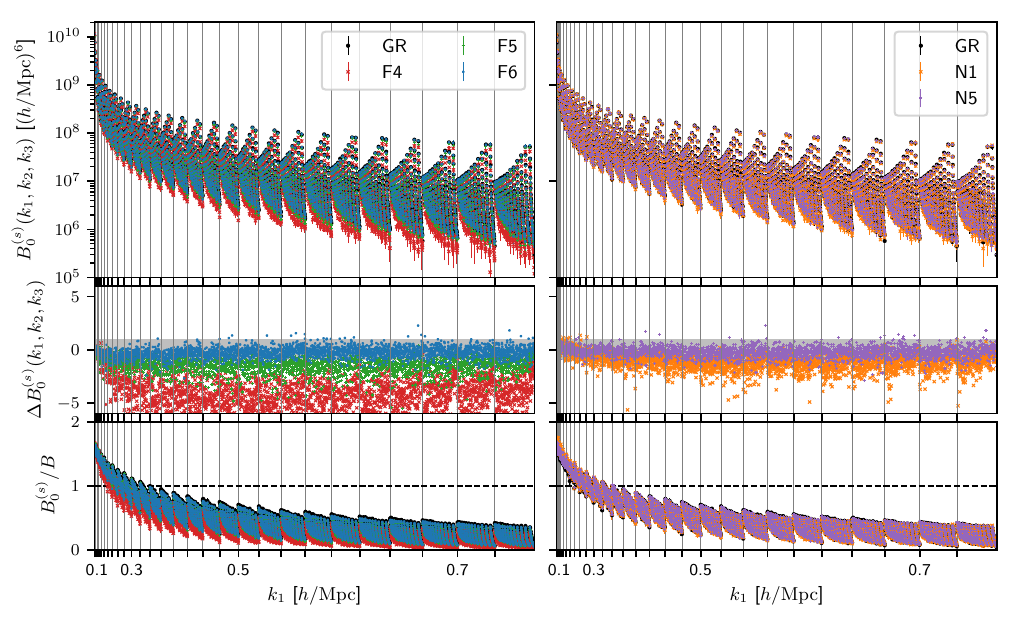}
  \caption{Measurements of the bispectrum monopole in redshift-space. The panels in the upper two rows are
    the same as those in Fig.~\ref{fig:bisp_rs}, while the bottom panels show the ratio between the
    redshift-space monopole and the real-space bispectrum.}
  \label{fig:bisp_zs}
\end{figure*}

In Fig.~\ref{fig:bisp_zs} we present the results of these measurements in a very similar fashion to Fig.~\ref{fig:bisp_rs}. Compared with the real-space case, the behavior of $\Delta B^{(s)}_0$ for the $f(R)$ models has changed qualitatively: a suppression now takes the place of the previously enhanced bispectrum relative to GR with the amplitude of deviations being strongest for F4, followed by F5. They are weakest for the F6 model, which is mostly consistent with GR within the combined 1-$\sigma$ scatter of the GR and F6 measurements. \revision{Furthermore, we find that the deviations from GR are no longer maximized for equilateral triangles, but display an overall reduced configuration dependence with only slightly larger differences for collinear shapes.} To understand the origin of this change in behaviour, we plot the ratio of the redshift-space monopole and real-space bispectrum in the bottom panels of Fig.~\ref{fig:bisp_zs}, which show an increased redshift-space signal on large scales that becomes heavily damped for triangles involving small-scale sides. This is simply a reflection of the two expected effects: on large scales the infall velocities of galaxies lead to a (configuration-space) squashing and thus enhanced clustering. On small scales, however, the FoG effect smears out the galaxy positions, which translates into a damping of the signal in Fourier space. In case of the $f(R)$ model, the FoG effect is significantly stronger than in GR, particularly for F4, \revision{which reverses the real-space enhancement in clustering and explains the reduced configuration dependence seen in the GR deviations of $B^{(s)}_0$. The fact that there is no appreciable dependence on triangle shape in differences of the ratio $B^{(s)}_0/B$ either, and that we find very similar differences in the damping of the power spectrum, further points to an increased velocity dispersion in the $f(R)$ models. We also note that in case of F6 the real-space enhancement and the FoG damping have nearly equal but opposite amplitudes, which makes the signal consistent with GR (a similar effect has been observed previously for the power spectrum, e.g., \cite{Bose2017}), and we have checked that this cancellation happens independently of whether we include satellite galaxies in the HOD sample or not. For the nDGP models, on the other hand, the ratio $B^{(s)}_0/B$} is mostly identical to GR on small scales, but we find a slight enhancement on large scales because of the increased growth rate, as already noted in Sec.~\ref{subsect:LRSD}. In total this means that the GR deviations of the bispectrum monopole tend to be similar to those observed in the real-space case, but with somewhat diminished significance because the reduced overall signal raises the relative importance of the shot noise.

In summary, our results indicate that the bispectrum can be a potentially powerful and complementary measure for discriminating theories of modified gravity. For the HOD galaxy catalogs used in this work we have found that the real-space clustering probed by the bispectrum can deviate significantly from GR on scales $k > 0.3\,h/\mathrm{Mpc}$ and these deviations tend to be at least a factor of a few larger than what an equivalent analysis implies for the power spectrum in all cases. In addition, we have seen that the difference between GR and modified gravity displays a clear dependence on triangle shape, which can further be exploited to break degeneracies with other cosmological and systematical effects. In redshift space, the small-scale bispectrum is heavily damped by the Finger-of-God effect, which leads to an increased impact of shot noise and thus less significant deviations from GR. This however is partly remedied for $f(R)$ models, which additionally differ from GR by having higher velocity dispersion.

Finally, we stress that all results shown here are based on uncertainties that correspond to the volume of a single simulation box, $\sim1\,(h/\mathrm{Gpc})^3$. DESI will observe a much larger volume and, moreover, will have a higher number density of tracers, so that we expect all of our reported deviations from GR to grow. However, we also  note that the analysis here is only based on a small number of realizations and a single HOD prescription. A future, more detailed, study is needed to assess the robustness of our results and to which degree they depend on the adopted HOD model. In addition, differentiating between GR and a given model of modified gravity using measurements such as those above will critically depend on our ability to make robust predictions in the non-linear regime. While recent progresses on the modelling of the bispectrum have been reported for chameleon and Vainshtein type models \cite{BosTar1018,Bose:2019wuz,Becker:2020}, further developments will be vital for a successful application.

\subsubsection{Hierarchical clustering}
\label{subsec:hierarchical_clustering}

The full information of all one-point correlations of cosmic density field (of matter or galaxies) is encoded in the shape and amplitude of the density probability distribution function, $pdf(\delta)$.  If the density field is a Gaussian random field, then the $pdf$ can be described simply by two numbers: the mean $\overline{\delta}$ and the variance $\sigma^2$. However, {as we have seen above,} in the gravitational instability scenario, the growth of cosmic structures gives rise to significant deviations of the evolved matter distribution from the initial Gaussianity. {An alternative 
\revision{summary statistic} to accurately capture this information is} the growth of higher-order central moments of the $pdf$ \cite{1980Peebles,Fry1984a,ber1992}. 

If we sample N-point correlations in a big enough volume, so the {\it fair-sample} hypothesis is satisfied, then  the central moments can be expressed as volume-averaged correlation functions \cite{1980Peebles}:
\begin{equation}
\label{eqn:xi_volume_av}
\overline{\xi}_n(R)\equiv\langle\delta_R^n\rangle_{\rm c} = \int \textrm{d}^3 x_1\ldots\textrm{d}^3 x_n\xi(\mathbf{x_1}\ldots\mathbf{x_n})W(x_1/R)\ldots W(x_n/R)\,.
\end{equation}
Here $W()$ is the smoothing window (usually a top-hat or a Gaussian) and $R$ is the smoothing scale.  By $\langle\delta_R^n\rangle_{\rm c}$ we denote here the $n$-th cumulants, which can be expressed in terms of  central moments, $\langle\delta^n\rangle$, of the underlying density $pdf(\delta_R)$. The first few cumulants are given by
\begin{eqnarray}
\label{eqn:cumulants}
\langle\delta\rangle_{\rm c} &=& 0,\,\,\textrm{(the mean)}\nonumber\\
\langle\delta^2\rangle_{\rm c} &=& \langle\delta^2\rangle\equiv\sigma^2,\,\,\textrm{(the variance)}\nonumber\\
\langle\delta^3\rangle_{\rm c} &=& \langle\delta^3\rangle,\,\,\textrm{(the skewness)}\nonumber\\
\langle\delta^4\rangle_{\rm c} &=& \langle\delta^4\rangle - 3\langle\delta^2\rangle_{\rm c}^2,\,\,\textrm{(the kurtosis)}\nonumber\\
\langle\delta^5\rangle_{\rm c} &=& \langle\delta^5\rangle - 10\langle\delta^3\rangle_{\rm c}\langle\delta^2\rangle_{\rm c}\, \text{(the hyperskewness)}  \,.
\end{eqnarray}
In reality it is very hard to reliably estimate the full density $pdf$ from a set of discrete field tracers (such as galaxies). However, it was established that the very first few cumulants already provide a robust characterization of the underlying density field and associated 1-point correlation statistics \cite{ber1992,ber1994B}. This was extensively exploited for the benefit of cosmological analysis, since the central moments of the $pdf$ can be readily extracted from observations.  In this context, the particularly useful concept is the {\it reduced cumulants } or the so-called {\it hierarchical amplitudes} $S_n$'s, which are given by \cite{Juszkiewicz1993,Lokas_kurt}  
\begin{equation}
\langle\delta^n\rangle_{\rm(c)} = S_n\langle\delta^2\rangle_{\rm(c)}^{n-1} = S_n\sigma^{2n-2}. 
\end{equation}
Considerations from cosmic perturbation theory (PT) have shown that in the classical gravitational clustering scenario the higher-order cumulants are strong functions of the field variance, $\sigma^2$. By rescaling the cumulants using the variance in a proper power to define the reduced cumulants, one can remove most of this dependence, and what is left is a statistic that is very sensitive to the higher-order and nonlinear effects of gravitational clustering. Calculations from {perturbation theory}  \cite{ber1994,Juszkiewicz1993} have shown that for smoothed fields (such as count-in-cell for galaxy counts) the reduced cumulants become weak functions of the smoothing scale $R$ and this effect is accurately captured by various combinations (set for a given cumulant order) of the logarithmic slope of the mass {field} variance, the so-called gamma-factors defined as
\begin{equation}
    \label{eqn:gamma_factors_PT}
    \gamma_n(R)\equiv \frac{\textrm{d}^n\log\sigma^2_M(R)}{\textrm{d}\log^n R}\,.
\end{equation}
\revision{For the first three cumulants the PT prediction yields}
\begin{eqnarray}
S_3=\frac{34}{7}+\gamma_1\,,\\
S_4=\frac{60712}{1323}+\frac{62}{3}\gamma_1+\frac{7}{3}\gamma_1^2+\frac{2}{3}\gamma_2\,,\\
S_5=\frac{200575880}{305613}+\frac{1847200}{3969}\gamma_1+\frac{6940}{63}\gamma_1^2+\frac{235}{27}\gamma_1^2+\frac{1490}{63}\gamma_2+\frac{50}{9}\gamma_1\gamma_2+\frac{10}{27}\gamma_3\,.
\end{eqnarray}
\revision{As we can see, the logarithmic slope of the matter variance field affects the predicted amplitude of cumulants. We expect that the effect of screening enhancing the nonlinearity of the matter filed, as illustrated, e.g., by the bottom panel of Fig.~\ref{fig:pofk}, should be reflected in the changed values of $\gamma$ factors.}

The above properties of the reduced cumulants in principle make them a very promising and suitable tool for testing the nature of gravitational instability \cite{Bernardeau:2001qr,SkewBAO} or the nature of the initial conditions \cite{Luo1993,Chodorowski1996,Durrer2000}. For the same reasons it was put forward that using $S_n$'s can be beneficial for testing GR and MG on cosmic scales \cite{Gaztanaga2001,Hellwing_npoint,Hellwing2013,Hellwing2017}.
\begin{figure*}[!tb]
\begin{center}
{\includegraphics[width=0.6\textwidth]{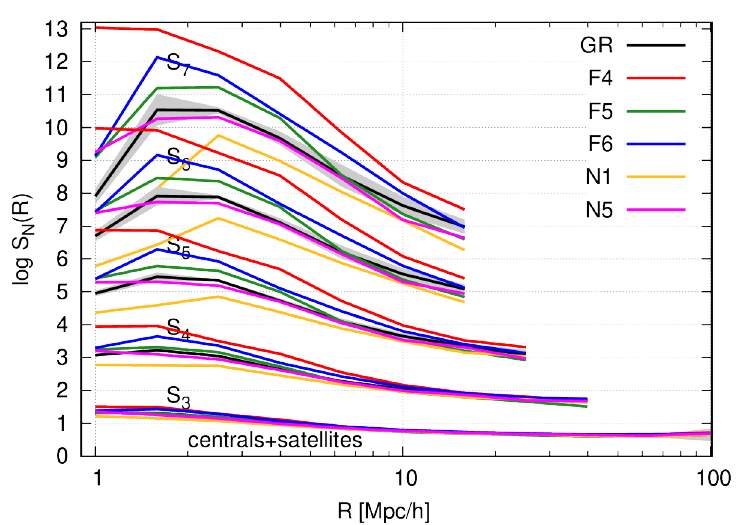}
}
\caption{The hierarchy of galaxy clustering reduced cumulants $S_n$'s for all models. The shaded regions represent $1\sigma$ scatters around the GR ensemble mean. The groups of lines from the bottom ($S_3$ or skewness) to the top ($S_7$) correspond to increasing cumulant order.
}
\label{fig:Sns_at05_compare}
\end{center}
\end{figure*}

\revision{In what follows we focus on the galaxy catalogs in configuration space split into two samples---all galaxies and centrals only---to be able to compare the magnitude of differences and scales at which they are attained between the samples. This will be useful for understanding the impact of satellites and one-halo term for the reduced cumulants.}

We use the count-in-cell method to measure the central moments of the HOD galaxy distributions, employing the algorithm presented in \cite{Hellwing2013,Hellwing2017}. The count-in-cell method gives good results when the expected number counts $\langle N\rangle$ in a cell of a given size is large. Given the number density of our HOD catalogs, $n_g\sim3.2\times10^{-4}(\Mpch)^{-3}$, this is guaranteed for a large enough smoothing scale, $R \gtrsim 8\Mpch$. At smaller scales we might expect to have a significant contamination by shot noise. To overcome this we performed the shot-noise correction for all our central moments. The  method uses the moment-generating function of the Poisson model to calculate the net contribution by discrete noise (see details in \cite{Gaztanaga1994,Hellwing_npoint}).
On the other hand, the higher-order moments are severely affected by finite volume effects \cite{Colombi1994}. The effect scales with the order $n$ of the moment. For the simulation box size we used, the cosmic variance becomes significant at $R\sim60\Mpch$ for the skewness ($S_3$), while for $S_7$ this scale drops to only $\sim 10\Mpch$. We expect that in a DESI-like survey both the shot noise and the cosmic variance will affect the measured cumulants to much less extent than can be noticed in our simulations.

In Fig.~\ref{fig:Sns_at05_compare} we plot the full hierarchy of reduced cumulants from $n=3$ (the skewness) up to seventh order ($S_7$). This gives us a general impression of the shapes and the amplitudes of these statistics. The solid black lines mark the fiducial GR case, while associated shaded regions delimit $1\sigma$ dispersion around the mean from the ensemble of five GR realizations. The scatters for all the other models have an amplitude and scale dependence that is very close to that of the GR case, and for clarity we do not plot them. This figure already illustrates a couple of interesting points. First, we can observe that the higher-order moments (starting from $n=5$) show a clear downturn of amplitudes at $R\lesssim2.5\Mpch$. This is possibly caused by the limitations of the HOD scheme, which is not designed to accurately capture the galaxy clustering in the one-halo term regime \cite{Kravtsov2004,Yang2005,Yang2008}. We shall not attempt to study this in greater details, since this is beyond the scope of this work. Reassuringly, the scale and the shape of this effect seems to be very similar in all the inquired models with F4 being exception, where the effect seems to much milder. Since this regime is strongly affected by both sampling noise and HOD accuracy, we shall stop at noting the exceptional behavior of F4 here. 

\begin{figure*}[!tb]
\begin{center}
{\includegraphics[angle=0, width=0.95\textwidth]{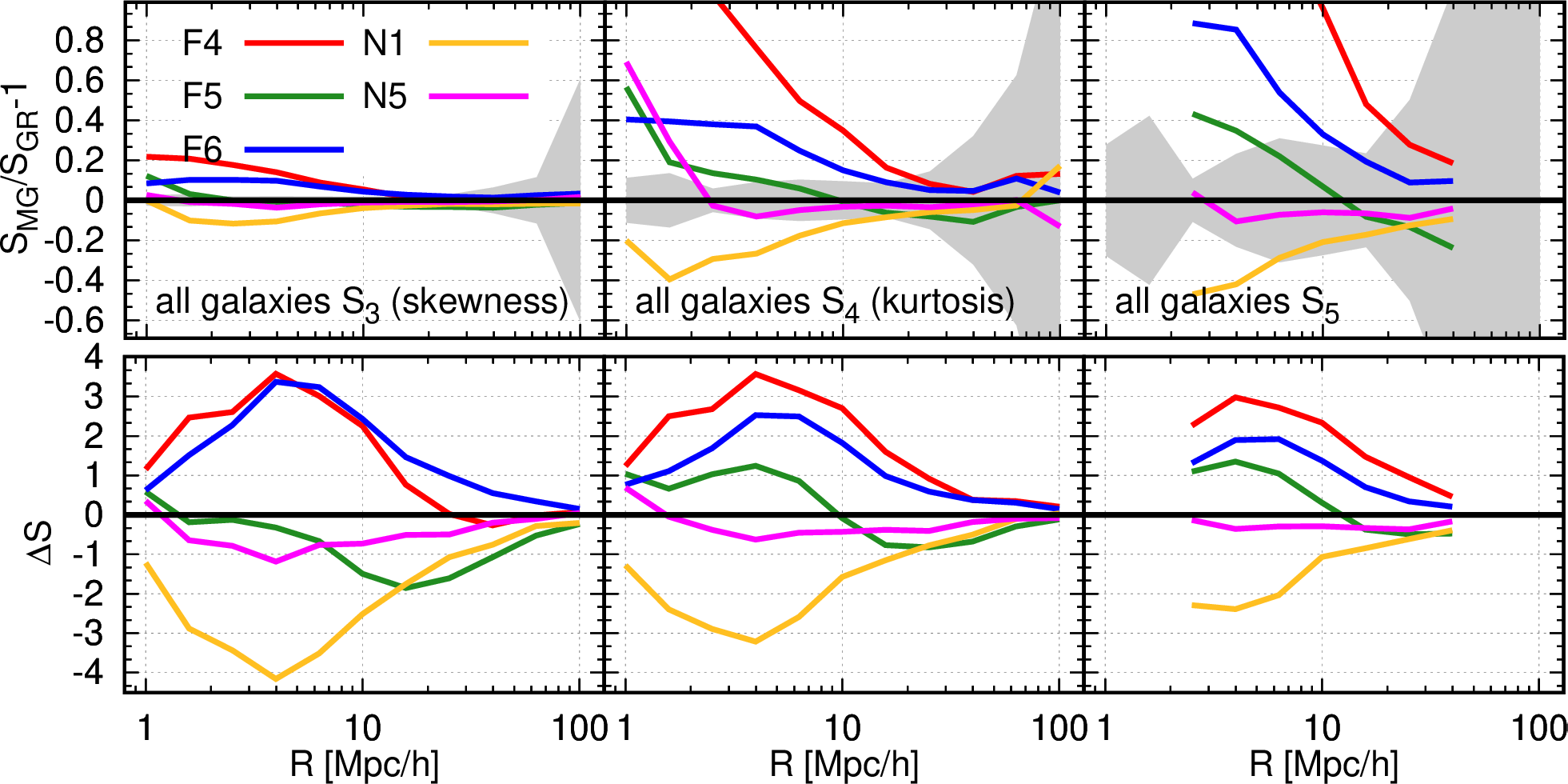}
}
\caption{Top row: The relative differences of the skewness (left panel), kurtosis (central panel) and hyperskenwess (right panel) taken with respect to the fiducial GR case. The shaded regions mark the error round the ratio of ensemble averages. \textcolor{black}{Bottom row: the equivalent error-weighted differences, $\Delta S$, similarly defined as in Eq.~\eqref{eq:bispectrum.DeltaB}, for each statistics.}}
\label{fig:s345_compare_gal}
\end{center}
\end{figure*}

\begin{figure*}[!tb]
\begin{center}
{\includegraphics[angle=0, width=0.95\textwidth]{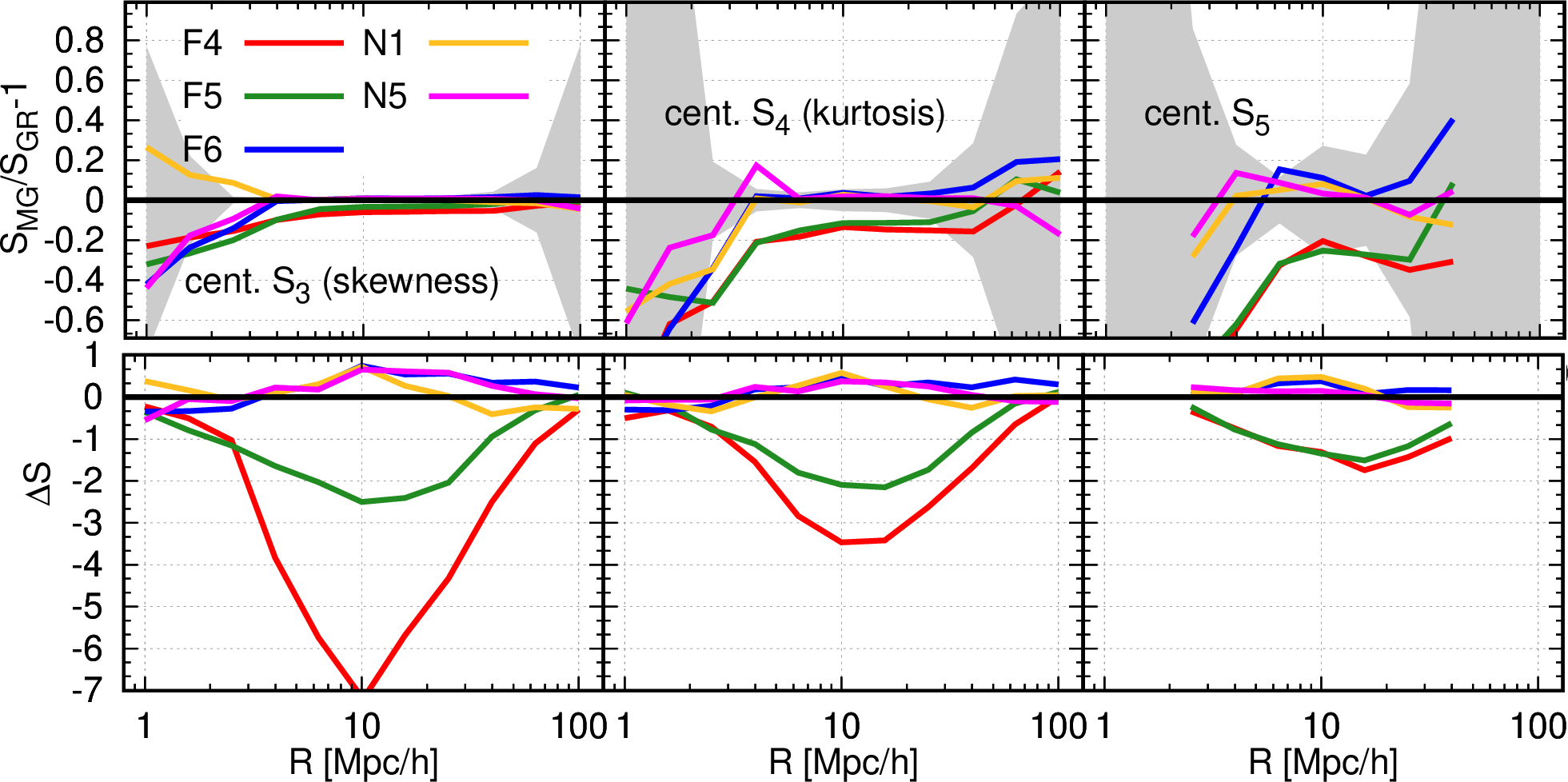}
}
\caption{The same as the previous figure, but for central galaxies only.}
\label{fig:s345_compare_centrals}
\end{center}
\end{figure*}

Another observation is that the deviations from the GR grow with the increasing statistic's order $n$. Therefore, the higher-order cumulants would appear as better observables to discriminate between the models. Alas, the stronger MG signal comes at a price of greatly increased scatter, which is illustrated by much larger $1\sigma$ contours at small and large scales for higher-order cumulants. In practice, we have assessed that reduced cumulants from $S_6$ and higher are subjects to a scatter too large to be suitable for our analysis. Thus in our further considerations we will limit ourselves to only the first three reduced cumulants.

\revision{Figure \ref{fig:s345_compare_gal} shows the differences for $S_3$, $S_4$, and $S_5$ in the ``all galaxies'' sample. The upper panels show the relative difference with respect to GR, $S_n/S_n^{\rm GR}-1$, for $n=3$, $4$, $5$, as a function of the smoothing scale $R$ The shaded regions delimit the $1\sigma$ scatters of this ratio.
The bottom row illustrated error-weighted cumulants difference, $\Delta S$ defined in an analogues way as it was for the bispectrum, $\Delta B$, in Eq.~\eqref{eq:bispectrum.DeltaB}}.
For nearly all scales, F4 appears to be the model deviating most strongly from GR, with the signal reaching $\sim20\%$ for the skewness, $\sim 100\%$ for the kurtosis, and up to $\sim200\%$ for $S_5$ at $R\lesssim10\Mpch$. The next two standing-out models are N1 and F6: for all three $S_n$'s their deviations from GR show similar shape and amplitude, but differ by the signs N1 promotes lower values of amplitudes than GR, while F6, like the other $f(R)$ models, fosters higher values. The absolute effects of these two models are $\sim10\%$ for the skewness, up to $40\%$ for the kurtosis and even $\gtrsim50\%$ for $S_5$. The remaining two MG models (F5 and N5)
exhibits considerably weaker deviations from GR, with N5 being mostly consistent with the latter.

The facts that F4 and N1 deviate most strongly from GR while N5 is close to it are as expected given that the screening effect is weaker in the former but stronger in the latter. The behavior of F6 and F5, on the other hand, is opposite to expectation, since the fifth force is more efficiently suppressed in F6. However, we note that, due to the additional complication of HOD modelling, the properties of the galaxy field may not follow exactly those of the matter field, making a physical interpretation more difficult: for example, Fig.~\ref{fig:HOD} shows that the galaxies may be hosted by different halo populations in these two models. Also note that the exceptional behavior of F5 here is consistent with the findings of the real-space galaxy bispectra in the previous subsection.

\revision{In Fig.~\ref{fig:s345_compare_centrals}, we present statistics analogues to the one just discussed above, but for central galaxies only. This plot demonstrates that, for centrals, the typical differences between GR and our MG models are much smaller than what we observed for the ``all galaxies'' catalogs; the associated scatter is also larger, as can be strongly visible at small scales ($R\leq 3h^{-1}{\rm Mpc}$). This might be somewhat surprising given that our HOD catalogs have small satellite fractions, $\simeq11\%$. However, due to halo-exclusion effect at these scales, the satellites make a significant contribution,
reducing the shot noise. In contrast to the ``all galaxies" sample, here F6 is characterized by minimal deviations from GR. Also, 
N1, which for the ``all galaxies" sample shows significant departures from GR, deviates weakly from GR for central galaxies. Only F4 and F5 remain as potentially detectable models. This results underlines the importance of the nonlinear regime described by the 1-halo term, as well as an accurate model for the halo-galaxy connection, for distinguishing between GR and MG scenarios. In future works, a more careful and detailed approach should be used for modelling galaxy satellites and their impact on the higher-order clustering statistics.}

To summarize, our results indicate that there is a great potential in using the reduced cumulants of galaxy distribution as strong discriminators of MG models.
\revision{This is provided we consider all galaxy sample that include satellites, which are necessary to probe the clustering in 1-halo regime.}
The relative differences we have found are quite large, from $\sim20\%$ in the case of the skewness to $\gtrsim100\%$ for $S_5$. Indeed, while their error bars are much larger, it seems that both $S_4$ and $S_5$ offers a better prospect for testing MG and GR, since the amplitude of the signal is much higher. There are, nevertheless, uncertainties associated with our analysis which are of a statistical nature, due to the relatively low number densities and small volume of our mock galaxy catalogs. For a DESI-like survey we can expect a 20 times larger volume to be probed and also higher number density of tracers \cite{Aghamousa:2016zmz}, which should reduce the statistical errors by a factor of at least 3 (see, {e.g.,} \cite{Bose2017} for more discussion). To match the specifications of such future data, larger-volume and higher-resolution simulations are needed to more accurately make the theoretical predictions for $S_n$.

In addition to the statistical uncertainties, like the other 
\revision{summary statistics}, there are a range of theoretical and observational systematics which should be better understood. One example is the HOD method to construct mock galaxy catalogs. We have repeated the analysis using the mock galaxy catalogs where the HOD parameters were tuned to match the real-space 3D galaxy 2PCFs amongst the different models, and the behaviors of $\Delta S_n$ are quite different, with much smaller deviations from GR in all models except N1. This suggests that galaxy modelling is an important uncertainty which should be addressed. A more accurate but also more expensive way to achieve this is to use hydrodynamical simulations which evolve the baryonic component as well. The latter can indeed be distributed in a different way than smooth dark matter at small scales \cite{skewness_barions}. Recently, however, there is a growing consensus that while nonlinear baryonic physics is crucial for understanding and modeling of galaxy internal properties, its impact on galaxy velocities and positions is very mild \cite{vanDaalen2014,Hellwing2016}. This offers a way to test whether simplistic HOD modelling like the one used here would cause a substantial bias in the predicted $S_n$'s.

Another potentially important source of systematic effects lies in the fact that we have neglected the effects of redshift space distortions in our analysis. In reality, the measured line-of-sight coordinate of a galaxy is affected by its peculiar velocity, and some studies have found that in MG scenarios there exists a degeneracy between increased spatial clustering caused by the fifth force and enhanced clustering damping precipitated in redshift space by enhanced dynamics (see, {e.g.}, \cite{Bose2017}). The situation can be to a large extent remedied by the fact that both the nominator and denominator of the reduced cumulants ratios are affected by the RSD to a similar magnitude and the overall effect is largely canceled out \cite{Hivon1995}. This, together with the fact that the damping is limited to only small scales, suggests that at the intermediate scales of $10\lesssim R/(\Mpch)\lesssim 60$ the signal we have measured is not severely affected by RSD effects. Therefore, we hope that with detailed studies using future simulations we will be able to more accurately quantify these effects and extract genuine MG signals on those scales.

While the MG signal unveiled in the clustering amplitudes appears to be strong and significant, in reality, the observational data is a subject of various selection effects. To foster robust data analysis one needs to model specific survey's radial (redshift), angular and luminosity (magnitude) selection functions. Since in our analysis we did not attempt to model any of such effects, our results 
should be taken as an optimistic best-case scenario for an idealized perfect survey. On the other hand, we are dealing here with volume-weighted central moments estimated by count-in-cells. In contrast to pair-weighted 
statistics such as 2PCF, the volume-weighting makes the central moments less prone to biases induced by specific selections (especially radial and angular). As shown for example by \cite{Cappi2015}, for the case of the VIPERS survey, careful modelling of the survey selection functions admits for a robust estimation of central moments and related hierarchical amplitudes. Thus, we expect that for a survey with a footprint like DESI, the observational selection effects, when modelled accurately, should not hinder the potential for obtaining strong MG constraints using the reduced cumulants of galaxy clustering.

\subsubsection{The Minkowski functionals of the density field}
\label{subsect:MF}

As discussed in \S~\ref{subsect:threept}, \S~\ref{subsec:bispectrum} and \S~\ref{subsec:hierarchical_clustering}, because the observed density field is not perfectly Gaussian due to the nonlinear evolution of the cosmic structure and possibly also to primordial non-Gaussianity, one cannot extract all the information from galaxy surveys that may be relevant for cosmological analysis using the commonly used two-point correlation function or power spectrum. It is true that higher-order correlation functions, or the corresponding high-order power spectra (referred to as N-point statistics hereafter), can be complementary, given the the computational challenges to measure those quantities from observations and theoretical difficulty to model them accurately, it is worthwhile to consider other probes that encode the same information. An example of such probes is cosmic voids (see \S~\ref{subsect:rsdvoid} and \S~\ref{subsect:voidlens}), the statistics of which can be related to the hierarchy of N-point correlation functions (see, e.g., \cite{White:1979,Fry:1986}). Other alternatives include topological and morphological measures of the density field; in this subsection we will briefly describe the Minkowski functionals (MFs, \cite{Mecke:1994ax}) as a potential test of gravity. 

Compared with the standard N-point statistics, the MFs are advantageous in several aspects. First, According to Hadwiger’s theorem, the morphological properties of a three-dimensional structure are completely specified by four MFs, namely the volume, the surface area, the integrated mean curvature and the genus. Second, MFs are independent of galaxy bias, which makes them an ideal tool for gravity tests. This is because galaxy bias can have a complicated scale dependence in general MG theories, which makes it challenging to build models for N-point statistics for MG theories. Last but not least, it is computationally much cheaper to measure than the N-point correlation functions from observational or mock galaxy catalogs. MFs have been applied to analyze density fields in galaxy surveys (e.g., \cite{Hikage:2003fc,Blake:2013noa,Wiegand:2013xfa}) and maps from CMB or weak lensing experiments (e.g., \cite{Schmalzing:1997uc,Hikage:2008gy,Fang:2013mxa,Ade:2015ava,Chenxiaoji:2014mxa,Munshi:2016lyb}).

\begin{figure}
  \centering
  \includegraphics[width=.9\linewidth]{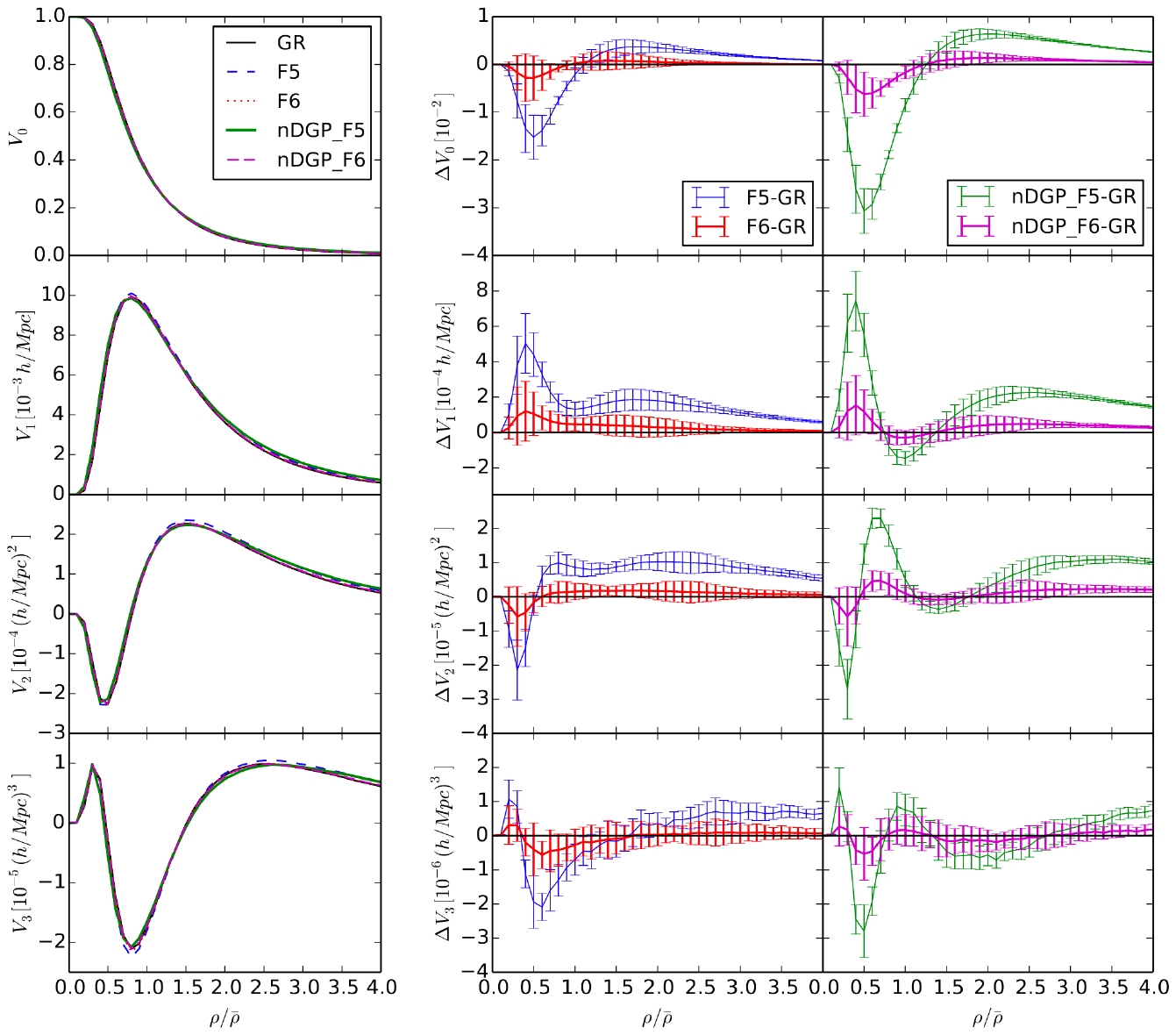}
  \caption{Left panel: The MFs computed from N-body simulations for different models of gravity at $z= 0$, with the dark matter density field smoothed with a spherical Gaussian filter with size $R_G= 5\Mpch$. The two $f(R)$ models -- F6 (red dotted lines), F5 (blue short dashed) -- and two nDGP models -- N5 (purple dashed), N1 (green solid) -- are compared with the GR case (black solid) for $V_0$ -- $V_3$ (from top to bottom). Right: The differences in the MFs between the four modified gravity models and GR, first column for $f(R)$ gravity, and second column for nDGP. The quantity $\rho/\bar{\rho}$ is the density threshold used for the MF calculations in ratio of the mean density. This plot is reproduced from \cite{MF} and so the colour scheme is different from the one adopted for most other plots in this paper. {As discussed in the text, while some features of the MFs are similar for GR and MG models, their amplitude is more pronounced and can be used to distinguish between the models.}}
 \label{fig:MF}
\end{figure}

A proof-of-concept study of using MFs for gravity tests was done in \cite{MF}, using the dark matter fields from the simulations described in \S~\ref{sec:simulations}, and the results are briefly summerized here. After smoothing the dark matter density field to suppress the shot noise, the MFs $V_0, V_1, V_2$ and $V_3$ were measured as a function of the density threshold, shown as $\rho/\bar{\rho}$. {Although the specific analytical form of the MFs is not needed here and can be found in, e.g.,   \cite{Schmalzing:1997aj}, their physical characterization is  informative: $V_0$ is the volume fraction of the excursion set (structure pattern); $V_1$ represents the area of the surface of the excursion; $V_2$ is its integrated mean curvature; and $V_3$ is its Euler characteristic per unit volume or the genus number describing the connectedness of the isodensity contours, see, e.g., \cite{Gott:1986uz,1996MNRAS.281.1375C,Matsubara:2000mw}}. In the left panel of Fig.~\ref{fig:MF} the four MFs of GR (black) and four MG models -- F6 (red), F5 (blue), N5 (purple) and N1 (green) -- are shown, while the right panels show the differences with respect to GR.

To get a sense on why MFs are useful for testing gravity, let us consider the results in more details. Taking $\Delta V_0$ (the difference in volume) to begin with, we can see that the volume fractions of the excursion set in MG are generally larger than that in GR, for densities above a sufficiently high threshold (i.e., $\rho>\bar{\rho}$). However, it is the opposite for under-dense regions, namely, the volume fraction above an under-dense threshold get smaller in MG, which is equivalent to having larger volume fraction below an under-dense threshold. This is as naturally expected as the halos and voids are more abundant and with larger sizes in $f(R)$ or nDGP than those in GR. Furthermore, for $\Delta V_1$ (the surface area), the overall trend is similar to that of $\Delta V_0$, except for the under-dense regions ($\rho<\bar{\rho}$). This can be understood as follows: if the excursion sets are all isolated regions, as is the case for high density threshold, it is expected that the change in their surface area follows that in the volume fraction they occupy. However, regions enclosed by the low iso-density contours are no longer the excursion sets, but under-dense regions with density below the threshold, with the volume fraction $1-V_0$. Thus, at the low density threshold region, $V_1$ changes in the opposite direction as $V_0$, and becomes larger in both $f(R)$ gravity and nDGP models. {Moreover, it can be seen from Fig.~\ref{fig:MF} that $V_3$ can help distinguish between GR and MG models as follows. In GR, the isodensity contours are more connected  with $V_3 < 0$ for the region $0.5 \lsim \rho/\bar{\rho} \lsim 1.5$, but more disconnected with $V_3>0$ in the other regions. For the $f(R)$ and the nDGP models, this feature is overall more pronounced and can thus serve to distinguish them from GR.} 

To summarize, the MFs $V_0$ -- $V_3$ capture information that is not available in the simple two-point statistics, and therefore can be useful for testing gravity and other cosmological models. However, the study in \cite{MF} was based on the dark matter, \revision{rather than galaxy fields}, which may have enhanced the model differences compared with the latter. The HOD tuning to match the (projected) galaxy 2PCFs, as used in most other probes of this paper, may further reduce the signal. Therefore, it will be useful to conduct a more detailed study using the MFs measured from realistic mock galaxy catalogs with a higher galaxy density and larger volume, allowing systematic effects from observations to be included, to fully assess the potential of this probe. 

\subsubsection{Stacked cluster phase spaces}
\label{subsect:stackedcl}

In the weak-field approximation of general relativity and at sub-horizon scales, a massive particle will still feel a force from the accelerated expansion of space \cite{Nandra2012}. The effective acceleration experienced by a massive particle with zero angular momentum in the vicinity of a galaxy cluster with gravitational potential ($\Psi$) is given by,
\begin{equation} 
\vec\nabla \Phi =  \vec\nabla \Psi + qH^2 r \hat{r}.
   \label{eq:acceleration_eq}
\end{equation} 
The effective potential $\Phi$ therefore takes into account both the effect produced by a {\it matter only} density field with potential $\Psi$ and the effect produced by the acceleration term $qH^2r$, with $q$ being the deceleration parameter $q\equiv-a\ddot{a}/\dot{a}^2$. From a Newtonian perspective, the latter term can be thought of as a repulsive force that opposes the inward pull of the cluster's mass distribution, caused by the accelerated expansion of space.

Given that the acceleration on a point mass is governed by the gradient of the gravitational potential, we define an equivalence radius as the point where the acceleration due to the cluster's gravity and the acceleration from the expanding space are equal to each other ($\vec\nabla \Phi= 0$), which yields, 
\begin{equation}
r_{\rm eq} = \left(\frac{GM}{-qH^2}\right)^{1/3},
\end{equation}
where $G$ is the gravitational constant and $M$ is the mass of the cluster. The escape velocity profile inferred from the observed phase space data can be modeled with a function of the mass distribution of a specific cluster, as specified by its gravitational potential, in our case we use the Einasto profile {with its standard three free parameters: $\alpha, r_{-2}, \rho_{-2}$ \cite{1965TrAlm...5...87E}}.  In the concordance $\Lambda$CDM cosmology, the escape profile is also a function of redshift $z$ and cosmological parameters, $\Omega_m, h$, etc.. Therefore in standard $\Lambda$CDM, the escape velocity radial profile is given by a function of the cosmology and cluster parameters combined,
\begin{equation} 
v_{\rm esc} = v_{\rm esc}\left(r,z,\Omega_m, h, \dots, \alpha, r_{-2},\rho_{-2}\right)
\end{equation}
where we include cosmological parameters like $\Omega_m$ as well as cluster observables like $\alpha$, $r_{-2}$ and $\rho_{-2}$.

In terms of escape speeds, we now recognize that the gravitational potential at $r_{\rm eq}$ plays an important role and that the distance to escape  a cluster is well-defined and finite in an accelerating Universe. In other words, we set the boundary condition such that the radial component of the escape velocity with respect to the cluster is zero at the equivalence radius, $-2\Phi\left(r_{\rm eq}\right) = v_{\rm esc}^2\left(r_{\rm eq}\right) = 0$. Using the Newtonian analogy that 
$v_{\rm esc}^2 = -2\Phi$, in Ref.~\citep{BehrooziEscape2013} it was noted that
\begin{equation} 
v_{\rm esc}(r,z) = \sqrt{-2 \big(\Psi(r) - \Psi(r_{\rm eq}) \big) - qH^2 \big(r^2 - r_{\rm eq}^2 \big)}.\label{eq:vesc}
\end{equation}
Eq.~\eqref{eq:vesc} yields a radial escape speed of 0, relative to the cluster center, at the equivalence radius given the gravitational potential profile $\Psi$. Note also that this equation applies to an accelerating universe for any choice of gravitational potential $\Psi$ \citep{Miller16}. For instance, modified gravity theories where the Poisson equation holds still satisfy Eq.~\eqref{eq:vesc} \cite{StarkMG}.

The observed escape velocity is readily identified by an edge in the radius-velocity phase-space data. This edge is typically suppressed statistically from under-sampling of the phase-space. Ref.~\citep{2020arXiv200302733H} showed that this suppression ($Z_v$) is dependent only on the number of galaxies which sample the phase space, such that when $\mathcal{O}(10^5)$ tracers are available, the statistical suppression term $Z_v \approx 1$ and a practically exact tracing of the 3D escape velocity profile can be made. In practice and even in stacked cluster phase-spaces, we typically have a few hundred (in well-sampled single clusters) to a few thousand (stacked) available tracers, and so $1.1 \lesssim Z_v \lesssim 1.5$. Thus, we measure the {line-of-sight} (LOS) escape velocity and correct for this statistical suppression to infer the 3D escape velocity: $v_{\rm esc} = Z_v{v}_{\rm esc,LOS}$.

Let us consider the $f(R)$ model as introduced in \S~\ref{sect:models}. The gravitational potential which massive particles experience is given by \citep{lombriserClusterDensity},
\begin{equation}
\Phi_{f(R)}(r) = \Phi_{\rm GR}(r) - \frac{1}{2} \delta f_{R}(r),
\end{equation}
where the $\delta$ signifies that the background has been subtracted from the scalar field: $\delta f_{R} = f_{R} - \bar{f}_{R}$. In an expanding chameleon $f(R)$ gravity Universe, instead of Eq.~\eqref{eq:acceleration_eq}, we have to modify the effective potential and its relation to the escape velocity as,
\begin{equation}
-2\Phi_{f(R)}(r) = v_{{\rm esc},f(R)}^2(r)
               =  v_{{\rm esc},{\rm GR}}^2(r)+ \big[\delta f_R(r)- \delta f_R(r_{\rm eq})\big].
\label{eq:vesc_fr}
\end{equation}
In practice, the gravitational potential in Eqs.~\eqref{eq:vesc} and \eqref{eq:vesc_fr} is constrained by weak lensing shear profiles around galaxy clusters. The LOS escape velocity surface can be observed using highly multiplexed spectroscopic instruments like DESI.

\begin{figure}
  \centering
  {	\includegraphics[width=0.520\textwidth]{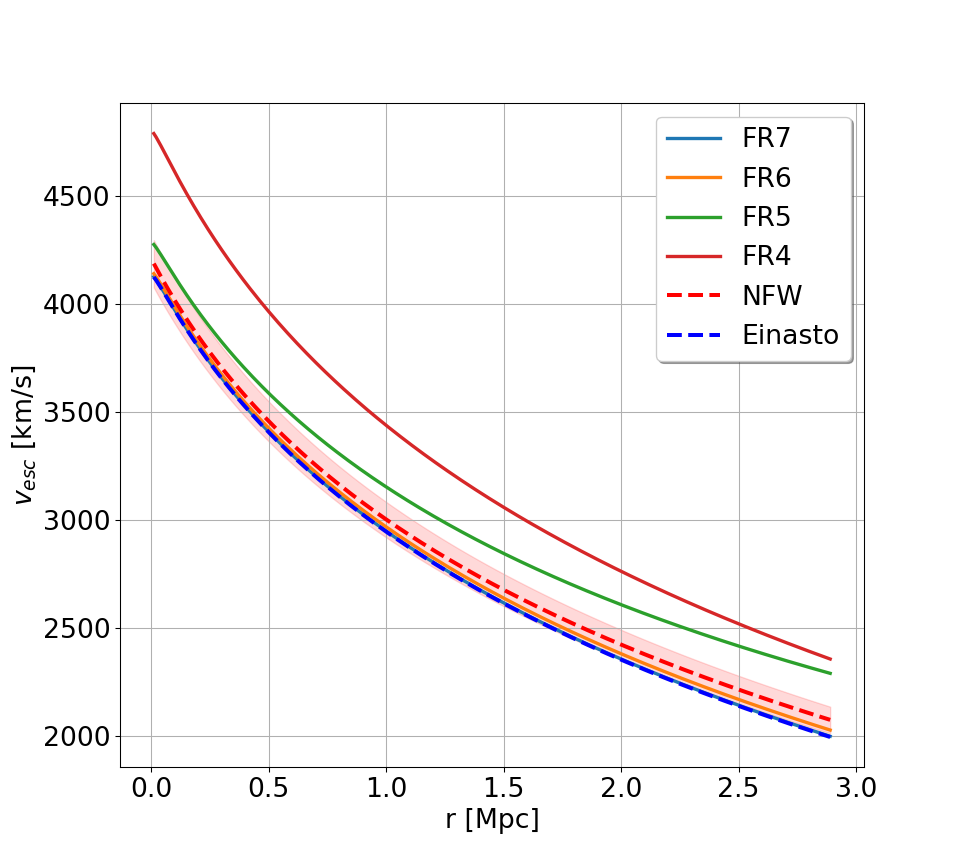}
	\includegraphics[width=0.462\textwidth]{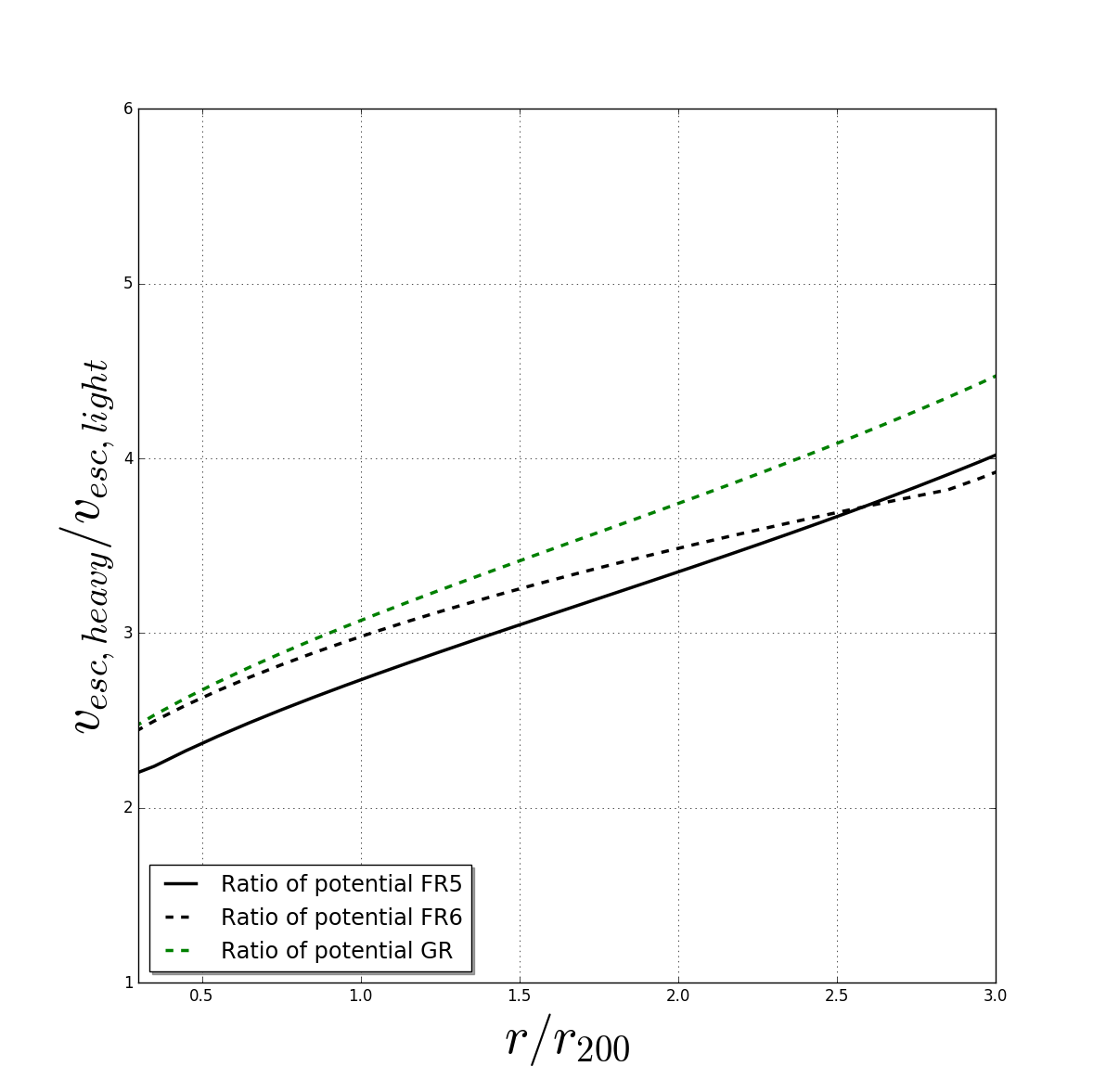}
}
  \caption{(Color Online) {\it Left}: The expected radial escape velocity profile for a stacked high mass ($10^{15}M_{\odot}$) cluster with $\sim 5000$ member galaxies. The shaded region includes nominal 5\% error from the weak-lensing inference of the gravitational potential. {\it Right}: The same as in the left panel, but shows the ratio between the high mass and low mass clusters escape velocities. This probe utilizes the ratio for two reasons: (1) we expect the inner $\sim0.5$ Mpc of high mass clusters to look like GR, which sets a lever-arm for the detection of $f(R)$ signals in lower mass systems; (2) this removes a primary systematic, the under-sampling which suppresses the true escape edge. Note that the color schemes are different in the two panels, as indicated in the legends.}
\label{fig:fr_theory}
\end{figure}

We expect the BGS data to provide spectroscopic redshifts for $\sim 25$ member galaxies per cluster in a typical $\sim 10^{15}M_{\odot}$ cluster, a value much lower than the actual cluster richnesses at this mass. This is because DESI suffers from significant fiber collisions in dense regions like the cores of galaxy clusters. However, due to the area and depth of the survey, we can expect $\sim 100$s of massive galaxy clusters to lie within the survey footprint. Therefore, DESI will provide useful line-of-sight velocities for a few thousand galaxies in a stacked cluster phase-space that has an average cluster mass of $\sim 10^{15}M_{\odot}$. The cluster masses themselves need to be measured via an independent weak lensing analysis. Note that unlike most of the other probes discussed in this paper which require some knowledge of the physical distances of the tracers, the escape velocity analysis is conducted purely in proper units, such as km/s and angular separations \cite{stark2017}. 

In the left panel of Fig.~\ref{fig:fr_theory}, \revision{we use Eq.~\eqref{eq:vesc_fr} to predict the} escape velocity for a stack of high mass ($10^{15}M_{\odot}$) clusters for different variants of HS $f(R)$ gravity including F4, F5, F6 and as well as F7 ($|f_{R0}|=10^{-7}$); note that the line colors are different from other plots of this paper. These profiles are not measured directly The dashed lines representing Einasto and NFW profiles are for GR and highlight the difference one would expect from an incorrect choice of mass profile when inferring the gravitational potential from the weak lensing data. The shaded region includes error from weak lensing (5\%). For F4 and F5, we find significant differences from GR in both the amplitude and the shape at all radii. For F6, it becomes difficult to differentiate against GR using a single stacked phase-space, and for F7 the result is practically indistinguishable from GR.

The right panel of Fig.~\ref{fig:fr_theory} shows the ratio between the escape edges of high- and low-mass clusters. Taking the ratio enables us to divide out the effect of the statistical suppression, on the assumption that the phase-space sampling is similar between them. In this case, we see that the difference between F6 and GR becomes appreciable at large radii. This is a result of the chameleon screening mechanism, which means that MG deepens the potential in the outskirts of low-mass galaxy clusters with respect to GR, but leaves the potential of high-mass clusters relatively unaffected. 

Therefore, this offers two ways to constrain gravity: using the radial escape velocity profile based on $v_{\rm esc}=Z_vv_{\rm esc,LOS}$, cf.~left panel of Fig.~\ref{fig:fr_theory}, or using the ratio between stacked high- and low-mass clusters, cf.~right panel of Fig.~\ref{fig:fr_theory}. We have found that the first approach, assuming a stack of massive ($\langle 10^{15}M_\odot\rangle$) clusters having $\sim5000$ tracers leads to a constraint of $|f_{R0}|\lesssim5\times10^{-6}$ when we take into account the uncertainty in $Z_v$ and the weak lensing mass errors. However, the latter approach which utilizes the ratio of the escape edge for a high and low mass stack produces superior constraints.

\begin{figure}
  \centering
  \includegraphics[width=1.0\linewidth]{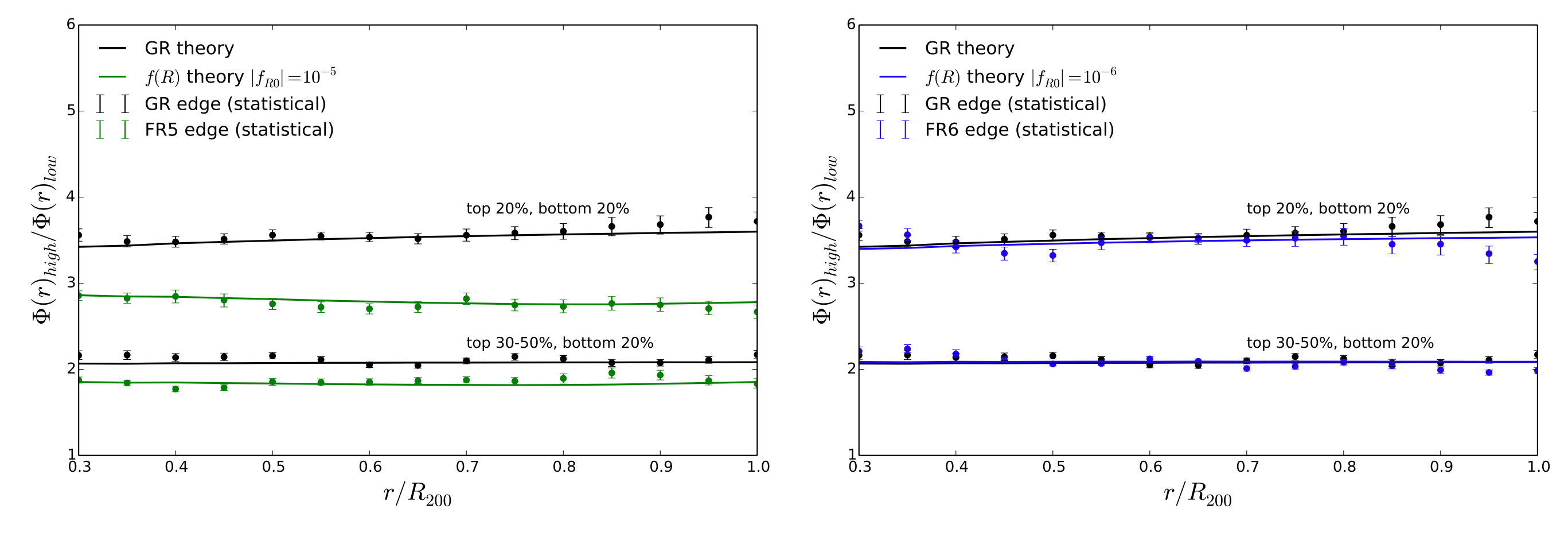}
  \caption{\small The $z=0$ gravitational potential ratio between high and low mass bins (of 20 halos) for GR (black), and the F5 (\textit{left}) (green) and  F6 (\textit{right}) parametrization of modified gravity (blue). The points are the averages of the square of the measured radial escape velocities using the simulation particles for each bin in radius and mass. The errors are 1$\sigma$ on the mean from boot-strap re-sampling. The solid lines represent the theoretical predictions using the NFW density parameter. The separation between GR and $f(R)$ potential ratios increases with increasing separation in the mass bins and we show two different mass binning schemes in each panel where the percentages denote the percentiles of cluster masses we keep in the sample. Note the large difference between the GR and F5 ratios at either mass binning scheme. Note the precise and accurate agreement between the theory and the measured escape profile (or $\Phi$) profile ratios. These figures justify our use of only theory in making predictions without the need for additional simulations at differing levels of $f_{R0}$. }
  \label{fig:fr_tests}
\end{figure}

Since the galaxy number in the mock HOD galaxy catalogs described in \S~\ref{sec:HOD} is too low, we use dark matter particles from the same simulations as described in \S~\ref{sec:simulations} to test the ratio of the high- and low-mass cluster potential ratios. Ref. \cite{Miller16} showed that \revision{semi-analytically modeled} galaxies are unbiased tracers of the dynamical escape velocity edge when compared to particles in a $\Lambda$CDM dark matter only simulation. Since our MG simulations are dark matter only, we assume galaxies in a MG universe will also be unbiased and that our constraints based on the particles will match expectations for real galaxies. \revision{However, we note that high volume and high resolution simulations are required to create large enough samples of massive halos which are well populated with sub-halos in order to conduct detailed phase-space explorations.}

We choose halos in the simulation such that the average masses in the high- and low-mass bins are nearly the same. For GR, the low mass bin is $9.10\times10^{13} \sim 1.96\times10^{14}h^{-1}M_{\odot}$ and the high mass bin is $7.48\times10^{14} - 1.58\times10^{15}h^{-1}M_{\odot}$. For F6, the low-mass bin is $9.13\times10^{13} - 1.97\times10^{14}h^{-1}M_{\odot}$ and the high-mass bin is $7.34\times10^{14} - 1.58\times10^{15}h^{-1}M_{\odot}$. For F5, the low-mass bin is $1.16\times10^{14} - 1.94\times10^{14}h^{-1}M_{\odot}$ and the high-mass bin is $7.49\times10^{14} - 1.58\times10^{15}h^{-1}M_{\odot}$. We then measure the radial escape velocity for each cluster in these bins and take the average, and calculate the error on the theory through bootstrap re-sampling of the clusters in each bin. As shown in Fig.~\ref{fig:fr_tests}. we find excellent agreement with theory, and this allows us to use the analytical prescriptions for making predictions for $f(R)$ gravity variants other than F5 and F6. We also note that there are detectable differences in the shape between GR and F6 in the outskirts, in agreement with our analytical result in Fig.~\ref{fig:fr_theory}.

We then utilize the probe
\begin{equation}
\frac{\Phi_{\rm high,LOS}(r)}{\Phi_{\rm low,LOS}(r)}  \equiv\frac{\langle{Z_v({\rm low})}\rangle}{\langle{Z_v({\rm high})}\rangle} \frac{\langle{v_{\rm esc,high}^2(r)}\rangle}{\langle v_{\rm esc,low}^2(r) \rangle} \approx \frac{\Phi_{\rm high}(r)}{ \Phi_{\rm low}(r)},
\label{eq:phiratio_proj}
\end{equation}
which was tested extensively on lightcone data from Ref.~\cite{henriques}. A forecast analysis was carried out in Ref.~\cite{StarkMG}, where it was found that using this probe, DESI BGS data could differentiate between $\Lambda$CDM (the null hypothesis) and $|f_{R0}|<4\times10^{-6}$ with a p-value of $3\times10^{-7}$ (e.g.,  $\sim 5\sigma$).

In reality, we expect to be able to make multiple independent measures of the escape velocity ratio, each with different average masses in their high- and low-mass bins. We will have enough clusters to choose from to ensure that there are no clusters residing in more than one binned measurement of the escape ratio. This additional power to differentiate between $f(R)$ and GR models lies in both more clusters and more mass bins. We have checked that the escape ratio linearly depends on the mass ratio of the high- and low-mass bins. Using the large sample from the GR simulations described in \S~\ref{sec:simulations}, we are able to create up to four more high- and low-mass bins, each with a different mass ratio. In practice, we treat each of these as independent measures as a function of radius, thus increasing our statistical sample by a factor of 5. In doing so, we are able to differentiate between GR and $f_{R_0}= 6 \times 10^{-7}$ at $>5\sigma$.

The observables for this test involve the galaxy redshifts and radial positions for the phase-spaces and the weak-lensing shear profiles. We assume that the systematics from these measurements are well-controlled and minimized in the data. For the phase-spaces, this should not be an issue, since the galaxy sky positions and redshifts should be determined at percent level accuracy compared to the escape surface. Recent advances in the methods to constrain weak lensing mass profiles claim control on the systematics at $<10\%$ \cite{applegate:2014}. Nominal cluster-based systematic issues like the selection function and mis-centering are not issues for this probe, simply because we require high signal-to-noise measurements of the phase-space density and the weak-lensing signal for individual clusters. 

It should be noted that while the ratio test exploits the unique properties of the f(R) model (e.g. the chameleon screening mechanism), the test of the radial escape velocity profile does not require any specific characteristics from the gravity models as it places constraints based on the difference in gravitational potential alone. That allows to utilize galaxy cluster's phase spaces to test other MG and alternative gravity models such as emergent gravity \cite{Verlinde:2016toy} in addition to constraints based on mass profiles of galaxy clusters
\cite{Halenka:2018qnj, Ettori:2018tus, 2019JCAP...05..053T}.

In the end, the primary observational systematic when using high signal-to-noise clusters and parameterized mass profiles is the weak lensing shear. It is likely that the DESI Bright Galaxy Survey will provide the required amount of phase-space sampling for over 1000 clusters. Many of these will have (albeit noisy) weak lensing profiles from the deep ground-based imaging from DES and DECals (as well as from the Hyper-SuprimeCam Survey).

The story is different when we stack phase-space data and weak-lensing shears. Stacking is the presumed scenario for this probe. DESI does a poor job of spatial sampling within high density regions. A concern is that weak-lensing profiles are hard to measure for clusters $\lesssim10^{14}$M$_{\odot}$, where the $f(R)$ signal becomes interesting. However, in order to make further in-roads into how well this probe will work for realistic DESI stacks, we require a new generation of mock galaxy catalogs in light-cones with estimated shear profiles and high enough galaxy sampling to measure phase-spaces with many 10s of galaxies per cluster. 

\subsection{Gravitational lensing statistics}
\label{subsect:lensing}

While the paper principally  focuses on utilizing  density and velocity statistics coming out of spectroscopic and photometric galaxy surveys, in this section we discuss the potential to use complementary lensing information to test the properties of gravity directly through the evolution of the gravitational potential. For a spectroscopic survey that has substantial sky coverage overlap with imaging surveys, lensing can be a useful probe for testing gravity. In this section we discuss two distinct environments that can be considered. In \ref{subsect:gglens} we consider galaxy-galaxy lensing, while in \ref{subsect:voidlens} we discuss the potential of lensing measurement in low density, void environments.

\subsubsection{Galaxy-galaxy lensing}
\label{subsect:gglens}

Galaxy-galaxy lensing (GGL) describes the distortions of images of background (source) galaxies around foreground (lensing) galaxies, and detects the matter distribution around the latter up to radii which typically go well beyond the dark matter halos of the lensing galaxies. It is an ideal probe to study properties of dark matter halos such as the mass profiles \cite{Mandelbaum:2006pw,vanUitert:2016}, galaxy-matter cross correlation and galaxy bias \cite{Sheldon:2003xj,Dvornik:2018}, and to constrain cosmological parameters \cite{Cacciato:2008hm}. It has been applied in multiple lensing surveys, such as {\sc cfhtlens} \cite{Gillis:2013qi,Velander:2013jga,Hudson:2013loa}, {\sc kids} \cite{Sifon:2015uca,Viola:2015csx,Dvornik:2017} and {\sc des} \cite{Clampitt:2016ljk}.

In regions well outside foreground galaxies, the screening mechanisms are expected to work less efficiently. Such regions can experience substantially stronger gravitational force, and consequently enhanced clustering of matter, in the MG models studied here, making GGL a potentially useful probe to test them. However, as in void lensing (see \S\ref{subsect:voidlens}), because individual galaxies generally do not produce strong enough lensing effect, one has to stack the tangential shear around many foreground galaxies to detect signals at high signal-to-noise. GGL in the context of MG models has been studied previously in, e.g., \citep[][]{Park:2014aga,Leauthaud:2016jdb,Li:2017}.

The tangential shear around a foreground galaxy is related to the average excess surface density profile, $\Delta\Sigma(r_p)$ \cite{1991ApJ...370....1M,2001ApJ...555..572W}, given by the following integration of the galaxy-matter cross correlation function, $\xi_{gm}(r)$:
\begin{eqnarray}\label{eq:excess}
\Delta\Sigma(r_p) &=& \rho_{\rm crit}\Omega_m\frac{2}{r_p^2}\int^{\infty}_{-\infty}{\rm d\chi}\int^{r_p}_0{\rm d}r\cdot r\xi_{gm}\left(\sqrt{r^2+\chi^2}\right) - \rho_{\rm crit}\Omega_m\int^{\infty}_{-\infty}{\rm d}\chi\xi_{gm}(r_p,\chi),
\end{eqnarray}
where $r$ and $r_p$ are respectively the 3D distance and projected distance from the lensing galaxy, and $\chi$ is the comoving distance from it. We have measured $\xi_{gm}$ using a modified version of the the publicly-available code {\sc cute} \citep{Alonso2012}, which counts pairs of HOD galaxies and simulation particles. To carry out the integrations in Eq.~(\ref{eq:excess}), we first interpolate $\xi_{gm}\left(r=\sqrt{r^2_p+\chi^2}\right)$ onto a grid in $\left({\rm log}(r_p),\chi\right)$ using cubic spline, and then do discrete summations of the integrands evaluated at the grid points. We use $\pm\chi_{\rm max}$, with $\chi_{\rm max}=90~h^{-1}$Mpc, as the integral limit and have checked that this choice leads to converged results. 

\begin{figure}[!tb]
\begin{center}
{\includegraphics[width=0.5\textwidth]{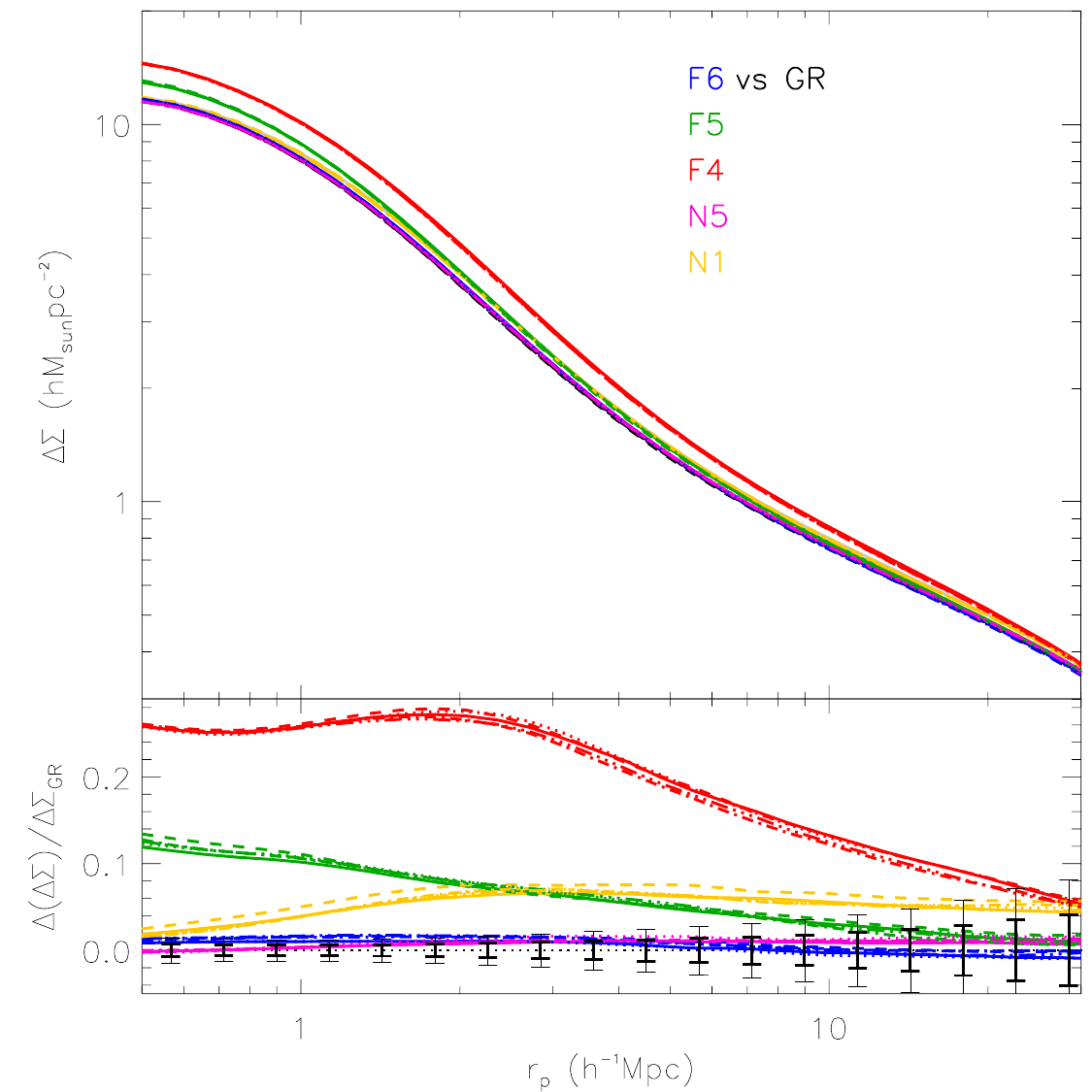}
}
\caption{(Color Online) The excess surface mass density profiles, $\Delta\Sigma$, measured at $z=0.5$, for the six models studied here: GR (black), F6 (blue), F5 (green), F4 (red), N5 (magenta) and N1 (orange). The lower sub-panel shows the relative differences with respect to GR, where the black horizontal dotted line is zero, and the thick and thin error bars indicate respectively the statistic uncertainties in an optimistic and a pessimistic case, in which an LSST-like imaging survey has $6,000$ and $1,500$ sq.~deg.~overlap with a DESI-like survey. We have used different lines styles to represent the five realisations, instead of showing their mean, to highlight the small scatter across the different realizations.
}
\label{fig:GGL}
\end{center}
\end{figure}

The results of $\Delta\Sigma$ and the relative differences between modified gravity models and GR are shown in Fig.  \ref{fig:GGL}, considering only lenses at $z=0.5$ for simplicity. The MG curves display the expected trends with respect to GR -- within the $f(R)$ family, F4 shows the largest deviation and F6 the smallest, and within the nDGP family, N1 shows stronger difference than N5; these are in the same order as decreasing screening efficiency and increasing strength of the fifth force. 
The differences between $f(R)$ gravity and GR decrease on larger scales, while for nDGP the deviation from GR remains scale-independent at $r_p$ greater than a few $h^{-1}{\rm Mpc}$. This behavior was expected since $f(R)$ gravity has scale-dependent linear growth, with the fifth force suppressed outside the Compton wavelength of the scalar field, while in DGP the linear growth rate is scale-independent. On the other hand, on small scales ($r_p\lesssim1h^{-1}$Mpc) the nDGP predictions become close to the GR ones, which is a result of the efficient Vainshtein screening near and inside dark matter halos.  

We have checked the results at $z=0$ and found similar qualitative and quantitative conclusions as $z=0.5$, with the maximum deviation from GR at the level of $\sim28\%$, $\sim12\%$ and $\sim9\%$ for F4, F5 and N1 respectively (the F6 and N5 models are, on the other hand, very close to GR). The model differences also have a weak dependence on the galaxy number density \cite{Li:2017}. The GGL results are not sensitive to the way in which the HOD catalogs are produced for the MG models; for example, we did a test by tuning the respective HOD parameters so that these models match the real-space 3D galaxy correlation function of GR, and found almost identical conclusions as in Fig.~\ref{fig:GGL}. 
Finally, note that in Fig.~\ref{fig:GGL}, instead of the average results from all 5 simulation realizations, we have shown the curve for each of them using different line styles, to see the scatter between them is very small even for the stronger MG models (F4 and N1); we also did a test by tuning the HOD parameters individually for each simulation realization instead of doing this for all 5 boxes together, and while the HOD parameters are now slightly different, the realization scatter was again negligible. This makes sample variance a lesser concern for testing models using GGL, especially for surveys like DESI which will give higher galaxy number densities in a larger volume than used in the analyses here.

To forecast the constraining power of GGL, we have calculated the signal-to-noise (S/N) of the distinguishability of the MG models from GR, which is defined as
\begin{eqnarray}\label{eq:SN_GGL}
\left({\rm S/N}\right)^2 \equiv \delta\Delta\Sigma^{T}(r_{p,i})\mathbb{C}^{-1}(r_{p,i},r_{p,j})\delta\Delta\Sigma(r_{p,j}),
\end{eqnarray}
in which $\delta\Delta\Sigma(r_{p,i})$ is the model difference of the excess surface mass density in the $i$th $r_p$ bin, and $\mathbb{C}(r_{p,i},r_{p,j})$ the covariance matrix between the $i$-th and $j$-th $r_p$ bins. The covariance matrix is calculated following the analytical prescription of \cite{Jeong:2009}, based on halo model predictions of the shear-shear, galaxy-galaxy and shear-galaxy correlation functions (for which we have used the same cosmological and HOD parameters as in the simulation and galaxy catalogs). The calculation takes into account contributions from cosmic variance, the Poisson noise of lens galaxies and source shape noise $\sigma_\gamma$, and assumes a single source redshift $z_{\rm S}=1.0$ and lens redshift $z_{\rm L}=0.5$. The original covariance matrix, $\mathbb{C}({\bf \theta},{\bf \theta}')$ is calculated for the tangential shear $\gamma_t(\theta)$, where $\theta$ is the angular separation from the lens, and then converted to the covariance matrix for $\Delta\Sigma$ using
\begin{eqnarray}\label{eq:ggl_cov}
\mathbb{C}(r_{p,i},r_{p,j}) &=& \Sigma^2_{\rm crit}(z_{\rm L})\mathbb{C}({\bf \theta},{\bf \theta}'),
\end{eqnarray}
where $\Sigma_{\rm crit}$ is the critical surface mass density, and $\gamma_t(\theta)=\Delta\Sigma\left[r_p=D_{\rm A}(z_{\rm L})\theta\right]/\Sigma_{\rm crit}$. We consider GGL measured using the synergy of a DESI-like spectroscopic survey and an LSST-like imaging survey, with two cases of overlapping sky areas -- an optimistic case of $6,000$  and a pessimistic case of $1,500$ squared degrees. In both cases we adopt a value $\sigma_\gamma=0.22$ and assume that the source galaxy number density is $n_{\rm S}=40~{\rm arcmin}^{-2}$. The error bars in the lower panel of Fig.~\ref{fig:GGL} show the square roots of the diagonal elements of the covariance matrix, $\mathbb{C}(r_{p,i},r_{p,i})$; even in the optimistic case the uncertainty is significantly larger than the model differences for F6 and N5, but for the other models the reverse is true.

In addition to the statistical uncertainties discussed above, in real observations the total error budget must also include contributions from a range of systematic effects. Following \cite{Clampitt:2016ljk}, we calculate the total covariance matrix as $\mathbb{C}^{\rm tot}_{ij} = \mathbb{C}^{\rm stat}_{ij}+\mathbb{C}^{\rm syst}_{ij}$, where the statistical contribution is as above, while the systematic contribution is given by 
\begin{eqnarray}\label{eq:GGL_systematics}
\mathbb{C}^{\rm syst}_{ij} &=& \mathbb{C}^{\rm syst}(r_{p,i},r_{p,j})\ =\ f_{\rm syst}^2\Delta\Sigma(r_{p,i})\Delta\Sigma(r_{p,j}),
\end{eqnarray}
where $\Delta\Sigma(r_{p,i})$ is the excess surface mass density for GR in the $i$-th $r_p$ bin, and $f_{\rm syst}$ is a multiplicative factor accounting for systematic uncertainties caused by photo-$z$ bias, shear calibration, stellar contamination, {intrinsic alignments of galaxies, etc., see, e.g., \cite{Clampitt:2016ljk,Mandelbaum:2017jpr}}. 

We have done the calculation of S/N in Eq.~(\ref{eq:SN_GGL}) using the full covariance matrix, with two values of $f_{\rm syst}$: $0.0$ (negligible systematic error) and $0.1$, and the results are summarized in Table \ref{table:SN_GGL}. 
Because of the relatively poor resolution of our simulations, on small scales the matter clustering and galaxy distribution they predict could be inaccurate; to be conservative, we only used the $\Delta\Sigma$ data within $r_p\in[2,30]h^{-1}$Mpc in the forecast (this will also make the result less affected by astrophysical uncertainties such as the impact of baryons on $\Delta\Sigma$). 
As one can see from Table \ref{table:SN_GGL}, including systematic uncertainties substantially reduces the power of GGL in distinguishing the various modified gravity models. However, GGL with a LSST-like imaging survey can still tell apart F5 and F4 from GR.

\begin{table}

\begin{tabular}{@{}lllllll}
\hline\hline
 & $f_{\rm syst}$ & F6 & F5 & F4 & N5 & N1 \\
\hline
 & $0.0$ & $1.0$ & $5.1$ & $18.9$ & $0.9$ & $5.3$ \\
Pessimistic case & $0.1$ & $0.4$ & $1.4$ & $4.8$ & $0.1$ & $0.8$ \\
\hline
 & $0.0$ & $2.0$ & $10.6$ & $39.2$ & $1.8$ & $11.1$ \\
Optimistic case & $0.1$ & $0.9$ & $2.7$ & $9.0$ & $0.2$ & $1.1$ \\
\hline\hline
\end{tabular}
\caption{The S/N values at which GGL from the synergy of a DESI-like spectroscopic and an LSST-like imaging survey can distinguish the different modified models studied in this paper from GR. See the main text for more details.}
\label{table:SN_GGL}
\end{table}

\subsubsection{Void lensing}
\label{subsect:voidlens}

We have discussed that while MG models are usually screened in halos and high-density regions, departures from GR can be substantial in underdense (void) regions. The fifth force present in many such models (including $f(R)$ gravity and nDGP) can lead to a more efficient evacuation of underdense regions, and therefore emptier voids than in GR \cite{Li:2010vo,Cai2014,Barreira:2015vra,Cautun2016a,Cautun:2017tkc,Falck2017}. Once the galaxy populations were matched to have the same two-point galaxy correlation function across models, voids in MG models such as $f(R)$ gravity, despite being emptier of matter, have essentially the same void abundances and void galaxy number density profiles as their GR counterparts \citep{Cai2014,Cautun:2017tkc}. However, the weak lensing signal of voids show significant differences with respect to GR, with voids in $f(R)$ and nDGP having a larger tangential shear signal than GR ones \cite{Cai2014,Cautun:2017tkc,Paillas2019:MNRAS.484.1149P,Davies2019:arXiv190706657D}. The weak lensing imprint of voids has already been measured by current surveys \cite{Clampitt2014,Gruen2015,Sanchez2016}, and future observational campaigns would greatly improve the quantity and quality of void lensing data. Voids have an additional advantage, namely their properties are largely insensitive to the baryonic and galaxy formation physics -- which is still a major uncertainty -- and are well reproduced by dark matter only simulations \cite{Paillas2017}. As a result, void lensing can be an appealing technique for testing MG models. {Voids are also highly versatile \cite{Cautun:2017tkc} in that there can be various different void definitions which trace different aspects of the cosmic web and can be tailored to maximize the potential of extracting certain features of a particular model.}

Here, we test the potential of two galaxy void finders to constrain the MG models studied in this paper. The first, the Watershed Void Finder \citep[][hereafter WVF]{Platen2007} identifies underdensities in the 3D distribution of galaxies. The voids are determined by the watershed basins of a given large-scale galaxy density field without imposing any constraints on the shape, size or underdensity of these objects. The method constructs a volume-filling galaxy density field using the Delaunay Tessellation Field Estimator \citep{Schaap2000,Cautun2011}, which is based on Delaunay triangulations. The resulting density is defined on a $1024^3$ regular grid with a grid cell size of $1\Mpch$. To reduce small-scale structures that could give rise to artificial voids, the method smooths the density field with a Gaussian filter of $2\Mpch$ radius -- this filter size corresponds to the typical width of the filaments and sheets that form the void edges \citep[e.g.][]{Cautun2013,Cautun2014a}. The smoothed density field is then segmented into watershed basins. This process is equivalent to following the path of a rain drop along a landscape: each volume element, in our case the voxel of a regular grid, is connected to the neighbor with the lowest density, with the same process repeated for each neighbor until a minimum of the density field is reached. Finally, a watershed basin is composed of all the voxels whose paths end at the same density minimum. The void centers are chosen as the volume-weighted barycentre of all the voxels associated to each void, and the void radius is the radius of a sphere with the same volume as the void volume.

The second method identifies tunnels, which are 2D underdensities in the distribution of galaxies projected onto the plane of the sky. The tunnels are defined as circular regions that are devoid of galaxies and consist of elongated line-of-sight regions that intersect one or more voids without passing through overdense regions \citep{Cautun:2017tkc}. To identify tunnels, we build a Delaunay tessellation of the projected galaxy distribution since, by definition, the circumcircle of every Delaunay triangle is empty of galaxies, with the closest galaxies being the ones that give the triangle vertices and that are found exactly on the circumcircle. The tunnels consist of the circumcircles whose centres are not inside a larger circumcircle. We are interested in the modified gravity signature of underdense regions, and so we select only the tunnels with radii above $1\Mpch$, which correspond to underdense regions in projection \cite{Cautun:2017tkc}. We project the entire simulation box, since its length roughly corresponds to the comoving distance between redshift $0.3$ and $0.7$.

To obtain the void tangential shear, we compute the mean excess surface density profile around each void. We have followed the procedure described in Ref.~\cite{Cautun:2017tkc}, where the $\Delta\Sigma(r_p)$ for the 3D underdensities, that is WVF objects, was computed similarly to GGL, that is using Eq.~(\ref{eq:excess}), but with the galaxy-matter cross correlation function replaced by the void-matter cross correlation one, $\xi_{vm}$. To average over the density profiles of voids of different sizes, we computed $\xi_{vm}$ as a function of the scaled radial distance, $r/R_{\rm void}$, with $R_{\rm void}$ the radius of each void. The mean excess surface density of WVF voids was computed as
\begin{eqnarray}\label{eq:excess_voids}
\Delta\Sigma(\eta_p) &=& \rho_{\rm crit}\Omega_m \bar{R}_{\rm void} \left[  \frac{2}{\eta_p^2} \int^{+3}_{-3}{\rm d\chi}\int^{\eta_p}_0{\rm d}\eta\cdot \eta\xi_{vm}\left(\sqrt{\eta^2+\chi^2}\right) - \int^{+3}_{-3}{\rm d}\chi\xi_{vm}\left(\sqrt{\eta^2_p+\chi^2}\right) \right] \;,
\end{eqnarray}
where $\bar{R}_{\rm void}$ is the mean void radius. The symbols $\eta$\footnote{{Note that here we use $\eta$ instead of $r_p$ for the distance transverse to the line of sight, to highlight the fact that the former is rescaled by $R_{\rm void}$.}} and $\chi$ denote the spatial coordinates {perpendicular to} and along the line-of-sight, respectively, with both coordinates representing scaled distances, that is in units of the void radius, $R_{\rm void}$. The value of $3$ in the ${\rm d\chi}$ line-of-sight integral comes from limiting the integral to three times the void radius, which is sufficient for calculating the void lensing signal \cite{Cautun:2017tkc,Cai2014}. In the case of the 2D underdensities, we compute the 2D void-matter cross correlation function, $\xi_{vm\;2D}(\eta)$, again as a function of the scaled distance, $\eta=r/R_{\rm void}$, by projecting the dark matter particle distribution along the same axis along which we projected the HOD galaxy distribution. Then, the excess surface density is given by
\begin{eqnarray}\label{eq:excess_voids_2D}
\Delta\Sigma(\eta_p) &=& \rho_{\rm crit}\Omega_m \left[ \frac{2}{\eta_p^2}\int^{\eta_p}_0{\rm d}\eta\cdot \eta\xi_{vm\;2D}\left(\eta\right) - \xi_{vm\;2D} \left(\eta_p\right) \right] \;.
\end{eqnarray}
We have measured the void-matter cross correlation functions using a brute force algorithm, similarly to \S\ref{subsect:gglens}.  

\begin{figure*}[!tb]
\begin{center}
{	\includegraphics[width=0.47\textwidth]{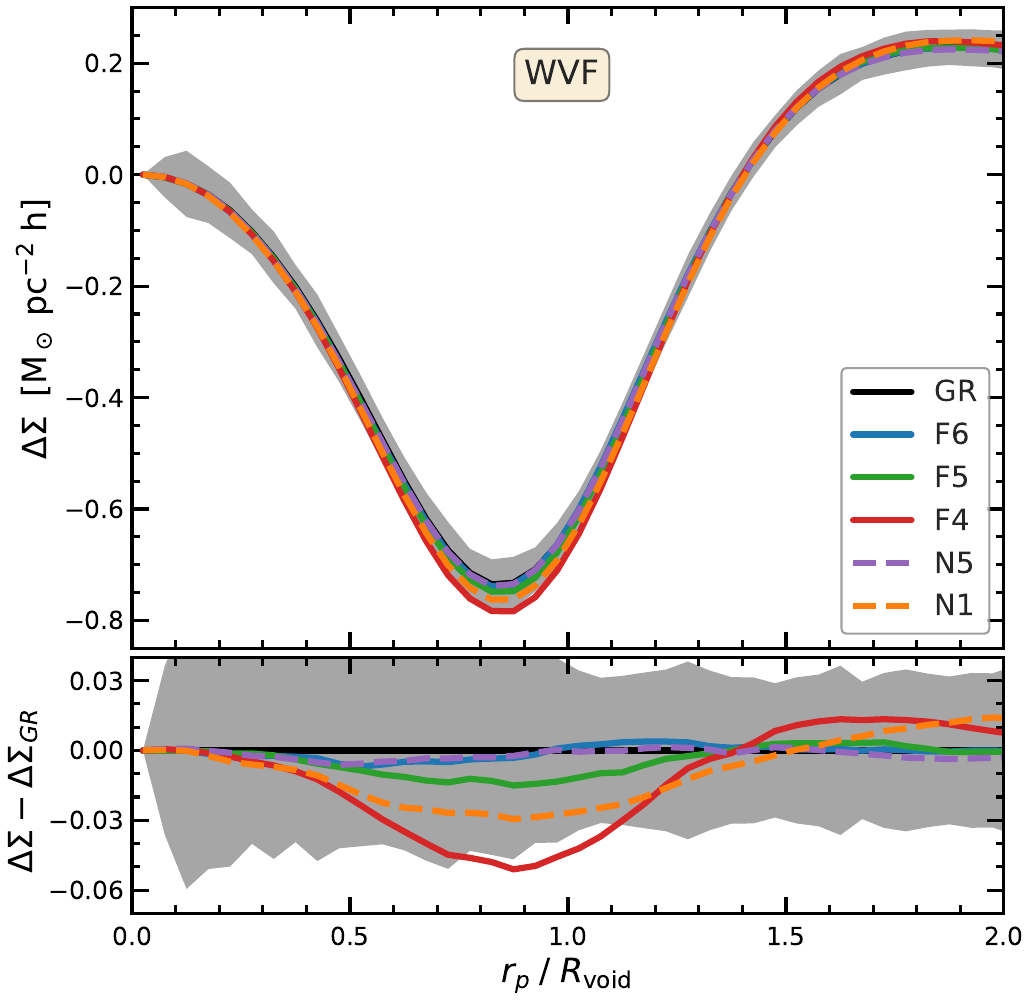}
	\includegraphics[width=0.462\textwidth]{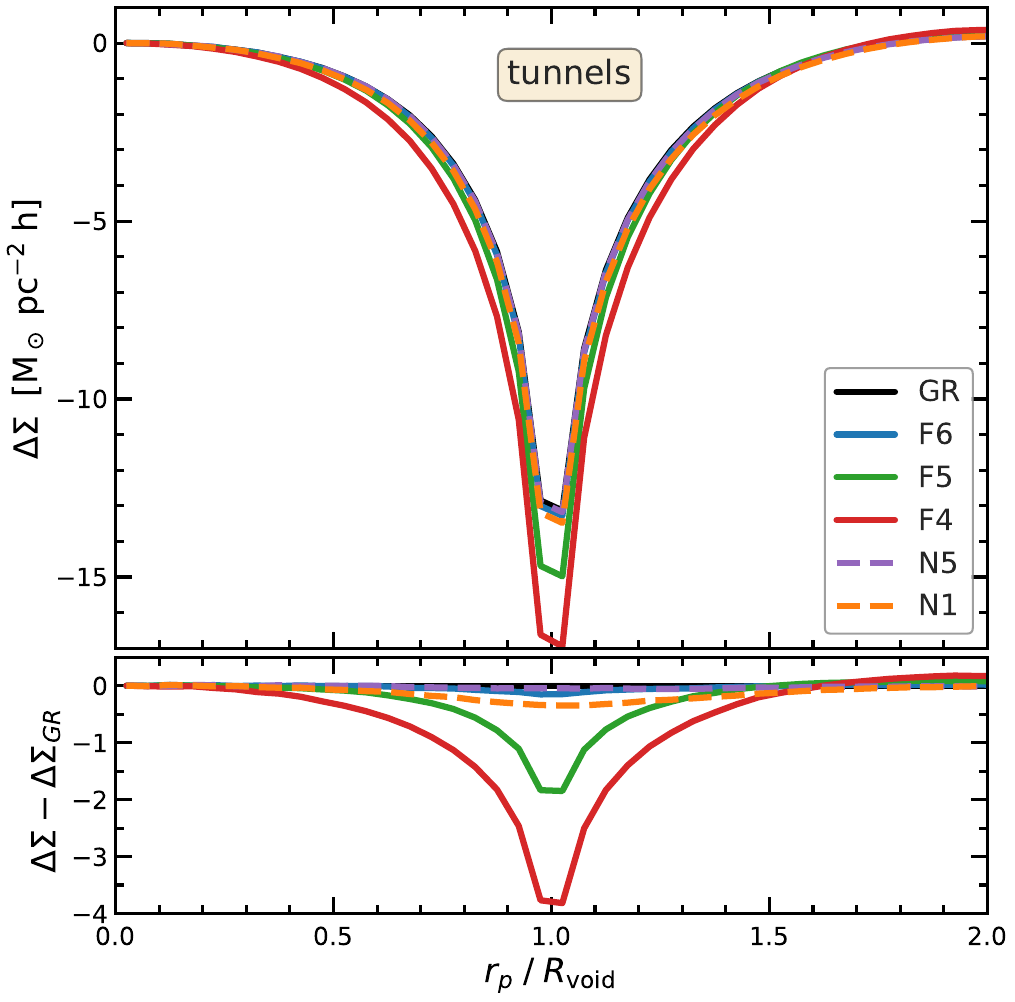}
}
\caption{(Color Online) The excess surface mass density profiles, $\Delta\Sigma$, of 3D voids (left) and 2D tunnels (right) at $z=0.5$, as a function of scaled radial distance, $r/R_{\rm void}$. The lines show the predictions for the six models studied here: GR (black), F6 (blue), F5 (green), F4 (red), N5 (cyan) and N1 (orange). The bottom panels show the difference between the MG models and GR, with the grey shaded region showing the $1\sigma$ sample variance uncertainties for a $6,000$ sq.deg. overlap with the LSST survey (for tunnels, the uncertainty range is very small and hence not easily visible).
}
\label{fig:void_weak_lensing}
\end{center}
\end{figure*}

The excess surface density profiles of WVF voids and tunnels at $z=0.5$ are shown in Fig.  \ref{fig:void_weak_lensing}. Since voids are underdense, they have negative $\Delta\Sigma$ values, which means that they give rise to a similar effect as a divergent lens. Of the two void finders, the tunnels have lensing signals that are nearly 20 times larger, demonstrating that selecting underdensities in projection results in a larger lensing signal than 3D underdensities. Void lensing would be even larger if we select the voids using the weak lensing convergence field, as shown in \cite{Davies2018:MNRAS.480L.101D,Davies2020:arXiv200411387D,Davies2020:arXiv201011954D}. Compared with GR, voids in the two MG models studied here have systematically larger lensing signals, i.e., show more divergent lensing effects. The differences are largest for F4 within the $f(R)$ family and for N1 in the nDGP family.
The differences vary with redshift (not shown here), being larger at lower $z$ \cite{Cautun:2017tkc}, because in both MG models studied here the screening becomes weaker at late times, leading to larger deviations from GR.

\revision{The significant difference in the $\Delta\Sigma$ profiles between 3D WVF voids and 2D tunnels is due to how these objects are selected. The 3D voids correspond to 3D underdensities, with the matter evacuated from within the voids being pilled up near their edges. This means that after projection 3D voids are on average only very mildly underdense, which results in a small tangential shear signal. The boundaries of 3D voids do not have uniform densities, but rather are composed from a few overdense regions with many lower density portions connecting them \cite{Cautun2014a,Falck2015:MNRAS.450.3239F}. The tunnels correspond to line-of-sight elongated portions of 3D voids selected such that they do not contain the overdense regions found on the edge of 3D voids (see Figure 3 in \cite{Paillas2019:MNRAS.484.1149P} for a sketch of the relation between tunnels and 3D voids). This means that tunnels correspond to significantly lower mean projected densities, which are reflected in a higher tangential shear signal than 3D voids.}

Interestingly, for 3D voids the F4 and N1 models show a similarly sized difference with respect to GR. However, for tunnels the N1 model -- which is the nDGP variant that deviates from GR more strongly -- is nearly as close to GR as the F6 model, while the F4 model shows a much larger difference. The tunnels are selected to maximize the transition contrast between underdense and overdense regions, with $\Delta\Sigma$ being proportional to the steepness of this transition. In contrast to $f(R)$ gravity, the overdense regions in nDGP models have a similar density as in GR thanks to the more efficient screening (see \S\ref{subsect:gglens}), and the differences between nDGP and GR lensing come only from underdense regions, which explains why tunnel lensing is better for testing $f(R)$ models than nDGP ones. This is another example to show how the different screening mechanisms can leave distinct imprints in the observed large-scale matter distribution, {which can in turn be picked up or maximized by a suitably-tailored void finding algorithm \cite{Paillas2019:MNRAS.484.1149P}.}

To estimate the power of WVF and tunnels to constrain modified gravity models, we have calculated the signal-to-noise (S/N) with which various models can be distinguished from GR. For this, we follow the same approach as in \S\ref{subsect:gglens} and compute the S/N using  Eq.~(\ref{eq:SN_GGL}). The covariance matrix contains the contribution from sample variance, shape noise and systematic effects. We estimate this by considering a single source redshift of $z_s=1.0$ and a distribution of void lenses between redshifts $z=0.3$ and $z=0.7$. Considering lower redshift lenses makes little difference since the $z<0.3$ volume is small; considering lenses with $z>0.7$ adds little constraining power due to a combination of decreasing lensing kernel and smaller differences between modified gravity models and GR. For simplicity, we consider that all void lenses between $0.3\leq z\leq 0.7$ have the same excess surface density given by the mean value at $z=0.5$. Similar to \S\ref{subsect:gglens}, we consider two cases in which the {DESI spectroscopic survey have different overlaps -- $6,000$ deg$^2$ for the {\it optimistic case} and $1,500$ deg$^2$ for the {\it pessimistic case} -- with the galaxy imaging survey LSST.} In both cases we adopt an intrinsic source shape noise $\sigma_{\gamma}=0.22$ {and a source galaxy number density $n_{\rm S}=40~{\rm arcmin}^{-2}$}. 

\begin{table}

\setlength{\tabcolsep}{5pt} 
\begin{tabular}{c @{\hskip 1.cm}c}
\\[-.2cm]

\begin{tabular}{@{}l @{\hskip .2cm}l lll ll}
\multicolumn{7}{c}{\bf 3D WVF voids} \\
\hline \\[-.3cm]
 & $f_{\rm syst}$ & F6 & F5 & F4 & N5 & N1 \\
\hline \\[-.3cm]
\multirow{ 2}{*}{Pessimistic case} 
 & $0.00$ & $0.1$ & $0.3$ & $1.0$ & $0.1$ & $0.7$ \\
 & $0.10$ & $0.1$ & $0.2$ & $0.6$ & $0.1$ & $0.4$ \\
\hline \\[-.3cm]
\multirow{ 2}{*}{Optimistic case}
 & $0.00$ & $0.3$ & $0.6$ & $2.0$ & $0.3$ & $1.3$ \\
 & $0.10$ & $0.3$ & $0.3$ & $0.7$ & $0.3$ & $0.6$ \\
\hline\hline
\end{tabular}

&

\begin{tabular}{@{}l @{\hskip .2cm}l lll ll}
\multicolumn{7}{c}{\bf 2D tunnels} \\
\hline \\[-.3cm]
 & $f_{\rm syst}$ & F6 & F5 & F4 & N5 & N1 \\
\hline \\[-.3cm]
\multirow{ 2}{*}{Pessimistic case} 
 & $0.00$ & $1.4$ & $15$ & $32$ & $0.8$ & $4.3$ \\
 & $0.10$ & $0.5$ & $3.1$ & $4.5$ & $0.7$ & $2.7$ \\
\hline \\[-.3cm]
\multirow{ 2}{*}{Optimistic case}
 & $0.00$ & $2.8$ & $30$ & $63$ & $1.7$ & $8.6$ \\
 & $0.10$ & $0.9$ & $5.8$ & $7.5$ & $1.5$ & $5.4$ \\
\hline\hline
\end{tabular}
 
\end{tabular}
\caption{The S/N values for using void tangential shear measurements to distinguish the various modified gravity models from GR. These are based on forecasts using the synergy between DESI and two overlapping imaging surveys as described in the text. We present results for two void identification methods: 3D WVF voids and 2D tunnels. }
\label{table:SN_VGL}
\end{table}

We estimate the sample variance using the 5 realizations of GR simulations. For each realization, we split the volume into $4^3$ non-overlapping regions and compute $\Delta\Sigma$ for each of these regions; we do so using Eq.~(\ref{eq:excess_voids_2D}) for both 3D voids and tunnels (see \cite{Cautun:2017tkc} for a discussion of why we cannot use Eq.~\ref{eq:excess_voids}). Then, we generate $100$ bootstrap samples over these regions and calculate the mean $\Delta\Sigma$ of each of the bootstrap samples. The procedure leads to $5\times100$ samples which we use to calculate the sample variance of $\Delta\Sigma$. The resulting uncertainties are shown as a grey shaded region in Fig.  \ref{fig:void_weak_lensing}. To estimate the sample variance for the survey, we scale it by the ratio of the box volume to the survey volume between $z=0.3$ and $0.7$. 
We estimate the shape noise covariance matrix, $\mathbb{C}_{\rm SN}$, by first calculating it for the tangential shear, $\gamma_t(\theta)$, using angular coordinates, $\theta$, and then converting it to a covariance matrix for $\Delta\Sigma$ similarly to Eq.~(\ref{eq:ggl_cov}):
\begin{eqnarray}\label{eq:VGL_covariance_transf}
\mathbb{C}_{\rm SN}(R_{\rm void}\eta_{p,i}, R_{\rm void}\eta_{p,j}) &=& \Sigma_{c;{\rm eff}}^2\mathbb{C}_{\rm SN}(\theta_i,\theta_j),
\end{eqnarray}
where $\Sigma_{c;{\rm eff}}$ is the effective critical surface density for lensing, computed using Eq.~(19) of \cite{Cautun:2017tkc} taking into account the variation of the lensing kernel across the considered redshift range. The total statistical uncertainty is given by the sum of the sample variance and shape noise covariance matrices. The lensing measurements can be affected by systematic effects, which we model as a fraction, $f_{\rm syst}$, of the total lensing signal, cf.~Eq.~(\ref{eq:GGL_systematics}), and which we add to the covariance matrix describing the statistical uncertainties similar to the case of GGL above.

The S/N values with which 3D WVF voids and 2D tunnels can discriminate the various MG models are given in Table \ref{table:SN_VGL}. A combination of 3D WVF voids and the pessimistic LSST overlap scenario is unable to constrain any of the studied MG models even before considering potential systematic errors in the lensing measurement. Having a survey with a larger sky coverage and deeper imaging, like the optimistic LSST overlap scenario, results in 3D voids being able to constrain only the F4 model to a modest $2\sigma$ level. However, including a 10\% systematic lensing error degrades again very much the constraining power of WVF voids and results in ${\rm S/N}\lesssim1$. The situation is much better for 2D tunnels, which can probe F5, F4 and N1 models even for the pessimistic overlap case. The constraining power of tunnels is even better for the optimistic overlap scenario, which would result in ${\rm S/N}\gtrsim3$ for all models studies here except N5 which has a ${\rm S/N}=1.7$. Including a $10\%$ systematic error leads to S/N values of ${\sim}6$ for F5, F4 and N1 (all models end up having the same S/N although they have very different S/N values in the absence of systematic errors), and S/N value of $1.5$ for N5 and $0.9$ for F6. 

\revision{When constraining modified gravity theories, the tunnels result in higher S/N value than WVF voids because of two aspects. Firstly, the tunnels' tangential shear signal is more sensitive to modified gravity models, e.g. the fractional change with respect to GR for the F4 and F5 models is ${\sim}2$ times higher for tunnels than for WVF voids (see bottom panels in Fig.  \ref{fig:void_weak_lensing}). This is due to the tunnels' interiors containing fewer overdense regions and thus probing preferentially the lowest density regions which are the least screened ones. Secondly, the tunnels tangential shear signal is ${\sim}20$ times higher than that of WVF voids, so similar fractional differences between GR and MG theories can be more robustly measured for tunnels than for 3D voids.}

Comparing the S/N values for GGL and tunnels, we notice that, amongst the $f(R)$ variants, tunnels give better constraints for F6 and F5; which is possibly because GGL probes the lensing of high-density regions while void lensing measures the effect of low-density regions where the effect of the fifth force is stronger. The case of F4 is special, in that chameleon screening is quite inefficient in this model so that the fifth force effect can be strong both in voids and near halos, and our result suggests that GGL is boosted more than void lensing in this case. Adding systematic errors downgrades constraints for all $f(R)$ models but does not qualitatively change this observation. In the case of nDGP, it is known that screening is strong in the vicinity of dark matter halos but weaker far away. Since GGL probes regions far beyond the halo virial radius, in the absence of systematic errors GGL and void lensing give similar S/N values, while the systematics seem to affect tunnel lensing more. We remark, however, that the GGL S/N values are calculated for distances larger than $2\Mpch$ and including the inner regions may boost the S/N values, especially for $f(R)$ models (though we would need to worry about uncertainties caused by the impact of baryons in that case), and that the ways to calculate the sample variance contributions to the statistic error budgets for GGL and void lensing are not exactly the same. A more detailed and rigorous comparison of the constraining powers of these two probes will be left for future work.

\section{Conclusions}
\label{sect:discussion}

In this paper, we have undertaken an initial study of 
\revision{summary statistics} that can potentially be employed to extract information about the properties of gravity from DESI large-scale structure data, leveraging the variety of environments DESI will observe, from voids to densely populated galaxy clusters. We have considered one, two and multi-point statistics of positional clustering and relative motions of galaxies in both configuration and Fourier spaces, utilizing mock galaxy samples from simulations that currently cover smaller volumes than will be covered in the full DESI survey. \newrevision{A brief summary of the summary statistics that we have looked at is given in Table \ref{tab:summary}.}

\begin{table*}
\centering
\newrevision{
\begin{tabular}{ccccccc} \hline \hline
           Summary statistics & Mock & Section & Information content & Physical signal & \makecell{Theoretical\\ prediction} & External data \\ 
\hline
Large-scale RSD  & HOD & \S~\ref{subsect:LRSD} & velocity field & 1 & PT/emulator & no \\
Small-scale RSD & SHAM/hydro & \S~\ref{subsect:subhalo} & velocity field & 1 & emulator & no \\
Void RSD & HOD & \S~\ref{subsect:rsdvoid} & \makecell{velocity field \\ environment dependence} & 2 & PT/emulator & no \\
Marked CF (density marks) & HOD & \S~\ref{sub:density} & environment dependence & 3 & PT/emulator & no \\
Marked CF (other marks) & HOD & \S~\ref{sub:potential} & environment dependence & 2 & emulator & yes \\
Clustering with density split & HOD & \S~\ref{sub:rho-wp} & environment dependence & 1 & emulator & no \\
3-point correlation & HOD & \S~\ref{subsect:threept} & non-Gaussianity & 2 & PT/emulator & no \\
Galaxy bispectrum & HOD & \S~\ref{subsec:bispectrum} & non-Gaussianity & 2 & PT/emulator & no \\
Hierarchical clustering & HOD & \S~\ref{subsec:hierarchical_clustering} & non-Gaussianity & 2 & PT/emulator & no\\
Minkowski functionals & HOD & \S~\ref{subsect:MF} & morphology & 2 & emulator & no \\
Cluster phase-space stacking & HOD & \S~\ref{subsect:stackedcl} & velocity field & 1 & emulator & no \\
Galaxy-galaxy lensing & HOD & \S~\ref{subsect:gglens} & matter clustering & 1 & emulator & yes \\
Void lensing & HOD & \S~\ref{subsect:voidlens} & \makecell{matter clustering\\ environment dependence} & 1 & emulator & yes \\
\hline\hline
\end{tabular}
\caption{A list of the summary statistics considered in this paper. The analyses are based on mock galaxy catalogs constructed with HOD (for all summary statistics other than small-scale RSD; cf.~\ref{sec:HOD}) or SHAM and hydrodynamical simulations (for small-scale RSD; cf.~\ref{sect:SHAM_SHYBONE}). Our HOD catalogs are at $z\simeq0.5$, with a number density of $\simeq3.2\times10^{-4}[\Mpch]^{-3}$, which is close to the specifications of DESI LRG targets; while the SHAM/hydro catalogs are at low redshift ($z=0$) with a number density of $10^{-2}[\Mpch]^{-3}$, close to DESI BGS. The 3rd column links to the relevant sections in the paper. The 4th column (`Information content') highlights the main information that is probed by a given summary statistic. In the 5th column we comment on the strength of the physical signal caused by MG: `1' means there is a strong and readily-interpretable effect of MG; `2' indicates there is a strong signal but the interpretation may be complicated, e.g., the signal strength and its relative ordering among different gravity models could be affected by the use of tracers such as halos and mock galaxies; `3' shows the signal is weak and inconclusive. The 6th column indicates how the theoretical predictions can be made: for some summary statistics (and on relatively large scales) it is possible to applied perturbation-theory (PT) based approaches, while for most others (and especially when tracers rather than the dark matter field itself are considered) this proves challenging, so that alternative methods, such as emulation, will be needed. The 7th column indicates if external data, e.g., lensing data or gravitational potential inferred elsewhere, is needed (which could bring further challenges or uncertainties in their application to DESI). For all summary statistics, further works using realistic mocks that include proper observational systematics will be needed, as discussed in the text.}
\label{tab:summary}
}
\end{table*} 

This work is a preparatory step to both assess the relative potential of different statistics to put constraints on the properties of gravity and to determine the requirements for future larger-scale cosmological simulations, in terms of both the necessary fidelity  in reproducing modified gravity phenomenology, and in reproducing the realities of the DESI survey extent, completeness and instrumental and astrophysical systematic effects.

To make accurate predictions for how gravity might be constrained by DESI, when fully leveraging all observed spatial scales, and linear to nonlinear clustering, numerical simulations and mock galaxy catalogs that include specific MG phenomenology are essential.

In this work, we have considered two representative examples of scalar-tensor theories, which have played an important role in the development of Solar System tests of GR, and serve as excellent case studies for this.  The chameleon $f(R)$ gravity and DGP braneworld models considered exhibit different `screening' mechanisms, that allow their predictions to mimic GR in the Solar System and pass existing local gravity tests, and both predict the same speeds of photons and gravitational waves, a key feature that makes them compatible with detections of gravitational waves with electromagnetic counterparts. These models exhibit subtle differences from GR and from each other, however, on cosmic scales, that one can hope to use to distinguish or constrain them with observations. Their features are expected to arise in a broader range of theories, making the simulations more broadly applicable. 
 
This paper primarily uses existing N-body simulations that, while fully incorporating the complex physics of the scalar-tensor theory, are over smaller volumes and with lower mass resolution than will ultimately be sought for a full simulation of DESI. We also briefly commented on faster, approximate, simulation methods such as {\sc MG}-{\sc cola}, which will be valuable complements to the full simulations, giving the capability of running a large number of realizations for covariance matrix estimates etc.. The dark matter particle simulations have been used to construct mock galaxy catalogs using simulated halo catalogs and applying simple HOD prescriptions. The HOD parameters have been tuned so that they have approximately the same galaxy number density and galaxy two-point clustering properties (equivalent to fixing the number density and projected galaxy 2-point correlation function of the galaxy sample to those from observations), and then studying other statistics to see if they show any appreciable difference between the different gravity models. \newrevision{For one particular summary statistic, small-scale RSD, we have used mock galaxy catalogs constructed using SHAM or from full-physics galaxy formation simulations, in order to get a high galaxy number density and to have better accuracy of the clustering signal on small scales.}

Of the statistics considered, we find that the redshift space distortions and higher order correlation functions offer the greatest immediate potential for distinguishing modifications to GR with the first two years of DESI data.

\begin{itemize}
\item Prospective large scale RSD constraints on the growth rate from DESI, parameterized by $\beta$, are expected to be markedly tighter than current constraints. This can be most useful in constraining models such as nDGP, which modifies the growth rate on large, linear, scales. The SHAM analysis presented here demonstrates a promising potential to differentiate between GR and some MG models, such as $f(R)$ gravity, whose effect is most prominent on small, nonlinear, scales; the inclusion of small-scale RSD offers complementary information that could improve the ability of differentiating the various models. Finally, void RSD is a relatively new probe complementary to RSD around galaxies or galaxy clusters, and is of particular interest since near voids the effect of modified gravity is usually stronger. We discussed the challenges facing these probes.

\item The 3PCF offers complementary information about the nonlinear evolution of matter to the traditional two-point statistics. We find deviations of the MG models from GR on all scales, with different degrees of strength depending on the triangular shape searched. On large scales, a multipole decomposition in one of the triangle angles allows for an efficient comparison of models, showing particular patterns in the two triangle sides that define the angle: the signal is relatively small on scales larger than $\sim 40\Mpch$, but the complexity in the patterns at each multipole is unique for each tested model, and cannot be easily reproduced by other mechanisms. On scales $\sim10\Mpch$, the full 3PCF can be calculated and the signal is strong enough to provide a robust 
\revision{summary statistic} to test gravity -- this is further confirmed by looking at the galaxy bispectrum at $k_1\gtrsim0.1h$Mpc$^{-1}$ and the higher order central moments with a smoothing scale of $\lesssim20\Mpch$.  However, detailed studies on the systematics involved in the different high-order statistics should be carried on, together with an appropriate modeling of the signal including effects such as RSD, either through perturbation theory or by using simulations, to assess the findings of these 
\revision{summary statistics} when applying to broader classes of MG models.
\end{itemize}

Other statistics appear to have a strong variation in outcomes depending on whether one uses the halo catalog or a HOD tuned galaxy catalog, or on the details of the HOD model. Our theoretical understanding of these 
\revision{summary statistics} is dependent on how well we can model their predictions and assumptions of the galaxy formation model. The utility of these will therefore require further analysis to assess if they offer strong constraining potential for DESI: 

\begin{itemize}
\item Marked correlation functions, with appropriate choices of marks, can in principle up-weight galaxies in environments where MG signals are stronger. Although the results depend on the details of the implementation, we find that if the mark is defined using the galaxy number density, which along with the galaxy two-point clustering has been tuned to match in the different models, the distinguishing power is degraded in general. Using additional information, e.g., the Newtonian potentials of the host halos of galaxies, is found to lead to increased model differences, which suggests that in future works other possibilities of marks should be explored. It is also important to understand, and properly account for, the systematic errors of the marks themselves, which are (directly or indirectly) related to observables.

\item Other measures of the information beyond two-point statistics, such as the Minkowski functionals, have the potential of offering complementary constraints on models or breaking parameter degeneracies. However, {their study} is not yet in a mature state and further efforts are needed to assess their model-constraining power.

\end{itemize}

In addition to statistics based solely on DESI data, we have also considered the potential to constrain gravity from combined results using lensing statistics in areas with DESI overlap. While the results shown are promising, further analyses using galaxy clustering simulations with concurrent lensing predictions are required to fully understand the constraining potential:  

\begin{itemize}
\item The MG models studied here predict stronger matter clustering around galaxies, with the enhancement of galaxy galaxy lensing with respect to GR predictions strongest near (far from) the galaxies for the $f(R)$ (nDGP) models as a result of their different screening mechanisms. We find that for a LSST-like imaging survey overlapping with DESI, with reasonably well-controlled systematics ($\lesssim10\%$), the enhanced galaxy-galaxy lensing signals can be used to distinguish F5, F4 and N1 from GR. 

\item Voids in modified gravity models are emptier and have a larger lensing signal. Of the two methods tested here, {i.e., the 3D Watershed Void Finder and the 2D tunnels, the latter} applied to a LSST-like deep lensing survey overlapping the DESI region can distinguish all the models investigated in this paper from GR (at more than $3\sigma$ level even when considering $10\%$ systematic errors).

\item Phase space statistics probe both position and velocity  data and photometric data for weak-lensing mass (or mass-richness relation). This probe compares the observations directly to the potential (GR or modified), as traced by the dynamics. If the modified gravity model affects the dynamics and not light-travel, this becomes a powerful probe. It can be used both to rule out specific non-GR models (e.g., by using $f(R)$ as the null hypothesis) or it can be a goodness-of-fit to GR (where GR is the null hypothesis). After removing the dominant systematic in our tests by using potential ratios and by using a large DESI-like sample of clusters, this probe can constrain $|\overline{f_{R0}}|= 6 \times 10^{-7}$ at $>5\sigma$.

\end{itemize}
\newrevision{In Table \ref{tab:summary} we have also presented a short summary of some of the information itemized above, for easier reference.}

To further assess the most promising statistics and obtain accurate forecasts of their constraining power with DESI data, it would be useful to have a single simulation with sufficient volume, resolution and redshift coverage to construct realistic mocks for all types of objects targeted by DESI. However, despite the latest technical developments, a full simulation of this kind with modified gravity effects is likely still too expensive. Therefore, based on the findings of this paper, we recommend the following alternatives:

(i) multiple simulations with higher mass and force resolutions that the ones used in this paper, e.g., $L_{\rm box}\sim500\Mpch$ and $N_p=1024^3$ or $L_{\rm box}\sim800$-$1000\Mpch$ and $N_p=2048^3$. These should be run for a large number of models enough for building emulators of various statistics in a higher-dimensional parameters space spanned by not just the modified gravity but also cosmological parameters.

(ii) even higher-resolution simulations with $L_{\rm box}\sim500\Mpch$ and $N_p=2048^3$. These are roughly of the same size and resolution as the Millennium Simulation \cite{Springel:2005mi}, and some of these simulations (for nDGP and $\Lambda$CDM) have already been running \cite{Hellwing:2021mi}. The high resolution will make such simulations useful for resolving even small halos which host galaxies such as ELGs, and the box size is still large enough to study clusterings on nonlinear and some linear scales. Their large cost means that they can not be run for a large number of models, unless a new generation of much more efficient full simulation codes come to existence in the near future.

At this point, we consider it important to prioritize resolution and coverage of the model/parameter space over volume, since most of the 
\revision{summary statistics} studied in this paper still lack reliable theoretical predictions, and for some of them simulations offer a more promising and accurate way.

These will be complemented by a large suite of fast approximate simulations using (appropriately tested and tuned versions of) {\sc mg}-{\sc cola} that can be used to estimate the covariance matrices for some of the 
\revision{summary statistics} studied here. In $\Lambda$CDM models, various approximate methods were compared in Ref.~\cite{Munari:2017aau} and their validity of creating galaxy mocks to estimate the covariance matrices established on BAO scales \cite{Kitaura:2015uqa}. In this paper, we are interested in 
\revision{summary statistics} that go into the non-linear scales and the validity of the {\sc cola} approach needs to be tested carefully in this regime. In $\Lambda$CDM, it was shown that by re-calibrating halo masses and/or velocities, it is possible to bring accuracy in clustering to one percent level on large scales \cite{Izard:2015dja}, and this analysis needs to be extended to low-mass dark matter halos and small-scale clustering. In addition, {\sc mg}-{\sc cola} uses an approximate method for screening based on spherically symmetric solutions \cite{Winther:2014cia}, which was shown to have a sufficient accuracy up to $k=1 h{\rm Mpc}^{-1}$ in describing deviations from $\Lambda$CDM in the matter power spectrum \cite{Winther:2017jof} but again a calibration might be required. Once it is calibrated and validated against full N-body simulations, {\sc mg}-{\sc cola} could provide a promising way to compute the covariance matrices for some of the 
\revision{summary statistics} considered in this paper.
 
The (purely dark matter) simulations mentioned above need to be populated with galaxies so that they can be compared with data. In this paper, we have primarily used mock galaxy catalogs based on HOD in our study, but other methods to make mock galaxies such as SHAM or using sub-sampled dark matter particles have also been used. As we pointed out when describing a few 
\revision{summary statistics}, uncertainties in such galaxy-halo connections may have a non-negligible impact on their theoretical predictions. A possible way to quantify this is by assessing the impact of different models of galaxy-halo connection on every 
\revision{summary statistic}; where hydrodynamics simulations of galaxy formation exist, it is also useful to check the effect of the subgrid baryonic models on galaxy clustering observables, or use them to calibrate the galaxy populating recipe. It is perhaps more advisable to directly use observational data, e.g., of the two-point galaxy clustering, to calibrate the HOD model (e.g., \cite{Zehavi:2011tzz,Smith:2017tzz}), a simplified variant of which is we have followed to create galaxy mocks in this paper.

\textcolor{black}{Moreover, during the past few years there has been a rise of interests in the use of emulators to constrain cosmological parameters directly from N-body simulations. For instances, the Aemulus \cite{McClintock:2018uyf,Zhai:2018plk} and Dark Quest \cite{Nishimichi:2018etk} projects presented accurate estimators for the halo mass function and galaxy clustering, among other summary statistics, as functions of the cosmological parameters in $\Lambda$CDM. In general, emulators learn to mimic the behavior of a physical model that is slow and expensive to evaluate. They provide predictions of the model outputs at input parameters where the expensive model has not been evaluated. By far, the most common emulation technique applied to cosmology has been Gaussian Process Emulation (see \cite{books/lib/RasmussenW06} for a comprehensive review), due to this model's ability to provide uncertainty estimations together with emulator outputs, although neural network approaches have also been used \cite{Kobayashi:2020zsw}. Directly leveraging N-body simulations to constrain gravity on intermediate to small scales can potentially play a big role in testing MG scenarios, where screening mechanisms introduce further nonlinearities that strongly affect the small scales. Although there has been some initial work done in this direction \cite{Ramachandra:2020lue}, a more extensive investigation including a variety of models will be material for future studies.}

Beyond the simulation creation, another important thing is to include critical systematic effects when constraining MG from small scales. Missing galaxies due to fiber collisions, i.e., the finite size of the fibers preventing their placement close to each other within the focal plane, generally affects all scales but -- relevant to our work --- leads to a factor of two discrepancy in the 2-point clustering on small scales if not accounted for properly.  We would want to incorporate fiber collision modeling into the mock data we analyze. One way to account for the effect of collisions is to introduce a probability based weighting scheme using the target algorithm itself \cite{Bianchi:2017saf}. While this method was developed within the framework of 2-point statistics, in principle it can be applied to other 
\revision{summary statistics} as well. Nonetheless, it is necessary to test for potential bias and efficiency when the method is applied to other statistics and in particular the interplay with statistics to detect the environment dependent effects of modified gravity.

Beyond this, one needs a pipeline that produces multiple mocks on a light-cone that incorporates the DESI survey geometry and imposes survey masks which incorporate information about the observing conditions, such as targeting (e.g., galaxy magnitude, color, surface brightness),  placement, galactic extinction, atmospheric extinction, seeing, cloud-cover, zodiacal light, and so on. Mitigation strategies for these observational systematics, including a forward modeling approach and weighting methods, will be developed by the DESI clustering working group and are in detail discussed in a companion paper elsewhere. We do not expect the measurement and observational effects, such as extinction or seeing, to particularly affect small scales, but rather mainly be important for large to intermediate scales. 

To close, DESI, a Stage IV state-of-the-art galaxy survey,  will open up a wide range of opportunities to test the theory of gravity on cosmological scales with unprecedented precision. This work provides a useful first step for the DESI collaboration, and the wider theoretical and observational communities, to plan joint efforts to ensure we can fully exploit the wealth of future observations and contribute to the tests of fundamental theories in physics.

\appendix

\begin{figure}
    \centering
    \includegraphics[width=0.9\textwidth]{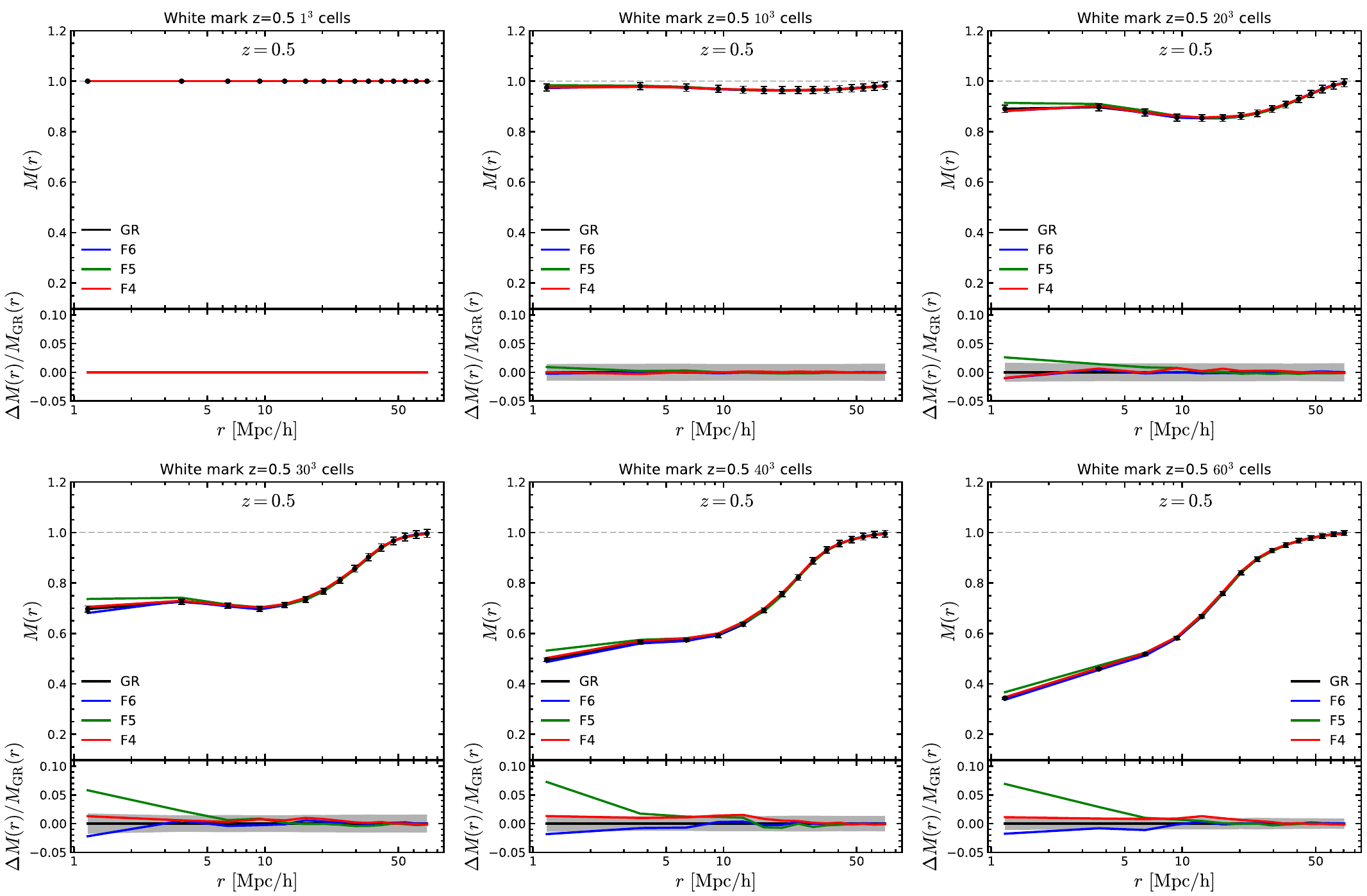}
    \caption{\newrevision{(Color Online) The marked correlation functions (mCF) with the White mark with $(\rho_\ast,p)=(4,10)$, for different numbers of cells used to calculate the density field.}}
    \label{fig:cells_mCF}
\end{figure}

\newrevision{\section{Impact of the cell-size on the marked correlation function measurements}
\label{App:A}} 

\newrevision{As we mentioned in Sec.~\ref{subsect:marked}, our galaxy number density is relatively low, therefore we cannot calculate the marks using cell sizes which are too small because that will cause a large number of cells to have no galaxies inside. Using a relatively large cell size helps to ensure the calculated mark values follow the large-scale density fluctuations, but apparently the cell size cannot be too large either, as that way the cells can no longer be considered as local environments of the tracers.}

\newrevision{For this reason, here we present a convergence test to show the impact of the cell size (or mean number of galaxy) on the mCF. Fig.~\ref{fig:cells_mCF} shows the mCFs with the White mark for different numbers of cells, in the top left panel we see that using $1$ cell, i.e., giving the same weight to the galaxies, the signal is identical in all models. When increasing the number of cells (or  reducing the cell size), we see that $f(R)$ models start to show some deviation from GR, and with $>30^3$ cells the relative differences from $\Lambda$CDM have stabilized. Note that the cell size of $\simeq17h^{-1}\textrm{Mpc}$ is only used to calculate the mark itself, and this does not mean that the (marked) correlation function is calculated for a galaxy density field after $17h^{-1}\mathrm{Mpc}$ smoothing.}

\newrevision{With a size of $17h^{-1}{\rm Mpc}$, the cells should still be a faithful representation of the local environments of galaxies. For example, when identified using galaxies as tracers,  cosmic voids are usually quite large, e.g., with radii of $\simeq10$--$100h^{-1}{\rm Mpc}$, which means that a cell of size $17h^{-1}{\rm Mpc}$ is not so large that the tracer densities in these cells are too close to the cosmic mean (i.e., cells of this size still capture the large-scale variations of the environment densities). This is supported by the convergence test in Fig.~\ref{fig:cells_mCF} described above.}

\newrevision{\section{Marked correlation function predictions for fixed halo number density}
\label{App:B}}

\newrevision{In this Appendix, we further investigate how the mCF results behave when halo catalogues are chosen to have the same number of haloes for all models, rather than a fixed halo mass cut as assumed in the main analysis. In particular, we repeat the analyses of Figs.~\ref{fig:marked1}, \ref{fig:mCF_negative_p} and \ref{fig:mCFg_phi} for halo samples with a fixed number density equal to the corresponding value of the BOSS-CMASS-DR9 sample, $n_h=3.2 \times 10^{-4}$ $h^{3}\mathrm{Mpc}^{-3}$.} 

\begin{figure}[!tb]
\begin{center}
\includegraphics[width=0.475\textwidth]{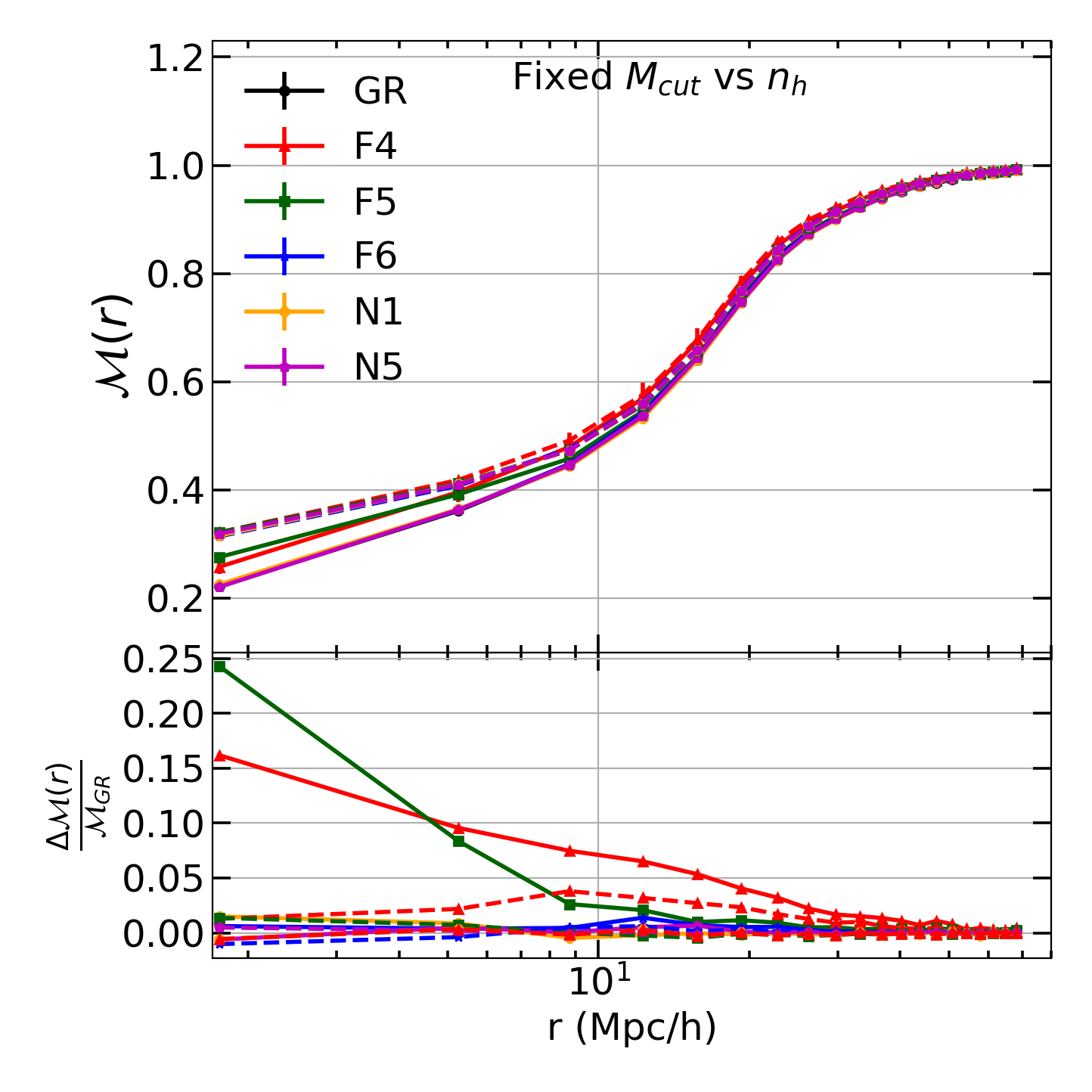}
\includegraphics[width=0.475\textwidth]{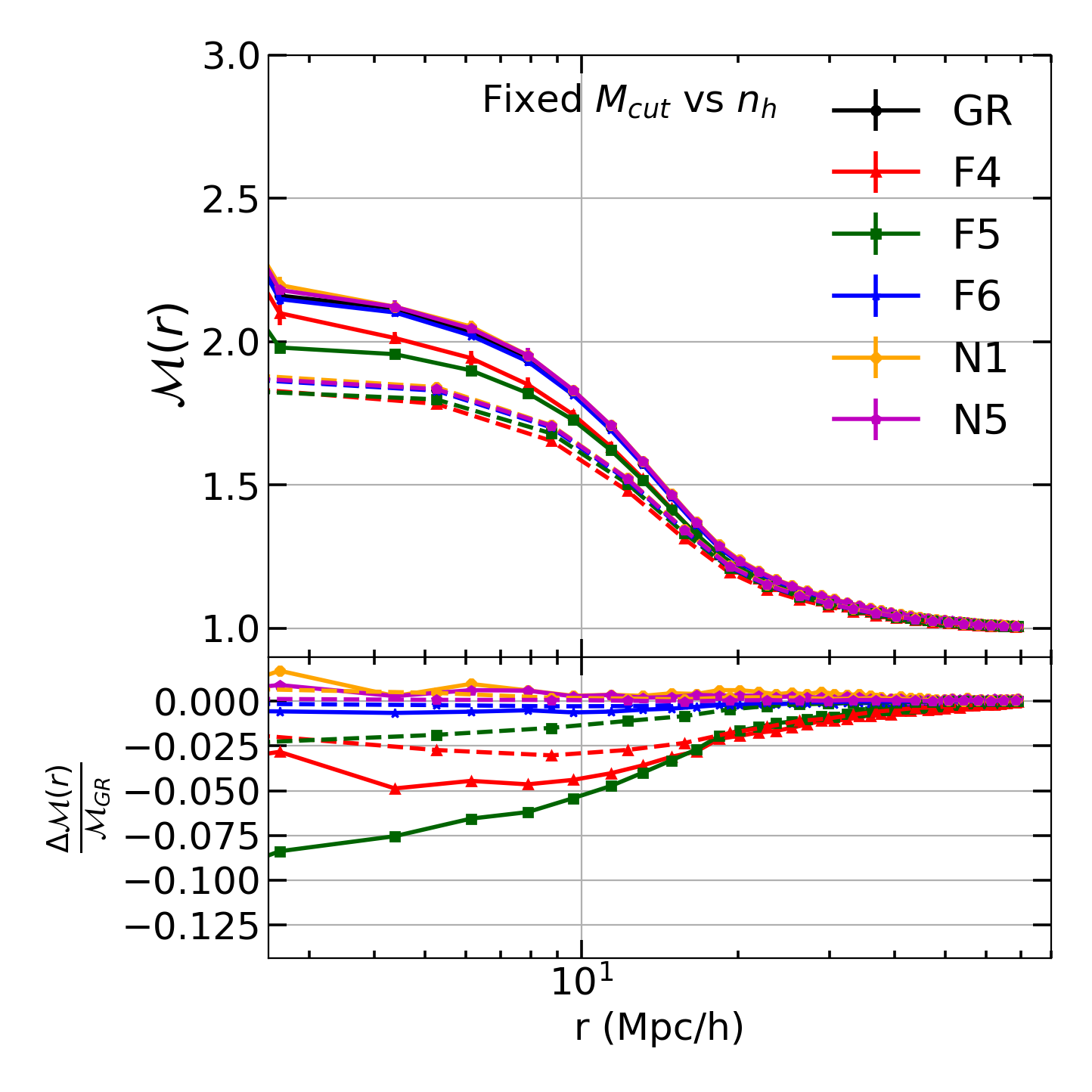}
\includegraphics[width=0.475\textwidth]{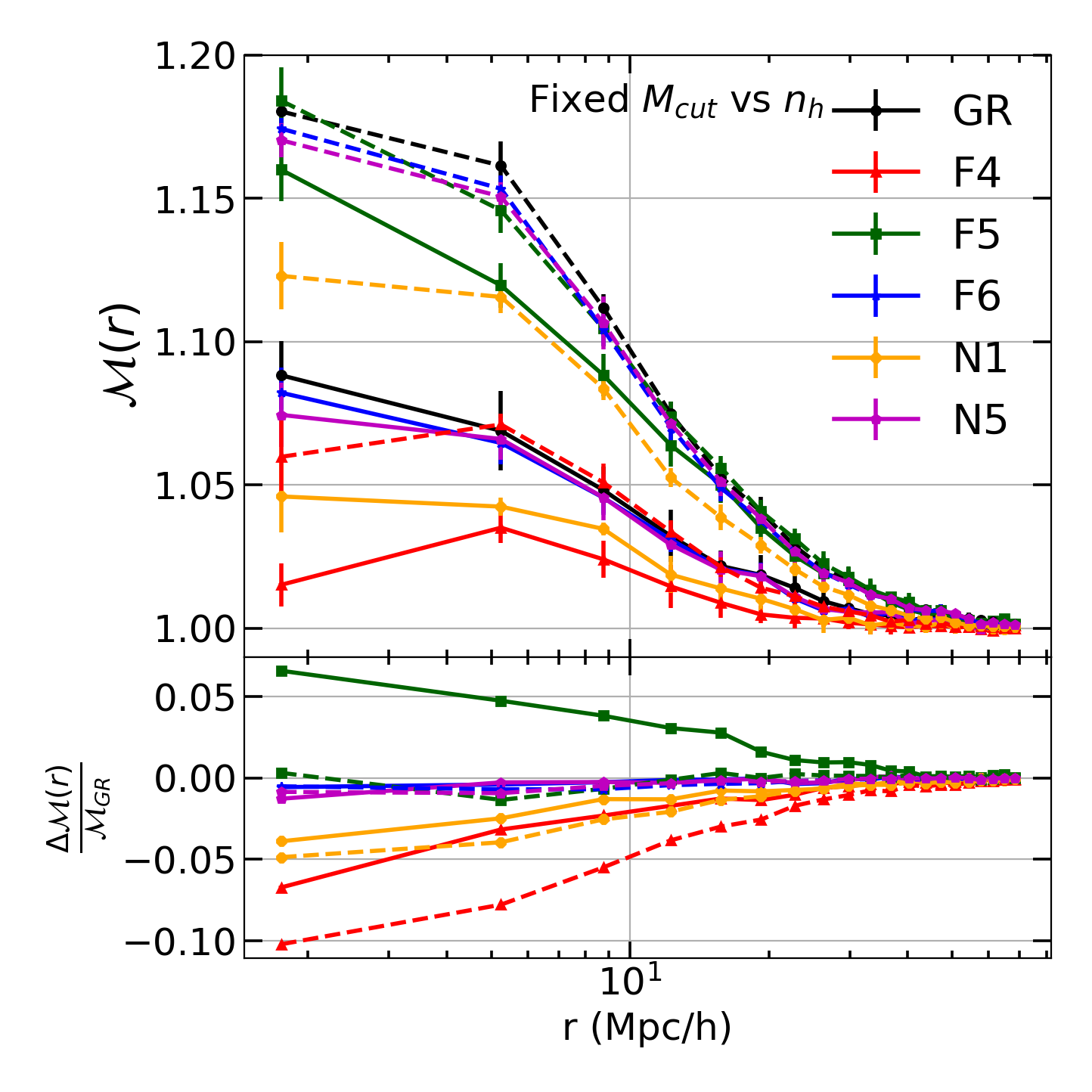}
\caption{\newrevision{(Color Online) A comparison of the marked correlation statistic obtained for dark matter halo samples designed with a fixed number density (dashed lines), as compared against the prediction using samples with a fixed mass cut (solid lines) equivalent to the original results of Figs.~\ref{fig:marked1} (Upper Left), \ref{fig:mCF_negative_p}  (Upper Right) and \ref{fig:mCFg_phi} (Bottom) of the main paper. All results are in real space.}}
\label{McutvsNh}
\end{center}
\end{figure}

\newrevision{The comparison is presented in Fig.~\ref{McutvsNh}. We notice that the fractional deviations with respect to the GR mCF prediction tend to become less pronounced in the case of the fixed number density samples (dashed lines), in particular for the two stronger $f(R)$ candidates, F4 and F5. For the rest of the models, the results for the fractional deviation do not change by more than $1\%$ compared to when using a fixed halo mass cut (solid lines).}

\newrevision{\section{Bispectrum measurements from halos and redshift-space galaxies}\label{App:C}}

\begin{figure}
    \centering
    \includegraphics{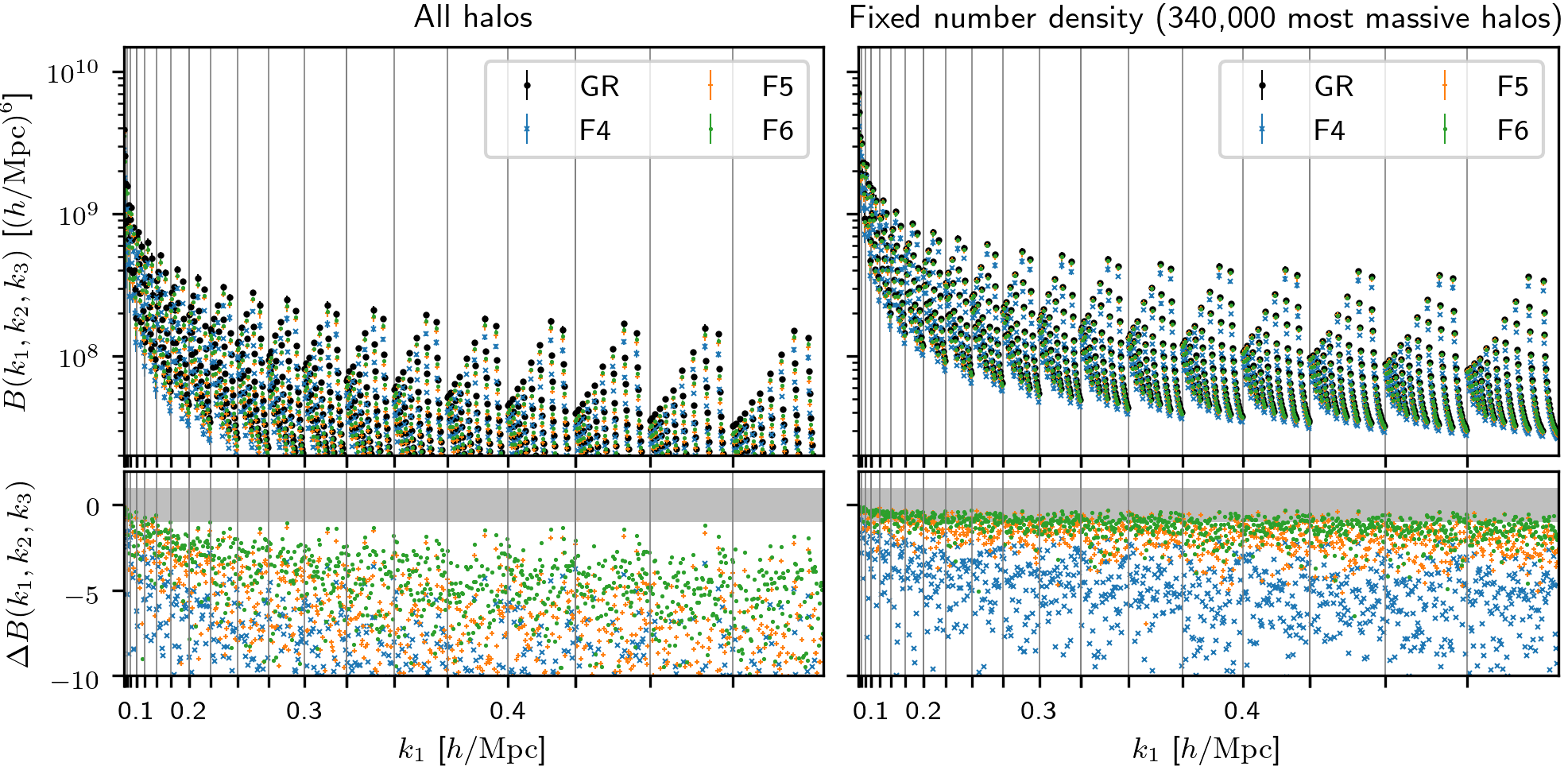}
    \caption{\newrevision{(Colour Online) Real-space bispectrum measured from halo catalogs using either all identified halos in the simulation box (left panels), or the 340,000 most massive ones (right panels).}}
    \label{fig:halo_bispectra}
\end{figure}

\newrevision{As a further test, we have measured the bispectrum using either all halos with over 20 particles from the simulation box, or halos with a constant number density threshold (the 340,000 most massive ones), and in both cases we find qualitatively similar results, see Fig.~\ref{fig:halo_bispectra}. In particular, the deviations from GR increase with increasing value of $|f_{R0}|$, as opposed to the results for the HOD samples (cf.~Fig.~\ref{fig:bisp_rs}), where F6 shows a stronger deviation from GR than F5.}

\begin{figure}
    \centering
    \includegraphics{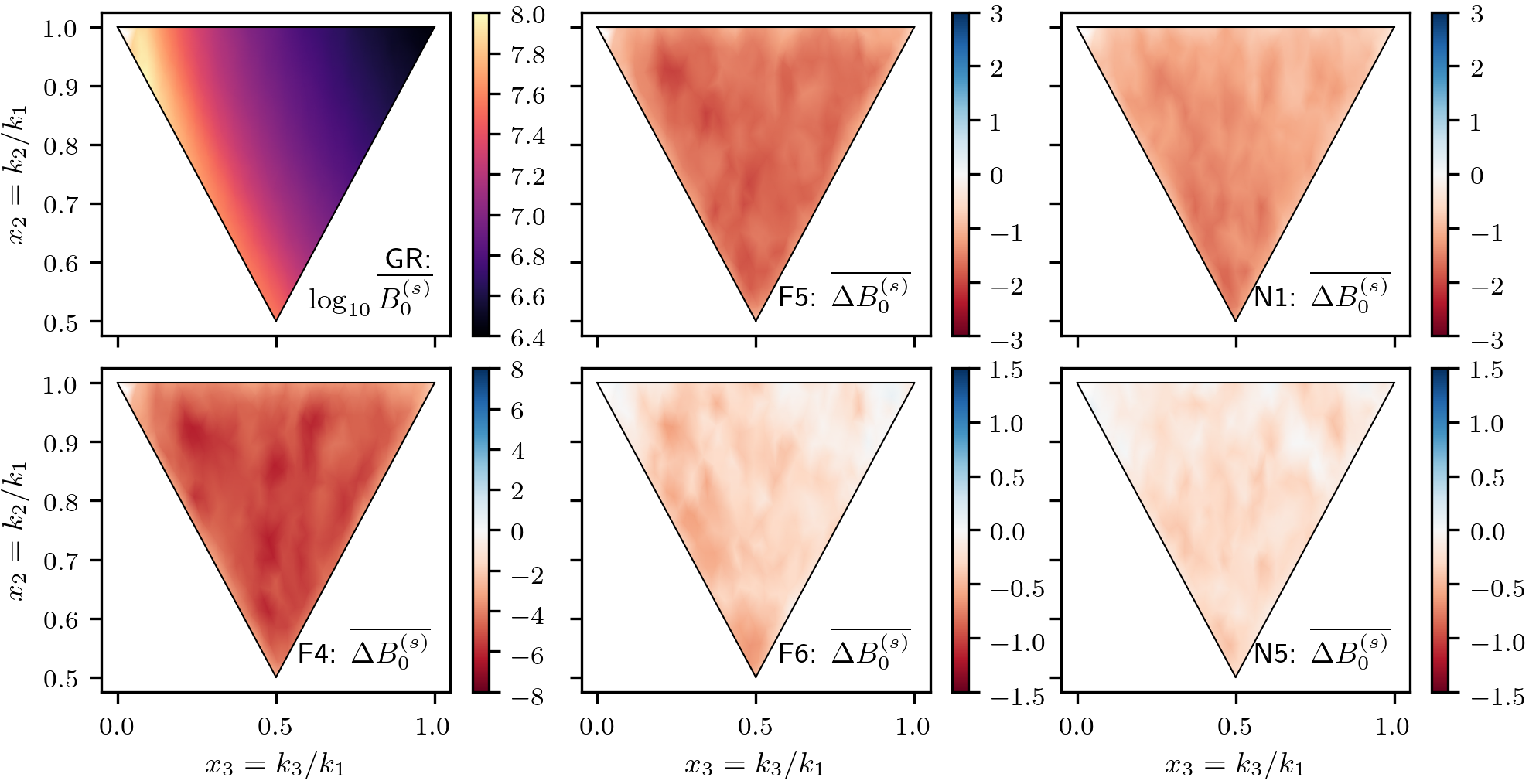}
    \caption{\newrevision{(Color Online) Configuration dependence of the monopole of the redshift space bispectrum.}}
    \label{fig:triangle_plots_bisp_zs}
\end{figure}

\newrevision{For illustration purposes, Fig.~\ref{fig:triangle_plots_bisp_zs} (the same as Fig.~\ref{fig:bisp_x2x3_rs}) shows the configuration dependence for the bispectrum monopole in redshift space. Unlike in real space, where the strongest deviations from GR occur for equilateral configurations, it shows that in redshift space the deviations have less configuration dependence, with only a slight increase towards collinear triangle shapes (lower left-hand side of the triangular plotting area). Since the bispectrum monopole signal is strongly affected by the Finger-of-God damping on small scales, the reduced configuration dependence is a consequence of the combination of this damping and the real space enhancement.}

\begin{acknowledgments}

We thank Peder Norberg and Martin White for useful comments on the early results of this work.

RB and GV's work is supported by DOE Grant DE-SC0011838, NASA ATP Grant NASA ATP Grants NNX14AH53G and 80NSSC18K0695G, NASA ROSES Grant 12-EUCLID12-0004 and NASA support for the Nancy Grace Roman Space Telescope High Latitude Survey Science Investigation Team.
MC was supported by Science and Technology Facilities Council (STFC) [ST/L00075X/1, ST/P000451/1], ERC Advanced Investigator Grant, DMIDAS [GA 786910], and the EU Horizon 2020 research and innovation programme under a Marie Sk{\l}odowska-Curie Grant agreement 794474 (DancingGalaxies).
JH is supported by a Durham co-fund Junior Research Fellowship, and the European Research Council (ERC-StG-716532-PUNCA).
CA is supported by the European Research Council (ERC-StG-716532-PUNCA).
WAH is supported by the Polish National Science Center
Grant No.~UMO-2018/30/E/ST9/00698 and UMO-2018/31/G/ST9/03388. This project has also benefited from numerical computations performed at the Interdisciplinary Centre for Mathematical and Computational Modelling (ICM) University of Warsaw under Grants Nos.~GA67-17 and GA65-30. 
CC-L is supported by a PhD Studentship from the Durham
Centre for Doctoral Training in Data Intensive Science, funded
by the UK Science and Technology Facilities Council (STFC,
ST/P006744/1) and Durham University. 
CH-A is supported by the Excellence Cluster ORIGINS which is funded by the Deutsche Forschungsgemeinschaft (DFG, German Research Foundation) under Germany's Excellence Strategy - EXC-2094 - 390783311. 
MI acknowledges support by the Department of Energy, Office of Science, under Award Number DE-SC0019206. 
KK is supported by the STFC Grant No.~ST/N000668/1. The work of KK and HAW has received funding from the European Research Council (ERC) under the European Union's Horizon 2020 research and innovation programme (Grant agreement 646702 ``CosTesGrav"). 
BL is supported by the European Research Council (ERC-StG-716532-PUNCA) under the H2020 Framework  of the European Commission, and STFC Consolidated Grants ST/L00075X/1, ST/P000451/1. 
CGS acknowledges support via the Basic Science Research Program from the National Research Foundation of South Korea (NRF) funded by the Ministry of Education (2018R1A6A1A06024977 and 2020R1I1A1A01073494).  
GN and AGM acknowledge support from CONACyT's Grants 179208, 286897 and Fronteras de la Ciencia 281, Instituto Avanzado de Cosmolog\'{\i}a (IAC), DAIP-UG and the computer infrastructure of the DCI-UG DataLab. 
MV is partially supported by Programa de Apoyo a Proyectos de Investigaci\'on e Innovaci\'on Tecnol\'ogica (PAPIIT) No.~IA102516, No.~IA101518, and from Proyecto LANCAD-UNAM-DGTIC-319 and LANCAD-UNAM-DGTIC-136. 
PN is funded by ‘Centre National d’\'Etudes Spatiales’ (CNES). 
SF is supported  by Programa de Apoyo a Proyectos de Investigaci\'on e Innovaci\'on Tecnol\'ogica (PAPIIT) No.~IA101619. 
AA and JLCC are supported by CONACyT project 283151.
GBZ is supported by the National Key Basic Research and Development Program of China (No.~2018YFA0404503), and by NSFC Grants 11720101004 and 11673025.  
CM and VH is partially supported by the National Science Foundation under Grant No.~181273.

O.V. and N. C. Devi are partially supported by a PAPIIT UNAM Grants IN112518 and AG101620. NCDevi acknowledges support from the European Commission’s Framework Programme 7,Marie Curie International Research Staff Exchange Scheme LACEGAL (PIRSES-GA-2010-269264). This study also used computer and human resources from the LAMOD supercomputing initiative.  

{This research used resources of the National Energy Research Scientific Computing Center (NERSC), a U.S. Department of Energy Office of Science User Facility operated under Contract No. DE-AC02-05CH11231.}
This research was partially supported through computational and human
resources provided by the LAMOD UNAM project through the clusters
Atocatl and Tochtli. LAMOD is a collaborative effort between the IA,
ICN and IQ institutes at UNAM.
The N-body simulations used in this work were carried out on two supercompuer systems: (1) the DiRAC Data Centric system at Durham University, operated by the Institute for Computational Cosmology (ICC) on behalf of the STFC DiRAC HPC Facility (www.dirac.ac.uk). This equipment was funded by BIS National E-infrastructure capital Grant ST/K00042X/1, STFC capital Grants ST/H008519/1 and ST/K00087X/1, STFC DiRAC Operations Grant ST/K003267/1 and Durham University. DiRAC is part of the National E-Infrastructure. (2) the Interdisciplinary Center for Mathematical and Computational Modelling (ICM) at University of Warsaw, under Grants No.~GA67-17 and No.~GA65-30.

\end{acknowledgments}

\bibliographystyle{prd}
\bibliography{references}

\end{document}